\documentclass[doublespacing]{utdthesis}

\usepackage{microtype}
\usepackage{amsmath,amssymb,amsthm}
\usepackage{graphicx}
\usepackage{url}
\usepackage[numbers]{natbib}
\let\cite=\citep
\usepackage{rotating}
\usepackage{ifpdf}
\ifpdf
  \usepackage{hyperref}
\fi

\usepackage[dvipsnames]{xcolor}
\usepackage{caption,subcaption}
\usepackage{algorithmic}
\usepackage[ruled,vlined,lined,commentsnumbered]{algorithm2e}
\usepackage{graphicx}
\usepackage{mathtools,bm,array}
\usepackage{enumitem}
\usepackage{multirow}

\newlength\mylen
\newcommand\myData[1]{%
  \settowidth\mylen{\KwData{}}%
  \setlength\hangindent{\mylen}%
  \hspace*{\mylen}#1}

\author{Josiah W. Smith}
\title{Novel Hybrid-Learning Algorithms for Improved \\ Millimeter-Wave Imaging Systems}
\thesistype{Dissertation}  
\degreefull{Doctor of Philosophy}
\degreeabbr{PhD}
\subject{Electrical Engineering}
\graduationmonth{May}
\graduationyear{2022}
\prevdegrees{BSEE} 

\committeemember*{Murat Torlak}
\committeemember{Naofal Al-Dhahir}
\committeemember{Carlos A. Busso-Recabarren}
\committeemember{Randall E. Lehmann}
\begin{document}

\frontmatter

\signaturepage
\copyrightpage{2022} 

\begin{dedication} 
Dedicated to my wife, Morgan. 
\end{dedication}

\maketitle

\begin{acks}{March 2022} 
First and foremost, I praise and thank God, the Almighty, from whom all blessings flow. His relentless pursuit of me during my time at The University of Texas at Dallas brought me life, strength, and ardor to persist through academic toil and learn to embody the love of Jesus to the world. 

I would like to express my utmost gratitude to my advisor, Professor Murat Torlak, who began mentoring me in 2017, while I finishing my undergraduate degree at The University of Texas at Dallas. 
He has continually challenged me in academic rigor, intellectual growth, and unrelenting commitment to achieve our goals. 
I owe much of my success to his patience, intuition, and plethora of experience in the field. 
The caring duality of his kindness and tenacity in problem-solving always provided the support and motivation necessary to achieve success. 
Working with him is a genuine blessing.

I am also grateful to my dissertation committee members, Professor Nafoal Al-Dhahir, Professor Carlos Busso, and Professor Randall Lehmann for generously offering their time and insight for both this dissertation and the influential courses I have had the honor of taken from them.

I would like to express my gratitude to Dr. Orges Furxhi at imec for his advice and mentorship during the summer of 2020, which was pivotal in my development as a researcher. 
I would like to thank my labmates at the Wireless Information Systems Laboratory (WISLAB) and Texas Analog Center of Excellence (TxACE) for their contributions to my research and constant friendship. 
I am thankful for the support of my research from Texas Instruments. 

I would like to thank my parents, brothers, and sisters, along with my spiritual family, for their influence on my life and their consistent love to accept me and challenge me to grow in Christ-likeness.

Finally, I am eternally indebted to my wife, my adventure partner, and my love for her endless love for me and support through the most challenging times. 
She is my inspiration and passion -- providing me with strength to overcome challenges and love to extend to others.
\end{acks}

\begin{abstract}
Increasing attention is being paid to millimeter-wave (mmWave), $30$ GHz to $300$ GHz, and terahertz (THz), $300$ GHz to $10$ THz, sensing applications including security sensing, industrial packaging, medical imaging, and non-destructive testing. 
Traditional methods for perception and imaging are challenged by novel data-driven algorithms that offer improved resolution, localization, and detection rates. 
Over the past decade, deep learning technology has garnered substantial popularity, particularly in perception and computer vision applications. 
Whereas conventional signal processing techniques are more easily generalized to various applications, hybrid approaches where signal processing and learning-based algorithms are interleaved pose a promising compromise between performance and generalizability. 
Furthermore, such hybrid algorithms improve model training by leveraging the known characteristics of radio frequency (RF) waveforms, thus yielding more efficiently trained deep learning algorithms and offering higher performance than conventional methods.

This dissertation introduces novel hybrid-learning algorithms for improved mmWave imaging systems applicable to a host of problems in perception and sensing. 
Various problem spaces are explored, including static and dynamic gesture classification; precise hand localization for human computer interaction; high-resolution near-field mmWave imaging using forward synthetic aperture radar (SAR); SAR under irregular scanning geometries; mmWave image super-resolution using deep neural network (DNN) and Vision Transformer (ViT) architectures; and data-level multiband radar fusion using a novel hybrid-learning architecture. 
Furthermore, we introduce several novel approaches for deep learning model training and dataset synthesis. 
Depending on the application, a varying balance of classical signal processing techniques and deep learning is applied to optimally leverage the advantages of each technique. 
To verify the proposed algorithms, we employ virtual prototyping via simulation and develop custom-built imaging testbeds for empirical testing. 
Our custom tools for algorithm development, dataset generation, system-level design, and deployment are made public to promote further innovation in this arena. 
The simulation and experimental results demonstrate the wide application space of hybrid-learning algorithms and the efficacy of joint signal processing data-driven algorithms for radar sensing, perception, and imaging.
\end{abstract}

\tableofcontents
\listoffigures 
\listoftables 

\mainmatter

\chapter{Introduction}
\label{ch:intro}

\section{Background and Motivation}
\label{sec:background}

Low-cost electromagnetic (EM) imaging systems have gained attention over the past decade as commercially available radar platforms have become increasingly affordable. 
Millimeter-wave (mmWave) radar has attracted exceptional interest for applications such as gesture recognition \cite{sang2018micro,kim2016hand,maragliulo2019foot,kim2017application,park2016_ir_uwb_gesture,kim2017staticgesture,smith2021sterile}, concealed threat detection \cite{sheen2016three,yanik2019near,zhuge2012automatic,carrer2014concealed}, and medical imaging \cite{chao2012millimeter,gao2016millimeter,di2017high,mirbeik2018synthetic,mirbeik2018ultra,fedeli2020microwave}, owing to its semi-penetrating non-ionizing nature and low power consumption. 
Although extensive studies have been conducted on deep learning for image processing and computer vision \cite{lim2017enhanced,glorot2011ReLU,bjorklund2017licenseplate_synthetic,wu2019_FaultSeg3D_synthetic} and conventional signal processing of radar signals \cite{sheen2016three,sheen1999real,sheen2001three,sheen2010near,sheen2018simulation,winkler2007range,kim2018joint,baral2021joint,soumekh1998wide,soumekh1999synthetic,lopez20003}, we examine an emerging field by leveraging the advantages of both approaches to develop novel hybrid-learning algorithms. 
The degree to which data-driven and conventional algorithms are employed varies by application, constraints, and requirements. 
As we approach several distinct applications, we apply the hybrid-learning approach of optimally leveraging the strengths of each domain to produce effective and efficient systems.

As machine learning algorithms are gaining significant attention for large-scale to edge applications, machine learning on radar data is gaining momentum particularly for classification and perception. 
Radar hand gesture recognition is of notable interest as increasing importance is placed on privacy and non-invasive sensing methods are generally preferred. 
In addition, sensing using optical or depth cameras often requires ideal lighting and temperature conditions \cite{lin2013_3D_hand_posture_RGBD,son2017image}. 
mmWave radar devices have recently emerged as a promising alternative offering low-cost system-on-chip sensors whose output signals contain precise spatial information even under non-ideal imaging conditions \cite{smith2021sterile}. 
However, proper handling of radar signals is essential for high-fidelity gesture sensing systems and can result in considerably varied performance. 

In addition to gesture sensing, data-driven perception on synthetic aperture radar (SAR) images is used in applications such as smartphone imaging \cite{alvarez2021towards}, UAV SAR \cite{garcia20203DSARProcessing}, and automotive imaging \cite{kan2020automotiveSAR}. 
However, high-resolution image reconstruction for irregular SAR scanning geometries using existing algorithms requires computationally prohibitive techniques. 
Efficient algorithms for common SAR patterns (linear \cite{paul2021systematic,maisto2021sensor}, rectilinear/planar \cite{sheen2016three,yanik2020development,mohammadian2019sar,zhuge2010sparse}, circular \cite{wu2020multilayered,gao2016efficient,demirci2011back,jia2014modifiedBPA}, and cylindrical \cite{smith2020nearfieldisar,amineh2019real,fortuny2001extension,detlefsen2005effective,laviada2017multiview}) have been investigated in the literature, but computationally tractable algorithms remain unexplored for irregular scanning geometries. 

On the other hand, while many studies on high-resolution near-field mmWave imaging have been conducted on the signal processing front \cite{sheen2016three,yanik2020development,smith2020nearfieldisar,yanik2019sparse,gao2018_1D_MIMO,gao2018cylindricalMIMO}, incorporating data-driven techniques into the imaging pipeline has received limited attention \cite{gao2018enhanced,wang2021tpssiNet}. 
Optical image super-resolution techniques have gained significant attention in recent years from the machine learning community \cite{lim2017enhanced,liang2021swinir}; however, SAR image super-resolution has largely been relegated to the far-field imaging regime \cite{zhu2021deep,wei2021sar,alver2019aNovel,hu2020inverse,liu2020sar,wu2020super,zhang2017complex}. 
Besides, near-field SAR super-resolution presents several unique challenges, particularly the availability of large, meaningful datasets for training convolutional neural network (CNN)-based algorithms.
Moreover, prior works assume simplistic targets consisting of only randomly placed point targets \cite{gao2016efficient,wang2021rmistnet,dai2021imaging,jing2022enhanced}.
To address these shortcomings, we propose a novel simulation platform for generating large meaningful datasets containing radar data from sophisticated objects and 3-D models.
With the flexibility of this toolbox, we address emerging problems in SAR image enhancement for irregular scanning geometries using data-driven techniques. 

Uniting our efforts towards preprocessing algorithms and improved data-driven imaging algorithms, we investigate a contactless gesture control framework that employs deep learning-aided super-resolution techniques in combination with classical radar signal processing methods to improve performance.
Existing studies on contactless gesture control, such as gesture radar, commonly apply machine learning and deep learning techniques.
Gurbuz \textit{et al.} employ a multi-frequency radio-frequency (RF) sensor to recognize American sign language patterns with high accuracy using several machine learning techniques such as support vector machines (SVM), random forest, linear discriminant analysis, and k-nearest neighbors \cite{gurbuz2021american}.
Gesture recognition algorithms have been examined for mmWave radar using SVM classifiers \cite{maragliulo2019foot}, CNN techniques \cite{kim2016hand,kim2017application,zhang2016dynamic,leem2020detecting}, LSTM networks \cite{suh2018_24GHz_gesture}, etc. \cite{sang2018micro}. 
However, these methods separately apply data-driven and signal processing techniques. 
By employing hybrid-learning algorithms, our proposed method combines the advantages of signal processing and deep learning to yield a performance gain over previous work \cite{joshi2015wideo}. 

Finally, interleaved hybrid-learning algorithms, particularly for near-field imaging, could significantly improve perception, as innate signal characteristics can be leveraged across spatial and spectral domains. 
Specifically, the dual relationships between RF signals and their spectral representations can be leveraged for superior resolution and image focusing. 
Towards this end, we investigate a novel hybrid-learning approach for multiband signal fusion to achieve 3-D SAR super-resolution. 
Compared to traditional signal processing approaches \cite{cuomo1999ultrawide,tian2014sparse,zou2016matrix,wang2018wavenumber,zhang2014coherent,zhang2017multiple,tian2013multiband,li2008mft,sarkar1995mpa}, our method enables technologies such as concealed weapon detection and occluded item classification as intricate targets can be recovered with high-resolution. 
Using this approach, we achieve 21 GHz bandwidth from two 4 GHz bandwidth radars operating at 60--64 GHz and 77--81 GHz. 
Extensive simulation and empirical experiments are provided to validate the robustness and generalizability of the proposed complex-valued CNN architecture \cite{jing2022enhanced}. 
Hybrid-learning algorithms require careful consideration of the mechanics of the problems but can offer considerable performance gains compared to signal processing or deep learning alone.

\section{Research Objectives and Previous Work}
\label{sec:research_objectives_and_previous_work}
The main objective of this dissertation is to present a framework through which to approach mmWave imaging problems by leveraging the advantages and trade-offs between conventional radar signal processing methods and modern data-driven algorithms.
The proposed technique is denoted as hybrid-learning as a hybrid approach that employs expertise and intuition in both the conventional radar signal processing domain and machine learning arena. 
To achieve this goal, we focus our efforts on several perception and imaging problems and develop novel data-driven techniques for improved classification, localization, and imaging. 

Towards machine learning classification of radar signals, we investigate various front-end signal processing techniques and their impact on perception systems. 
We explore data preprocessing techniques for static (stationary) and dynamic (moving) hand gestures using mmWave radar, considering both fidelity and computational load. 
By developing a thorough understanding of the challenges and opportunities in gesture recognition, we demonstrate a novel training technique by employing ``sterile'' data during the model training process. 
Additionally, we develop a model to decompose irregular SAR scanning geometries in the near-field to develop an efficient image reconstruction technique that overcomes the excessive computational burden required in previous studies \cite{alvarez2021towards,alvarez2019freehand,alvarez2021system}. 

As near-field SAR image super-resolution is gaining increasing attention \cite{wang2021tpssiNet,wang2021rmistnet,dai2021imaging,jing2022enhanced,long2019terahertz,li2020adaptive,han2020terahertz,wang2020csrnet,wang2021fista}, there is a significant need for large quantities of meaningful high-fidelity SAR data.
To this end, we design an open-source software platform for high-fidelity near-field dataset generation. 
Similar existing software implementations in the literature \cite{jost2005matlabToolkitSAR,Auer2016RaySAR,kedzierawski2011MATLAB,deo2012MATLAB} address only the simplified far-field scenario and cannot produce data relevant for near-field imaging. 
Using the custom framework to generate large, meaningful datasets, we consider several methods for improved imaging using hybrid-learning.
First, a SAR super-resolution algorithm is detailed to overcome image distortion caused by positioning errors common to near-field mmWave systems \cite{yanik2020development}. 
The proposed framework leverages a mobile-friendly Vision Transformer (ViT) architecture \cite{mehta2021mobilevit,sandler2018mobilenetv2} for image-to-image super-resolution.
Additionally, an image enhancement network is developed to perform spatial super-resolution on near-field SAR images with irregular sampling geometries common to emerging applications \cite{alvarez2021towards,smith2022efficient}.
The proposed algorithm is the first of its kind to pair mobile-suitable neural network architectures and efficient irregular SAR near-field imaging algorithms, thereby enabling several applications constrained to arbitrary scanning patterns and low computational load.

Based on  our expertise in radar gesture recognition and imaging, we propose a hybrid-learning framework for contactless musical interface. 
Our algorithm leverages high-fidelity spatiotemporal signatures embedded in the radar signal to provide a responsive, precise interface for a host of human computer interaction (HCI) tasks.
A hybrid deep learning, signal processing, and computer vision approach yields spatial resolution exceeding the theoretical bound and outperforms state-of-the-art localization performance compared with previous methods.

Finally, we propose an end-to-end interleaved hybrid-learning approach for near-field imaging employing deep learning super-resolution and regression techniques \cite{lim2017enhanced,liang2021swinir} throughout the imaging signal chain.
Fully embracing hybrid-learning, this technique allows a network to learn the various characteristics of the signal across the spatial and spectral domains. 
The proposed hybrid-learning algorithms yield significant performance gains for both image fidelity and computational efficiency. 

\section{Contributions and Proposed Work}
\label{sec:contributions}
In response to the challenges and opportunities of hybrid-learning algorithms, we present several studies on data-driven solutions to radar imaging problems and propose novel methods for sensing, tracking, imaging, super-resolution, and multiband radar fusion to achieve the following contributions: 

\begin{enumerate}
    \item We investigate static and dynamic gesture recognition using a small-platform MIMO-FMCW mmWave radar and CNN classifiers. 
    We perform an extensive study of the challenges and opportunities for static gesture recognition by examining several datasets and data preprocessing techniques.
    We explore the trade-offs of CNN classifiers for dynamic hand gesture recognition. 
    This contribution is based on the following publication:
    \begin{itemize}
        \item J. W. Smith, S. Thiagarajan, R. Willis, Y. Makris, and M. Torlak, ``Improved static hand gesture classification on deep convolutional neural networks using novel sterile training technique,'' \textit{IEEE Access}, vol. 9, pp. 10893–10902, Jan. 2021.
    \end{itemize}
    
    \item We address another aspect of static gesture recognition to improve the robustness of the classifier given the challenges of static gesture recognition. 
    We propose an efficient data collection approach and a novel technique for deep CNN training by introducing ``sterile'' data which aid in distinguishing distinct features among the static gestures and subsequently improve the classification accuracy. 
    We provide experimental results demonstrating the ability of this method to improve the classification accuracy of real human hand gestures. 
    This contribution is based on the following publication:
    \begin{itemize}
        \item J. W. Smith, S. Thiagarajan, R. Willis, Y. Makris, and M. Torlak, ``Improved static hand gesture classification on deep convolutional neural networks using novel sterile training technique,'' \textit{IEEE Access}, vol. 9, pp. 10893–10902, Jan. 2021.
    \end{itemize}
    
    \item In addition, we extend the work of \cite{alvarez2021towards,alvarez2021system,alvarez2021freehand} by proposing a novel imaging algorithm to enable efficient near-field irregular SAR. 
    This work addresses the need for efficient imaging algorithms for edge applications such as smartphone imaging and automotive SAR. 
    Subsequent deep learning for classification or super-resolution requires high-fidelity SAR images under computational constraints.
    Whereas conventional mmWave imaging relies on high-precision systems \cite{yanik2020development}, many edge applications, such as freehand imaging, necessitate both irregular array geometries and low computational complexity.
    The proposed reconstruction algorithm efficiently projects irregularly sampled multistatic data onto a virtual planar monostatic array achieving image resolution consistent with the computationally prohibitive backprojection algorithm (BPA) with equivalent efficiency to the range migration algorithm (RMA).
    This contribution is founded on the following publication:
    \begin{itemize}
        \item J. W. Smith and M. Torlak, ``Efficient \mbox{3-D} near-field MIMO-SAR imaging for irregular scanning geometries,'' \textit{IEEE Access}, vol. 10, pp 10283-10294, Jan. 2022.
    \end{itemize}
    
    \item To enable data-driven algorithms for near-field SAR, we develop a novel software framework for system prototyping, imaging algorithm development, and dataset generation. 
    The proposed software is implemented as an open-source MATLAB toolbox capable of efficiently generating high-fidelity SAR data that can be used for a host of applications.
    The contribution will be based on the following publications: 
    \begin{itemize}
        \item J. W. Smith and M. Torlak, ``Survey of emerging systems and algorithms for near-field THz SAR imaging,'' \textit{Proc. IEEE}, to be submitted.
    \end{itemize}
    
    \item To overcome the positioning errors common in many near-field SAR systems \cite{yanik2020development}, we propose a novel Vision Transformer (ViT) approach for SAR image super-resolution and artifact mitigation. 
    Using data generated from the software toolbox, we train our algorithm on images generated from SAR scenarios with image distortion and defocusing caused by array perturbations.
    The proposed algorithm employs a mobile-friendly image-to-image enhancement architecture \cite{mehta2021mobilevit,sandler2018mobilenetv2} suitable for a host of applications from laboratory environments to edge implementations. 
    We validate the proposed method using both simulations and empirical studies.
    This contribution is detailed in the following publication:
    \begin{itemize}
        \item J. W. Smith, Y. Alimam, G. Vedula, and M. Torlak, ``A vision transformer approach for efficient near-field SAR super-resolution under array perturbation,'' in \textit{Proc. IEEE Tex. Symp. Wirel. Microw. Circuits Syst. (WMCS)}, Waco, TX, Apr. 2022, pp. 1--6.
    \end{itemize}
    
    \item We then propose extending our work in \cite{smith2022efficient} to develop the first CNN-based SAR super-resolution algorithm for mobile applications.
    Emerging applications for mobile SAR imaging in the near-field are constrained to irregular sampling geometries and low computational complexity. 
    Previous studies on near-field SAR super-resolution algorithms are limited to conventional SAR geometries and are unsuitable for mobile applications \cite{wang2021tpssiNet,wang2020csrnet}.
    Using the software toolbox we developed, we generate large synthetic datasets to train a neural processor to perform 3-D SAR image super-resolution. 
    The proposed algorithm employs a generative adversarial network (GAN) architecture for image super-resolution using a patch discriminator technique \cite{isola2017image} and efficient depth-wise convolution implementation \cite{liu2022convnext}. 
    A thorough discussion of this contribution is provided in the following publication: 
    \begin{itemize}
        \item C. Vasieleiou, J. W. Smith, S. Thiagarajan, M. Nigh, Y. Makris, and M. Torlak, ``Efficient CNN-based super resolution algorithms for mmWave mobile radar imaging,'' in \textit{Proc. IEEE Int. Conf. Image Process. (ICIP)}, Bourdeaux, France, Oct. 2022, pp. 3803--3807.
    \end{itemize}
    
    \item Additionally, we developed a novel framework for human-computer interaction using a fully convolutional neural network (FCNN) for localization super-resolution in real-time.
    Our system offers unprecedented high-resolution tracking of hand position and motion characteristics by leveraging spatial and temporal features embedded in the reflected radar waveform. 
    By employing a hybrid-learning approach, we developed a novel spatial super-resolution technique that exceeds the theoretical limitations and a modified tracking algorithm to optimally leverage the inherent characteristics of radar signatures.
    This contribution is based on the following publication:
    \begin{itemize}
        \item J. W. Smith, O. Furxhi, M. Torlak, ``An {FCNN}-based super-resolution {mmWave} radar framework for contactless musical instrument interface,'' \textit{IEEE Trans. Multimedia}, vol. 24, pp. 2315--2328, May 2021.
    \end{itemize}
    
    \item Finally, we propose a novel hybrid-learning technique for multiband radar image fusion. 
    Using off-the-shelf 4 GHz bandwidth radars at 60-64 GHz and 77-81 GHz, we develop a high-fidelity testbed for collecting multiband radar images.
    The proposed algorithm achieves an effective bandwidth of 21 GHz and outperforms previous methods, particularly on high-bandwidth targets, in terms of image fidelity and computation time. 
    By leveraging a novel dual-domain architecture, the proposed hybrid-algorithm demonstrates super performance compared to conventional techniques \cite{zou2016matrix,wang2018wavenumber,li2008mft} in both simulation and empirical studies.
    This contribution is based on the following publication:
    \begin{itemize}
        \item J. W. Smith and M. Torlak, ``Deep learning-based multiband signal fusion for 3-D SAR super-resolution,'' in \textit{IEEE Trans. Aerosp. Electron. Syst.}, Apr. 2023.
    \end{itemize}
\end{enumerate}
Through these investigations, we also developed an advanced imaging system and a novel algorithm for near-field cylindrical MIMO-ISAR, which appeared in the following publication
\begin{itemize}
    \item J. W. Smith, M. E. Yanik and M. Torlak, ``Near-field MIMO-ISAR millimeter-wave imaging,'' \textit{Proc. IEEE Radar Conf. (RadarConf)}, Florence, Italy, Sep. 2020, pp. 1-6.
\end{itemize}

\section{Outline of the Dissertation}
\label{sec:outline}
The rest of the dissertation is organized as follows:
\begin{itemize}
    \item Chapter \ref{ch:fmcw_signal_model} details the FMCW signal model employed at length throughout this dissertation. 
    \item Chapter \ref{ch:signal_processing} presents an investigation of the impact of front-end signal processing techniques on deep learning perception algorithms and details a novel training technique for gesture sensing using mmWave radar. 
    \item Chapter \ref{ch:data_driven} investigates the inclusion of data-driven approaches in the near-field imaging pipeline; details the development and implementation of a software framework for near-field SAR imaging simulation, prototyping, and dataset generation; and presents near-field SAR image super-resolution and restoration algorithms using hybrid-learning techniques. 
    \item Chapter \ref{ch:rmi} details a real-time deep learning-based framework for contactless musical interface using mmWave radar. 
    \item Chapter \ref{ch:dual_radar} presents a multiband radar imaging system built from off-the-shelf 60 GHz and 77 GHz radars and a multiband fusion algorithm that leverages a novel hybrid-learning, dual-domain technique to provide an equivalent bandwidth of 21 GHz from the two 4 GHz bandwidth radars.
    \item Conclusions, summary, and discussion of proposed work are detailed in Chapter \ref{ch:conclusion}.
\end{itemize}
\chapter{Preliminaries of FMCW Signaling}
\label{ch:fmcw_signal_model}

In this chapter, we detail the fundamentals of frequency-modulated continuous-wave (FMCW) radar to be utilized extensively throughout this dissertation.
Over the past several decades, FMCW radars have emerged as an inexpensive option for high-bandwidth systems \cite{yanik2020development}. 
FMCW signals contain precise spatial information of the illuminated target and are used for a wide array of applications from gesture recognition \cite{kim2016hand,kim2017application,smith2021sterile}, concealed threat detection \cite{sheen2016three,yanik2019near}, and medical imaging \cite{fedeli2020microwave}. 
Throughout this dissertation, we will leverage the characteristics of FMCW signaling for high-fidelity perception and imaging. 

\section{FMCW Signal Model}
\label{sec:fmcw_signal_model}
We begin by considering a single bistatic FMCW transceiver, whose transmitter and receiver are positioned at the points ($x_T$,$y_T$,$Z_0$) and ($x_R$,$y_R$,$Z_0$) in a three-dimensional (\mbox{3-D}) space, respectively, and one stationary ideal point reflector in the scene with reflectivity $\sigma$ located at the point ($x_0$,$y_0$,$z_0$). 
The radar transceiver is positioned on the $x'$-$y'$ plane, located at $z = Z_0$.
\begin{figure}[h]
    \centering
    \includegraphics[width=\textwidth]{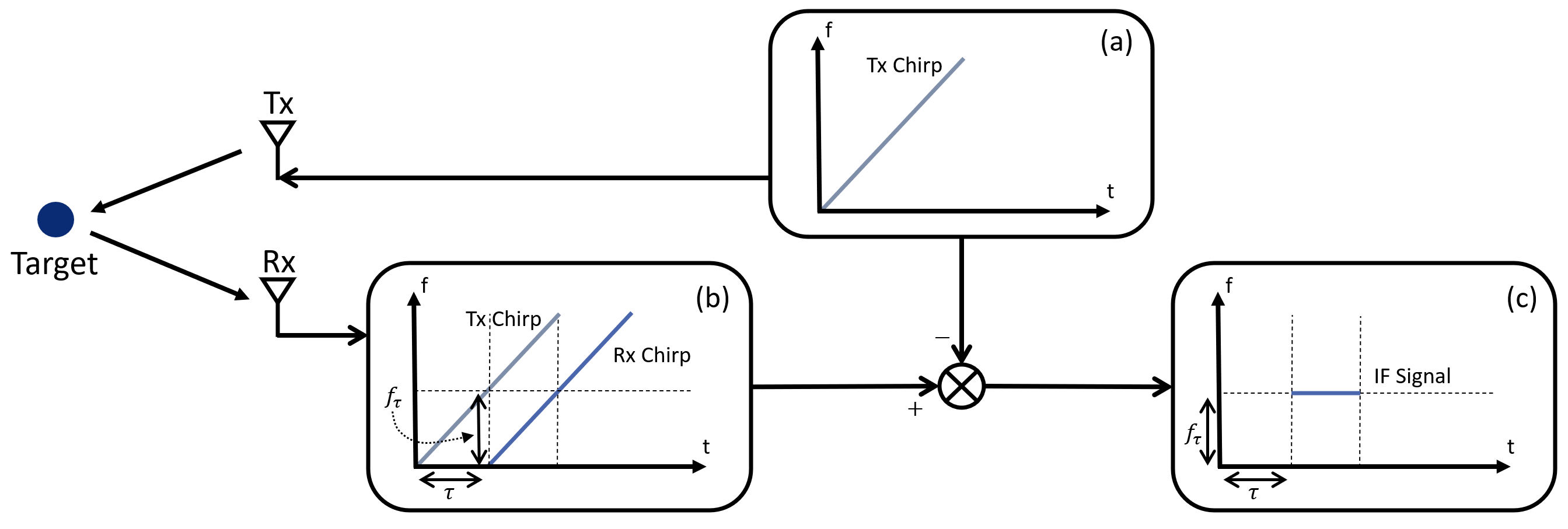}
    \caption{FMCW signal chain. (a) FMCW pulse generation. (b) Received signal from single ideal point scatterer. (c) FMCW beat signal after dechirping.}
    \label{fig:fmcw_signal_chain}
\end{figure}

As shown in Fig. \ref{fig:fmcw_signal_chain}, the FMCW device first generates what is known as a chirp signal, which can be modeled as a complex sinusoidal signal whose frequency increases linearly with time as
\begin{equation}
    m(t) = e^{-j2\pi(f_0t + \frac{1}{2}Kt^2)}, \quad 0 \leq t \leq T,
\end{equation}
where $f_0$ is the instantaneous frequency at time $t=0$, $K$ is the chirp slope, and $T$ is the chirp duration. 
The chirp bandwidth can be computed easily using $B = KT$ \cite{yanik2019near,smith2020nearfieldisar,wang2020AnEfficientCSA}.

The chirp signal $m(t)$ is transmitted by the transmit antenna, reflects off the ideal point reflector, and returns to the receive antenna as a scaled and time-delayed version of the transmitted signal. 
Consider the round-trip amplitude decay, the received signal can be modeled as
\begin{equation}
    \hat{m}(t) = \frac{\sigma}{R_T R_R}m(t-\tau) = \frac{\sigma}{R_T R_R} e^{-j2\pi(f_0(t-\tau) + \frac{1}{2}K(t-\tau)^2)},
\end{equation}
where $\tau$ is the round-trip time delay \cite{yanik2018millimeter} and the values $R_T$ and $R_R$ (see Fig. \ref{fig:fmcw_signal_chain}) are given by
\begin{gather}
    R_T = \left[ (x_0-x_T)^2 + (y_0-y_T)^2+(z_0-Z_0)^2 \right]^{\frac{1}{2}}, \\
    R_R = \left[ (x_0-x_R)^2 + (y_0-y_R)^2+(z_0-Z_0)^2 \right]^{\frac{1}{2}}.
\end{gather}

Therefore, the round trip time delay $\tau$ can be computed by
\begin{equation}
    \tau = \frac{R_T+R_R}{c},
\end{equation}
where $c$ is the speed of light.

The received signal $\hat{m}(t)$ is demodulated with the transmitted signal $m(t)$ yielding what is known as the IF signal or FMCW beat signal, written as
\begin{equation}
\label{eq:fmcw_beat_signal}
    s_0(t) = \frac{\sigma}{R_T R_R}e^{-j2\pi(f_0\tau +K\tau - \frac{1}{2}K\tau^2)}.
\end{equation}
The last phase term in (\ref{eq:fmcw_beat_signal}) is called the residual video phase (RVP) term and is known to be negligible \cite{yanik2019sparse}. 
Finally, the beat signal can be simplified to the expression
\begin{equation}
\label{eq:fmcw_beat_signal_k_multistatic}
    s_0(x_T,x_R,y_T,y_R,k) = \frac{\sigma}{R_T R_R} e^{-jk(R_T + R_R)},
\end{equation}
where $k = 2\pi f/c$ is the wavenumber corresponding to the instantaneous frequency $f = f_0 + Kt$ for $t \in [0,T]$.

The continuous-time signal (\ref{eq:fmcw_beat_signal_k_multistatic}) is sampled with sampling frequency $f_S$ by the radar analog-to-digital converter (ADC) and can be written in discrete time as
\begin{equation}
	\label{eq:mimo_beat_signal_discrete}
	s(x_T,x_R,y_T,y_R,n_k) = \frac{\sigma}{R_T R_R} e^{-j(k_0 + \Delta_k n_k)(R_T + R_R)},
\end{equation}
where $n_k$ is the wavenumber index, $k_0 = 2\pi f_0/c$ is the starting wavenumber corresponding to the starting frequency $f_0$, and $\Delta_k = 2\pi K /(c f_S)$ is the wavenumber step size.

\begin{figure}[h]
\centering
    \begin{subfigure}[b]{0.5\textwidth}
         \centering
         \includegraphics[width=\textwidth]{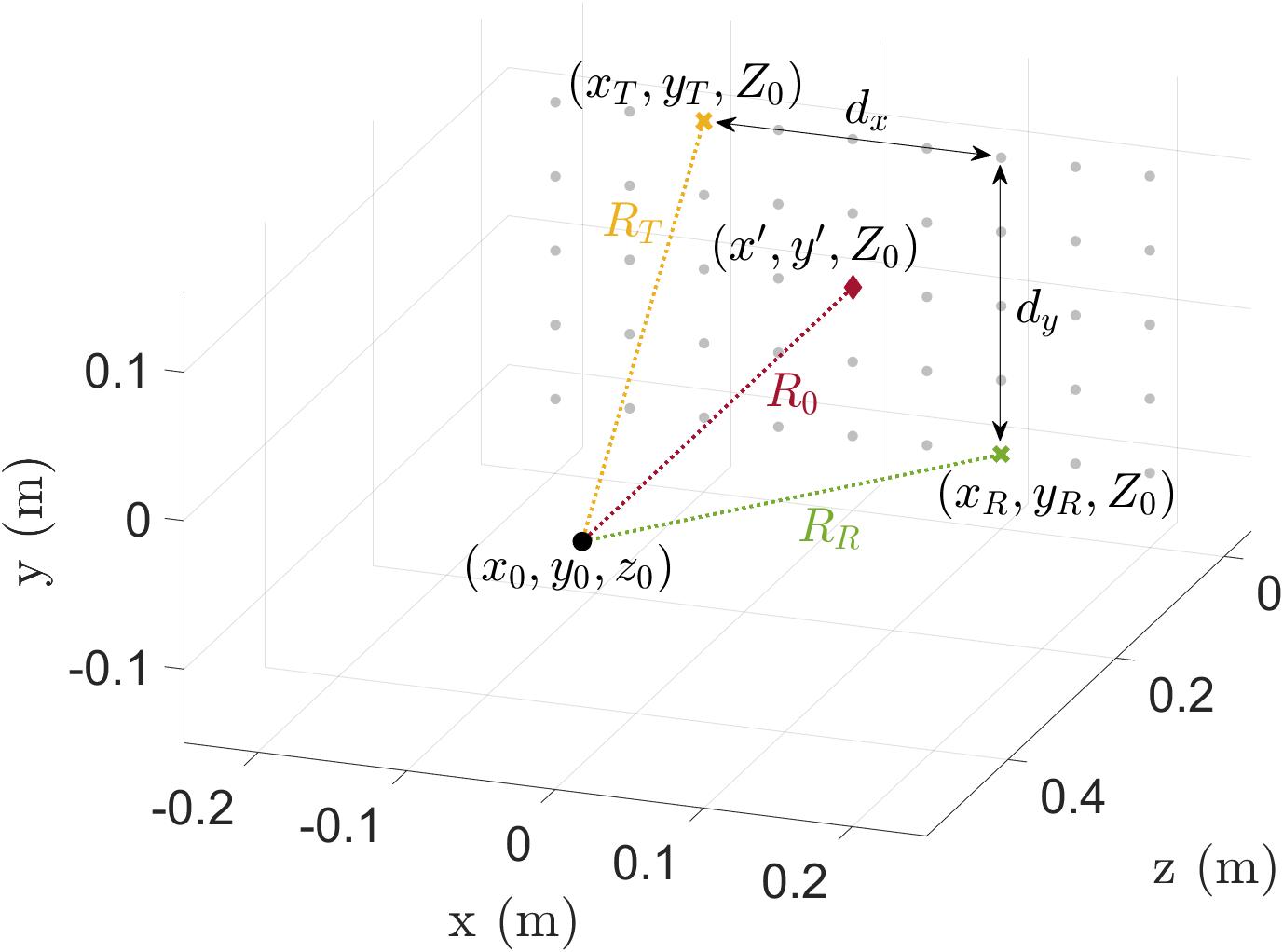}
         \caption{}
         \label{fig:fmcw_scenario}
    \end{subfigure}%
    ~
    \begin{subfigure}[b]{0.45\textwidth}
         \centering
         \includegraphics[width=\textwidth]{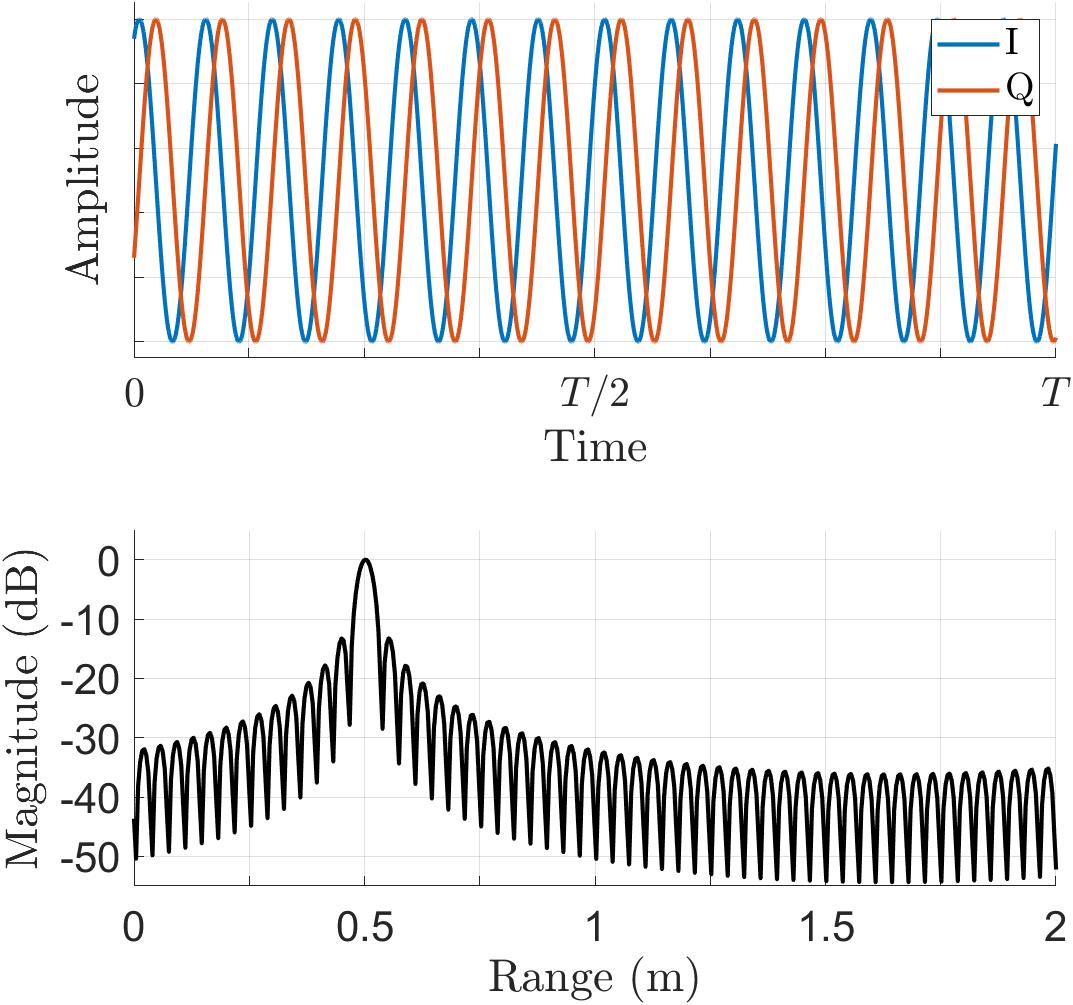}
         \caption{}
         \label{fig:fmcw_signal}
    \end{subfigure}
\caption{(a) Geometry for near-field FMCW radar scenario with Tx and Rx elements located at $(x_T,y_T,Z_0)$ and $(x_R,y_R,Z_0)$, respectively, and a point scatterer at $(x_0,y_0,z_0)$, using $Z_0 = 0$ m and $z_0$ = 0.5 m. (b) FMCW beat signal (top) and corresponding spectral representation using range Fourier transform.}
\label{fig:fmcw_example}
\end{figure}

To ease the subsequent signal processing, it is desirable to approximate the multistatic MIMO beat signal, represented in (\ref{eq:mimo_beat_signal_discrete}) as its corresponding monostatic equivalent using the approximation developed in \cite{yanik2020development,smith2020nearfieldisar,yanik2019sparse} as
\begin{equation}
	\label{eq:mult-to-mono}
	\hat{s}(x',y',n_k) = s(x_T,x_R,y_T,y_R,n_k) e^{j(k_0 + \Delta_k n_k)\frac{d_x^2 + d_y^2}{4 \Tilde{Z}}},
\end{equation}
valid only for small values of $d_x$ and $d_y$, the distances between the transmitter and receiver elements along the $x$- and $y$-directions, respectively, where $\Tilde{Z}$ is a reference plane typically given as the center of the target scene. 
Taking $(x',y',Z_0)$ as the location of the virtual element located at the midpoint between the transceiver pair, as shown in Fig. \ref{fig:fmcw_scenario}, and $R_0$ as the corresponding distance from the virtual element to the point reflector, the resulting monostatic beat signal is approximately 
\begin{equation}
	\label{eq:mono_beat_signal}
	\hat{s}(x',y',n_k) \approx \frac{p}{R_0^2}e^{-j2(k_0 + \Delta_k n_k)R_0}.
\end{equation}
From (\ref{eq:mono_beat_signal}), the spatial location, $(x_0,y_0,z_0)$, of the target is embedded in frequency of the radar beat signal, in the form of the radial distance $R_0$, which can be expressed as 
\begin{equation}
    \label{eq:monostatic_R}
    R_0 = \left[ {(x_0-x')^2 + (y_0-y')^2 + (z_0-Z_0)^2} \right]^{\frac{1}{2}}.
\end{equation}

An example geometry is given in Fig. \ref{fig:fmcw_scenario}, with a single transceiver pair located at $Z_0 = 0$ m and a point scatterer at $(0,0,0.5)$. 
From (\ref{eq:mono_beat_signal}), the FMCW beat signal, shown in Fig. \ref{fig:fmcw_signal}, is a single tone sinusoidal signal. 
Taking the Fourier transform of the FMCW beat signal yields the range profile, which shows a dominant peak at the distance from the radar to the point scatterer. 

\section{Range-Doppler Processing}
\label{sec:range_doppler}

The relative velocity of a target can be extracted from the beat signal expressed in (\ref{eq:mono_beat_signal}) by exploiting the Doppler effect.
As discussed in \cite{winkler2007range}, by transmitting a series of chirp waveforms at a known pulse repetition interval (PRI), $T_{PRI}$, the velocity of a moving target can be identified as the frequency component along the chirp index dimension given by
\begin{equation}
	\label{eq:doppler_final}
	\hat{s}(x',y',n_k,n_c) = \frac{p}{R_0^2} e^{-j(2(k_0 + \Delta_k n_k)R_0 + \frac{4\pi v T_{PRI}}{\lambda_0}n_c)},
\end{equation}
where $R$ is the initial range of the target, $v$ is the velocity of the target, $\lambda_0$ is the wavelength corresponding to $f_0$, and $n_c$ is the chirp index,

Thus, the beat signal sampled across time is a \mbox{2-D} complex sinusoidal signal with frequencies corresponding to the range and velocity of the target in the first and second dimensions, respectively. 
Subsequently, to extract the range and velocity, traditional methods perform a \mbox{2-D} fast Fourier transform (FFT) over a matrix whose rows or columns consist of subsequent chirps. 
This analysis is known as range-Doppler processing and is commonly applied to many radar signal processing problems \cite{soumekh1998wide,winkler1995making}.

\section{FMCW Response to Distributed Target}
\label{sec:fmcw_distributed_target}
Assuming a distributed target occupying volume $V$ in Cartesian $x$-$y$-$z$ space and the same transceiver pair discussed in Section \ref{sec:fmcw_signal_model}, the FMCW beat signal can be expressed as
\begin{equation}
    \label{eq:fmcw_beat_signal_distributed_target_multistatic}
    s(x_T,x_R,y_T,y_R,k) = \iiint_V \frac{p(x,y,z)}{R_T R_R} e^{-j k (R_T + R_R)} dx dy dz,
\end{equation}
where $p(x,y,z)$ is known as the reflectivity function of the target representing the intensity of reflection from each point of the target throughout volume $V$ and $R_T$ and $R_R$ are the radial distances from the target to the transmitter and receiver, respectively, as
\begin{gather}
    R_T = \left[ (x-x_T)^2 + (y-y_T)^2+(z-Z_0)^2 \right]^{\frac{1}{2}}, \\
    R_R = \left[ (x-x_R)^2 + (y-y_R)^2+(z-Z_0)^2 \right]^{\frac{1}{2}}.
\end{gather}

By applying the multistatic-to-monostatic conversion in (\ref{eq:mult-to-mono}) \cite{yanik2019sparse}, the virtual monostatic response can be written as
\begin{equation}
    \label{eq:fmcw_beat_signal_distributed_target_monostatic}
    s(x',y',k) = \iiint_V \frac{p(x,y,z)}{R_0^2} e^{-j 2 k R_0} dx dy dz,
\end{equation}
where $R_0$ is the distance between the virtual monostatic element and the target as
\begin{equation}
    R_0 = \left[ (x-x')^2 + (y-y')^2 + (z-Z_0)^2 \right]^\frac{1}{2}.
\end{equation}

The virtual monostatic response can be written in discrete-time as 
\begin{equation}
    \label{eq:fmcw_beat_signal_distributed_target_monostatic_discrete}
    s(x',y',n_k) = \iiint_V \frac{p(x,y,z)}{R_0^2} e^{-j 2 (k_0 + \Delta_k n_k) R_0} dx dy dz.
\end{equation}

In many applications, it is desirable to extract the reflectivity function $p(x,y,z)$ from the radar beat signal. 
This process is known as imaging and requires inversion of the integral in (\ref{eq:fmcw_beat_signal_distributed_target_multistatic}).
However, to achieve this, the radar must be sampled throughout space by utilizing a large array of radar transceivers, known as real array radar (RAR) \cite{wang20203}, or the concept of synthetic aperture radar (SAR), in which a small radar platform is scanned throughout space to synthesize a larger array. 
In this dissertation, orthogonality is leveraged across time by operating the MIMO radar using the time-division multiplexing (TDM) MIMO technique such that each Tx/Rx pair is activated sequentially. 
Hence, the MIMO-SAR operation involves performing TDM-MIMO at each location in space, but involves its own challenges \cite{yanik2020development,yanik2019sparse}.
High-resolution near-field SAR and MIMO-SAR imaging algorithms and systems for multiple modalities are discussed throughout this dissertation. 

FMCW signaling enables low-cost ultra-wideband radar systems for a host of applications. 
With precise spatial information embedded in the frequency content of the signal, FMCW radars are suitable for many sensing tasks. 
Extracting and leveraging the spatial information for applications such as classification and perception will be addressed at length in the remainder of this dissertation. 
Additionally, as radars operate with limited bandwidth, causing a sinc effect in the range domain, improving spatial resolution using machine learning techniques is a promising solution to overcoming system limitations. 
Throughout this dissertation, we will introduce novel techniques using signal processing, machine learning, and hybrid-learning algorithms leveraging the characteristics of FMCW signals. 

\chapter{Impact of Front-End Signal Processing Techniques on Deep Learning Perception}
\label{ch:signal_processing}

In this chapter, we explore various front-end signal processing techniques for improving perception using data-driven algorithms. 
We investigate signal processing algorithms to extract and process spatiotemporal signatures embedded in the FMCW radar signals, as discussed in Chapter \ref{ch:fmcw_signal_model}.
Applications including gesture recognition, SAR image segmentation for concealed weapon detection, and SAR image super-resolution using deep learning display variable performance depending on the signal processing techniques applied to the data prior to the learning algorithm. 
Here, we explore the impact of front-end signal processing methods by optimizing the fidelity and computational load of hybrid-learning algorithms for perception and imaging. 
Part of the following work was previously published in \cite{smith2021sterile}\footnote{\copyright 2021 IEEE. Reprinted, with permission, from J. W. Smith, S. Thiagarajan, R. Willis, Y. Makris, and M. Torlak, ``Improved static hand gesture classification on deep convolutional neural networks using novel sterile training technique,'' \textit{IEEE Access}, vol. 9, pp. 10893–10902, Jan. 2021.} and \cite{smith2022efficient}\footnote{\copyright 2022 IEEE. Reprinted, with permission, from J. W. Smith and M. Torlak, ``Efficient \mbox{3-D} near-field MIMO-SAR imaging for irregular scanning geometries,'' \textit{IEEE Access}, vol. 10, pp. 10283-10294, Jan. 2022.} and will be presented in \cite{smith2023ThzToolbox}.

\section{Gesture Recognition with mmWave Radar}
\label{sec:gestures_background}
Accurately classifying human hand gestures has recently received significant attention as non-contact human-computer interaction (HCI) sensors have become increasingly prevalent and desirable. 
Many efforts have been made to classify moving (dynamic) hand gestures and non-moving (static) hand gestures using optical cameras and many different classifiers \cite{baek2013comparison}. 
Applications of static gesture classification include augmented/virtual reality (AR/VR) \cite{son2017image}, human-computer interaction \cite{matilainen2016OUHANDS}, and even medical applications for range of motion and therapeutic applications \cite{anaz2018classification}. 
Such optical systems offer high-resolution two-dimensional (\mbox{2-D}) images but have innate drawbacks, such as requiring specific lighting conditions and lacking depth information. 
Some solutions have investigated the use of an RGB-D depth camera \cite{lin2013_3D_hand_posture_RGBD}, but these devices suffer from sunlight, restricting their usage to exclusively indoor applications \cite{son2017image}. 
On the other hand, small form-factor mmWave frequency-modulated-continuous-wave (FMCW) radar offers high-resolution depth information but does not have the cross-range resolution of an optical camera. 
mmWave radars are advantageous over optical solutions because of the semi-penetrative nature of EM radiation at wavelengths in the mmWave frequency range and independence from ambient temperature effects, allowing for fine measurements in non-ideal lighting and temperature environments including occlusion, fog, indoor/outdoor, etc. 
Additionally, FMCW mmWave radars enable simultaneous gesture classification and localization. 
High-resolution spatial information reflected from the human hand is embedded in the FMCW return signal. 
However, owing to the nature of FMCW radar as a time-of-flight (ToF) sensor and hardware size limitations, an off-the-shelf radar device cannot reconstruct an image reminiscent of the human hand or meaningful to the human eye without employing time-consuming SAR techniques. 
Thus, a deep convolutional neural network (CNN) approach is adopted to classify dynamic gestures from radar return signals \cite{kim2017application}. 

\begin{figure}[h]
\centering
    \begin{subfigure}[b]{0.25\textwidth}
         \centering
         \includegraphics[width=\textwidth]{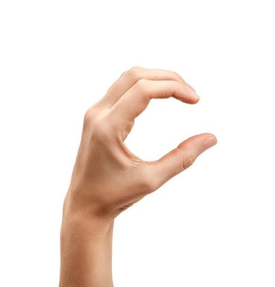}
         \caption{}
         \label{fig:c}
    \end{subfigure}
    \hfill
    \begin{subfigure}[b]{0.25\textwidth}
         \centering
         \includegraphics[width=\textwidth]{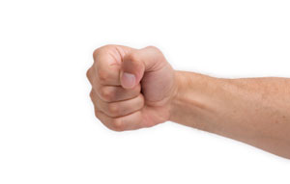}
         \caption{}
         \label{fig:fist}
    \end{subfigure}
    \hfill
    \begin{subfigure}[b]{0.25\textwidth}
         \centering
         \includegraphics[width=\textwidth]{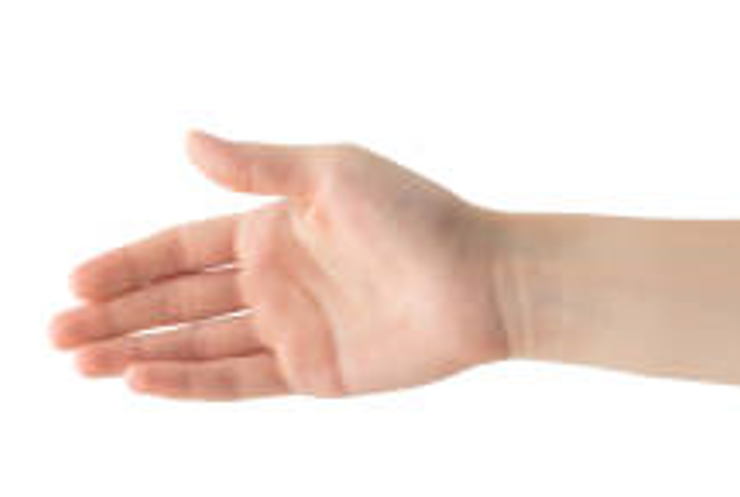}
         \caption{}
         \label{fig:palm}
    \end{subfigure}
    
    \vskip\baselineskip
    
    \begin{subfigure}[b]{0.25\textwidth}
         \centering
         \includegraphics[width=\textwidth]{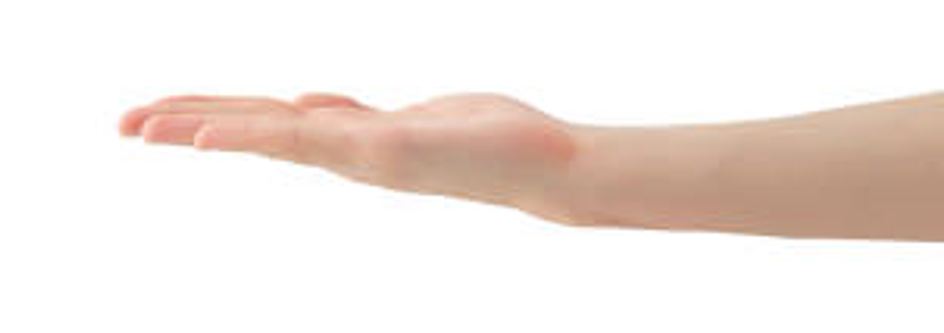}
         \caption{}
         \label{fig:perp}
    \end{subfigure}
    \hfill
    \begin{subfigure}[b]{0.25\textwidth}
         \centering
         \includegraphics[width=\textwidth]{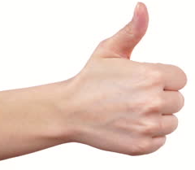}
         \caption{}
         \label{fig:tu}
    \end{subfigure}
    \hfill
    \begin{subfigure}[b]{0.25\textwidth}
         \centering
         \includegraphics[width=\textwidth]{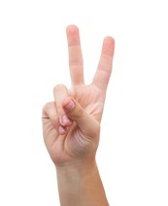}
         \caption{}
         \label{fig:two}
    \end{subfigure}
    
\caption{Static hand gestures: (a) ``c'', (b) ``fist'', (c) ``palm'', (d) ``perpendicular'', (e) ``thumbs up'', (f) ``two''.}
\label{fig:static_gestures}
\end{figure}

\subsection{Static Gesture Recognition with mmWave Radar}
\label{subsec:static_gestures}
In this section, we explore the application of hybrid-learning algorithms to the static (stationary) gesture recognition problem. 
Rather than employing SAR or ISAR to capture images of the hand from many locations using imaging algorithms to recover the reflectivity, we propose classifying a hand gesture using only a single, stationary MIMO-FMCW radar. 
Similar applications have been employed in commercial products for gesture-based HCI. 
The most notable example is the use of a 60 GHz radar in the Google Pixel 4 \cite{bernardo2017_o_soli_mio}. 
Fig. \ref{fig:static_gestures} shows the six gestures employed in the following studies, which shed light on the various degrees of complexity and opportunities for innovation in mmWave radar hand gesture recognition. 

\begin{figure}[h]
\centering
    \begin{subfigure}[b]{0.35\textwidth}
         \centering
         \includegraphics[width=\textwidth]{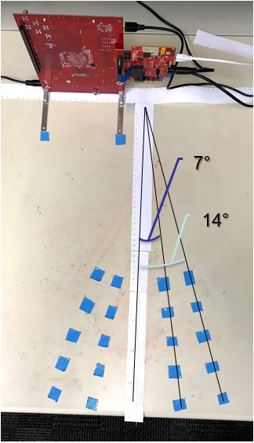}
         \caption{}
         \label{fig:static_setup0}
    \end{subfigure}%
    ~
    \begin{subfigure}[b]{0.45\textwidth}
         \centering
         \includegraphics[width=\textwidth]{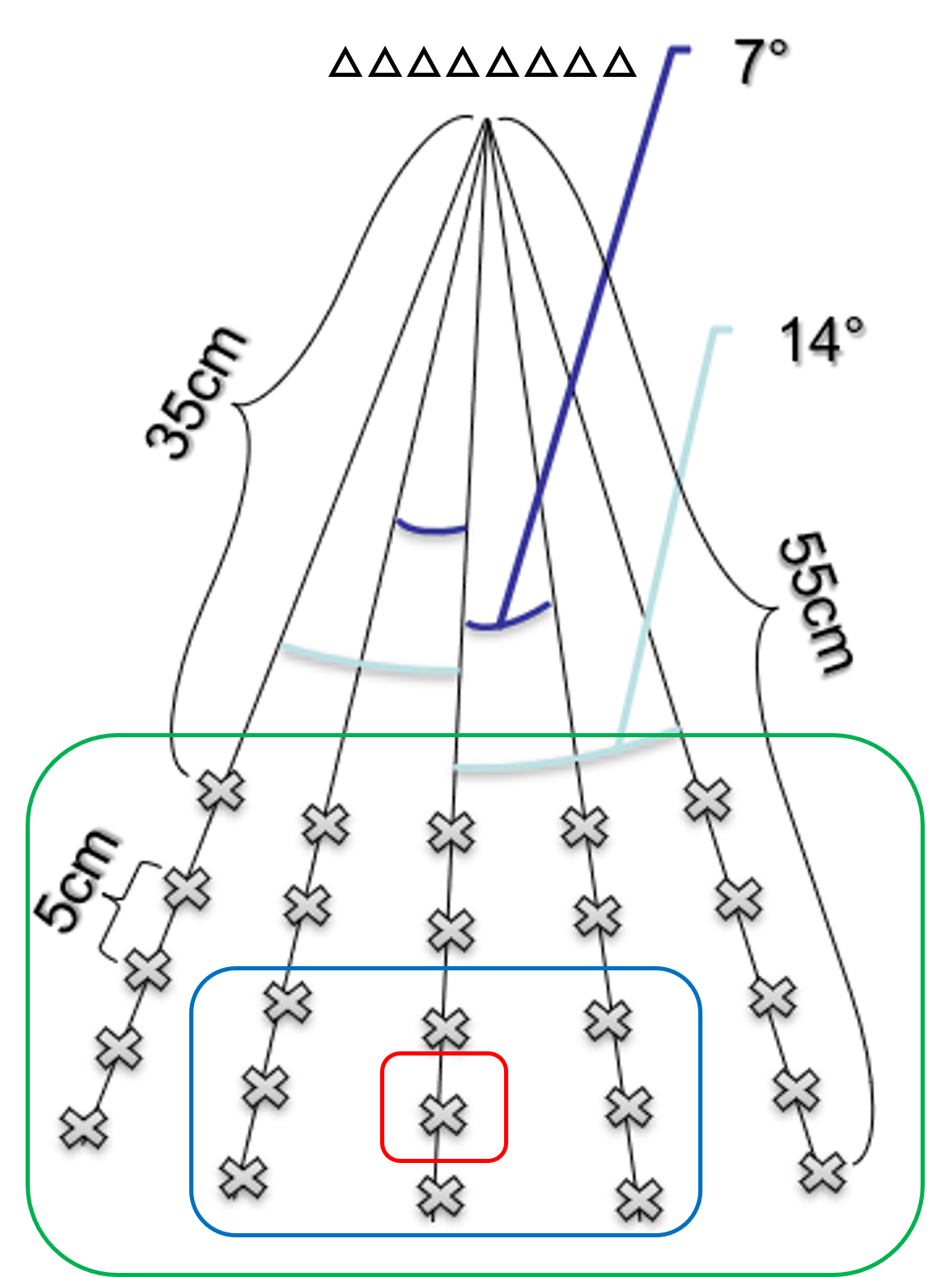}
         \caption{}
         \label{fig:static_datasets_setup}
    \end{subfigure}
\caption{(a) Hardware setup consisting of a TI 77 GHz radar and visual indicators for the test subjects. (b) Locations of data captured for each dataset: \textcolor{red}{Dataset 1}, \textcolor{ForestGreen}{Dataset 2}, \textcolor{blue}{Dataset 3}.}
\label{fig:static_setup}
\end{figure}

Data for subsequent experiments were collected from a diverse set of five participants. 
The subjects were instructed to position their hand angled forward, backward, left, and right by $30^\circ$, resulting in nine Degrees of Freedom (DoF) at each sampling location.

First, a dataset was collected at a single location 50 cm from the radar boresight. 
Each gesture was captured with the 9 DoFs detailed previously, resulting in a total of 2250 gestures per class, with six gesture classes, as shown in Fig. \ref{fig:static_gestures}. 
After the preliminary results were promising, a diverse dataset is collected using the setup in Fig. \ref{fig:static_setup0}. 

As indicated by the green box in Fig. \ref{fig:static_datasets_setup}, Dataset 2 comprises 25 locations from 35--55 cm, spanning a $28^\circ$ field of view (FOV), with a total of 4500 captures per class. 
Finally, a third dataset was collected within a smaller region, indicated by the blue box in Fig. \ref{fig:static_datasets_setup}, spanning 45--55 cm and a $14^\circ$ FOV. 
A summary of these datasets is presented in Table \ref{table:static_datasets}

\begin{table}[h]
    \centering
    \caption{Summary of static hand gesture datasets.}
    \begin{tabular}{c||c|c|c}
         & Dataset 1 & Dataset 2 & Dataset 3  \\
         \hline \hline
         Captures/class & 2250 & 4500 & 405 \\
         \hline
         Ranges & 50 cm & 35 cm -- 55 cm & 45 cm -- 55 cm \\
         \hline
         FOV & $0^\circ$ & $28^\circ$ & $14^\circ$ \\
         \hline \hline
    \end{tabular}
    \label{table:static_datasets}
\end{table}

\begin{figure}[h]
    \centering
    \includegraphics[width=0.6\textwidth]{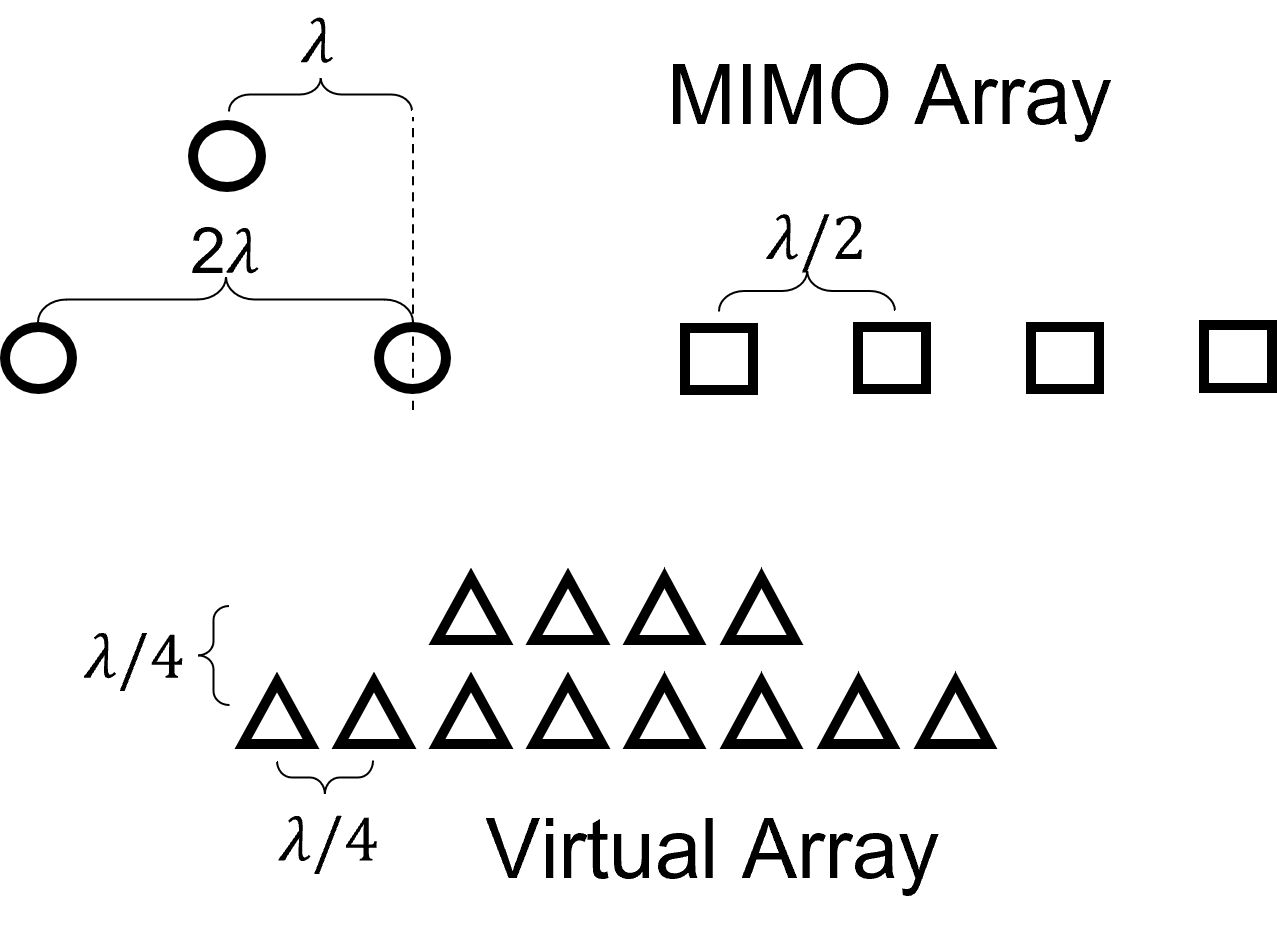}
    \caption{TI AWR1443BOOST MIMO antenna array and virtual monostatic array.}
    \label{fig:awr1243}
\end{figure}

The data are collected using a Texas Instruments (TI) AWR1443BOOST radar with \mbox{4 GHz} bandwidth from \mbox{77 GHz} to \mbox{81 GHz} is mounted on a TI mmWave-Devpack and TSW1400 data capture card to store the data and transfer it to the PC, where the samples are manipulated in MATLAB.
The TI AWR1443BOOST is equipped with a MIMO array consisting of two Tx elements spaced by $2\lambda_c$, one Tx element vertically displaced by $\lambda/2$, and four Rx elements spaced by $\lambda_c/2$ \cite{yanik2019sparse}, as shown in Fig. \ref{fig:awr1243}.
By orienting the radar in the horizontal direction, a virtual array exists consisting of a row of eight antennas underneath a row of four antennas, as shown in Fig. \ref{fig:awr1243}.
It should be noted that although the radar setup is mounted on a desk, the reflections of the desk are negligible, according an empirical study, owing to the narrow beamwidth of the radar along the vertical direction. 

After the datasets were collected, preprocessing techniques are applied to investigate the optimal presentation of data to a CNN. 
CNNs of varying dimensionality are implemented with three hidden layers consisting of a convolution layer with kernel sizes of 5, 5 $\times$ 5, or 5 $\times$ 5 $\times$ 5, a batch normalization layer, and a Rectified Linear Unit (ReLU) \cite{glorot2011ReLU}. 
After the three convolution layers, a fully connected layer is employed for the six classes and cross-entropy loss is used to train the network using an ADMM optimizer. 
The real and imaginary parts of the sample were layered to leverage the signatures embedded in the phase of the data.  

64 samples are taken over the 4 GHz bandwidth of the radar; hence, each sample is an array of size 64 $\times$ 12, owing to the 12 virtual channels.
To obtain a baseline, we apply a simple \mbox{1-D} CNN to the samples of Dataset 1 vectorized as 768 $\times$ 1 vectors. 
Even for a simple set of data in Dataset 1, the classification rate is 83\%. 
However, given the underlying mechanics of the problem and format of the data, the samples are not presented to the network in a meaningful way.  
A range-FFT is performed across the first dimension of the 64 $\times$ 12 array, as described in Section \ref{sec:fmcw_signal_model}. 
After selecting the range bins of interest, a process known as ``range-gating,'' a \mbox{2-D} CNN is trained using the range-FFT data from Dataset 1, yielding a classification accuracy of 95\%. 

\begin{figure}[h]
\centering
    \begin{subfigure}[b]{0.45\textwidth}
         \centering
         \includegraphics[width=\textwidth]{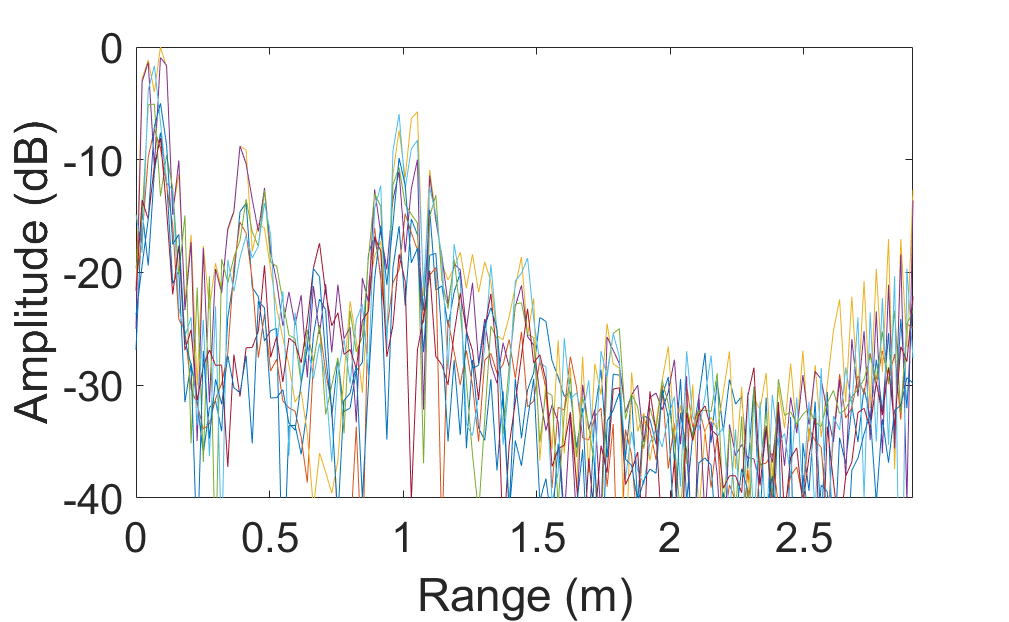}
         \caption{}
         \label{fig:c_range_fft}
    \end{subfigure}
    \hfill
    \begin{subfigure}[b]{0.45\textwidth}
         \centering
         \includegraphics[width=\textwidth]{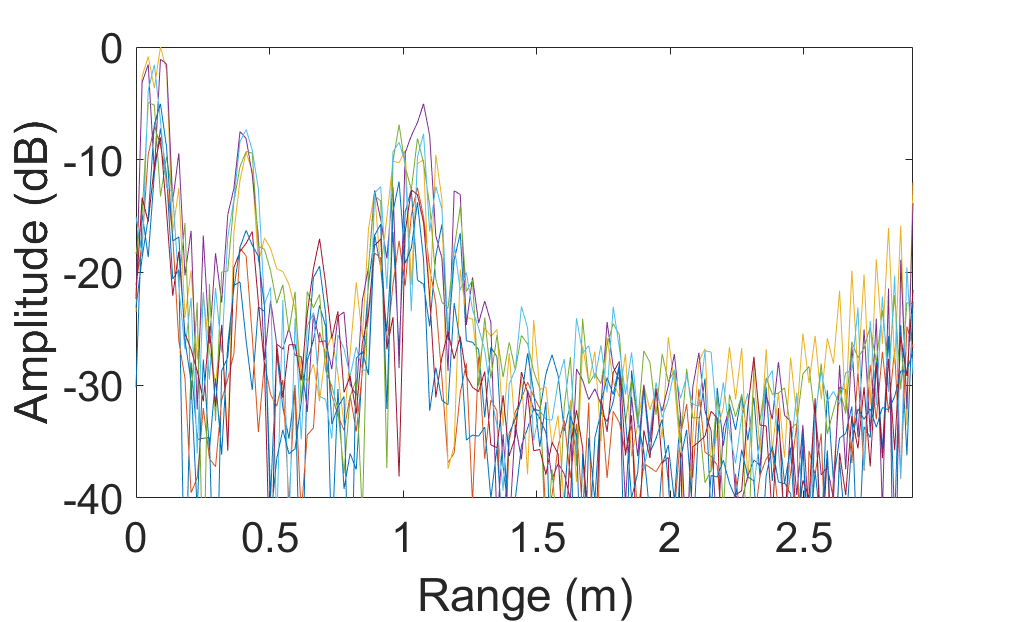}
         \caption{}
         \label{fig:fist_range_fft}
    \end{subfigure}
\caption{(a) Range-FFT of a ``c'' sample with the hand at 45 cm. (b) Range-FFT of a ``fist'' sample with the hand at 45 cm.}
\label{fig:range_ffts}
\end{figure}

Here, we note the behavior of the mmWave radar gesture data in the range and angle domains. 
Two sample range-FFT spectra are computed from random data points selected from the ``c'' and ``fist'' classes with the hand at 45 cm and shown in Fig. \ref{fig:range_ffts}. 
Because the frequency content of the FMCW signal corresponds to the range of the targets, the range-FFT spectrum represents the magnitude (and phase) of the reflection of the target at a given distance. 
As expected, a significant reflection is observed at approximately 1 m because of the human torso. 
However, while a reflection from the hand is visible around 45 cm, there is no distinguishing characteristic, to the human eye between the two classes. 

Next, an autocorrelation strategy is applied to the radar data using the eight collinear channels.
The \mbox{2-D} autocorrelation matrix is computed from the 25 $\times$ 8 range-FFT array. 
The autocorrelation method leverages the spatial relationships between adjacent channels to provide a more learnable representation of the samples; however for Dataset 1, the classification rate remained at 95\% using this approach.

Finally, an angle-FFT technique is applied to the samples along with the range-FFT, referred to as ``range-angle FFT.'' 
After the range-FFT is performed, the sample is rearranged and zero-padded according to the geometry in Fig. \ref{fig:awr1243}, with a size of 25 $\times$ 8 $\times$ 2.
An angle-FFT of size 16 is performed across the second dimension yielding a data cube of size 25 $\times$ 16 $\times$ 2. 
A \mbox{3-D} CNN is trained using the range-angle-FFT data, yielding a classification accuracy of 99\% for Dataset 1. 

\begin{figure}[h]
\centering
    \begin{subfigure}[b]{0.45\textwidth}
         \centering
         \includegraphics[width=\textwidth]{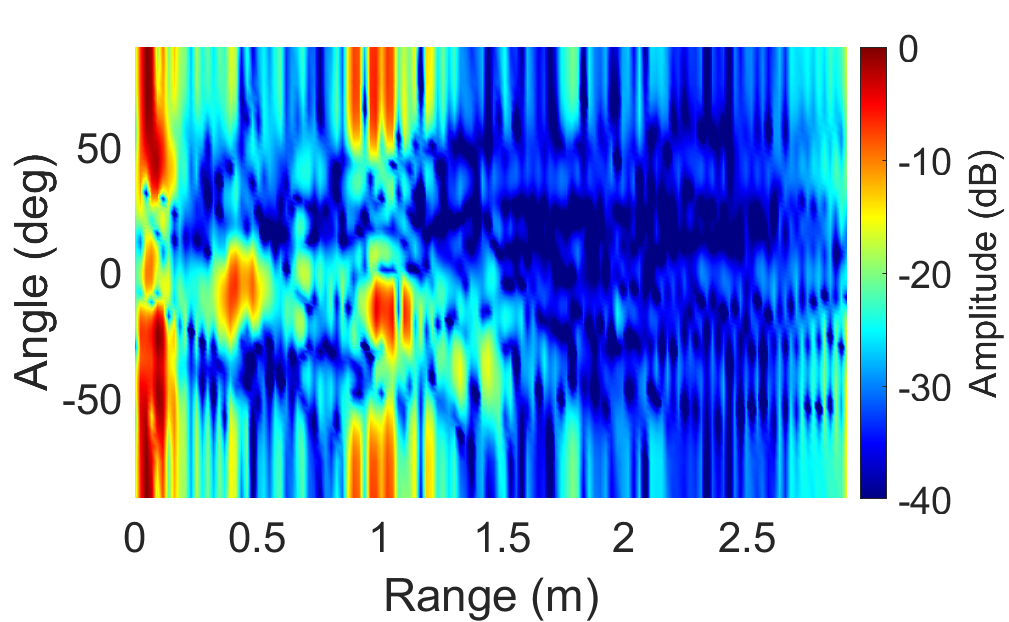}
         \caption{}
         \label{fig:c_range_angle_fft}
    \end{subfigure}
    \hfill
    \begin{subfigure}[b]{0.45\textwidth}
         \centering
         \includegraphics[width=\textwidth]{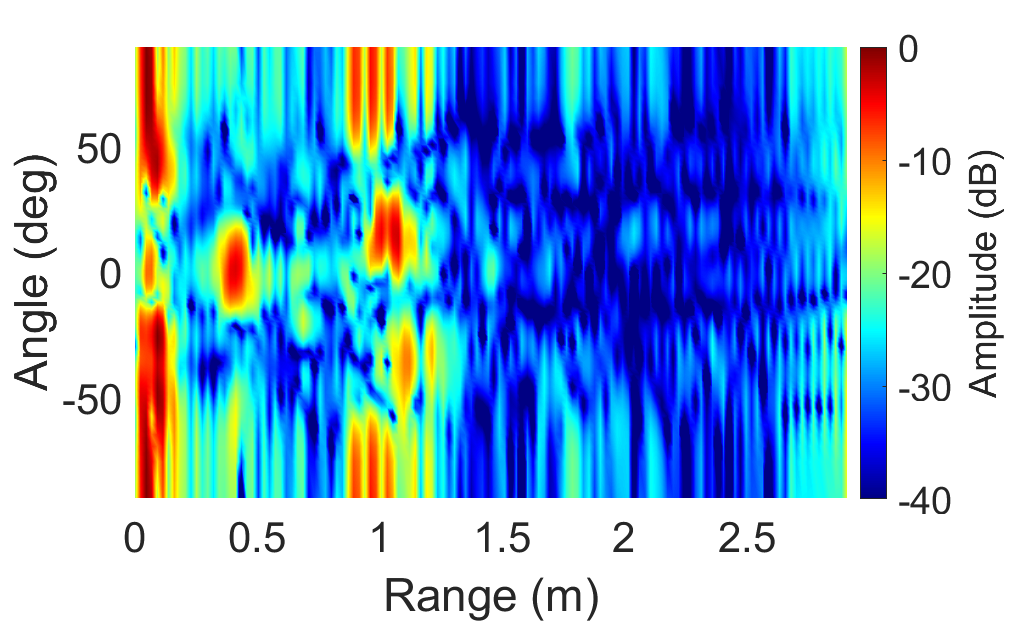}
         \caption{}
         \label{fig:fist_range_angle_fft}
    \end{subfigure}
\caption{(a) Range-Angle-FFT of a ``c'' sample with the hand at 45 cm. (b) Range-Angle-FFT of a ``fist'' sample with the hand at 45 cm.}
\label{fig:range_angle_ffts}
\end{figure}

Similarly, in Fig. \ref{fig:range_angle_ffts}, we examine the \mbox{2-D} range-angle-FFT spectra of the same two samples, as shown in Fig. \ref{fig:range_ffts}. 
We again note the reflection of the torso at approximately 1 m and the hand at approximately 45 cm, close to the center of the FOV; however, the reflections have negligible meaning to the naked eye. 

Similarly, the aforementioned preprocessing techniques are applied to Datasets 2 and 3, and the results are summarized in Table \ref{table:static_results}. 
As expected, Dataset 2, which is the most diverse dataset, is the most difficult to classify. 
Upon closer inspection, because the test subjects are seated in front of the radar with their hand in front of them, many of the samples in Dataset 2 do not contain a meaningful reflection from the hand as the hand reflection is obscured by the sidelobes from the much stronger torso reflection. 
Hence, a nulling strategy is employed to project the collected data onto the null space of the peak along the range-FFT corresponding to the torso.
However, this method yields a minimal increase in the classification rate for Dataset 2 and reduces the classification accuracy for Datasets 1 and 3. 
This phenomenon is likely due to the nulling process unintentionally reducing the learnable information about the hand gesture and the proposed nulling procedure is discarded. 

However, the autocorrelation and range-angle-FFT methods yield performance increases for all datasets. 
The autocorrelation technique results in classification rates of 62\% and 87\% for Datasets 2 and 3, respectively.
The range-angle-FFT strategy results in classification rates of 75\% and 91\% for Datasets 2 and 3, respectively.

\begin{table}[h]
    \centering
    \caption{Classification results for various preprocessing techniques of static gesture data.}
    \begin{tabular}{c||c|c|c}
         & Dataset 1 & Dataset 2 & Dataset 3  \\
         \hline \hline
         Raw (Vectorized) & 83\% & - & - \\
         \hline
         Reformatted Range-FFT & 95\% & 61\% & 86\% \\
         \hline
         Nulled & 91\% & 61\% & 80\% \\
         \hline
         Autocorrelation & 95\% & 62\% & 87\% \\
         \hline
         Range-Angle-FFT & 99\% & 75\% & 91\% \\
         \hline \hline
    \end{tabular}
    \label{table:static_results}
\end{table}

\subsection{Dynamic Gesture Recognition with mmWave Radar}
\label{subsec:dynamic_gestures}

\begin{figure}[h]
\centering
    \begin{subfigure}[b]{0.25\textwidth}
         \centering
         \includegraphics[width=\textwidth]{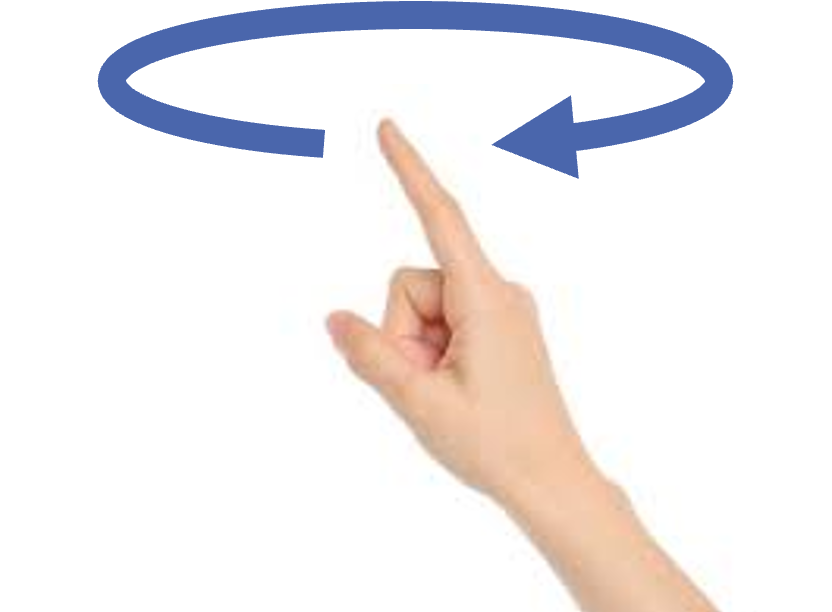}
         \caption{}
         \label{fig:circle}
    \end{subfigure}
    ~
    \begin{subfigure}[b]{0.25\textwidth}
         \centering
         \includegraphics[width=\textwidth]{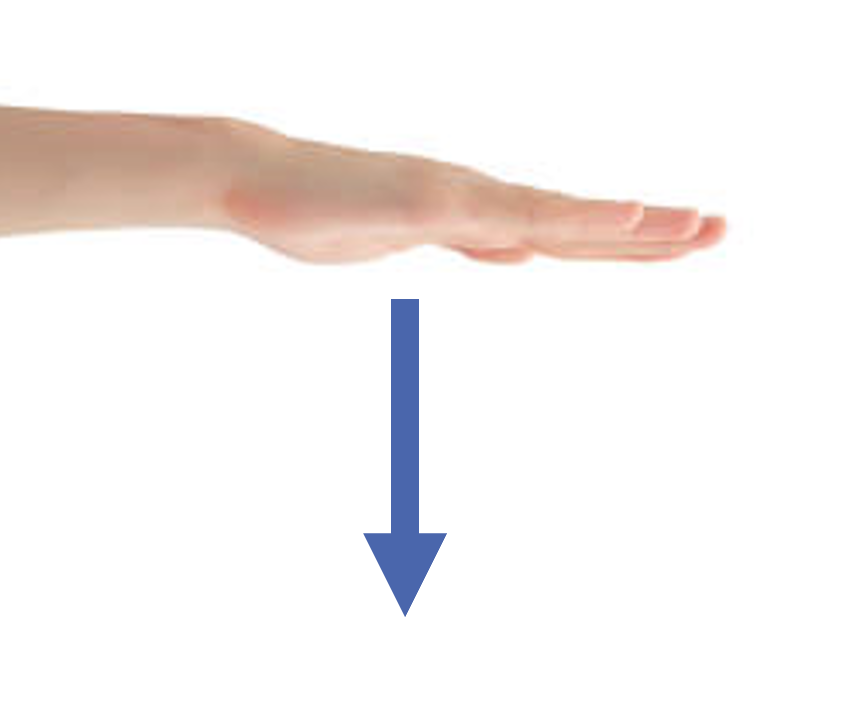}
         \caption{}
         \label{fig:push}
    \end{subfigure}
    ~
    \begin{subfigure}[b]{0.25\textwidth}
         \centering
         \includegraphics[width=\textwidth]{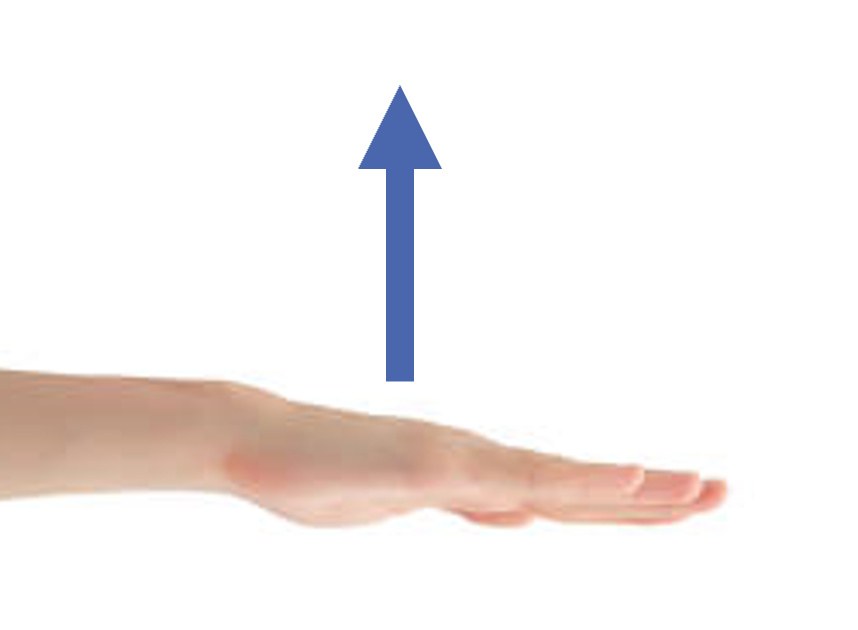}
         \caption{}
         \label{fig:pull}
    \end{subfigure}
    
    \vskip\baselineskip
    
    \begin{subfigure}[b]{0.25\textwidth}
         \centering
         \includegraphics[width=\textwidth]{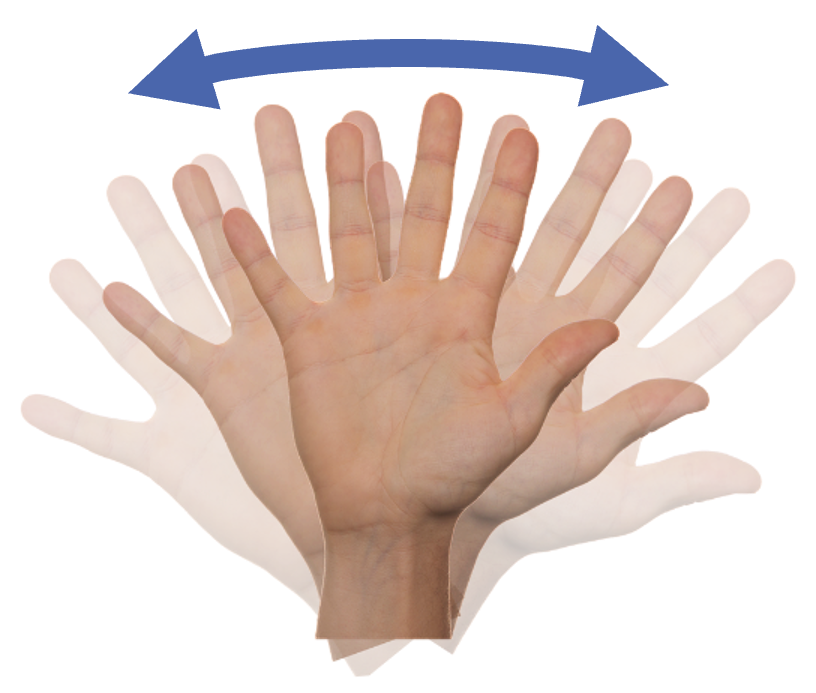}
         \caption{}
         \label{fig:wave}
    \end{subfigure}
    ~
    \begin{subfigure}[b]{0.25\textwidth}
         \centering
         \includegraphics[width=\textwidth]{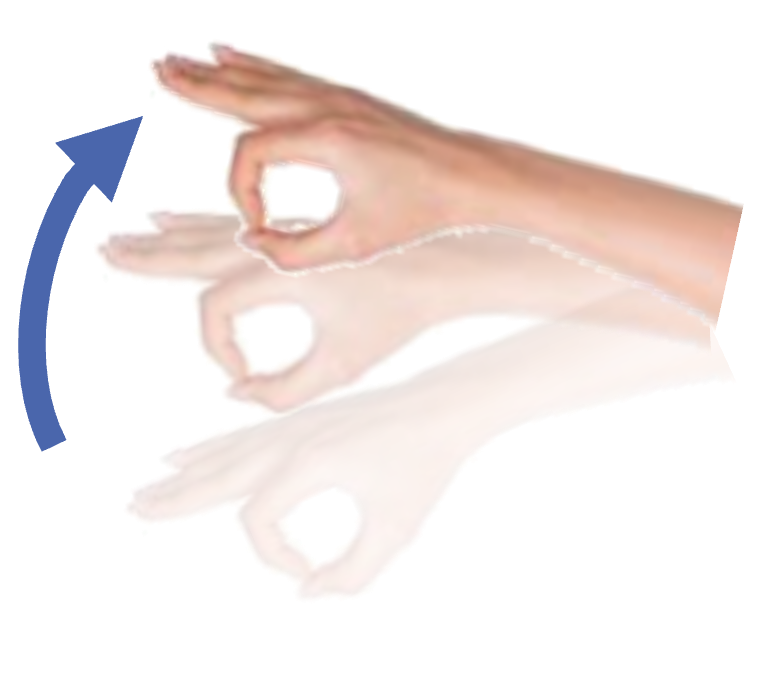}
         \caption{}
         \label{fig:whoosh}
    \end{subfigure}
    
\caption{Dynamic hand gestures: (a) ``circle'', (b) ``push'', (c) ``pull'', (d) ``wave'', (e) ``whoosh''.}
\label{fig:dynamic_gestures}
\end{figure}

Similarly, a study is conducted on dynamic (moving) gestures to investigate the impact of data presentation on classification rate. 
Five dynamic hand gestures are employed, as shown in Fig. \ref{fig:dynamic_gestures}, requiring the user to move their hand in a circle around the boresight of the radar, push towards the radar, pull away from the radar, wave at the boresight of the radar, or perform the University of Texas at Dallas ``whoosh'' spirit symbol, pulling their hand from their waist to face level. 
Five test subjects collect a single dataset while seated at a distances of \mbox{1 m} from the radar, consisting of 600 captures per class. 
Each capture consists of 512 FMCW pulses, known as frames, across 2.56 s; hence, the pulse repetition interval (PRI) is 5 ms. 

\begin{figure}[h]
    \centering
    \includegraphics[width=\textwidth]{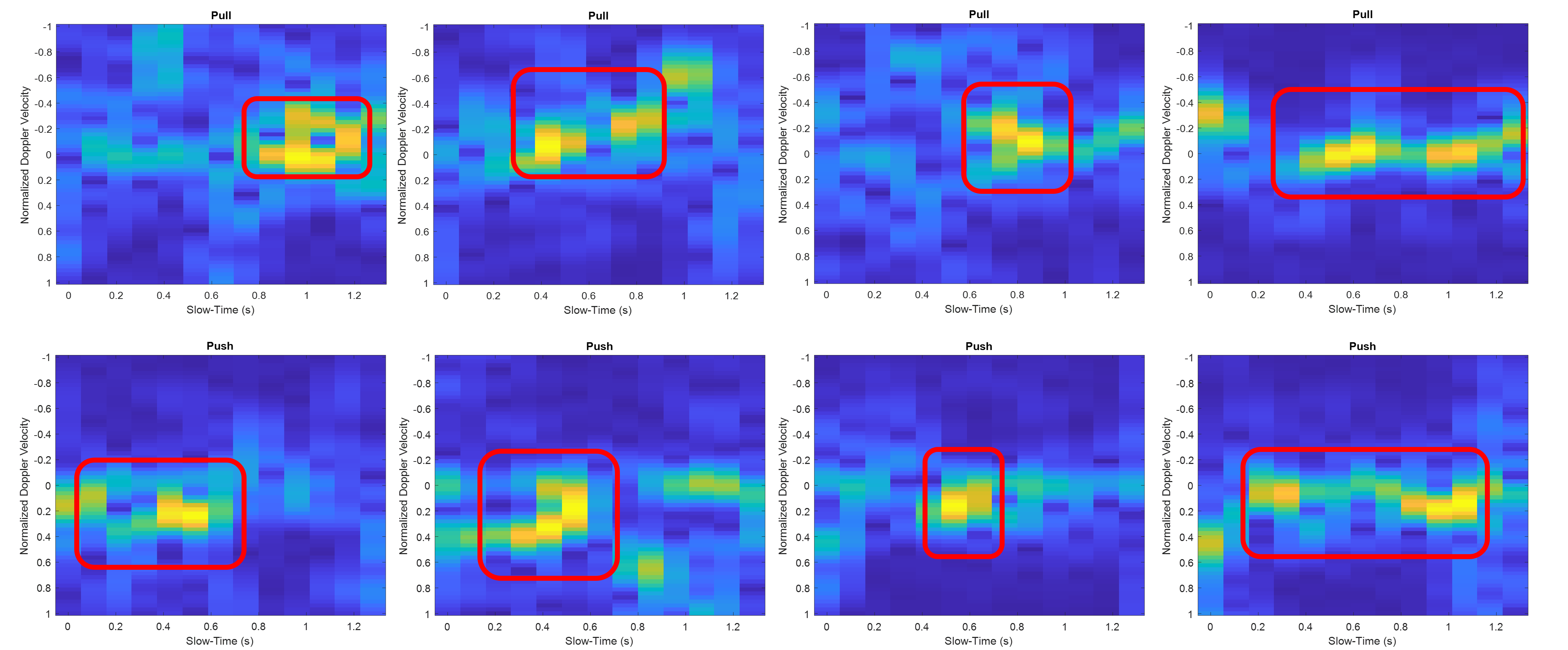}
    \caption{Doppler-STFT for pull and push gestures demonstrating where the gesture is performed during the 2.56 s capture.}
    \label{fig:dynamic_stft}
\end{figure}

The additional dimension of time allows for several new ways of presenting data to the CNN classifier. 
First, the conventional range-FFT and range gating are applied to the region of interest in which the hand and torso are both located. 
After the range-FFT, the network can be trained on the range-time data or range-Doppler data, using the Doppler-FFT detailed in Section \ref{sec:range_doppler}. 
Alternatively, as is commonly employed in speech processing, the short-time Fourier Transform (STFT) can be applied along the time dimension to yield a velocity versus time mapping of the data \cite{griffin1984stft}.
In this sense, the CNN will observe the velocity as it changes across the 2.56 s of the capture, as shown in Fig. \ref{fig:dynamic_stft}. 
However, the Doppler-STFT increases the dimensionality of the problem necessitating greater computational power. 

\begin{figure}[h]
    \centering
    \includegraphics[width=\textwidth]{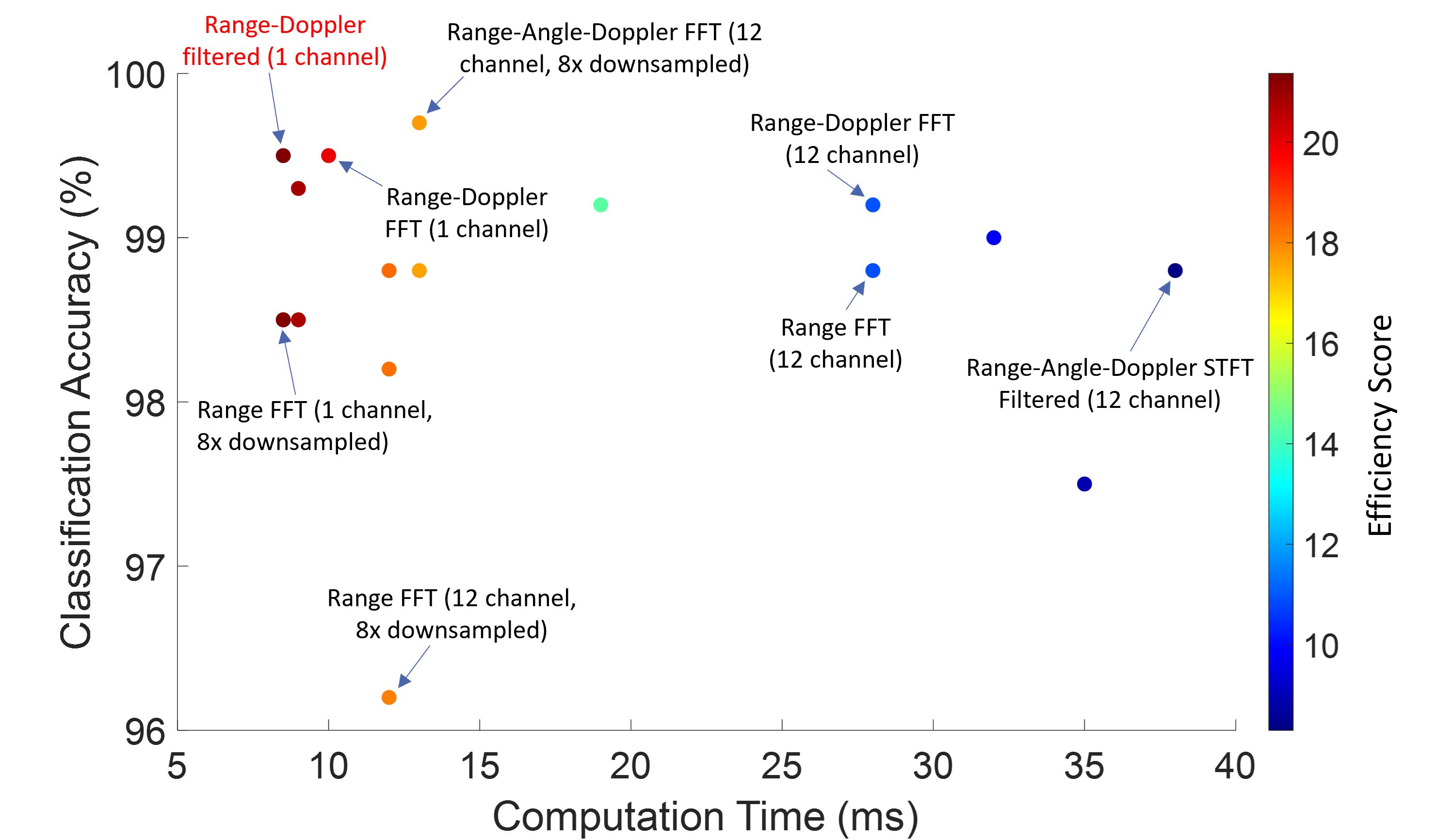}
    \caption{Doppler STFT for pull and push gestures demonstrating where the gesture is performed during the 2.56 s capture.}
    \label{fig:dynamic_comparison}
\end{figure}

As expected, given the considerable differences among the gestures over time, the classifier outperforms the static gesture case, in terms of classification accuracy. 
All combinations of the following preprocessing techniques are compared to evaluate the performance of the CNN: range-FFT, angle-FFT, Doppler-FFT/Doppler-STFT, x8 downsampled in time, 12 channels, and only 1 channel. 
The x8 downsampling operation is employed to compare the relative classification and computational performance if the gesture is sampled less frequently across time.
Utilizing only 1 channel rather than the full 12 channels reduces the dimensionality of the classifier and hence the computational load. 
Since the most notable variation between classes is in the range-time or range-Doppler domains, employing only a single channel may still capture enough information for robust classification. 
A comparison of the classification accuracy and required computation time is provided in Fig. \ref{fig:dynamic_comparison}, where the ``Efficiency Score,'' $\eta$, is computed by 
\begin{equation}
    \label{eq:efficiency_score}
    \eta = 20 \log_{10} \frac{\alpha}{T},
\end{equation}
where $\alpha$ is the classification accuracy and $T$ is the computation time.

Based on this analysis, the most efficient classifier is the range-Doppler, with a filter in the Doppler domain to only velocities near zero, using only 1 channel. 
From this result, we can infer that the most meaningful parameters for the neural network to learn are along the range-Doppler domains and the spatial/channel domain offers little insight into classifying dynamic gestures. 
Additionally, although the same information is present in the range-FFT and range-Doppler-FFT signals, the CNN learns noticeably different features that have a significant impact on algorithm performance. 

\section{Improved Static Gesture Classification using Novel Sterile Training Technique}
\label{sec:sterile}
Upon closer inspection of the mechanics of the problem, previous results on static and dynamic gesture classification become more apparent. 
For gesture recognition, a human hand can be mathematically modeled as a distributed target consisting of a continuously varying reflectivity across space. 
Understanding how radar captures such target scenes provides insight into the difficulty of hand gesture recognition using mmWave signaling.

\begin{figure}[h]
\centering
    \begin{subfigure}[b]{0.5\textwidth}
         \centering
         \includegraphics[width=\textwidth]{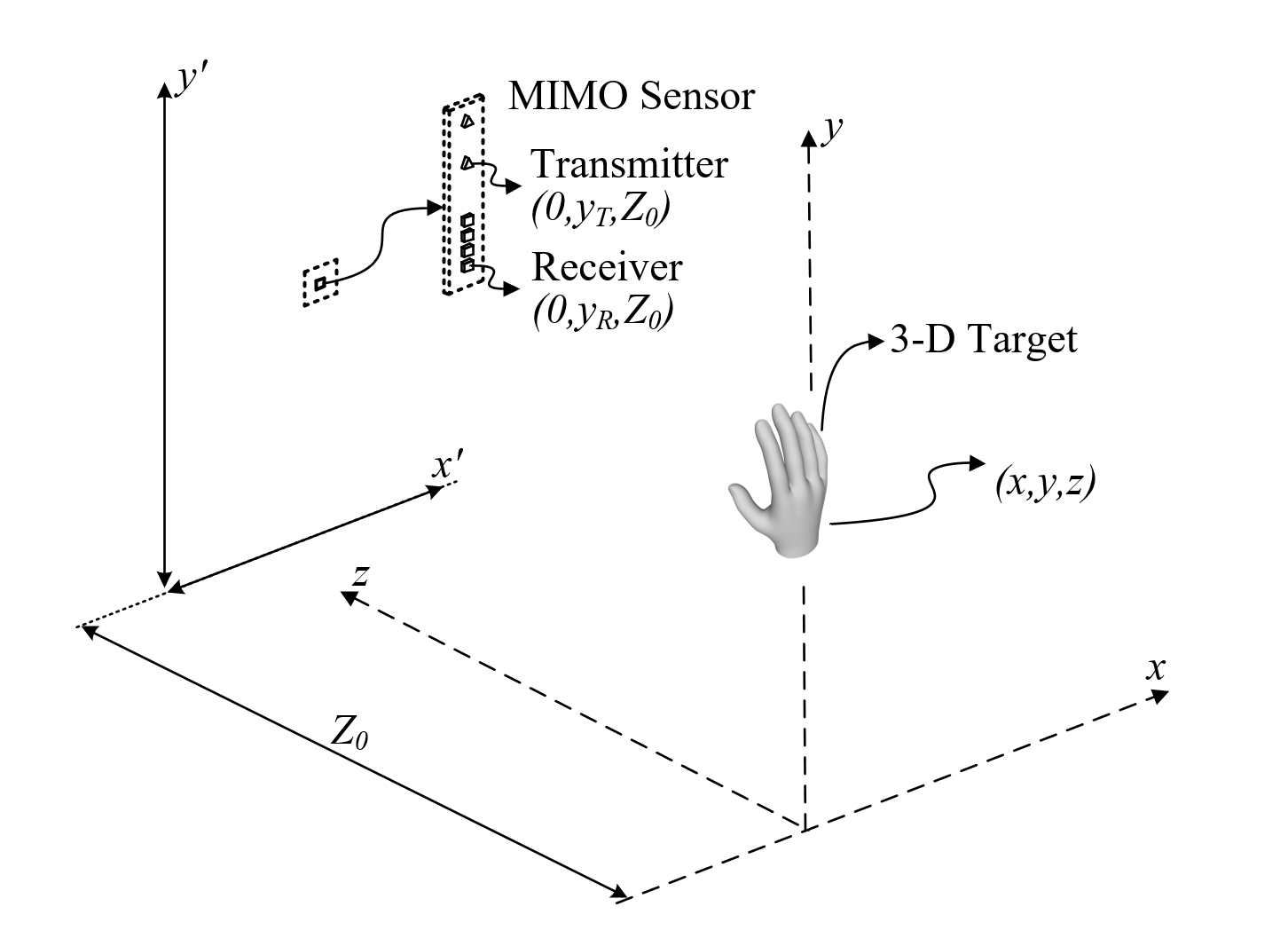}
         \caption{}
         \label{fig:hand_scenario}
    \end{subfigure}
    \begin{subfigure}[b]{0.225\textwidth}
         \centering
         \includegraphics[width=\textwidth]{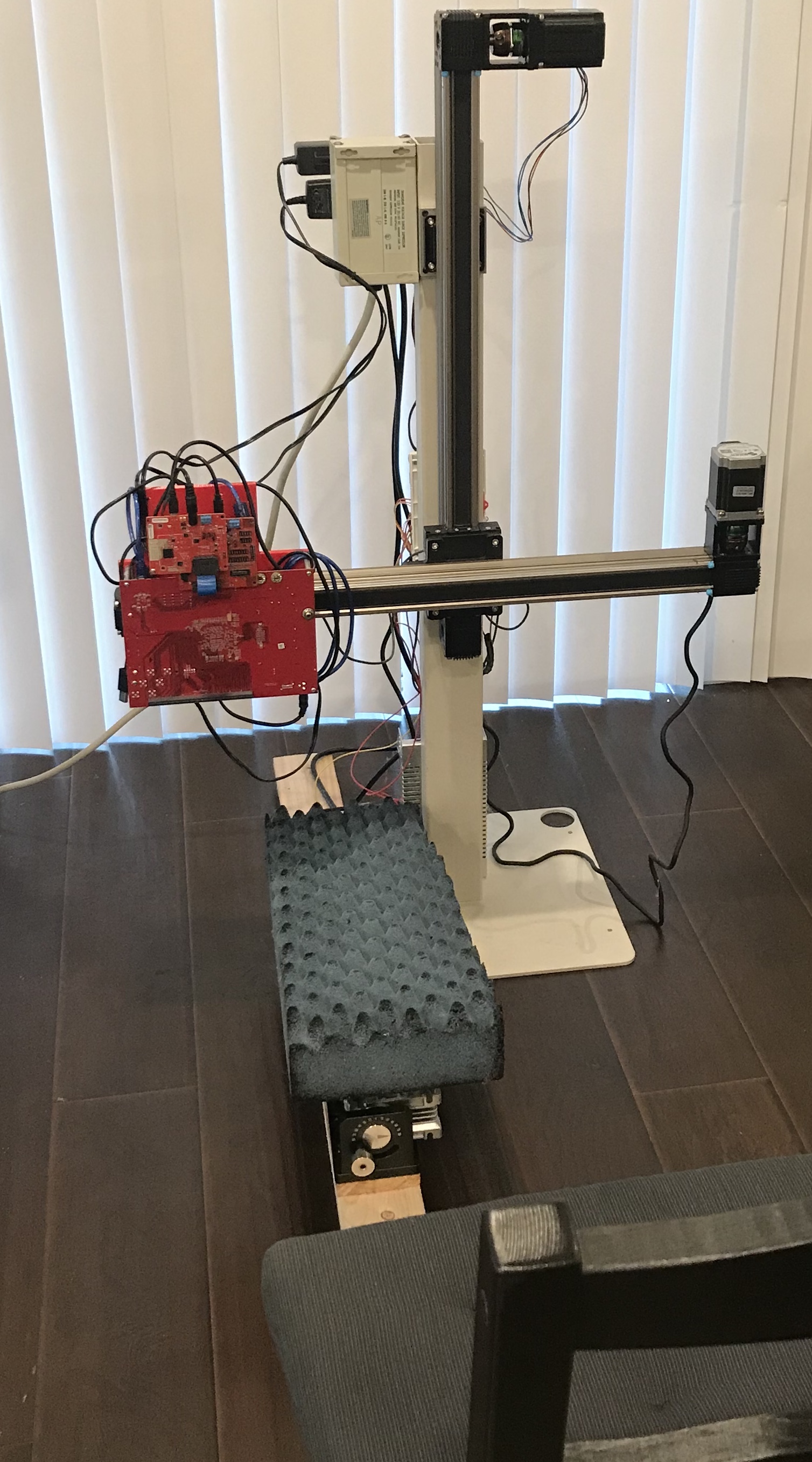}
         \caption{}
         \label{fig:xy_scanner}
    \end{subfigure}
\caption{(a) A MIMO radar sensor with transmitter and receiver antenna elements located at ($0$,$y_T$,$Z_0$) and ($0$,$y_R$,$Z_0$), respectively captures the return signal from a three-dimensional (3-D) target whose reflectivity function is $p(x,y,z)$. (b) Two-dimensional $x$-$y$ rectangular scanner system with chair for test subject to sit.}
\label{fig:hand_imaging}
\end{figure}

Assuming a simple linear MIMO array along the $y$-axis, such as the depiction in Fig. \ref{fig:hand_scenario}, and applying the multistatic to monostatic conversion in (\ref{eq:mult-to-mono}), the return signal from a distributed target can be modeled as the superposition of the echo signals from each of the target coordinates scaled by the target's reflectivity function $\sigma(x,y,z)$. 
The beat signal from each virtual monostatic transceiver at the positions $y'$ can be expressed as
\begin{equation}
\label{eq:distrubted_target}
    s(y',k) = \iiint \frac{p(x,y,z)}{R^2}e^{-j2kR}dxdydz.
\end{equation}
where $R$ is the radial distance from each virtual monostatic element located at the positions $y'$ to each point in the distributed target domain as 
\begin{equation}
    R = \left[{x^2 + (y-y')^2 + (z-Z_0)^2}\right]^\frac{1}{2}.
\end{equation}

If samples are taken throughout the $x'$-$y'$ plane, the reflectivity function can be reconstructed by inverting (\ref{eq:distrubted_target}); however, for applications such as hand gesture recognition, the transceiver elements span only a small space along the $y'$-axis. 
This model provides insight into the simultaneous plausibility and difficulty of the static gesture recognition problem using FMCW radar. 

Embedded in the beat signal are high-resolution spatial features describing the shape of the target or static gesture being performed, meaning that different hand poses or static gestures have distinct echo signals unique to that gesture. 
However, the target scene or hand cannot be analytically reconstructed as a three-dimensional (3-D) image and can be used to easily classify gestures using traditional optical image approaches. 
Thus, classifying static hand gestures involves attempting to learn a high-dimensional pattern (hand pose in three dimensions) from low-dimensional radar data. 

Another issue inherent to the hand gesture problem is the small radar cross-section (RCS) of the human hand, which results in a low signal-to-noise ratio (SNR). 
Even with a large amount of data, because the RCS of the hand is low, the features unique to each gesture class are not pronounced. 
As a result, the CNN has difficulty discerning meaningful features for static gestures. 

To overcome these deficiencies, we propose a novel data collection strategy and training technique that employs ``sterile'' data during network training to improve classification accuracy. 
First, we employ a \mbox{2-D} $x$-$y$ SAR scanner, as shown in Fig. \ref{fig:xy_scanner}, to capture data from numerous perspectives, both vertically and horizontally. 
In this manner, while the user remains stationary, many different views of the hand are captured quickly.

As mentioned previously, the RCS of the human hand is problematically small in comparison to noise and propagation effects. 
Comparing the range profiles of the different gestures, the differences are mostly indistinguishable to the human eye, as shown in Fig. \ref{fig:range_ffts}.
Even though a peak exists in the range FFT at a distance corresponding to the human hand, the features of the gesture reflected back to the radar are not sharply defined and are centered at different places on the human hand. 

\begin{figure}
     \centering
     \begin{subfigure}[b]{0.45\textwidth}
         \centering
         \includegraphics[width=\textwidth]{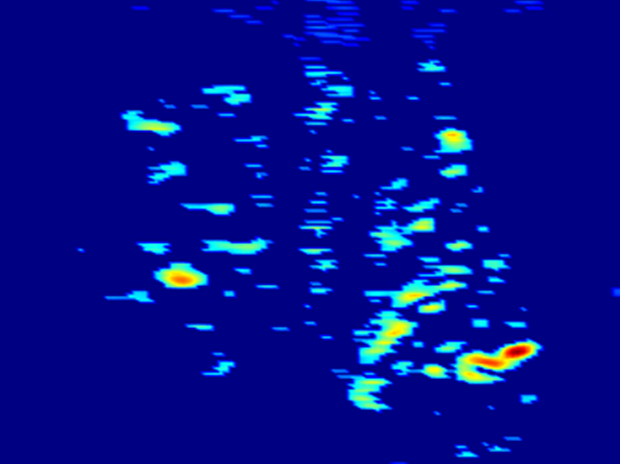}
         \caption{}
         \label{fig:sar_real_hand}
     \end{subfigure}
     \begin{subfigure}[b]{0.45\textwidth}
         \centering
         \includegraphics[width=\textwidth]{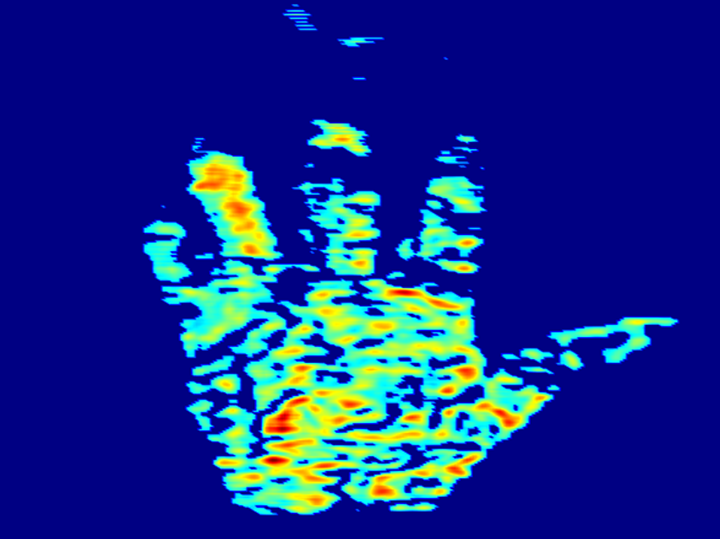}
         \caption{}
         \label{fig:sar_foil_hand}
     \end{subfigure}
        \caption{Comparison of the reconstructed SAR images from the (a) real human hand and the (b) aluminum cutout of the human hand demonstrating the low RCS of the human hand.}
        \label{fig:sar_images}
\end{figure}

To demonstrate this phenomenon, a SAR approach is temporarily adopted to reconstruct an image of the human hand using the methods described in \cite{sheen2010near,yanik2020development}. 
It is important to note that the images shown in Fig. \ref{fig:sar_images} are not the data used to train and validate the CNN. 
These images require all the data (thousands of samples) from the entire horizontal and vertical scan, which takes approximately $5$ min to complete. 

The reconstructed image of the human hand (Fig. \ref{fig:sar_real_hand}) shows a poor image of the hand owing to low RCS and SNR.
Comparatively, a SAR image is also reconstructed using an aluminum cutout in the shape of the hand to demonstrate an ideal hand target, as shown in Fig. \ref{fig:sar_foil_hand}. 

This empirical analysis reveals the innate difficulty in classifying hand gestures from radar beat signals. 
Even when employing thousands of radar return signals to construct the SAR image, the hand is barely visible and the gesture is difficult to recognize. 
From these images, we can infer that the features from a human hand contained in a single beat signal reflected are not pronounced and have a relatively low magnitude compared to the surroundings, noise, etc. 
In contrast, as shown in Fig. \ref{fig:sar_foil_hand}, the aluminum cutout demonstrates a high SNR, implying that the features of the gesture are much more prominent and consistent for each static gesture. 
The novel technique proposed in this section consists of capturing data from many perspectives using a \mbox{2-D} mechanical scanner from both ``real'' human hands and ``sterile'' aluminum cutouts, to improve classification accuracy.

To validate our technique, we collected data from eight participants for three gesture classes: ``palm'' (Fig. \ref{fig:palm}), ``perm'' (Fig. \ref{fig:perp}), and ``thumbs up'' (Fig. \ref{fig:tu}). 
Similarly, mmWave radar data were collected from the aluminum cutout for each gesture class using a SAR scanner. 
To compare against a control, we first train two networks using only real human hand data with range and range-angle preprocessing, respectively. 
For these networks, we use $8000$ set aside captures as the validation dataset, making the split between training and validation $80\%$ to $20\%$. 
The networks used to classify hand gestures vary based on the preprocessing applied to the dataset. 
For the range dataset, convolutional layers with kernel sizes of $13 \times 2$, each with $16$ filter,s are each followed by a Rectified Linear Unit \cite{glorot2011ReLU}. 
\begin{figure}[h]
    \centering
    \includegraphics[width=0.65\textwidth]{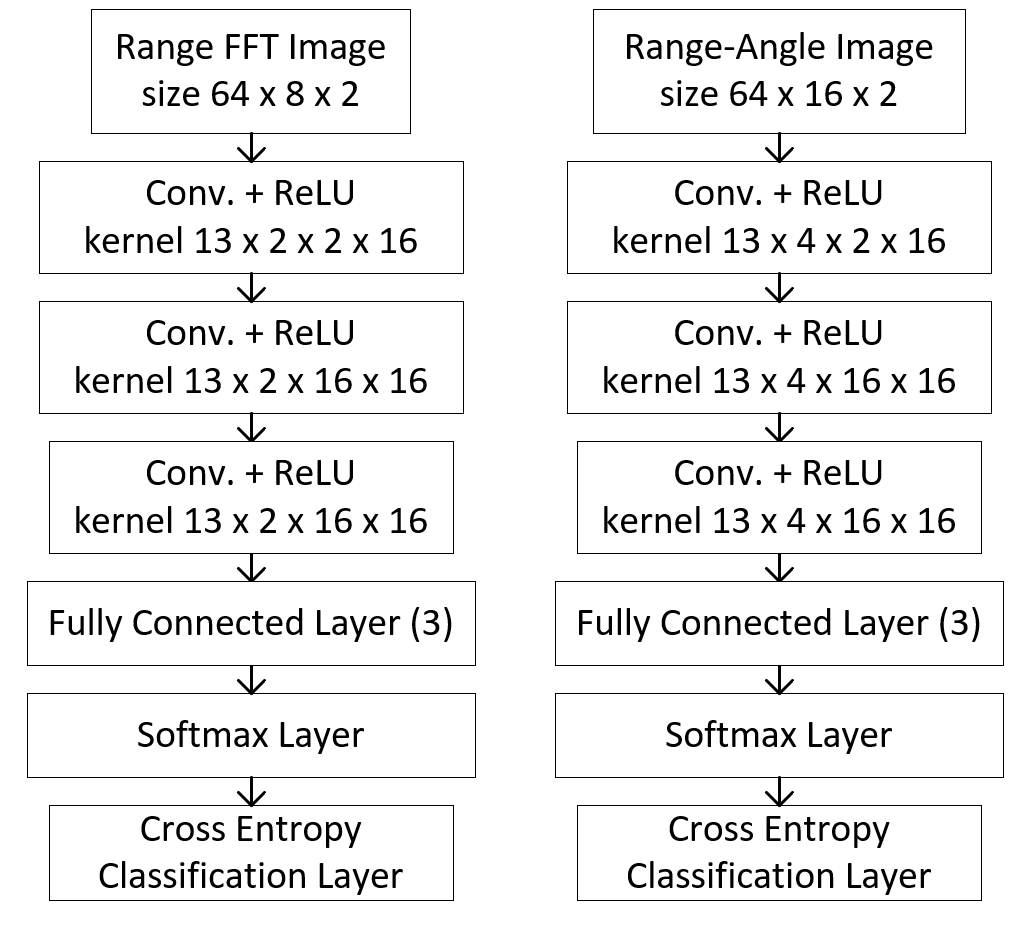}
    \caption{The network architecture for the Range FFT DNN and Range-Angle FFT DNN.}
    \label{fig:cnn_architectures}
\end{figure}
These are connected in series, followed by a fully connected layer with three output neurons, softmax layer, and final classification layer using the cross-entropy loss function. 
The range-angle dataset employs a network with the same architecture, changing only the size of the convolutional layers to $13 \times 4$ to account for larger image sizes. 
The key to both networks is the complex-valued layering and network architectures. 
Considering the real and imaginary parts of the radar range data as distinct layers of the image allows the network to identify pixel-to-pixel and layer-to-layer relationships, which correspond to the phase information of each complex-valued pixel. 
However, other complex-valued neural network architectures have been explored in the literature \cite{gao2018enhanced,wang2021tpssiNet,jing2022enhanced} and are investigated in Chapter \ref{ch:dual_radar}. 
The architectures of both networks are chosen after close inspection of the feature sizes in the observation image domain in both range and channel/angle, in addition to extensive testing to optimize the real-time implementation efficiency and classification rate. 
Both network architectures are shown in Fig. \ref{fig:cnn_architectures}. 

After training each network with only real human hand data, the range CNN and range-angle CNN yield classification rates of $84.9\%$ and $90.2\%$, respectively. 
These networks are named ``Human Only'' in Table \ref{table:sterile_results} since they are trained with only the range and range-angle profiles from human hands. 
Next, two new networks with identical architectures are trained using the complete datasets, consisting of real human hand data supplemented by ``sterile'' data from aluminum cutouts. 
These networks are dubbed ``Combined'' since they are trained with both real and ``sterile'' images. 
It is important to note that the ``Combined'' networks are validated with the same validation data as the ``Human Only;'' the only difference being the training dataset used for each network.
These results corroborate our hypotheses on training with ``sterile'' data, as the classification rates improve to $93.1\%$ and $95.4\%$ for the range and range-angle datasets, respectively. 

\begin{table}[h]
    \centering
    \caption{Comparison of classification rate between networks trained with only human hand data (Human Only) and networks trained using sterile data to supplement the real human hand data (Combined).}
    \large
    \begin{tabular}{c||c|c}
         & Human Only & Combined \\
         \hline \hline
         Range & 84.9\% & 93.1\% \\
         \hline
         Range-Angle & 90.2\% & 95.4\% \\
         \hline \hline
    \end{tabular}
    \label{table:sterile_results}
\end{table}

Compared to prior work in the literature, our proposed method improves upon gesture recognition by using sterile data while offering a solution to the difficult classification problem of static gestures under three-dimensional spatial translation. 
Kim \textit{et al.} \cite{kim2017staticgesture} employ a time-domain gesture recognition approach on an ultra-wideband (UWB) impulse-radio (IR) radar. 
The approach in \cite{kim2017staticgesture} considers two scenarios separately: (1) six gestures using human hands $15$ cm away from the transceiver and (2) three plaster model gestures rotated at $10^\circ$ increments. 
For scenario (1), the hand is kept at a constant position for all captures. 
Both training and testing are performed using human hand data resulting in a classification rate of $91\%$ using a CNN classifier. 
In scenario (2), plaster models of each gesture are captured from different perspectives by rotating the plaster model. 
For this scenario, Kim \textit{et al.} record classification accuracies of more than $90\%$ for three gestures and validated the models using data from the plaster model. 
Comparatively, our method yields a more robust classifier by including both real human hand reflections and ``sterile'' reflections in the training processes and validating them with only human hand data. 
Rather than creating two distinct classifiers for human and sterile data separately, as discussed in \cite{kim2017staticgesture}, the technique proposed in this section unites human and sterile data to construct a robust classifier. 
Furthermore, our approach investigates more diverse scenarios by capturing data from multiple test subjects at many locations relative to the hand position. 

Extensive work has been carried out towards dynamic gesture recognition using mmWave radar, Doppler radar, and IR-UWB sensors \cite{kim2016hand,kim2017application,park2016_ir_uwb_gesture,zhang2016dynamic,leem2020detecting,suh2018_24GHz_gesture,dekker2017gesture}; however, this is an entirely separate problem from the problem addressed in this section as the dynamic gesture case considers only motion.
This reduces the dimensionality of the classification to temporal motion features, whereas static gesture recognition on mmWave radar involves classification of a three-dimensional structure using lower-dimensional data, as discussed in Section \ref{subsec:dynamic_gestures} previously. 
Thus, our model is trained for the more difficult problem of static gesture classification under spatial translation and demonstrates superior classification accuracy compared to prior static gesture classification studies \cite{kim2017staticgesture}.

These studies on static and dynamic gesture classification approach hybrid-learning from the perspective of improving deep learning classification techniques by leveraging expertise in signal processing. 
Similarly, we extend our analysis by examining a similar preprocessing problem in high-resolution imaging. 

\section{Efficient \mbox{3-D} Near-Field MIMO-SAR Imaging for Irregular Scanning Geometries}
\label{sec:ffh}

With the emergence of fifth-generation (5G) and sixth-generation (6G) technologies, UWB mmWave transceivers are enabling unprecedented sensing and communications feats \cite{alvarez2021towards,li2021Integrated,basrawi2021reverse}.
Small form-factor multiple-input-multiple-output (MIMO) radars are becoming increasingly popular owing to their low cost and power consumption \cite{yanik2019near,smith2021An}. 
In addition to emerging 5G communications, mmWave radar has already been realized for high-resolution sensing on the Google Pixel 4 \cite{basrawi2021reverse}.
Of particular interest, recent studies have enabled freehand mmWave imaging by employing positioning sensors commonly employed in smartphones and virtual reality (VR) sensor suites \cite{alvarez2021towards,alvarez2019freehand,alvarez2021system,alvarez2021freehand,alvarez2021freehandsystem}.
Sub-wavelength localization accuracy was previously unachievable by conventional techniques such as 5G mmWave \cite{wymeersch2017mmWavePositioning} or Bluetooth low-energy (BLE) ranging \cite{hajiakhondi2020bluetooth}.
Freehand mmWave imaging is a high-resolution imaging technique that relies on conventional synthetic aperture radar (SAR) principles \cite{lopez20003,yanik2020development,smith2020nearfieldisar,yanik2019sparse,yanik2019cascaded} and precise tracking of a handheld radar device as it is moved by a human user throughout space \cite{alvarez2021towards,baumgartner2017sononet,blackall2005alignment,gilbertson2015force}.
Whereas traditional mmWave SAR imaging requires precise motion systems to achieve near-ideal synthetic arrays \cite{yanik2020development}, the scanning geometry employed by freehand imaging systems is generally irregular and does not conform to the typical array geometries required for efficient image reconstruction algorithms \cite{smith2023ThzToolbox}.

While a recent investigation proposes a fast imaging algorithm for irregular SAR geometries using array linearization \cite{zeng2021aperturelinearization}, the proposed technique adopts a simplistic model of the array displacement and does not explore near-field multistatic effects, both of which are addressed in this study. 
However, efficient algorithms for near-field MIMO-SAR operation under irregular scanning geometries have not been explored in the literature. 

Extensive research on freehand mmWave imaging has been conducted by Laviada \textit{et al.} at the University of Oviedo \cite{alvarez2021towards,garcia20203DSARProcessing,wu2020multilayered,alvarez2019freehand,alvarez2021system,alvarez2021freehand,alvarez2021freehandsystem}.
High-precision localization systems that enable freehand SAR imaging have been investigated using an infrared camera network to accurately track device location over time and recover EM images \cite{alvarez2019freehand}.
Their work was extended to employ an inertial measurement unit (IMU) and depth camera sensors to achieve standalone freehand imaging with promising results \cite{alvarez2021towards,alvarez2021system}.
In each of these efforts, the subject attempted to move the hand in a raster pattern to synthesize an approximately rectangular planar aperture using a linear frequency-modulated (LFM) handheld radar \cite{alvarez2021towards,alvarez2019freehand,alvarez2021freehand}. 
Owing to the subject's inability to move their hand in an ideal planar trajectory and the sensitivity of the mmWave signal to sub-millimeter perturbations, the image was reconstructed using the generalized back-projection algorithm (BPA). 

Similar irregular and non-cooperative scanning geometries have been observed in unmanned aerial vehicle (UAV) SAR imaging \cite{garcia20203DSARProcessing}, nonuniform NDT \cite{wu2020multilayered}, and automotive SAR imaging \cite{kan2020automotiveSAR}. 
However, for many edge and mobile applications, limitations on power consumption and computational complexity cannot be overcome using existing approaches for irregularly sampled SAR.
Although image reconstruction algorithms have been thoroughly investigated in the literature for cooperative synthetic array geometries \cite{yanik2019near,sheen2001three,lopez20003,yanik2020development,smith2020nearfieldisar,yanik2019sparse,smith2023ThzToolbox,yanik2019cascaded,fan2020linearMIMOArbitraryTopologies}, widely applicable efficient near-field imaging algorithms for applications such as freehand smartphone imaging, UAV imaging, and automotive SAR imaging have not been thoroughly addressed in the existing literature. 
Furthermore, while MIMO arrays, commonly employed in commercially available radar devices, offer spatially efficient small array sizes, the MIMO-SAR operation introduces a handful of complications in the image reconstruction process and proper handling of the multistatic array is necessary to avoid imaging artifacts \cite{yanik2019sparse}.
While progress has been made towards projecting MIMO-SAR radar data to virtual single-input-single-output (SISO) monostatic data \cite{yanik2019sparse,smith2023ThzToolbox}, the analysis is performed on a coplanar assumption that does not generally hold for irregular scanning geometries.

In this section, we propose a novel image reconstruction technique for efficient near-field imaging with irregular scanning geometries, such as those present in freehand imaging, UAV SAR, and automotive scenarios.
We examine the system and signal models for UWB MIMO-SAR and develop a multi-planar multistatic approach to mathematically decompose the irregularly sampled synthetic array such that an equivalent virtual planar monostatic array can be constructed. 
This technique is the first to extend the range migration algorithm (RMA) such that non-cooperative SAR scanning and multistatic effects are simultaneously mitigated. 
The analysis in subsequent sections provides a novel framework for decomposing irregular SAR scenarios and efficiently projecting irregular MIMO-SAR samples to a virtual planar monostatic equivalent. 
The proposed algorithm is validated through simulations and empirical experiments, demonstrating robustness to arbitrary scanning patterns and low computational complexity.
A thorough study of the relationship between the array irregularity and image resolution of the proposed algorithm is provided. 
The proposed technique demonstrates high-fidelity focusing comparable to the traditional planar RMA, even under array perturbation on the order of 10s of wavelengths. 
Our solution enables the development of emerging technologies that require non-ideal SAR scanning geometries, MIMO multistatic radar, and efficient image reconstruction.

\begin{figure}[h]
    \centering
    \includegraphics[width=0.6\textwidth]{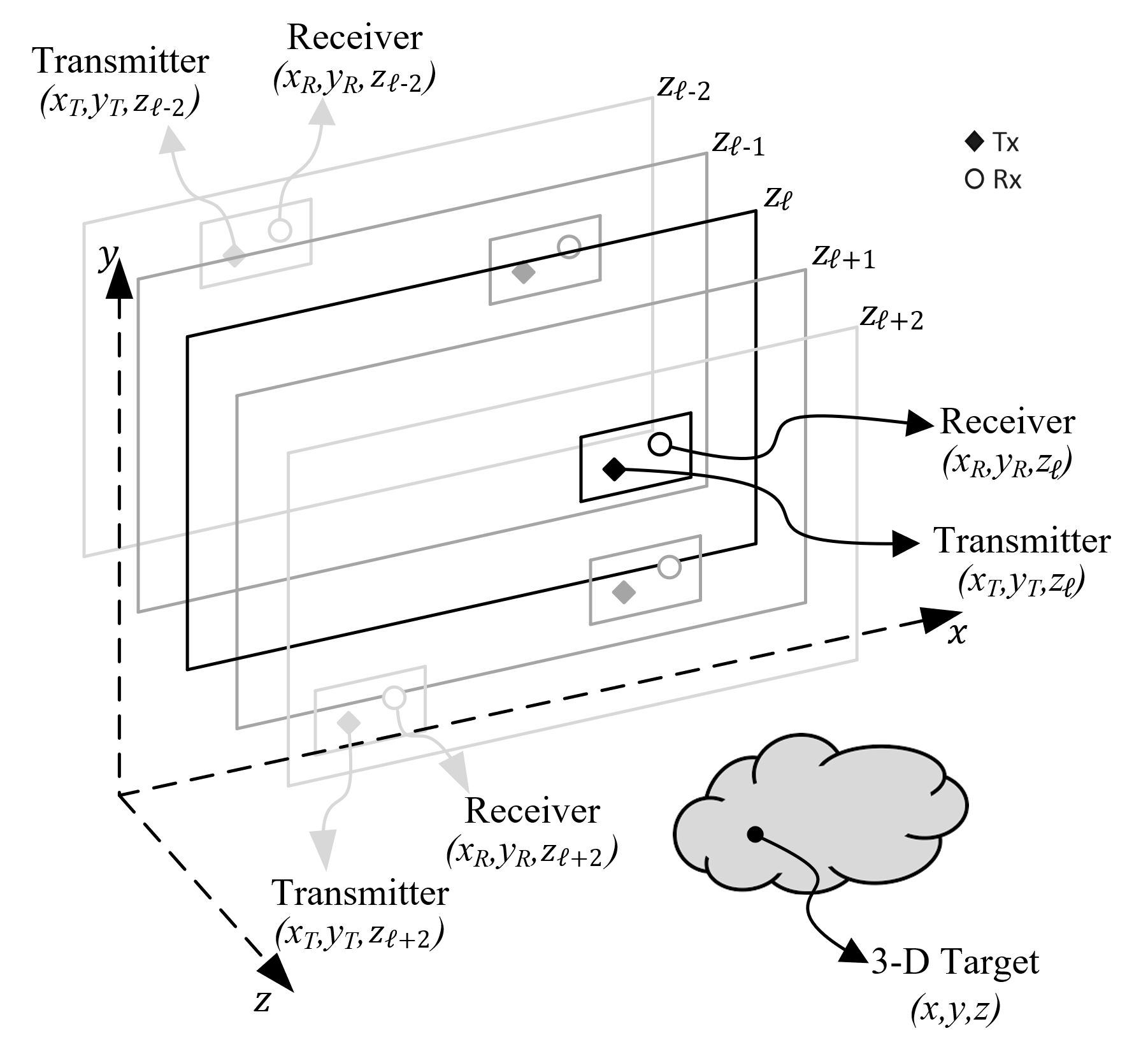}
    \caption{Geometry of the multi-planar SAR irregular scanning geometry with a multistatic array.}
    \label{fig:multiplanar}
\end{figure}

The remainder of this section is organized as follows.
Section \ref{sec:system_model} introduces the system model, including the multi-planar multistatic SAR concept, signal model, and a novel compensation technique for planar monostatic SAR. 
In Section \ref{sec:image_reconstruction}, efficient imaging methods and implementation details are discussed and the Efficient Multi-Planar Multistatic (EMPM) algorithm is proposed.
Section \ref{sec:prototype} details the hardware and software implementation for collecting multi-planar multistatic SAR data.
The results of the simulation and empirical studies are presented and discussed in Section \ref{sec:results}.

\subsection{Near-Field Irregular SAR System Model}
\label{sec:system_model}
In this section, we propose the characterization of irregular or arbitrary three-dimensional (\mbox{3-D}) MIMO-SAR sampling geometry using the multi-planar multistatic scenario shown in Fig. \ref{fig:multiplanar}, where data are collected along different $z$-planes by a MIMO multistatic radar with respect to a stationary \mbox{3-D} target.

\subsubsection{Multi-Planar MIMO-SAR Configuration}
\label{subsec:multiplanar_SAR}
For many emerging SAR applications, as the radar is moved throughout \mbox{3-D} space, it is generally oriented in the same direction towards some target; however, the samples are taken across several $z$-planes. 
Because the data are collected during an arbitrary SAR scanning path, the resulting synthetic aperture does not conform to standard scanning regimes, such as rectilinear/planar \cite{sheen2016three,yanik2020development}, circular \cite{wu2020multilayered,gao2016efficient}, or cylindrical \cite{smith2020nearfieldisar,amineh2019real,smith2023ThzToolbox}.
Hence, the image reconstruction process must consider the irregularity of the spatial sampling, the geometry of which is detailed in Fig. \ref{fig:multiplanar}.

Compared with planar MIMO-SAR, which requires a multistatic MIMO array to be scanned across a planar track \cite{yanik2020development,yanik2019sparse,fan2020linearMIMOArbitraryTopologies}, multi-planar MIMO-SAR allows the multistatic array to be scanned across a \mbox{3-D} space. 
For freehand imaging or automotive SAR, a MIMO array is fixed to a smartphone or vehicle, respectively, and is moved throughout space, generating a multi-planar MIMO-SAR irregular aperture. 
As shown in Fig. \ref{fig:multiplanar}, because the multistatic array is scanned in an irregular pattern spanning multiple $z$-planes, the locations of the transmit (Tx) and receive (Rx) elements are spatially translated by the movement of the MIMO array. 
The analyses in the subsequent sections present an efficient solution for irregular MIMO-SAR imaging, such that the position of the radar is known throughout the scan and the planar array assumption does not hold. 
This scenario is common to many of the aforementioned applications and necessitates both irregular scanning geometries and efficient image recovery.

\subsubsection{The \mbox{3-D} Multi-Planar Virtual Array Response in Near-Field Imaging}
\label{subsec:virtual_array}
\begin{figure}[h]
    \centering
    \includegraphics[width=0.6\textwidth]{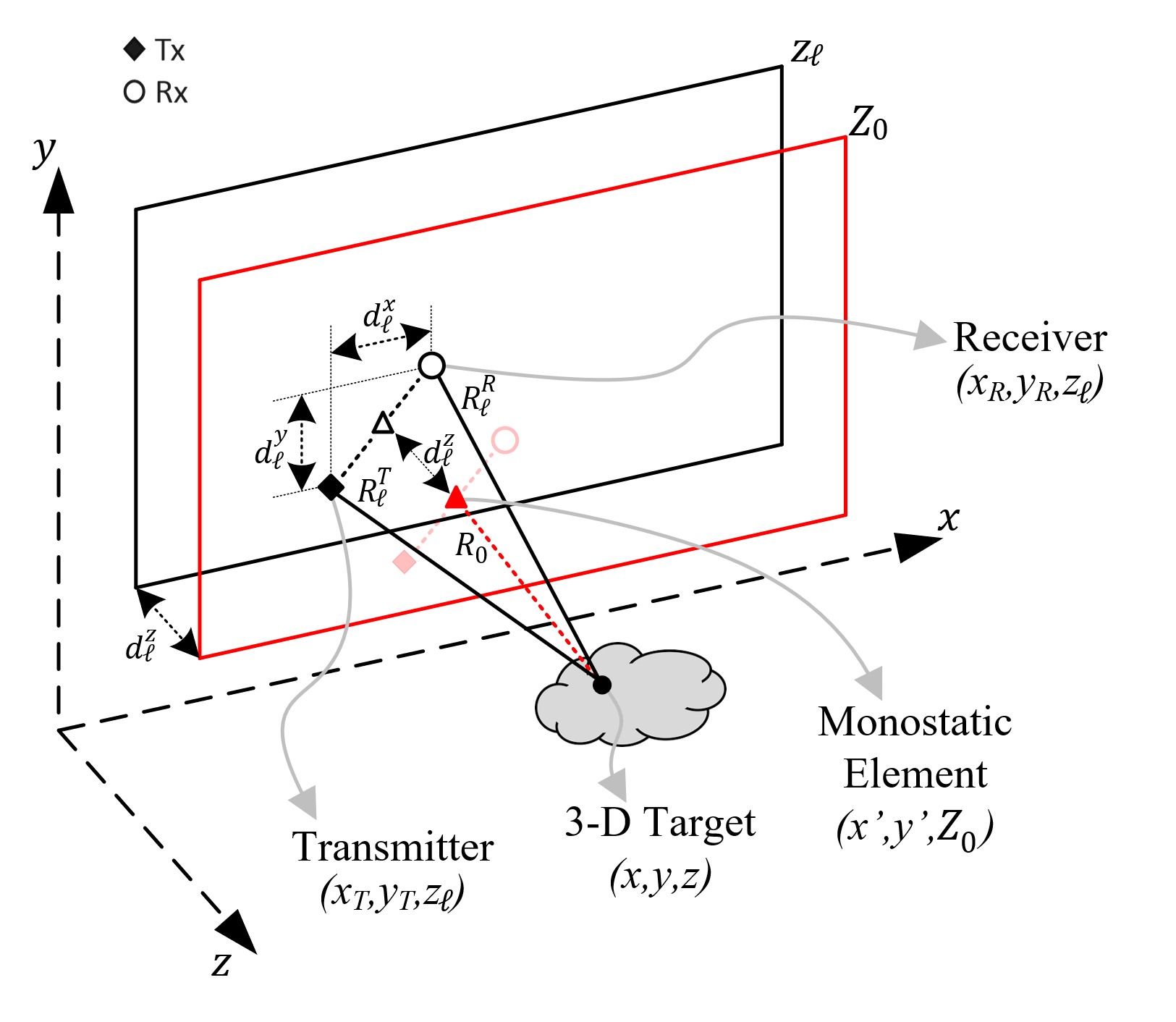}
    \caption{Relationship between the multi-planar multistatic elements and virtual planar monostatic elements.}
    \label{fig:virtual_array_compensation}
\end{figure}

By the analysis of \cite{yanik2019sparse,smith2023ThzToolbox,ender2009systemMIMOSAR} for the \mbox{2-D} case, a multistatic MIMO array can be approximated by a monostatic virtual element located at the midpoint of the Tx and Rx elements under the far-field assumption for a small fraction $\epsilon$ as 
\begin{equation}
\label{eq:far_field_assumption}
    \sqrt{(d_\ell^x)^2 + (d_\ell^y)^2} \leq \sqrt{4 \epsilon \lambda R},
\end{equation}
where $d_\ell^x$, $d_\ell^y$ are the distances between the Tx and Rx elements along the $x$- and $y$-directions, respectively, as shown in Fig. \ref{fig:virtual_array_compensation}, $\lambda$ is the wavelength of the carrier frequency, and $R$ is the distance from the midpoint of the antenna elements to a reference point in the scene. 

However, under the multi-planar multistatic framework, it is desirable to approximate each Tx/Rx pair using its virtual element located on a $Z_0$ plane in the near-field.
Thus, multi-planar data can be projected onto a virtual planar array to ease the subsequent image reconstruction process.
As shown in Fig. \ref{fig:virtual_array_compensation}, the $\ell$-th Tx/Rx pair located on the $z_\ell$ plane can be approximated by the element located at the midpoint between the Tx and Rx elements migrated to the $Z_0$ plane.

For near-field SAR, the assumption in (\ref{eq:far_field_assumption}) is invalid and the approximation must be handled more delicately. 
Hence, we derive an efficient compensation algorithm to approximate the multistatic multi-planar array as a monostatic planar array for near-field imaging scenarios.

The transmitter (Tx) and receiver (Rx) of the $\ell$-th multistatic MIMO array are located at $(x_T,y_T,z_\ell)$ and $(x_R,y_R,z_\ell)$, respectively, and the target scene is assumed to be a distributed target whose coordinates are given by $(x,y,z)$. 
In this study, orthogonality is leveraged across time by operating a MIMO radar using the time-division multiplexing (TDM) MIMO technique such that each Tx/Rx pair is activated sequentially. 
The round-trip distance between the $\ell$-th Tx/Rx pair and the point scatter located at $(x,y,z)$ can be written as 
\begin{align}
\label{eq:Rl_of_xT_xR_yT_yR}
    \begin{split}
        R_\ell^{RT}&= R_\ell^T + R_\ell^R, \\
        R_\ell^T &= \left[(x_T - x)^2 + (y_T - y)^2 + (z_\ell - z)^2 \right]^{\frac{1}{2}}, \\
        R_\ell^R &= \left[(x_R - x)^2 + (y_R - y)^2 + (z_\ell - z)^2 \right]^{\frac{1}{2}}.
    \end{split}
\end{align}

Denoting the virtual antenna element locations as $(x',y',Z_0)$, the $x$- and $y$-coordinates of the Tx/Rx pair can be expressed as
\begin{align}
\label{eq:xT_xR_yT_yR_to_virtual}
    \begin{split}
        x_T &= x' - d_\ell^x/2, \quad y_T = y' - d_\ell^y/2, \\
        x_R &= x' + d_\ell^x/2, \quad y_R = y' + d_\ell^y/2.
    \end{split}
\end{align}

Similarly, denoting $d_\ell^z$ as the distance between the $Z_0$ plane and the $z_\ell$ plane, as shown in Fig. \ref{fig:virtual_array_compensation}, the $z$-coordinate of the Tx and Rx elements can be expressed with respect to $Z_0$ as
\begin{equation}
\label{eq:zl_to_virtual}
    z_\ell = Z_0 + d_\ell^z.
\end{equation}

As described in Appendix \ref{app:taylor_series_ffh}, substituting (\ref{eq:xT_xR_yT_yR_to_virtual}) and (\ref{eq:zl_to_virtual}) into (\ref{eq:Rl_of_xT_xR_yT_yR}) and applying the third-order Taylor series expansion of $R_\ell$ for small values of $d_\ell^x$, $d_\ell^y$, and $d_\ell^z$ yields
\begin{align}
\label{Rl_approximation1}
\begin{split}
    R_\ell^{RT} &\approx 2R_0 + \frac{2(Z_0-z)d_\ell^z}{R_0} + \frac{(d_\ell^x)^2 + (d_\ell^y)^2 + 4(d_\ell^z)^2}{4R_0} \\
    &- \frac{\left[(x'-x)d_\ell^x + (y'-y)d_\ell^y\right]^2 + 4(Z_0-z)^2 (d_\ell^z)^2}{4R_0^3},
\end{split}
\end{align}
where $R_0$ is the distance between the virtual monostatic element located at $(x',y',Z_0)$ and the point scatterer at $(x,y,z)$, expressed as
\begin{equation}
\label{eq:R0}
    R_0 = \left[ (x'-x)^2 + (y'-y)^2 + (Z_0-z)^2 \right]^{\frac{1}{2}}.
\end{equation}

Centering the target to the origin of the $(x,y,z)$ coordinate system and considering $(x'-x),(y'-y) \ll Z_0$, we can acquire the improved approximation of the round-trip distance between the $\ell$-th Tx/Rx pair and the point scatterer as
\begin{equation}
\label{eq:Rl_best_approximation}
    R_\ell^{RT}= R_\ell^T + R_\ell^R \approx 2 R_0 + 2 d_\ell^z + \frac{(d_\ell^x)^2 + (d_\ell^y)^2}{4 Z_0}.
\end{equation}

\subsection{Multi-Planar Multistatic Signal Model}
\label{subsec:signal_model}
Consider a multi-planar multistatic array whose Tx and Rx elements are located at $(x_T,y_T,z_\ell)$ and $(x_R,y_R,z_\ell)$, respectively, and a distributed target occupying volume $V$ at locations $(x,y,z)$ in \mbox{3-D} space with a continuous reflectivity function given by $p(x,y,z)$.
The radar beat signal can be written as
\begin{equation}
    \label{eq:received_k}
    s(x_T,x_R,y_T,y_R,z_\ell,k) = \iiint_V \frac{p(x,y,z)}{R_\ell^T R_\ell^R} e^{-jk(R_\ell^T + R_\ell^R)} dx dy dz,
\end{equation}
where $k = 2\pi f/ c$ denotes the instantaneous wavenumber. 
Image recovery requires the inversion of (\ref{eq:received_k}) to produce $p(x,y,z)$.
However, given arbitrary sampling locations, the image cannot be computed efficiently using existing techniques \cite{alvarez2021towards,garcia20203DSARProcessing,wu2020multilayered,alvarez2019freehand,alvarez2021system,alvarez2021freehand,alvarez2021freehandsystem}.
The frequency-domain model of the received signal (\ref{eq:received_k}) is valid for any UWB radar signaling scheme, including frequency-modulated continuous-wave (FMCW), phase-modulated continuous wave (PMCW), and orthogonal frequency-division multiplexing (OFDM), which is commonly employed in 5G and IoT applications \cite{roos2019radar}. 
Furthermore, prior research on freehand imaging and similar IoT applications has employed a purely stepped-frequency FMCW signal model \cite{alvarez2019freehand,alvarez2021system,alvarez2021freehand,alvarez2021freehandsystem}. 
Similarly, Google Pixel 4 utilizes a Google Soli 60 GHz mmWave FMCW radar for sensing \cite{basrawi2021reverse}. 

However, the derivation of (\ref{eq:Rl_best_approximation}) enables efficient compensation of multistatic multi-planar data by careful handling of the phase.
To achieve the proposed compensation, we express the frequency response of the virtual planar monostatic array, whose elements are located at $(x',y',Z_0)$, as
\begin{equation}
    \label{eq:virtual_planar}
    \hat{s}(x',y',k) = \iiint_V \frac{p(x,y,z)}{R_0^2} e^{-j2kR_0} dx dy dz,
\end{equation}
where $R_0$ is given by (\ref{eq:R0}), $x'$ and $y'$ are the midpoints between each Tx/Rx pair, and $Z_0$ is the plane on which the samples are projected. 
From the analysis in Section \ref{subsec:virtual_array}, the relationship between the multi-planar multistatic response and the virtual monostatic array response is given by
\begin{equation}
    \label{eq:multiplanar_compensation}
    \hat{s}(x',y',k) \approx s(x_T,x_R,y_T,y_R,z_\ell,k) e^{j k\beta_\ell},
\end{equation}
where
\begin{equation}
    \beta_\ell = 2 d_\ell^z + \frac{(d_\ell^x)^2 + (d_\ell^y)^2}{4 Z_0},
\end{equation}
is the near-field residual phase term owing to the arbitrary scanning and MIMO effects in the near-field, as derived in (\ref{eq:Rl_best_approximation}). 
Hence, the virtual planar monostatic response can be efficiently acquired from the irregular samples by removing the residual phase to simultaneously account for the multi-planar scanning geometry and near-field multistatic effects. 
The novel phase compensation technique derived in this section efficiently reduces the dimensionality of the MIMO-SAR imaging problem and projects multi-planar samples onto a single plane to enable computationally tractable algorithms for image reconstruction.

\subsection{Efficient Image Reconstruction Algorithms for Near-Field Planar SAR}
\label{sec:image_reconstruction}
In this section, we review traditional planar SAR image reconstruction methods that employ efficient Fourier-based solutions to recover EM images \cite{smith2023ThzToolbox} and propose a novel technique for multi-planar multistatic SAR. 
Existing research on irregularly sampled SAR imaging problems employs the gold-standard back-projection algorithm (BPA) \cite{alvarez2021towards,garcia20203DSARProcessing,wu2020multilayered,alvarez2019freehand,alvarez2021system,alvarez2021freehand,alvarez2021freehandsystem}.
However, this approach is computationally infeasible for most edge and mobile applications.
To overcome this challenge, we employ the approximation in (\ref{eq:multiplanar_compensation}) to project multi-planar data to a planar-sampled scenario to satisfy the requirements for efficient image reconstruction.
The Fourier-based algorithm detailed in the subsequent analysis is known as the range migration algorithm (RMA) or \mbox{$f$-$k$} algorithm, and has been explored in greater detail elsewhere \cite{sheen2001three,gao2018_1D_MIMO,smith2023ThzToolbox,yanik2019cascaded,fan2020linearMIMOArbitraryTopologies}.

The key step to efficiently invert the integral in (\ref{eq:virtual_planar}) is to represent the spherical wave term as a superposition of plane waves using the method of stationary phase (MSP) \cite{yanik2019sparse,smith2023ThzToolbox}, such that
\begin{equation}
\label{eq:MSP_expansion}
    \frac{e^{-j2 k R_0}}{R_0} \approx \iint_A \frac{e^{-j(k_x'(x-x') + k_y'(y-y') + k_z(z-Z_0))}}{k_z} dk_x' dk_y',
\end{equation}
where
\begin{equation}
    k_z^2 = 4k^2 - (k_x')^2 - (k_y')^2,
\end{equation}
and $A$ is the region in $k_x'$-$k_y'$ space occupied by the spherical wavefront.

Following the analysis in \cite{yanik2020development,smith2023ThzToolbox}, substituting (\ref{eq:MSP_expansion}) into (\ref{eq:virtual_planar}) and rearranging the phase terms to leverage the Fourier relationships yields
\begin{equation}
    P(k_x,k_y,k_z) = \hat{S}(k_x',k_y',k) k_z e^{-j k_z Z_0},
\end{equation}
where $P(k_x,k_y,k_z)$ and $\hat{S}(k_x',k_y',k)$ are the spatial spectral representations of the reflectivity function $p(\cdot)$ and array response $\hat{s}(\cdot)$, respectively.
Because the primed and unprimed coordinate systems are coincident, the distinction can be dropped for the remaining analysis.
Hence, the RMA image recovery process can be summarized as
\begin{equation}
\label{eq:RMA_final}
    p(x,y,z) = \text{IFT}_{\text{3D}}^{(k_x,k_y,k_z)}\left[ \mathcal{S} \left[ \text{FT}_{\text{2D}}^{(x',y')} \left[ \hat{s}(x',y',k) \right] k_z e^{-j k_z Z_0} \right] \right],
\end{equation}
where $\text{FT}[\cdot]$ and $\text{IFT}[\cdot]$ are the forward and inverse Fourier transform operators, respectively, $\mathcal{S}$ is the Stolt interpolation operator required to compensate for the spherical wavefront \cite{smith2023ThzToolbox}, and $\hat{s}(\cdot)$ is obtained from (\ref{eq:multiplanar_compensation}).
The spatial resolution along each dimension of the recovered image is given by

\begin{equation}
\label{eq:spatial_resolution}
    \delta_x = \frac{\lambda_c Z_0}{2 D_x}, \quad
    \delta_y = \frac{\lambda_c Z_0}{2 D_y}, \quad
    \delta_z = \frac{c}{2B},
\end{equation}
where $D_x$ and $D_y$ are the sizes of the aperture along the $x$- and $y$-directions, respectively; $B$ is the system bandwidth; and $\lambda_c$ is the wavelength of the center frequency \cite{yanik2019sparse,gao2018_1D_MIMO,smith2023ThzToolbox}.

Although (\ref{eq:RMA_final}) provides an efficient solution for planar array imaging problems, its application to irregular scanning geometries requires a discussion of several key issues.
Applying the compensation technique in (\ref{eq:multiplanar_compensation}) for irregularly sampled data, the multi-planar data can be approximately projected to planar sampling; however, they are likely non-uniform at positions $(x',y',Z_0)$ along the $x$- and $y$-directions.
Traditional efficient implementations rely on the common fast Fourier transform (FFT) algorithm; however, recent work on non-uniform planar MIMO-SAR \cite{gao2018_1D_MIMO,fan2020linearMIMOArbitraryTopologies} and irregular MIMO real aperture radar (MIMO-RAR) \cite{wang20203} imaging has produced solutions using a non-uniform FFT (NUFFT) approach employing fast Gaussian gridding (FGG), as discussed in \cite{greengard2006nufft}, for the Fourier transforms and Stolt interpolation step in (\ref{eq:RMA_final}). 
The sampling criteria for the nonuniform planar case are discussed in detail in \cite{gao2018_1D_MIMO,wang20203,fan2020linearMIMOArbitraryTopologies} and apply correspondingly to irregular scanning scenarios after multi-planar compensation. 
Similarly, the FGG-NUFFT technique is employed in this study to perform the proposed RMA efficiently on irregularly sampled planar data.

For the multi-planar sampling scenario discussed in Section \ref{subsec:multiplanar_SAR}, the RMA cannot be applied directly without multi-planar compensation because the data are sampled on different $z$-planes, as discussed in Section \ref{sec:results}.
If the RMA is applied to the raw multi-planar data, the forward Fourier transform in (\ref{eq:RMA_final}) is invalid because the data along the $x'$ and $y'$-directions are not coplanar and the resulting image will suffer from significant distortion, rendering the resulting images unusable in most cases.

\begin{figure}[h]
    \centering
    \includegraphics[width=0.4\textwidth]{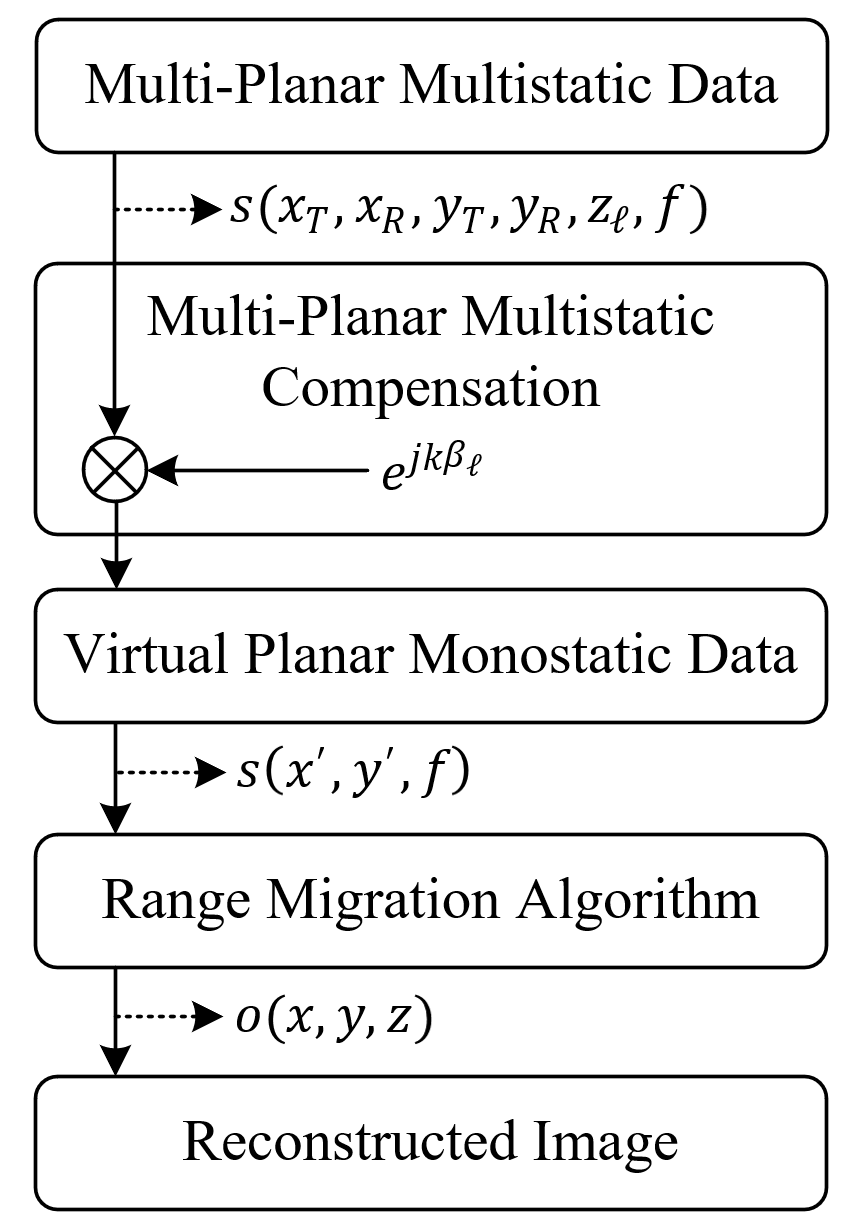}
    \caption{The complete EMPM image reconstruction process from irregular sampling compensation to RMA image recovery.}
    \label{fig:image_reconstruction_process_complete}
\end{figure} 

Sampling considerations for image reconstruction remain identical to those in analyses elsewhere \cite{sheen2001three,yanik2019sparse} after the multi-planar compensation algorithm.
Baseband frequency sampling criteria can be determined using the maximum range for a given application.
As given in \cite{sheen2010near}, the maximum frequency sampling interval is given by $\Delta_f < c/(2R_\text{max})$, where $R_\text{max}$ is the maximum target range. 
Although spatial sampling criteria are not guaranteed for irregular SAR scanning, if the relationship between the capture rate of the radar and the velocity of the radar platform is tuned appropriately during system design, undersampling artifacts are typically minimal \cite{alvarez2019freehand,alvarez2021system,alvarez2021freehand,alvarez2021freehandsystem}. 
To avoid spatial undersampling, the lower bound of the pulse repetition frequency (PRF) can be computed using $\text{PRF} > 4v_\text{max}/\lambda_c$, where $v_\text{max}$ is the maximum velocity for a certain application. 
For example, assuming that the maximum velocity of the human hand for a freehand SAR is 1 m/s and a center frequency of 79 GHz, the lower bound of the PRF is approximately 1.06 kHz. 
It is important to note that the number of captures increases proportionally with the PRF; hence, at high velocities, a large number of samples are captured. 
The computational performance of traditional techniques employing the BPA degrades substantially when many samples are captured. 
On the other hand, the signal-to-noise ratio can be improved by increasing the number of samples, at the cost of increasing the computational burden.
Hence, an efficient algorithm for multi-planar MIMO-SAR imaging is required to enable many such technologies. 

In terms of computational complexity, the EMPM algorithm offers a significant advantage over existing techniques in the literature \cite{alvarez2021towards,garcia20203DSARProcessing,wu2020multilayered,alvarez2019freehand,alvarez2021system,alvarez2021freehand,alvarez2021freehandsystem}, which employ the BPA, whose computational complexity is on the order of $O(N^6)$ \cite{yanik2020development,fan2020linearMIMOArbitraryTopologies}. 
The time complexity of the RMA and its FGG-NUFFT variants has been investigated in the literature \cite{wang20203,fan2020linearMIMOArbitraryTopologies} and the multi-planar compensation step proposed in this section presents negligible computational expense to the RMA, which is on the order of $O(N^3 \log{N})$ \cite{sheen2001three,lopez20003}.
Hence, as discussed in Section \ref{sec:results}, the EMPM algorithm offers comparable imaging performance to the BPA with tractable execution time for mobile platforms, similar to the RMA.

The EMPM reconstruction process for efficient near-field SAR imaging with irregular scanning geometries is illustrated in Fig. \ref{fig:image_reconstruction_process_complete}.
Using the analysis in Section \ref{sec:system_model}, irregular scanning geometries can be modeled as multi-planar sampling scenarios, as shown in Fig. \ref{fig:multiplanar}, and compensated by removing the residual phase due to the multi-planar multistatic conditions. 
The key difference between the traditional RMA and the EMPM is the alignment of the multi-planar multistatic (MIMO-SAR) data to virtual planar monostatic data. 
This crucial step compensates for both the sampling irregularities and multistatic MIMO effects simultaneously, while significantly reducing the dimensionality, from \mbox{6-D} $(x_T,x_R,y_T,y_R,z_\ell,f)$ to \mbox{3-D} $(x',y',f)$, and subsequently the computational complexity. 
Finally, virtual planar monostatic data are used to efficiently recover the image using the RMA.
In simulation and empirical studies on irregular SAR scanning geometries, the EMPM algorithm is applied to efficiently produce high-resolution \mbox{3-D} images previously infeasible due to algorithmic deficiencies.

\subsection{Multi-Planar Multistatic Imaging Hardware Prototype}
\label{sec:prototype}
\begin{figure}[h]
    \centering
    \includegraphics[width=0.8\textwidth]{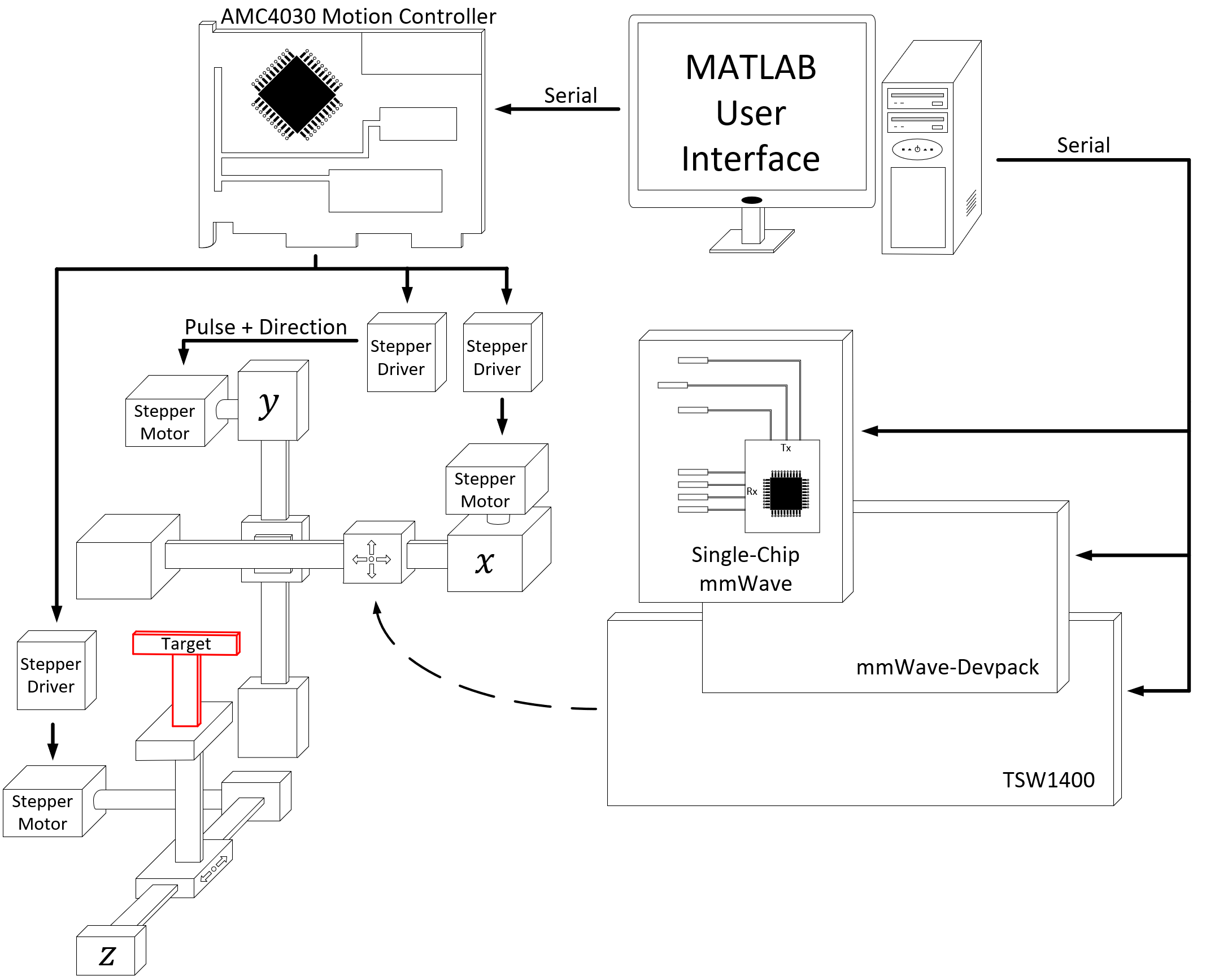}
    \caption{System design for \mbox{3-D} scanner with radar mounted on planar $x$-$y$ rails and target mounted on a linear $z$ rail. The TI radar, data capture card, and mechanical scanner are controlled by MATLAB via USB serial interface.}
    \label{fig:scanner_xyz}
\end{figure}

In this section, we discuss the hardware prototype implementation for empirically validating the proposed imaging algorithm by collecting multi-planar multistatic SAR data.
The hardware architecture of the mmWave imaging system is illustrated in Fig. \ref{fig:scanner_xyz}.

A Texas Instruments (TI) mmWave MIMO radar is mounted on an $x$-$y$ planar scanner. 
The TI AWR1443BOOST radar with a bandwidth of \mbox{4 GHz} from \mbox{77 GHz} to \mbox{81 GHz} is mounted on a TI mmWave-Devpack and TSW1400 data capture card to store the data from the SAR scan and transfer it to the PC, where the image recovery algorithm is implemented in MATLAB.
The TI AWR1443BOOST is equipped with a MIMO array consisting of two Tx elements spaced by $2\lambda_c$ and four Rx elements spaced by $\lambda_c/2$ \cite{yanik2019sparse}.
Although 5G and IoT applications commonly employ an OFDM modulation scheme, the TI radar employed for the following experiments utilizes FMCW signaling. 
However, FMCW radar has been utilized for smartphone applications, notably the Google Pixel 4, which is equipped with a Google Soli FMCW radar \cite{basrawi2021reverse}. 
The proposed range migration-based algorithm is applicable to both OFDM and FMCW radars; hence, the results discussed in the following section are relevant for a wide array of 5G, IoT, smartphone, and automotive applications \cite{zhang2015ofdm}.

Additionally, a linear rail is used to move the target along the $z$-direction to collect multi-planar multistatic data under the geometry discussed in Section \ref{subsec:multiplanar_SAR}.
All three $x$-$y$-$z$ rails are driven by stepper motors controlled by an AMC4030 motion controller, and the scanning process and radar set up are handled in MATLAB.
Additional details on system development and device calibration can be found in \cite{yanik2020development}.
The images are reconstructed using MATLAB implementations running on a desktop PC equipped with a 12-core AMD Ryzen 9 3900X running at 4.6 GHz with 64 GB of memory.
Using this hardware prototype, data can be collected for many target scenarios under the multi-planar multistatic scenario by performing multiple planar SAR scans with the target at different $z$-locations.
To emulate irregular scanning geometries, data collected throughout the $x$-$y$-$z$ space are subsampled, as discussed in Section \ref{sec:results}.
Implementations in the literature employ multi-camera infrared camera systems to track the radar as it is moved through space by the user \cite{alvarez2019freehand,alvarez2021freehand}.
Other studies on irregular scanning geometries explored freehand imaging using a stereo camera with an IMU for positioning estimation \cite{alvarez2021towards} and UAV near-field imaging with a laser rangefinder and a real-time kinematic (RTK) system for localization \cite{garcia20203DSARProcessing}. 
These implementations, among others \cite{kan2020automotiveSAR,wu2020multilayered,alvarez2021system,alvarez2021freehandsystem}, demonstrate the viability of high-resolution sensors for precise positioning to enable novel imaging techniques using UWB mmWave radars.
Hence, this study focuses on improving the computational efficiency of the imaging technique and assumes that the radar position is known across an irregularly sampled geometry.

\subsection{Measurement Results and Discussion}
\label{sec:results}
In this section, we validate the EMPM algorithm derived in Sections \ref{sec:system_model} and \ref{sec:image_reconstruction}, as illustrated in Fig. \ref{fig:image_reconstruction_process_complete}.
Irregular scanning geometries were simulated using the simulation platform developed in \cite{smith2023ThzToolbox}, and the image reconstruction results are shown by comparing our enhanced method with the gold standard BPA and RMA without multi-planar multistatic compensation.
Similarly, using the \mbox{3-D} mechanical system detailed in Fig. \ref{fig:scanner_xyz}, irregular scanning geometries were emulated by collecting planar scans with the target at different $z$-locations and subsampling the collected data.
Imaging results, comparing the EMPM with the BPA and RMA, demonstrate the computational advantage of our technique while achieving nearly identical spatial resolution.
The EMPM algorithm achieves image quality comparable to that of the BPA while offering time and space complexity on par with the RMA.

\subsubsection{Simulated Irregular Geometry SAR Imaging Results}
\label{subsec:sim}
\begin{figure}[h]
\centering
    \begin{subfigure}[b]{0.45\textwidth}
         \centering
         \includegraphics[width=\textwidth]{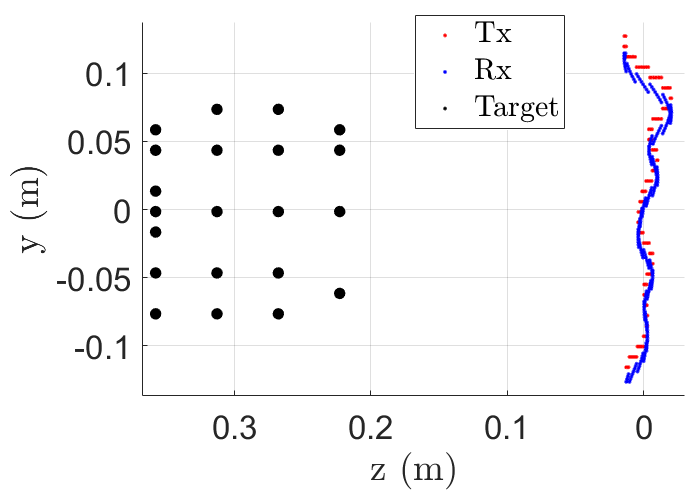}
         \caption{}
         \label{fig:sim1_UTD_scenario}
    \end{subfigure}
    \begin{subfigure}[b]{0.45\textwidth}
         \centering
         \includegraphics[width=\textwidth]{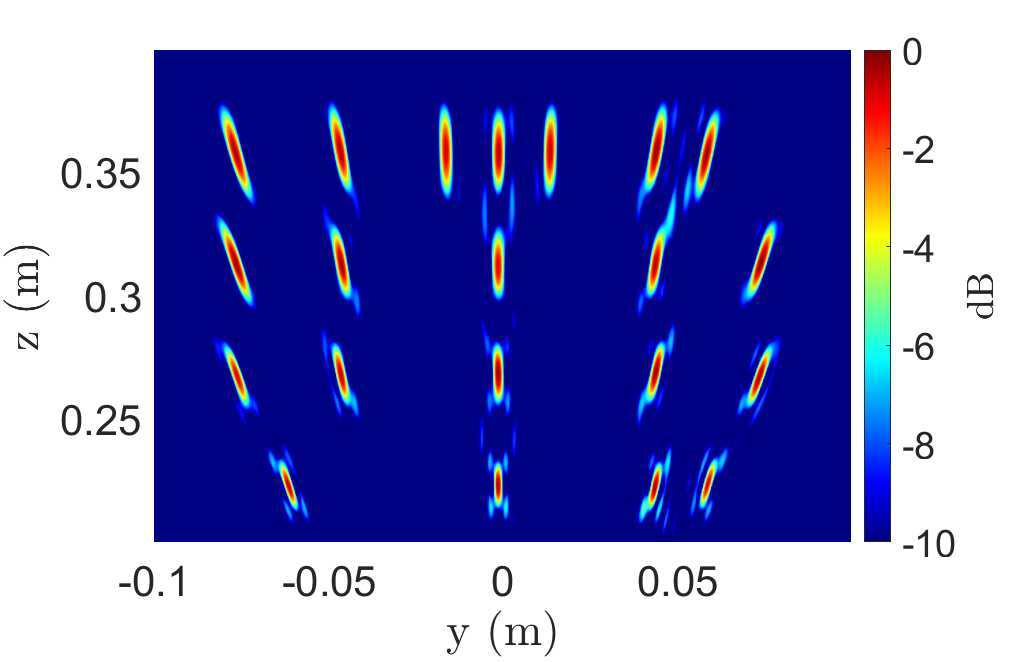}
         \caption{}
         \label{fig:sim1_UTD_BPA}
    \end{subfigure}
    \vskip\baselineskip
    \begin{subfigure}[b]{0.45\textwidth}
         \centering
         \includegraphics[width=\textwidth]{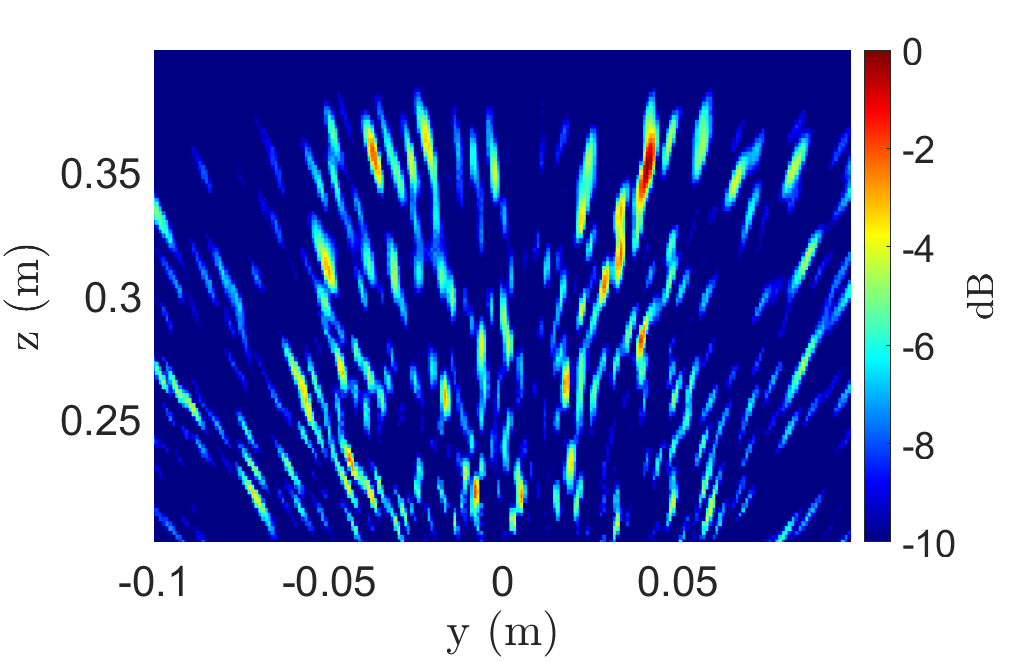}
         \caption{}
         \label{fig:sim1_UTD_RMA}
    \end{subfigure}
    \begin{subfigure}[b]{0.45\textwidth}
         \centering
         \includegraphics[width=\textwidth]{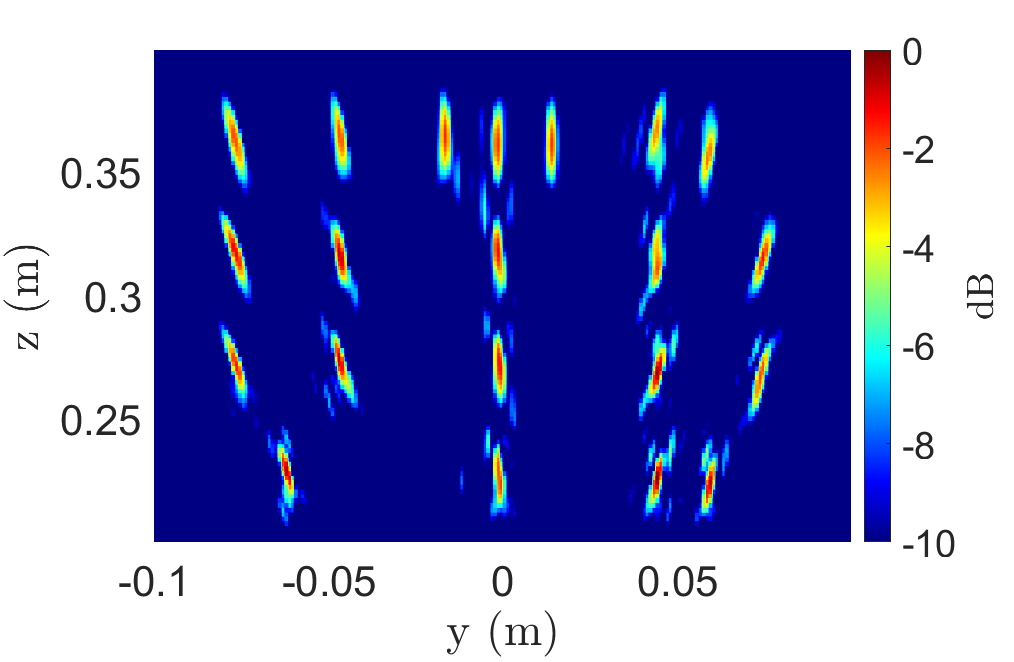}
         \caption{}
         \label{fig:sim1_UTD_RMA_FFH}
    \end{subfigure}
\caption{(a) Irregular scanning geometry for ``UTD'' scenario consisting of a multi-linear array in the $y$-direction at \mbox{$x$ = 0 m} and corresponding imaging results using the (b) BPA \mbox{(296.3 s)}, (c) RMA without multi-planar multistatic compensation \mbox{(29 ms)}, and (d) EMPM \mbox{(30 ms)}.}
\label{fig:sim1_UTD}
\end{figure}

To validate our proposed algorithm in simulation, we consider three distinct scenarios.
First, we investigate the impact of array irregularities on image resolution.
We consider the point spread functions (PSFs) of several multi-planar MIMO-SAR scenarios and compare them with an ideal planar scanning scenario to analyze the range and cross-range resolution of the EMPM algorithm. 
We assume a single ideal point target located at \mbox{$(0, 0, 0.5$ m$)$} in \mbox{3-D} space for the PSF simulation. 
For comparison, an ideal linear MIMO-SAR pattern is generated along with several irregular SAR scanning patterns with increasing irregularity.
Each non-cooperative motion track is generated by a semi-smooth, random curve spanning \mbox{$y' \in [-12.5, 12.5]$ cm} with varying $z_\ell$ around \mbox{$Z_0 = 0$ m} with 256 sampling locations, as shown in Fig. \ref{fig:sim1_UTD_scenario}. 
\begin{figure}[h]
    \begin{subfigure}[b]{0.5\textwidth}
         \centering
         \includegraphics[width=\textwidth]{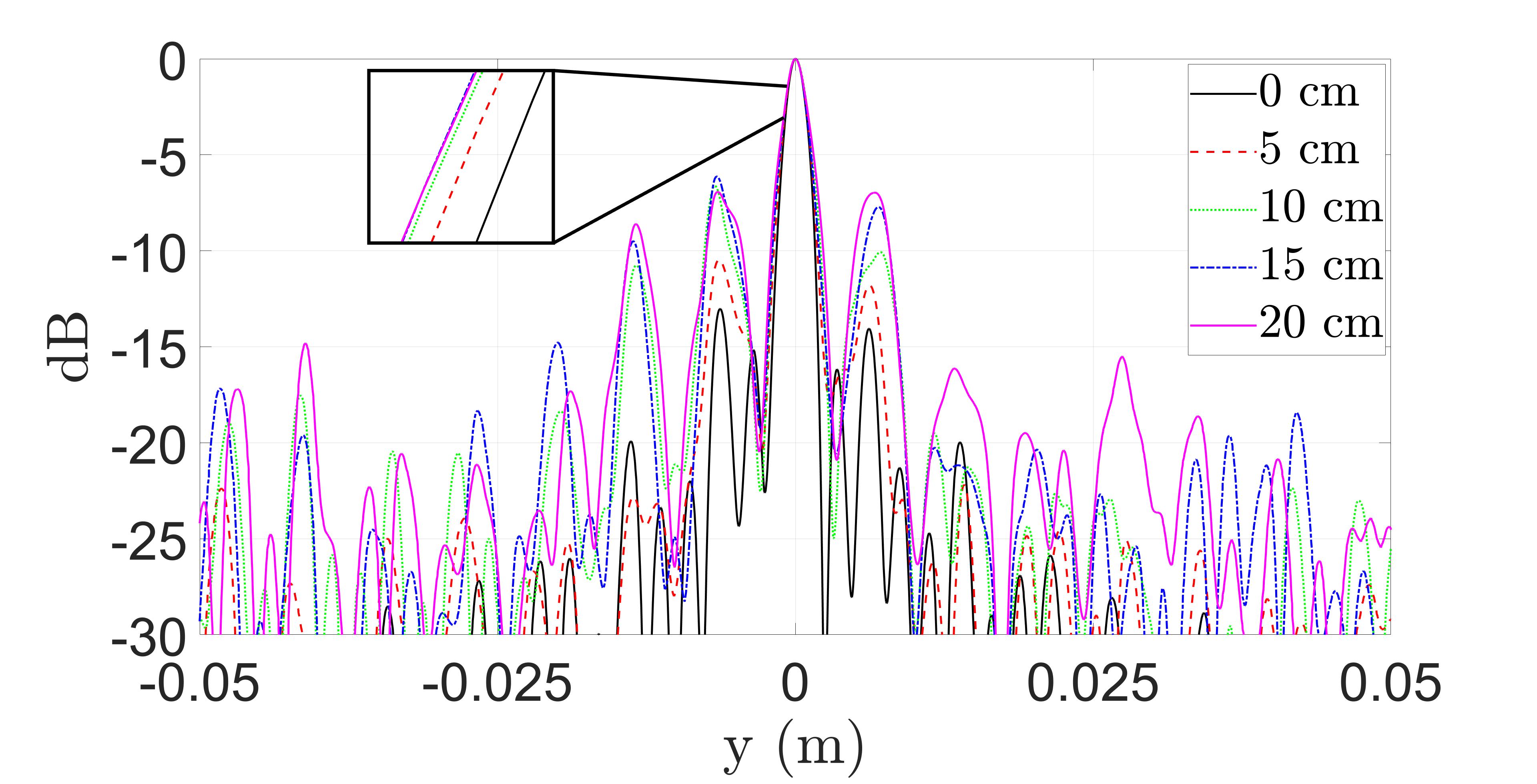}
         \caption{}
         \label{fig:sim0_PSF_y}
    \end{subfigure}
    \begin{subfigure}[b]{0.5\textwidth}
         \centering
         \includegraphics[width=\textwidth]{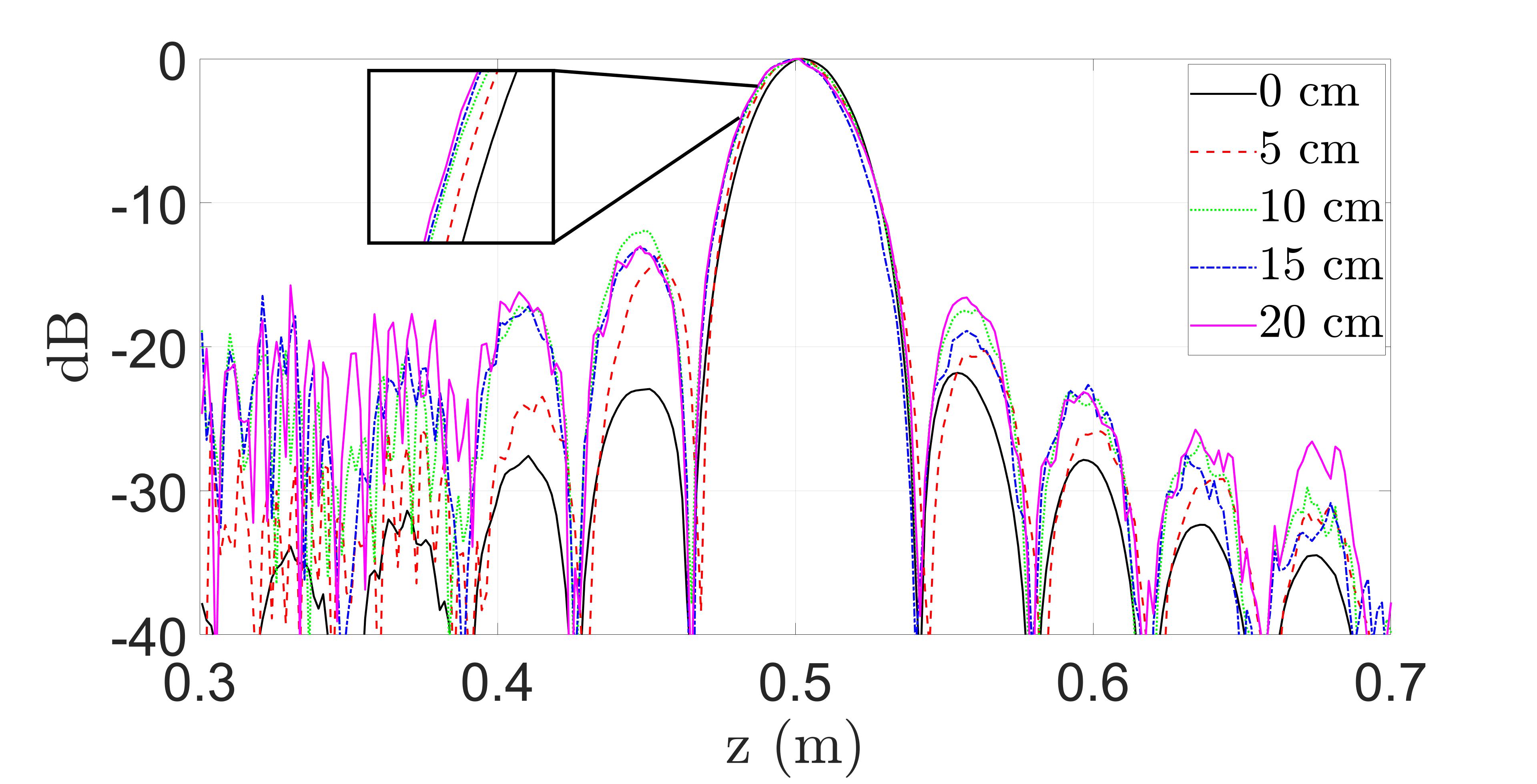}
         \caption{}
         \label{fig:sim0_PSF_z}
    \end{subfigure}
\caption{Comparison of point spread function (PSF) resolution along the (a) $y$- and (b) $z$-direction with varying maximum distance from the reference plane, $Z_0$, to the samples at $z_\ell$. The distance between the samples and reference plane, $\Delta_z^\text{max}$, is varied from \mbox{0 cm} (linear) to \mbox{20 cm} with a step size of \mbox{5 cm}. The linear case (0 cm) is computed with the conventional RMA. Each of the remaining PSFs are computed using the EMPM.}
\label{fig:sim0_PSF}
\end{figure}
To analyze the impact of $d_\ell^z = z_\ell - Z_0$, the distance between the reference plane at $Z_0$, and the samples at $z_\ell$, we simulate several multi-planar multistatic with increasing variance of $d_\ell^z$, as shown in Fig. \ref{fig:sim0_PSF}. 
The absolute maximum distance, $\Delta_z^\text{max} \triangleq \max |d_\ell^z|$, varies from \mbox{0 cm}, the linear case, to \mbox{20 cm} with a step size of \mbox{5 cm}. 
At a center frequency of \mbox{79 GHz}, \mbox{$\Delta_z^\text{max} =$ 20 cm} is more than 50 times the wavelength, $\lambda_c =$ \mbox{3.79 mm}. 
Prior work on freehand smartphone imaging system design assumes deviations on the order of several centimeters \cite{alvarez2021towards}. 
Along the cross-range dimension, which is symmetric along both the $x$- and $y$-directions, the resolution is minimally affected by the irregular scanning geometry when the algorithm is applied, as shown in Fig. \ref{fig:sim0_PSF_y}. 
However, a direct relationship between $\Delta_z^\text{max}$ and the main beamwidth is observed along with decreased sidelobe suppression compared with the ideal linear case, where the traditional RMA can be employed directly. 
Along the $z$-direction, the resolution of the EMPM suffers as $\Delta_z^\text{max}$ increases but remains quite similar to the resolution of the linear case, as shown in Fig. \ref{fig:sim0_PSF_z}. 
Hence, the EMPM achieves a focusing performance comparable to that of the ideal linear or planar RMA case while allowing for irregular scanning geometries with large deviations from the reference plane in the $z$-direction. 

To evaluate the performance of the algorithm for more complex targets, a linear array along the $y$-axis is simulated as shown in Fig. \ref{fig:sim1_UTD_scenario} with 21 point scatterers arranged as the letters ``UTD.''
Again, irregular array locations are generated by a semi-smooth, random curve spanning \mbox{$y' \in [-12.5, 12.5]$ cm} and \mbox{$z_\ell \in [-2.5, 2.5]$ cm} with 256 sampling locations.
As shown in Fig. \ref{fig:sim1_UTD_BPA}, the gold standard BPA recovers each point scatterer without artifacts; however, computing the BPA image requires 296.3 s on our machine. 
The RMA without multi-planar multistatic compensation and the EMPM are considerably more efficient, requiring only 30 ms for computation.
However, while the RMA image in Fig. \ref{fig:sim1_UTD_RMA} is significantly distorted and the target shape is lost, the EMPM resolves the point targets comparably to the BPA and requires a fraction of the computation time.

\begin{figure}[h]
    \centering
    \includegraphics[width=0.6\textwidth]{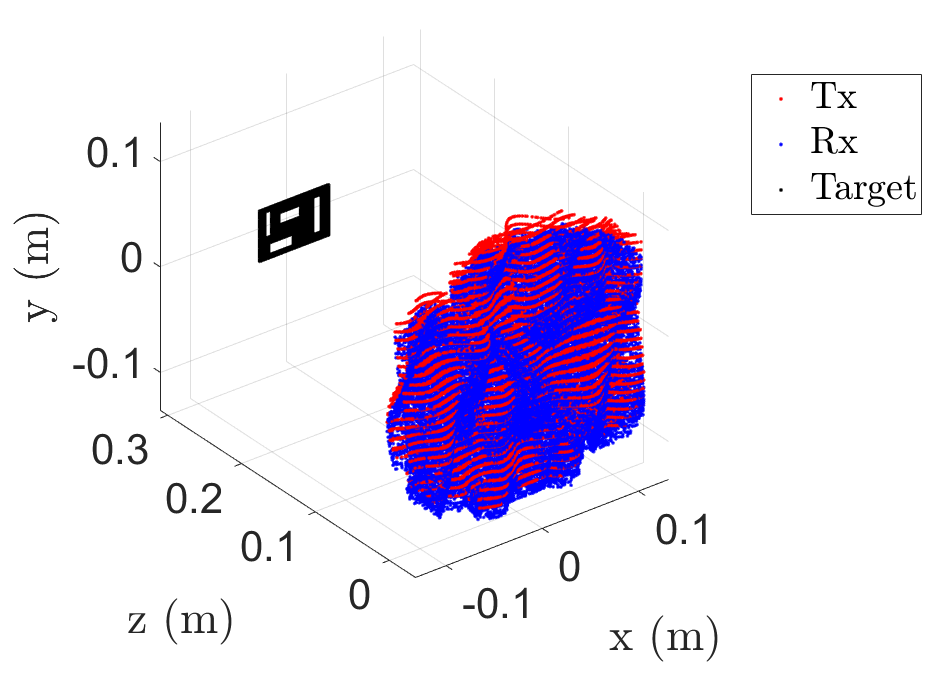}
    \caption{Irregular scanning geometry for cutout consisting of a multi-planar array along the $x$- and $y$-directions.}
    \label{fig:sim2_cutout2_scenario}
\end{figure}

\begin{figure}[h]
\centering
    \begin{subfigure}[b]{0.45\textwidth}
         \centering
         \includegraphics[width=\textwidth]{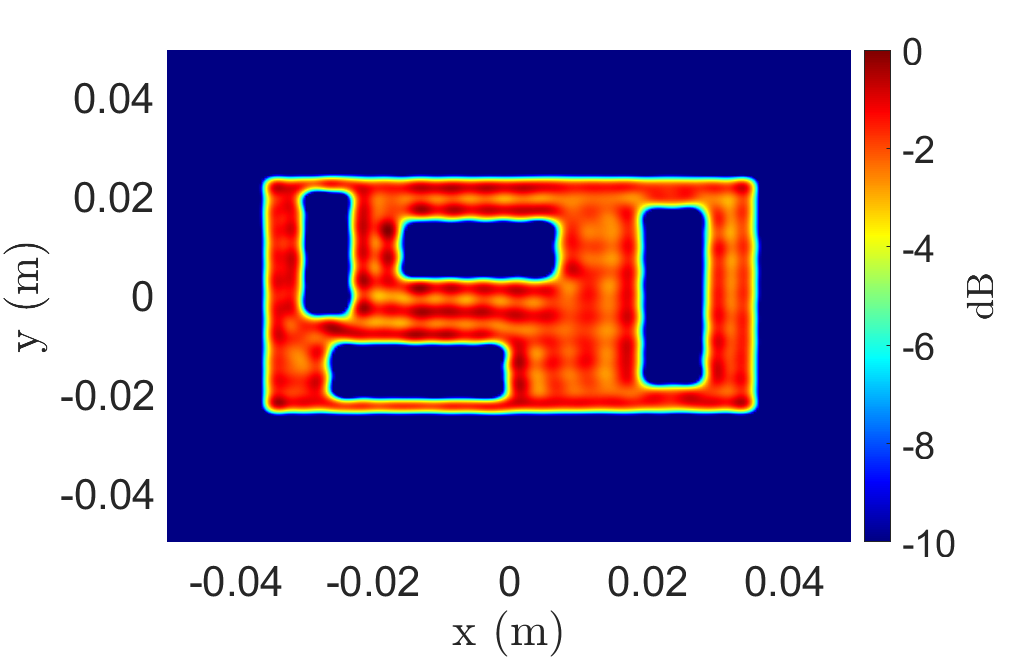}
         \caption{}
         \label{fig:sim2_cutout2_BPA}
    \end{subfigure}
    \begin{subfigure}[b]{0.45\textwidth}
         \centering
         \includegraphics[width=\textwidth]{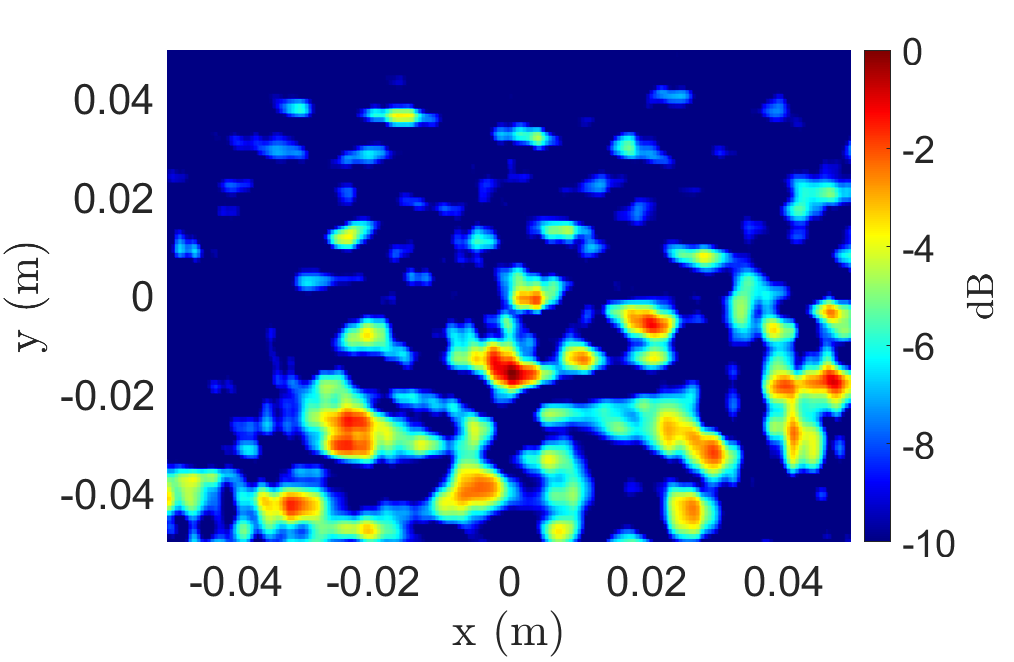}
         \caption{}
         \label{fig:sim2_cutout2_RMA}
    \end{subfigure}
    \vskip\baselineskip
    \begin{subfigure}[b]{0.45\textwidth}
         \centering
         \includegraphics[width=\textwidth]{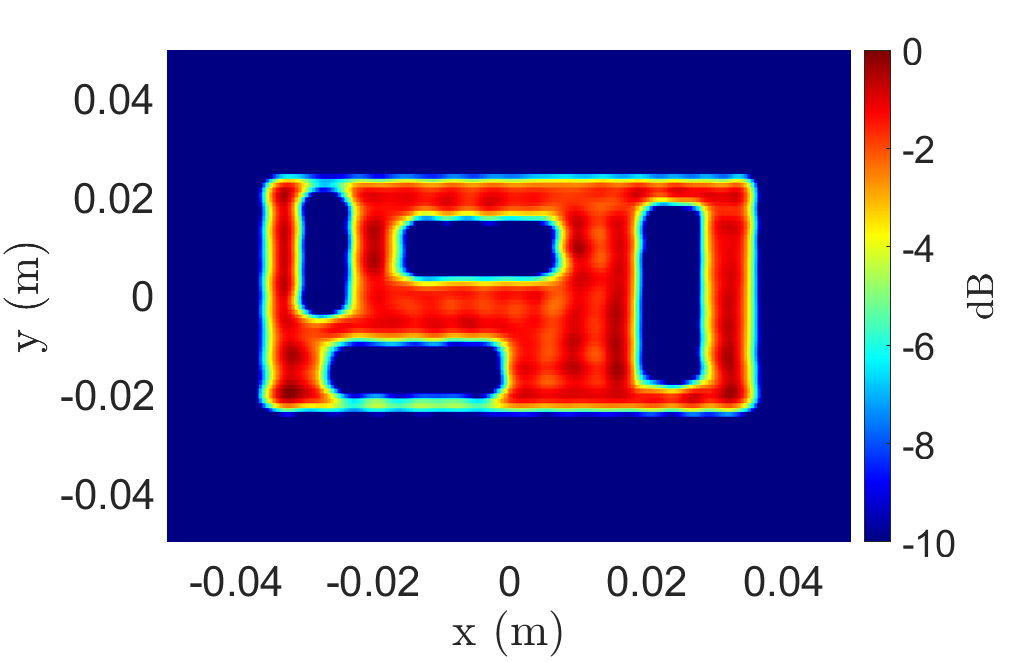}
         \caption{}
         \label{fig:sim2_cutout2_RMA_FFH}
    \end{subfigure}
    \begin{subfigure}[b]{0.45\textwidth}
         \centering
         \includegraphics[width=\textwidth]{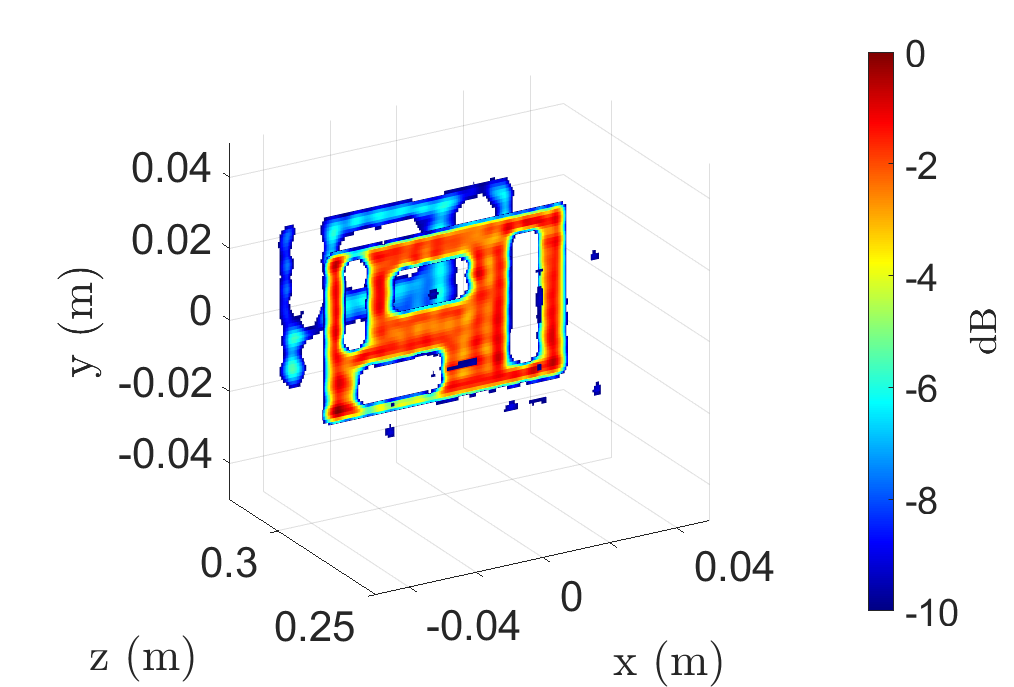}
         \caption{}
         \label{fig:sim2_cutout2_RMA_FFH_3D}
    \end{subfigure}
\caption{Imaging results for the scenario in Fig. \ref{fig:sim2_cutout2_scenario} using the (a) BPA \mbox{(1324.8 s)}, (b) RMA without multi-planar multistatic compensation \mbox{(1.1 s)}, (c) EMPM at the \mbox{$z$ = 300 mm} plane \mbox{(1.1 s)}, and (d) the \mbox{3-D} reconstructed image using the EMPM \mbox{(4.8 s)}.}
\label{fig:sim2_cutout2}
\end{figure}

Considering the more broadly applicable \mbox{2-D} scanning case, a \mbox{2-D} multi-planar multistatic scenario is simulated with a solid target located at \mbox{$z$ = 300 mm}, as shown in Fig. \ref{fig:sim2_cutout2_scenario}.
The target is a rectangular strip with cutouts of various sizes and the irregular sampling geometry is generated as a \mbox{2-D} semi-smooth random curve occupying \mbox{$x' \in [-12.5, 12.5]$ cm}, \mbox{$y' \in [-12.5, 12.5]$ cm}, and \mbox{$z_\ell \in [-2.5, 2.5]$ cm} with 102956 sampling locations. 
The number of sampling locations is selected to approximate a virtual array spanning $[-12.5, 12.5]$ cm along the $x$- and $y$-directions, which is a realistic aperture size for many applications, while satisfying the sampling condition, such that the distance between subsequent sampling points is always less than $\lambda_c/4$.
Because the target is located on a single $z$-plane parallel to the planar projection after our compensation technique, a \mbox{2-D} $x$-$y$ image is recovered at \mbox{$z$ = 300 mm}. 
Again, while the BPA yields a robust reconstruction, the computation time is excessive for most applications, requiring 1324.8 s on a desktop machine.
On the other hand, the EMPM algorithm outperforms the RMA significantly in terms of image quality, nearly matching that of the BPA with only slight artifacting, while demonstrating superior efficiency to the BPA computing a high-resolution \mbox{2-D} image in only 1.1 s.
Similarly, \mbox{3-D} images can be reconstructed using these methods.
The \mbox{3-D} reconstructed image using the EMPM is shown in Fig. \ref{fig:sim2_cutout2_RMA_FFH_3D}, requiring 4.8 s to compute, whereas the RMA and BPA are computed in 4.8 s and 339159.2 s, respectively. 

Comparing the results in Figs. \ref{fig:sim1_UTD} and \ref{fig:sim2_cutout2}, aberrations appear to be more pronounced along the $z$-dimension or depth. 
This phenomenon is expected given the analysis in Section \ref{subsec:virtual_array}, where $d_\ell^z$ and the size of the target in the $z$-direction are assumed to be small. 
Hence, for targets of significant size in the $z$-direction, such as the target in Fig. \ref{fig:sim1_UTD_RMA}, the proposed compensation suffers from slight artifacting compared with the BPA. 
However, for many applications, the considerable time savings achieved using our technique is a necessary trade-off compared with the prohibitively slow BPA. 

\subsubsection{Empirical Irregular Geometry SAR Imaging Results}
\label{subsec:real}
\begin{figure}[h]
\centering
    \begin{subfigure}[b]{0.4\textwidth}
         \centering
         \includegraphics[width=\textwidth]{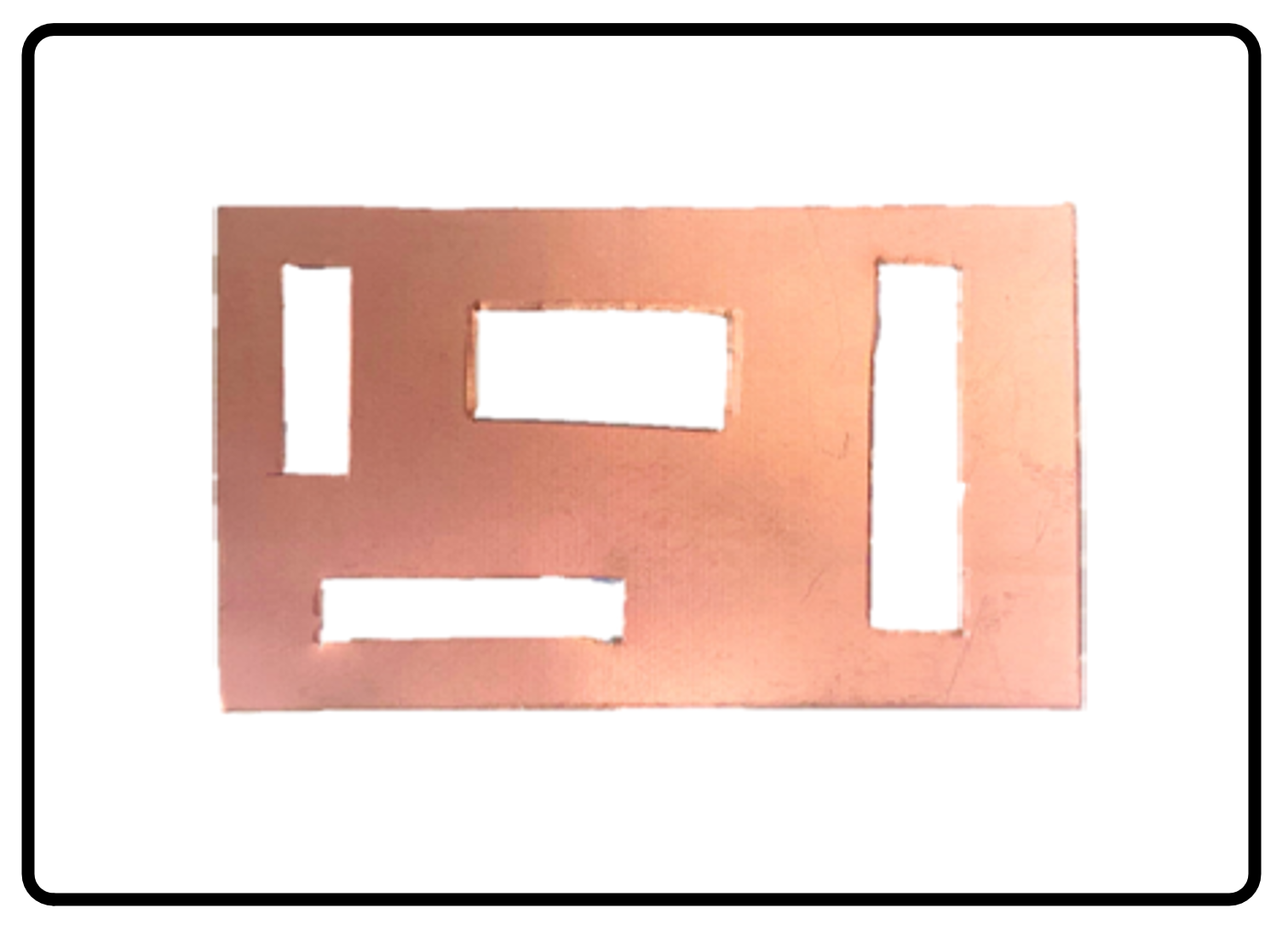} 
         \caption{}
         \label{fig:cutout2}
    \end{subfigure}
    \begin{subfigure}[b]{0.4\textwidth}
         \centering
         \includegraphics[width=\textwidth]{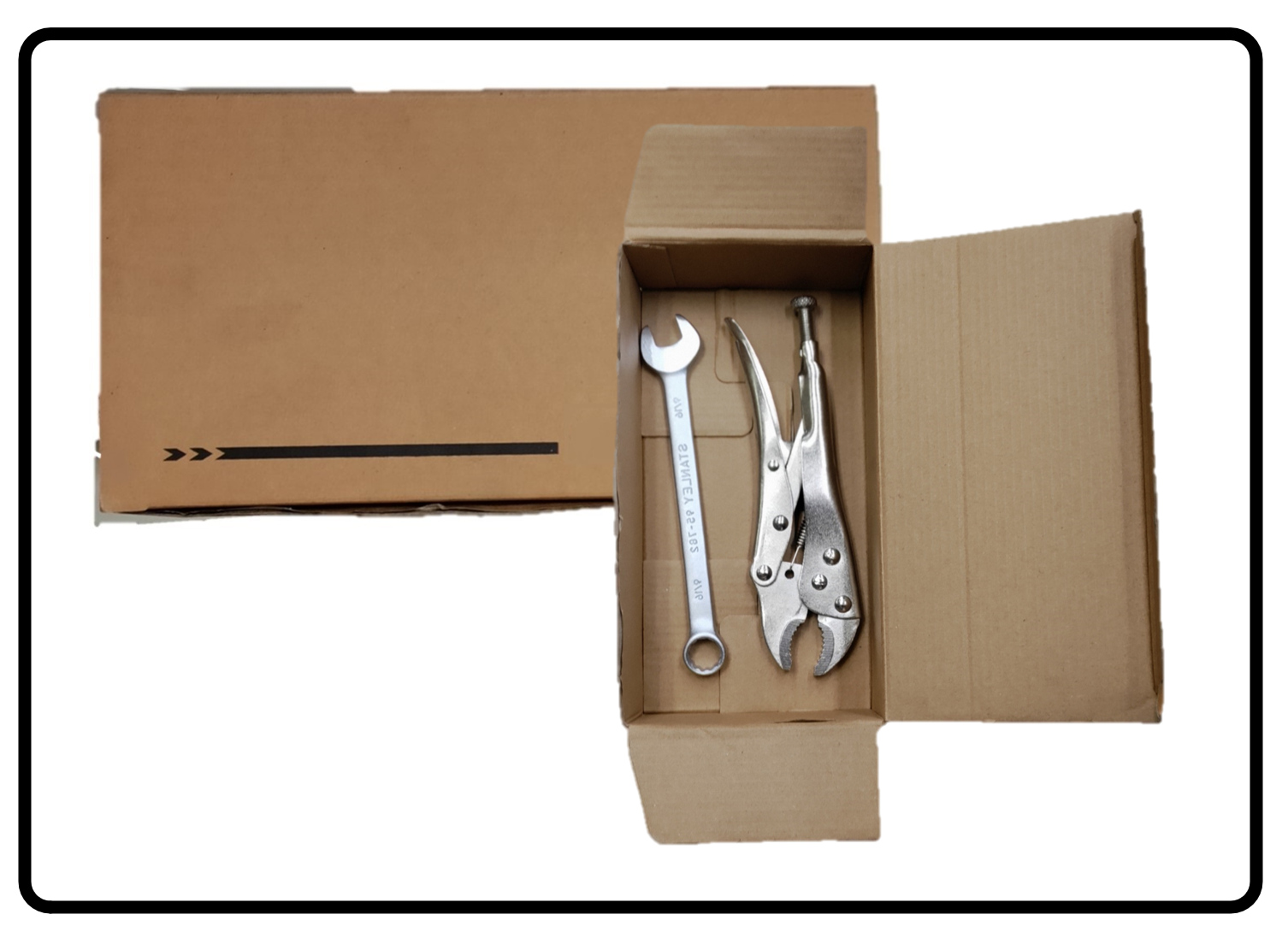} 
         \caption{}
         \label{fig:hiddentools}
    \end{subfigure}
    \vskip\baselineskip
    \begin{subfigure}[b]{0.4\textwidth}
         \centering
         \includegraphics[width=\textwidth]{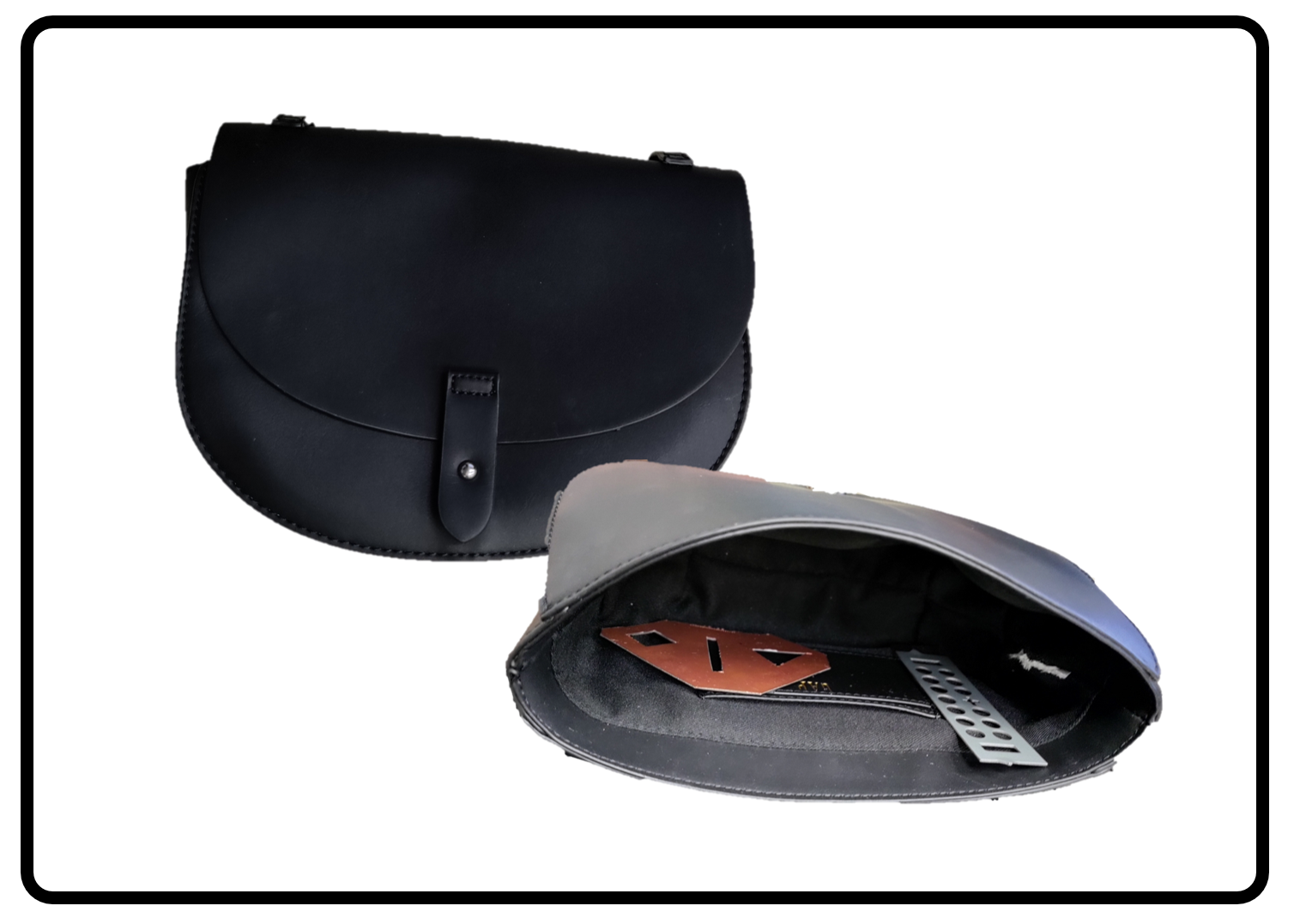} 
         \caption{}
         \label{fig:purse}
    \end{subfigure}
\caption{Various experimental targets: (a) copper-clad laminate test target, (b) tools hidden inside box, and (c) purse containing metal cutouts.}
\label{fig:targets}
\end{figure}

The proposed EMPM imaging technique and system prototype are validated experimentally by capturing SAR data of various target scenes, as shown in Fig. \ref{fig:targets}.
The reconstructed images obtained using each method are compared and discussed.

\begin{figure}[h]
\centering
    \begin{subfigure}[b]{0.4\textwidth}
         \centering
         \includegraphics[width=\textwidth]{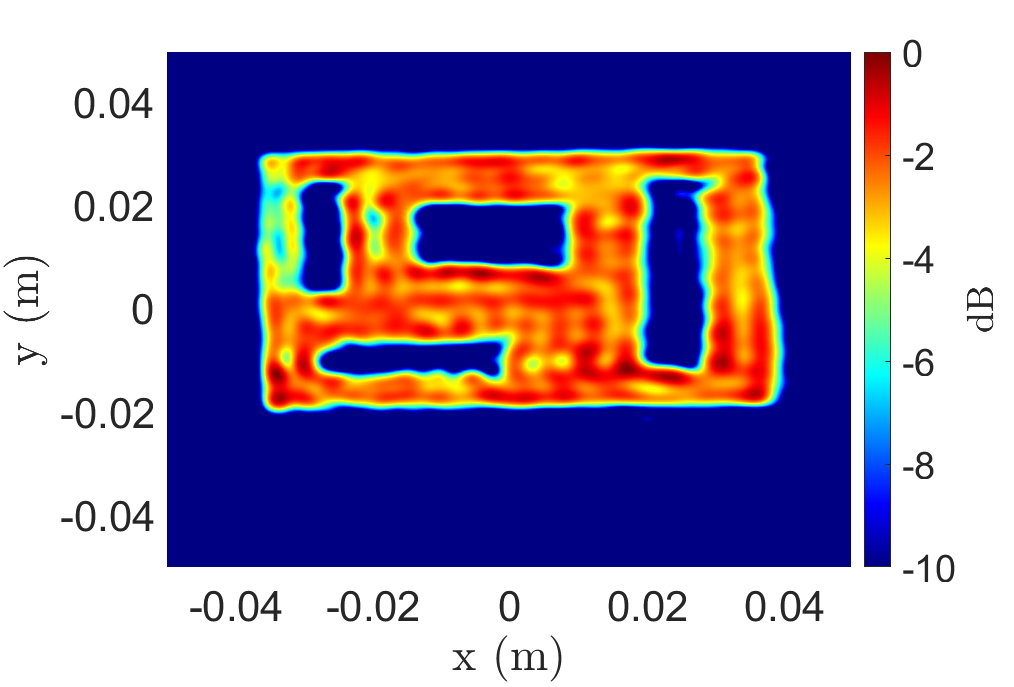} 
         \caption{}
         \label{fig:exp1_cutout2_BPA}
    \end{subfigure}
    \begin{subfigure}[b]{0.4\textwidth}
         \centering
         \includegraphics[width=\textwidth]{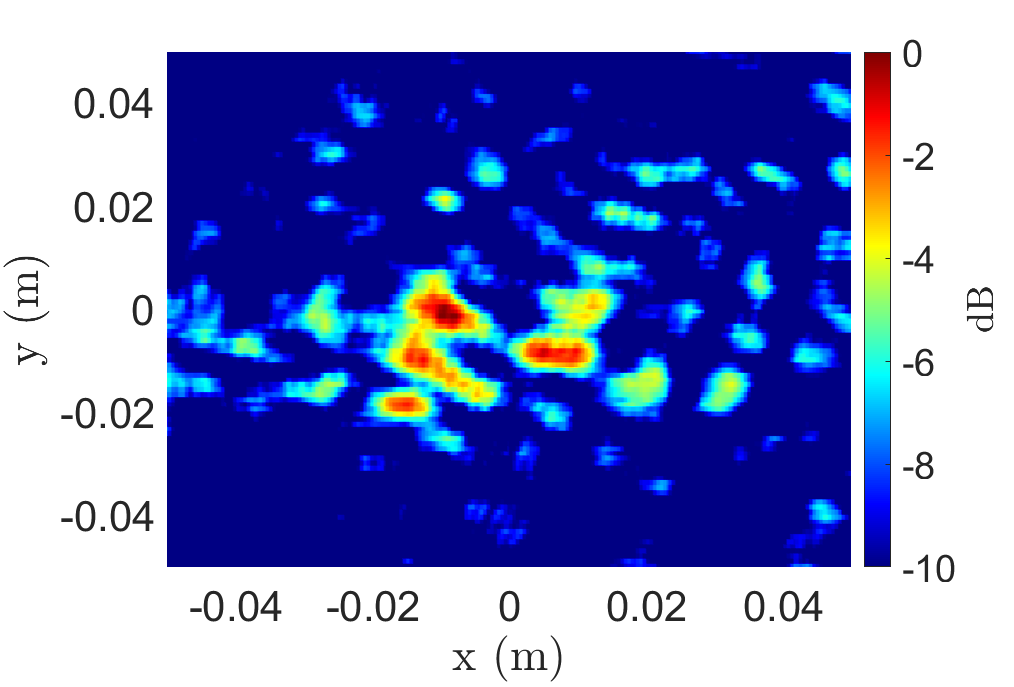} 
         \caption{}
         \label{fig:exp1_cutout2_RMA}
    \end{subfigure}
    \vskip\baselineskip
    \begin{subfigure}[b]{0.4\textwidth}
         \centering
         \includegraphics[width=\textwidth]{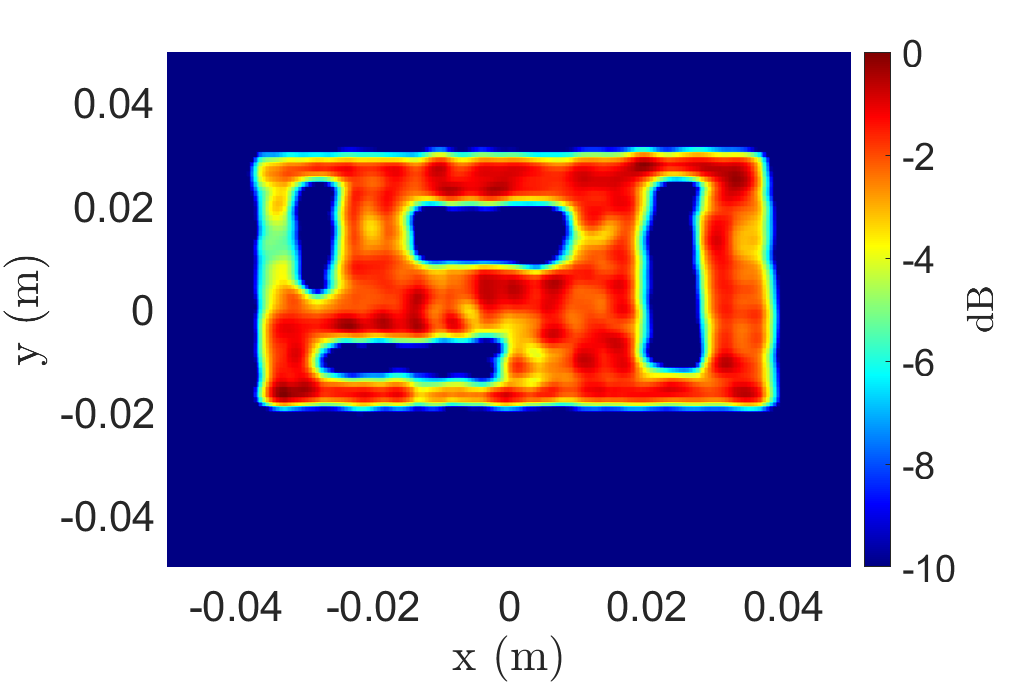} 
         \caption{}
         \label{fig:exp1_cutout2_RMA_FFH}
    \end{subfigure}
\caption{Imaging results for the copper test target using the (a) BPA \mbox{(1324.8 s)}, (b) RMA without multi-planar multistatic compensation \mbox{(1.1 s)}, and (c) EMPM \mbox{(1.1 s)}.}
\label{fig:exp1_cutout2}
\end{figure}

The test target with several horizontal and vertical rectangular cutouts made from copper-clad laminate (Fig. \ref{fig:cutout2}) is illuminated by the $x$-$y$ scanner at the planes $z \in [275, 324]$ mm with a separation of \mbox{1 mm}. 
Hence, data are collected throughout the same region discussed previously such that $x' \in [-12.5,12.5]$ cm, $y' \in [-12.5,12.5]$ cm, and $z_\ell \in [-2.5,2.5]$ cm with 102762 sampling locations. 
After the data were collected, the 50 planar scans were subsampled using a similar random \mbox{2-D} curve as shown in Fig. \ref{fig:sim2_cutout2_scenario} to emulate the multi-planar irregular sampling scenario.
The imaging results and corresponding computation times for each reconstruction algorithm are shown in Fig. \ref{fig:exp1_cutout2}.
The EMPM imaging technique demonstrates robustness in projecting the irregular scanning geometry to a planar scenario for more efficient image recovery, as the cutout is recovered cleanly without significant artifacting, as shown in Fig. \ref{fig:exp1_cutout2_RMA_FFH}.
In contrast, the image recovered using the BPA requires nearly 30 min to compute, and although the RMA is computed efficiently, the RMA without multi-planar compensation cannot resolve the target scene, as shown in Fig. \ref{fig:exp1_cutout2_RMA}.
Furthermore, when the target location is unknown in the $z$-direction, \mbox{3-D} imaging offers an improved solution with a slightly higher computational cost, using the proposed algorithm.

\begin{figure}[h]
\centering
    \begin{subfigure}[b]{0.24\textwidth}
         \centering
         \includegraphics[width=\textwidth]{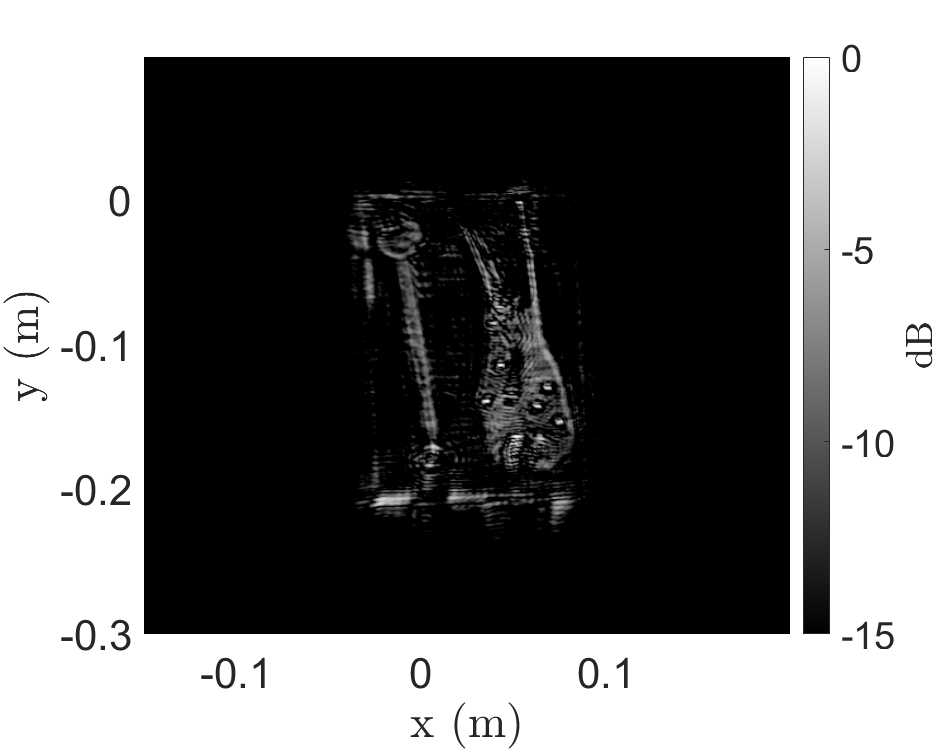} 
         \caption{}
         \label{fig:exp2_hiddentools_BPA_2D}
    \end{subfigure}
    \begin{subfigure}[b]{0.24\textwidth}
         \centering
         \includegraphics[width=\textwidth]{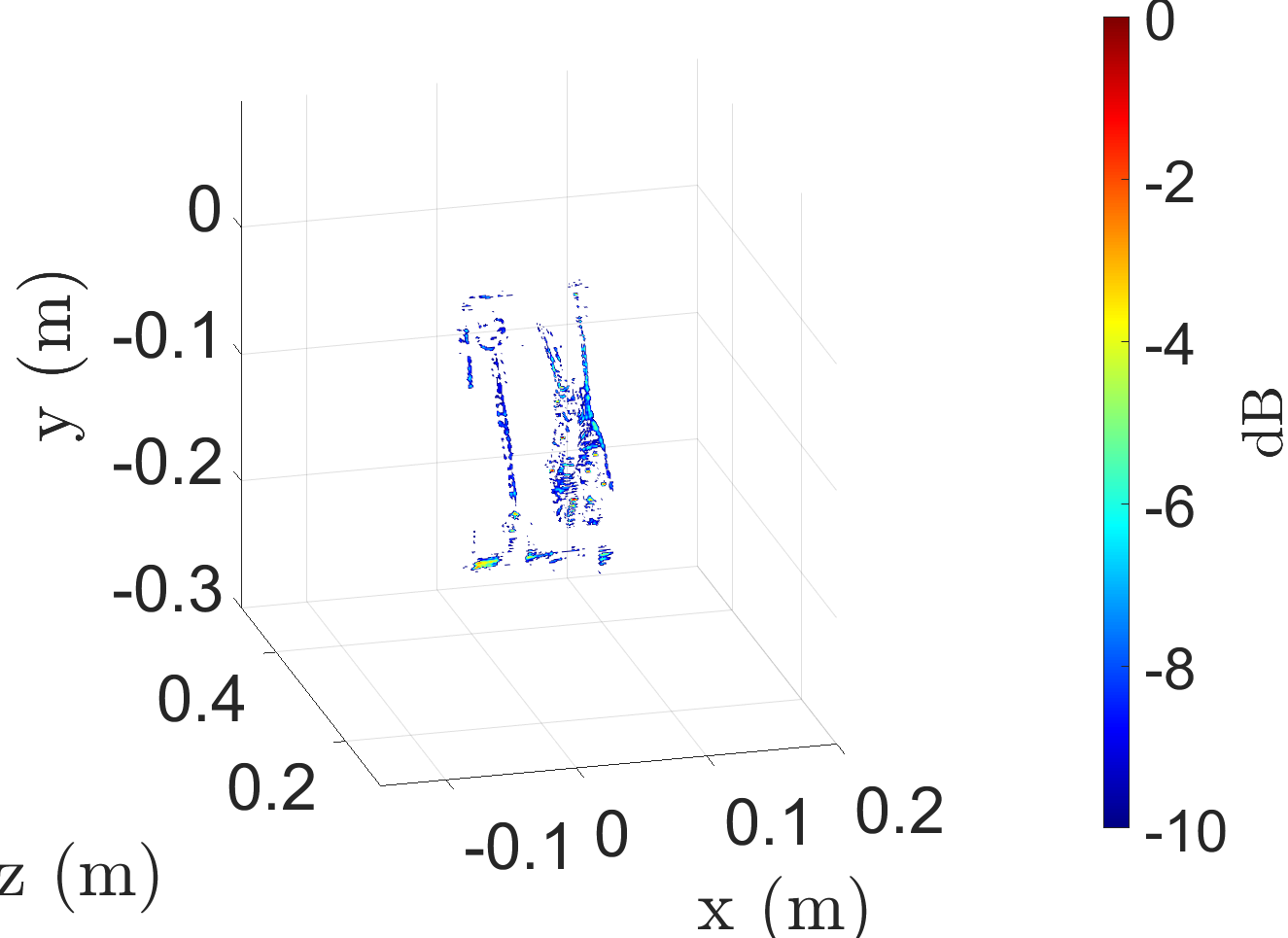} 
         \caption{}
         \label{fig:exp2_hiddentools_BPA_3D}
    \end{subfigure}
    \begin{subfigure}[b]{0.24\textwidth}
         \centering
         \includegraphics[width=\textwidth]{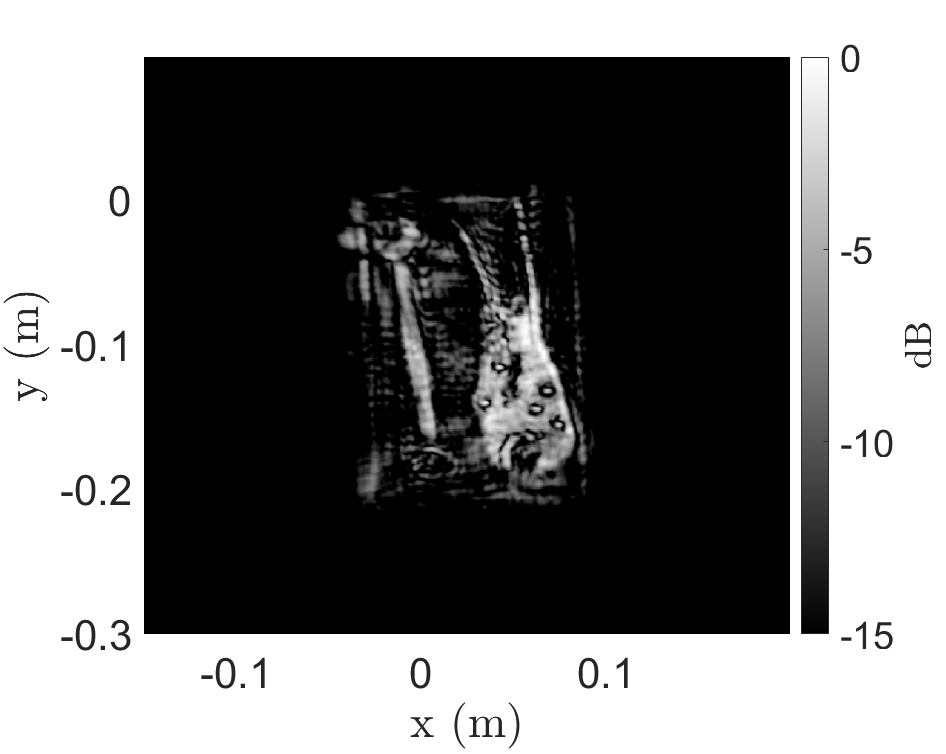} 
         \caption{}
         \label{fig:exp2_hiddentools_RMA_FFH_2D}
    \end{subfigure}
    \begin{subfigure}[b]{0.24\textwidth}
         \centering
         \includegraphics[width=\textwidth]{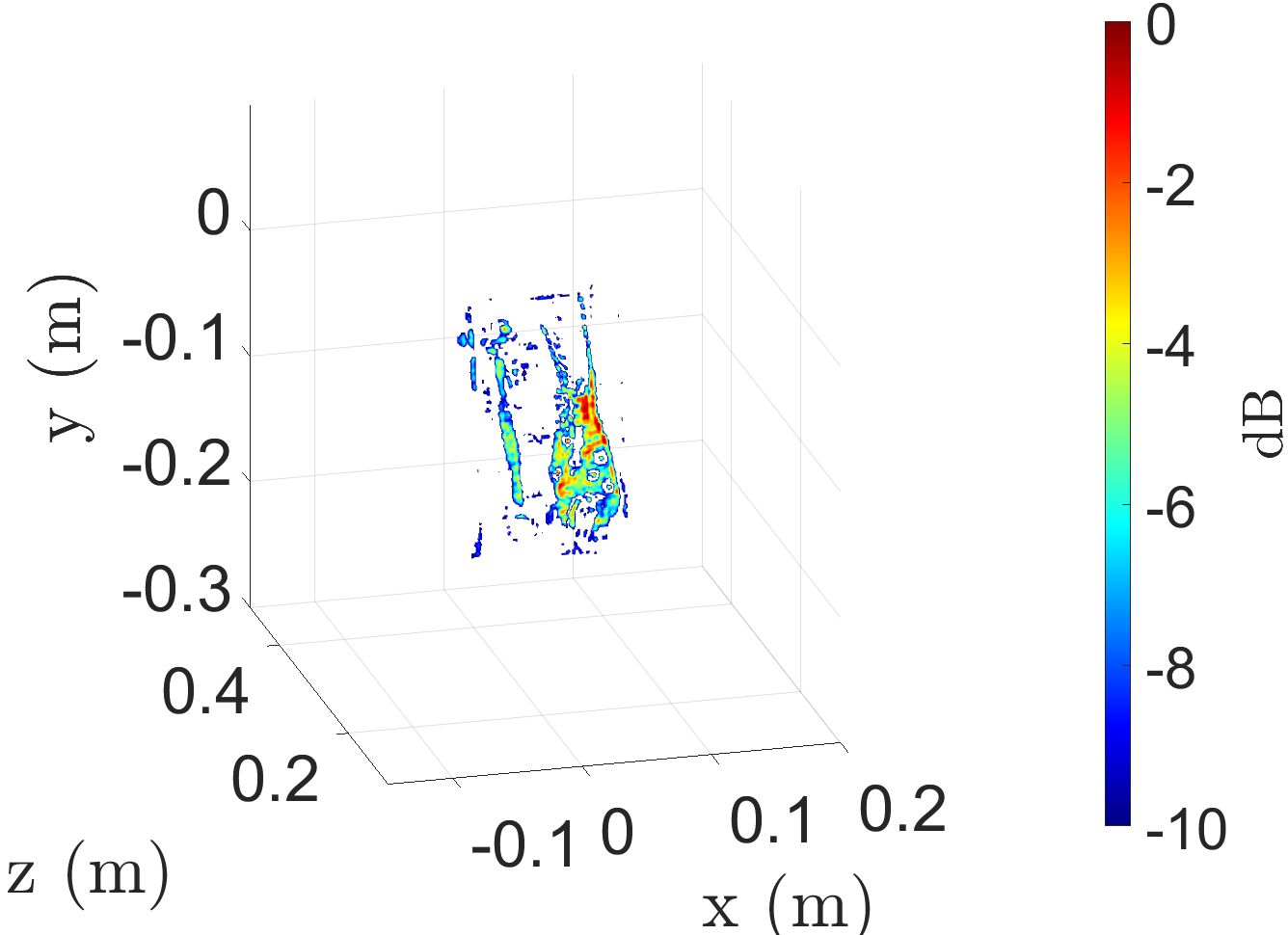} 
         \caption{}
         \label{fig:exp2_hiddentools_RMA_FFH_3D}
    \end{subfigure}
    \vskip\baselineskip
    \begin{subfigure}[b]{0.24\textwidth}
         \centering
         \includegraphics[width=\textwidth]{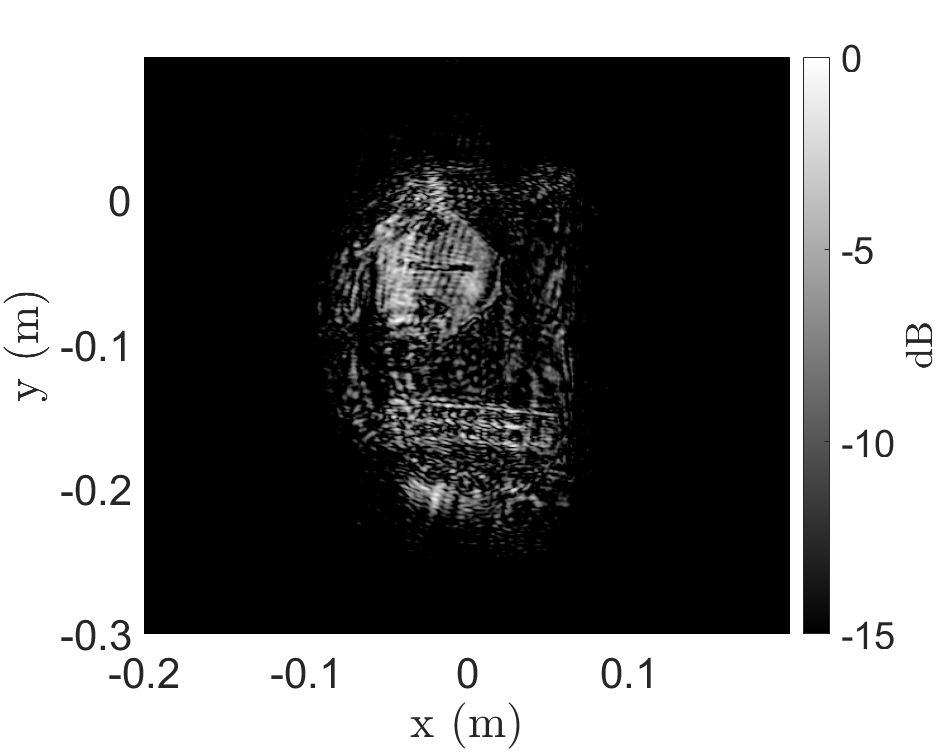} 
         \caption{}
         \label{fig:exp3_purse_BPA_2D}
    \end{subfigure}
    \begin{subfigure}[b]{0.24\textwidth}
         \centering
         \includegraphics[width=\textwidth]{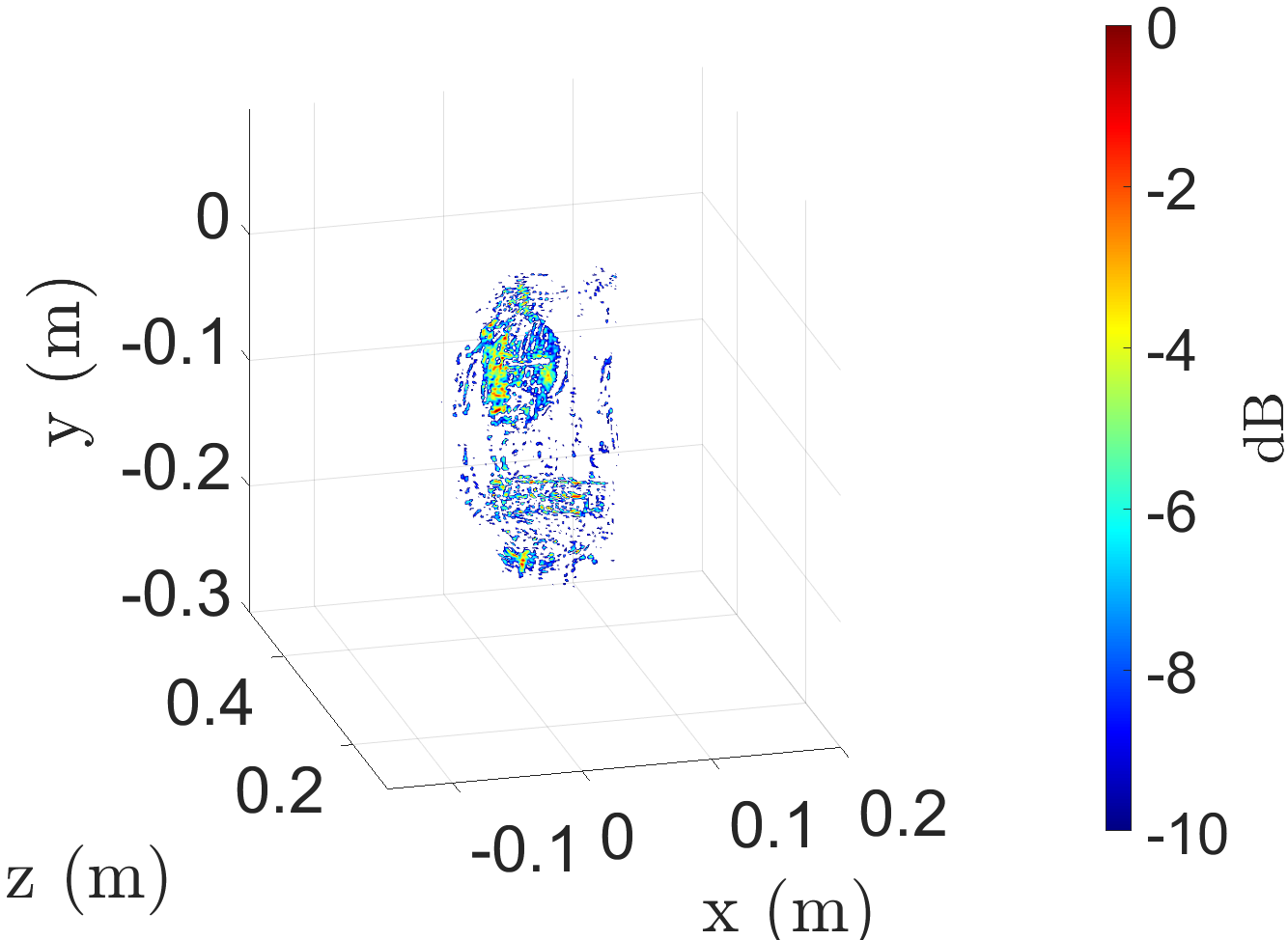} 
         \caption{}
         \label{fig:exp3_purse_BPA_3D}
    \end{subfigure}
    \begin{subfigure}[b]{0.24\textwidth}
         \centering
         \includegraphics[width=\textwidth]{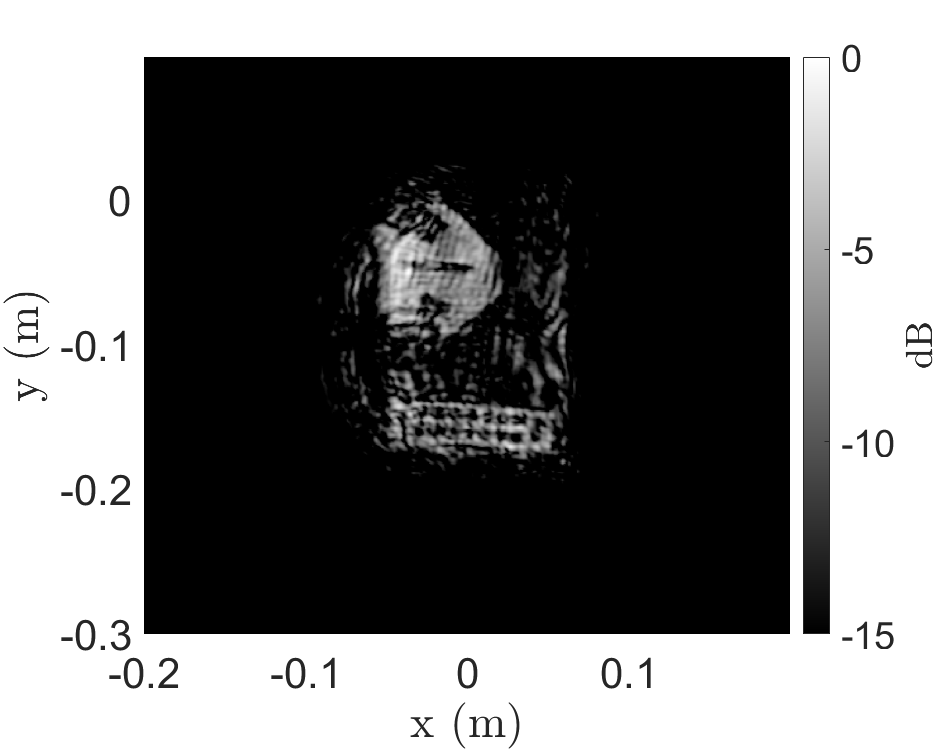} 
         \caption{}
         \label{fig:exp3_purse_RMA_FFH_2D}
    \end{subfigure}
    \begin{subfigure}[b]{0.24\textwidth}
         \centering
         \includegraphics[width=\textwidth]{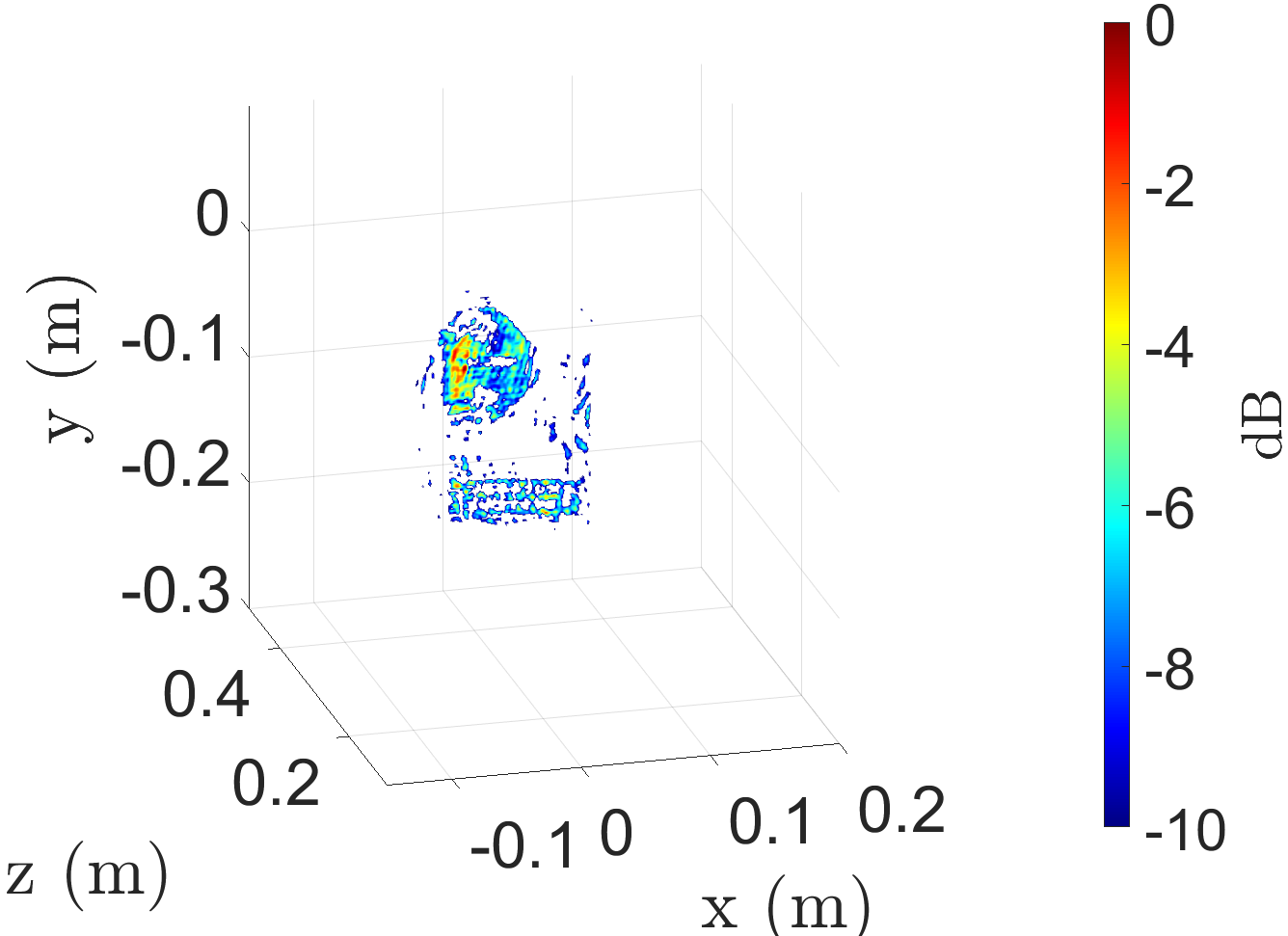} 
         \caption{}
         \label{fig:exp3_purse_RMA_FFH_3D}
    \end{subfigure}
\caption{Imaging results for the hidden tools target, as shown in Fig. \ref{fig:hiddentools}, using the (a) \mbox{2-D} BPA (\mbox{5299.4 s}), (b) \mbox{3-D} BPA (\mbox{1356636.9 s}), (c) \mbox{2-D} EMPM (\mbox{4.3 s}), and (d) \mbox{3-D} EMPM (\mbox{10.7 s}). Imaging results for the concealed items in a purse, as shown in Fig. \ref{fig:purse}, using the (e) \mbox{2-D} BPA (\mbox{5299.4 s}), (f) \mbox{3-D} BPA (\mbox{1356636.9 s}), (g) \mbox{2-D} EMPM (\mbox{4.3 s}), and (h) \mbox{3-D} EMPM (\mbox{10.7 s}).}
\label{fig:exp2_and_exp3}
\end{figure}

The second target screened by the prototype to demonstrate a hidden target scenario consists of two wrenches (a combination wrench and a vise grip) hidden inside a cardboard box, as shown in Fig. \ref{fig:hiddentools}.
SAR scans of the target are performed with the target at the $z$-planes \mbox{$z \in [275, 324]$ mm} with a separation of \mbox{1 mm}.
To accommodate a larger target size, the aperture is increased to \mbox{$x' \in [-25,25]$ cm}, \mbox{$y' \in [-25,25]$ cm}, and \mbox{$z_\ell \in [-2.5,2.5]$ cm} with 102545 sampling locations.
Similarly, the $x$-$y$-$z$ data are sampled to emulate the multi-planar multi-static scenario using a semi-smooth random curve, as shown in Fig. \ref{fig:sim2_cutout2_scenario}.
The \mbox{2-D} and \mbox{3-D} implementations of the BPA and proposed algorithm are applied to the nonuniform data under the irregular scanning geometry, and the recovered images are shown in Figs. \ref{fig:exp2_hiddentools_BPA_2D} -- \ref{fig:exp2_hiddentools_RMA_FFH_3D}.
Both wrenches are visible in the reconstructed images; however, while the \mbox{2-D} image from the BPA and proposed algorithm provide high-fidelity reconstructions of the hidden tools, the \mbox{2-D} $z$-plane must be carefully selected to obtain such images.
The presence and location of targets are generally unknown for concealed item detection problems. 
Hence, \mbox{3-D} imaging is preferable for such scenarios and is primarily constrained by computational expense.
Our proposed algorithm offers an elegant compromise between the efficiency of the RMA and the image quality of the BPA.
In Fig. \ref{fig:exp2_hiddentools_RMA_FFH_3D}, the \mbox{3-D} image is computed by the proposed algorithm with an image quality comparable to that of the BPA, with a significantly reduced computational cost. 

A third experiment is conducted with several metal cutouts concealed in a purse to emulate a scenario wherein a suspicious personal item is quickly screened with an irregular scanning geometry, such as freehand SAR or drone imaging.
Fig. \ref{fig:purse} shows the purse and two hidden items: a triangularly shaped metal plate with different cutout shapes and a rectangular metal plate with circular holes.
The target is scanned by the multi-planar multistatic prototype discussed in Section \ref{sec:prototype}, and the data are employed to emulate an irregular sampling scenario.
Scans are performed with the target at the $z$-planes \mbox{$z \in [275, 324]$ mm} with a separation of \mbox{1 mm} and an aperture is synthesized within \mbox{$x' \in [-25,25]$ cm}, \mbox{$y' \in [-25,25]$ cm}, and \mbox{$z_\ell \in [-2.5,2.5]$ cm} with 102821 sampling locations.
The reconstructed images and corresponding computation times are shown in Figs. \ref{fig:exp3_purse_BPA_2D} -- \ref{fig:exp3_purse_RMA_FFH_3D}.
Both metal cutouts are resolved using our algorithm, with an image quality comparable to that of the BPA. 
Again, assuming that the contents of the purse are generally unknown, computing the \mbox{3-D} image is preferable for concealed item detection.
To efficiently recover a \mbox{3-D} image with irregularly sampled data, existing inversion techniques require excessive computation time and memory capacity, as shown in Figs. \ref{fig:exp3_purse_BPA_2D} and \ref{fig:exp3_purse_BPA_3D}.
However, the proposed EMPM (Figs. \ref{fig:exp3_purse_RMA_FFH_2D} and \ref{fig:exp3_purse_RMA_FFH_3D}) offers an efficient solution that does not compromise image quality.

These experiments demonstrate the advantages of the EMPM and the limitations of the RMA and BPA.
A comparison of the computation times required for each algorithm is presented in Table \ref{tab:computation_times}.
Applying the RMA directly to the multi-planar data, as shown in Figs. \ref{fig:sim1_UTD_RMA}, \ref{fig:sim2_cutout2_RMA}, and \ref{fig:exp1_cutout2_RMA}, yields significant aberrations to the point of failed reconstruction. 
Therefore, the RMA images are not shown in the other examples. 
When the target is known to be \mbox{2-D} and located at a single known $z$-plane, the 2-D BPA implementation can be computed somewhat efficiently in certain instances by employing a graphics processing unit (GPU) and parallelizing the computation \cite{alvarez2021towards,alvarez2019freehand,alvarez2021system,alvarez2021freehand,alvarez2021freehandsystem}.
However, particularly for mobile applications, access to high-capacity GPUs is rare or size-prohibitive, and such acceleration is infeasible.
Moreover, as the BPA is scaled up to three dimensions, the time and space complexities increase exponentially, requiring excessive computational power and memory.
In many emerging applications, efficient \mbox{3-D} image computation on low-power devices is preferable, if not mandatory, as the precise location of the target is generally unknown.
However, efficient algorithms, such as the RMA, require monostatic, planar assumptions that are unachievable by these applications. 
To enable such technologies, the EMPM imaging algorithm efficiently compensates for the irregular scanning geometry by carefully handling the phase of each sample.
This enables image reconstruction under dynamic conditions with computational complexity identical to that of the RMA and image quality comparable to that of the BPA.

\begin{table}[h]
    \centering
    \caption{Computation time, in seconds, required by the various algorithms for each experiment.}
    \begin{tabular}{c||c|c|c}
         & Metal Cutout & Hidden Tools & Purse \\
         \hline
         \hline
         \mbox{2-D} {BPA} & 1324.8 & 5299.4 & 5299.4 \\
         \mbox{3-D} {BPA} & 339159.2 & 1356636.9 & 1356636.9 \\
         \hline
         \mbox{2-D} {RMA} & 1.1 & 4.3 & 4.3 \\
         \mbox{3-D} {RMA} & 4.8 & 10.7 & 10.7 \\
         \hline
         \mbox{2-D} {EMPM} & 1.1 & 4.3 & 4.3 \\
         \mbox{3-D} {EMPM} & 4.8 & 10.7 & 10.7 \\
         \hline
         \hline
    \end{tabular}
    \label{tab:computation_times}
\end{table}

In this section, we presented a novel approach for high-resolution, efficient \mbox{3-D} near-field SAR imaging for irregular scanning geometries. 
We proposed a multi-planar multistatic framework applicable to a diverse set of applications, including freehand imaging, UAV SAR, and automotive imaging. 
A novel algorithm, the EMPM, is proposed to efficiently compensate for irregularly sampled multi-planar multistatic data to equivalent planar monostatic mmWave radar data. 
Our technique extends the traditional RMA by presenting an algorithm for efficiently aligning multi-planar multistatic data to a virtual planar monostatic scenario. 
By projecting the data onto a virtual planar monostatic equivalent array, the RMA is extended to account for both irregular scanning and MIMO-SAR effects, resulting in high-fidelity focusing. 
The EMPM is valid for common radar signaling techniques in 5G, IoT, smartphones, and automotive applications. 
The simulation results demonstrate the robustness of our approach in the presence of significant spatial deviations among the samples along the $z$-direction. 
Furthermore, we empirically validated the EMPM by using a custom prototype to capture multi-planar multistatic data for several concealed and obscured scenarios.

In both simulation and experimental studies, our algorithm achieves efficient image reconstruction matching the focusing quality of existing techniques while reducing computational complexity by a considerable margin. 
The EMPM is valuable for enabling efficient medium-fidelity image recovery; however, additional image enhancement can be pursued via data-driven approaches, such as those discussed in Chapter \ref{ch:data_driven}. 
Without the EMPM imaging algorithm, deep learning-based SAR super-resolution on irregularly sampled SAR images remains infeasible.
Herein, we observe the interaction between front-end signal processing techniques and deep learning approaches. 

For many problems in perception and imaging, a parallel design of signal processing and data-driven techniques yields impressive results.
By examining static and dynamic gestures, we studied the effects of various preprocessing techniques on classification accuracy and proposed a novel ``sterile'' training technique to improve CNN robustness \cite{smith2021sterile}.
We extended our analysis of preprocessing algorithms to the near-field SAR imaging domain for irregular scanning geometries.
Many emerging applications require efficient SAR image recovery with non-cooperative arrays in the near-field.
The proposed EMPM algorithm achieves image quality comparable to that of the computationally prohibitive BPA with the efficiency of the planar array-constrained RMA \cite{smith2022efficient}. 
The EMPM enables rapid medium-fidelity image reconstruction for irregularly sampled near-field SAR scenarios. 
Further investigation into data-driven methods for enhanced imaging is discussed in subsequent chapters. 

\chapter{Enhanced High-Resolution Imaging Algorithms using Data-Driven Methods}
\label{ch:data_driven}

In the previous chapter, we employed signal processing expertise to improve deep learning sensing and perception capabilities. 
In this study, we employ data-driven techniques to enhance high-resolution near-field SAR image reconstruction algorithms.
Part of the following work was previously published in \cite{smith2022ffh_vit}\footnote{\copyright 2022 IEEE. Reprinted, with permission, from J. W. Smith, Y. Alimam, G. Vedula, and M. Torlak, ``A vision transformer approach for efficient near-field SAR super-resolution under array perturbation,'' in \textit{Proc. IEEE Tex. Symp. Wirel. Microw. Circuits Syst. (WMCS)}, Waco, TX, USA, Apr. 2022, pp. 1--6.} and \cite{vasileiou2022efficient}\footnote{\copyright 2022 IEEE. Reprinted, with permission, from C. Vasileiou, J. W. Smith, S. Thiagarajan, M. Nigh, Y. Makris, and M. Torlak, ``Efficient {CNN}-based super resolution algorithms for {mmWave} mobile radar imaging,'' in \textit{Proc. IEEE Int. Conf. Image Process. (ICIP)}, Bourdeaux, France, Oct. 2022, pp. 3803--3807.
} and will be presented in \cite{smith2023ThzToolbox}.

\begin{figure}
     \centering
     \begin{subfigure}[b]{0.45\textwidth}
         \centering
         \includegraphics[width=\textwidth]{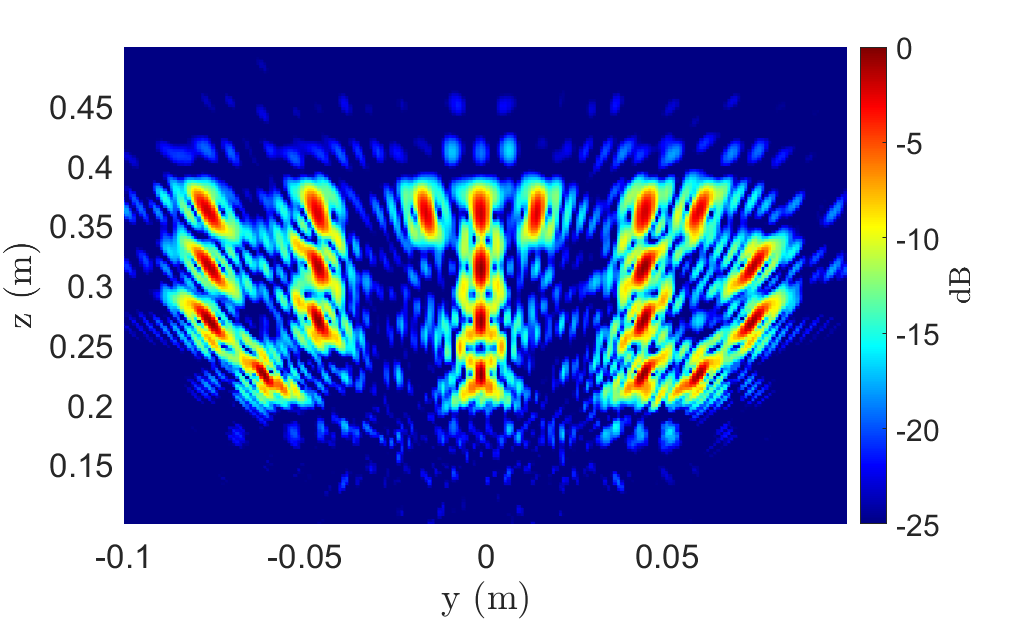}
         \caption{}
         \label{fig:utd_blurry}
     \end{subfigure}
     \begin{subfigure}[b]{0.45\textwidth}
         \centering
         \includegraphics[width=\textwidth]{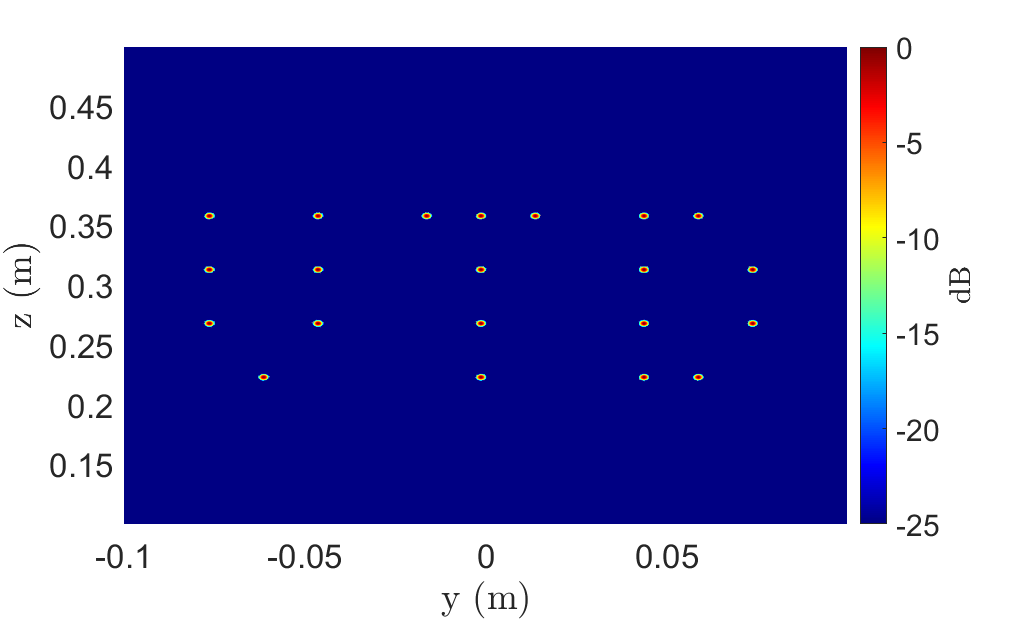}
         \caption{}
         \label{fig:utd_ideal}
     \end{subfigure}
        \caption{21 ideal point scatters arranged in the letters ``UTD.'' (a) SAR image. (b) Ideal image.}
        \label{fig:utd_images}
\end{figure}

Near-field SAR and ISAR imaging systems are becoming increasingly popular for numerous applications, from security sensing to industrial packaging. 
Such systems suffer from distortions owing to assumptions and approximations in the image reconstruction process \cite{yanik2019sparse} in addition to system limitations \cite{smith2021An} and non-idealities \cite{yanik2020development}. 
To illustrate these phenomena, we compare the reconstructed SAR image from the points in the shape of the letters ``UTD'' with its ideal counterpart, as shown in Fig. \ref{fig:utd_images}.
Even in the noiseless case, distortion around the point targets in Fig. \ref{fig:utd_blurry} is present owing to physical limitations and assumptions of the imaging algorithm. 
Because system cross-range and range resolution are limited by the effective length of the aperture and bandwidth, respectively, any resolution improvement through data processing can enable technologies that are otherwise infeasible due to cost or size constraints. 
Although some attempts have been made towards SAR image super-resolution using sparse coding and compressed sensing (CS) techniques \cite{guo2019millimeter,qiao2015compressive}, near-field SAR has only recently received attention for image super-resolution \cite{wang2021tpssiNet,wang2021rmistnet,dai2021imaging,wang2020csrnet,wang2021fista}.
Additionally, deep learning-based techniques have been shown to outperform CS algorithms for far-field SAR image super-resolution \cite{gao2018enhanced}.
Prior deep learning-based SAR super-resolution methods are trained exclusively on simulated SAR data consisting of randomly placed points and then tested on real SAR data.
Thus, there is an increasing need for meaningful near-field SAR data containing sophisticated targets beyond the simple random points employed in previous studies \cite{jing2022enhanced}.

\section{High-Fidelity Near-Field SAR Simulation Toolbox}
\label{sec:sar_simulator}
Given that a major obstacle to deep learning on mmWave imaging data is the availability of meaningful data for model training and prototyping, we develop an open-source MATLAB toolbox for system prototyping, imaging algorithm development, and high-fidelity dataset generation. 

Many array topologies (real or synthetic) have been investigated for near-field array imaging. 
The Pacific Northwest National Laboratory (PNNL) is a front-runner in the field of array imaging algorithm development. 
Sheen \textit{et al.} investigated rectilinear (planar) array imaging \cite{sheen2016three,sheen2001three} and cylindrical SAR (ECSAR) \cite{sheen2018simulation} using a switched antenna array time-division multiplexing (TDM) approach by sequentially operating transmitters and receivers and approximating the switched array as a monostatic array. 
Yanik \textit{et al.} have driven work towards efficient algorithms for near-field and sparse array image reconstruction \cite{yanik2019near,yanik2019sparse} as well as system-level design \cite{yanik2020development,smith2020nearfieldisar,yanik2019cascaded}.
Similarly, the National University of Defense Technology (NUDT), China, has presented extensive work on MIMO and MIMO-SAR imaging algorithms. 
Their work included efforts on \mbox{3-D} imaging using \mbox{1-D} scanning of a \mbox{1-D} MIMO array \cite{gao2018_1D_MIMO,fan2020linearMIMOArbitraryTopologies}, \mbox{2-D} imaging using circular SAR in the near-field \cite{gao2016efficient}, and cylindrical MIMO-SAR \cite{gao2018cylindricalMIMO}. 
In addition, extensive progress towards image reconstruction has been made using the range migration algorithm (RMA) for rectilinear patterns \cite{lopez20003,paul2021systematic,maisto2021sensor,yanik2020development,yanik2019cascaded,guo2019millimeter,shin2007range,zhuge2012three,zhu2017range,ding2013thz,batra2021Submm,batra2021short}, polar formatting algorithm (PFA) for cylindrical scanning geometries \cite{amineh2019real,fortuny2001extension,detlefsen2005effective,yang2018three}, and back-projection algorithm (BPA) for any array geometry \cite{fedeli2020microwave,soumekh1998wide,mohammadian2019sar,zhuge2010sparse,jaeschke20143d,moll2012towards}. 
The toolbox presented in this section offers tools for simulating and reconstructing images from \mbox{1-D} linear, \mbox{2-D} rectilinear, \mbox{1-D} circular, and \mbox{2-D} cylindrical array geometries using the provided BPA, RMA, and PFA image reconstruction algorithms.

Prior work on array imaging simulation tools has largely been limited to two domains: far-field simplified imaging simulation  \cite{soumekh1999synthetic,jost2005matlabToolkitSAR,Auer2016RaySAR,kedzierawski2011MATLAB,deo2012MATLAB,gorham2010sar} and near-field antenna simulation software \cite{sheen2018simulation}.
Existing simple far-field packages ignore spherical wave phenomena, thus drastically reducing the simulation and image reconstruction complexity, whereas near-field antenna simulators employ application-specific implementations using computational electromagnetic (CEM) solvers such as HFSS, Xpatch, or FEKO. 
Efforts falling into the former camp are applicable to many remote sensing applications typical of classical strip-map and spotlight SAR imaging modes but fall short for near-field imaging scenarios where the wavefront curvature must be considered \cite{soumekh1999synthetic,jost2005matlabToolkitSAR,Auer2016RaySAR,gorham2010sar}. 
On the other hand, despite being quite robust, computational EM solvers require a tedious process for setting up each SAR scenario and can be prohibitively slow, requiring a computation time on the order of $10$s of hours \cite{sheen2018simulation}. 
The proposed toolbox leverages the advantages of both software types, enabling custom waveform design, complex antenna gain patterns, scanning scenario design, and image reconstruction.
This drastically increases the accessibility of imaging simulation and reconstruction to researchers across numerous fields.
In particular, this contribution is key to enabling rapid high-fidelity dataset generation.
Although the proposed software platform is designed for high-fidelity THz SAR imaging, it can easily be operated at lower frequencies in the mmWave region.
The toolbox user interface and API provide a complete solution for numerous SAR imaging tasks and promote data and benchmark standardization, which is a vital step in the evolution of data-driven signal processing algorithms.

\begin{figure}[h]
    \centering
    \includegraphics[width=\textwidth]{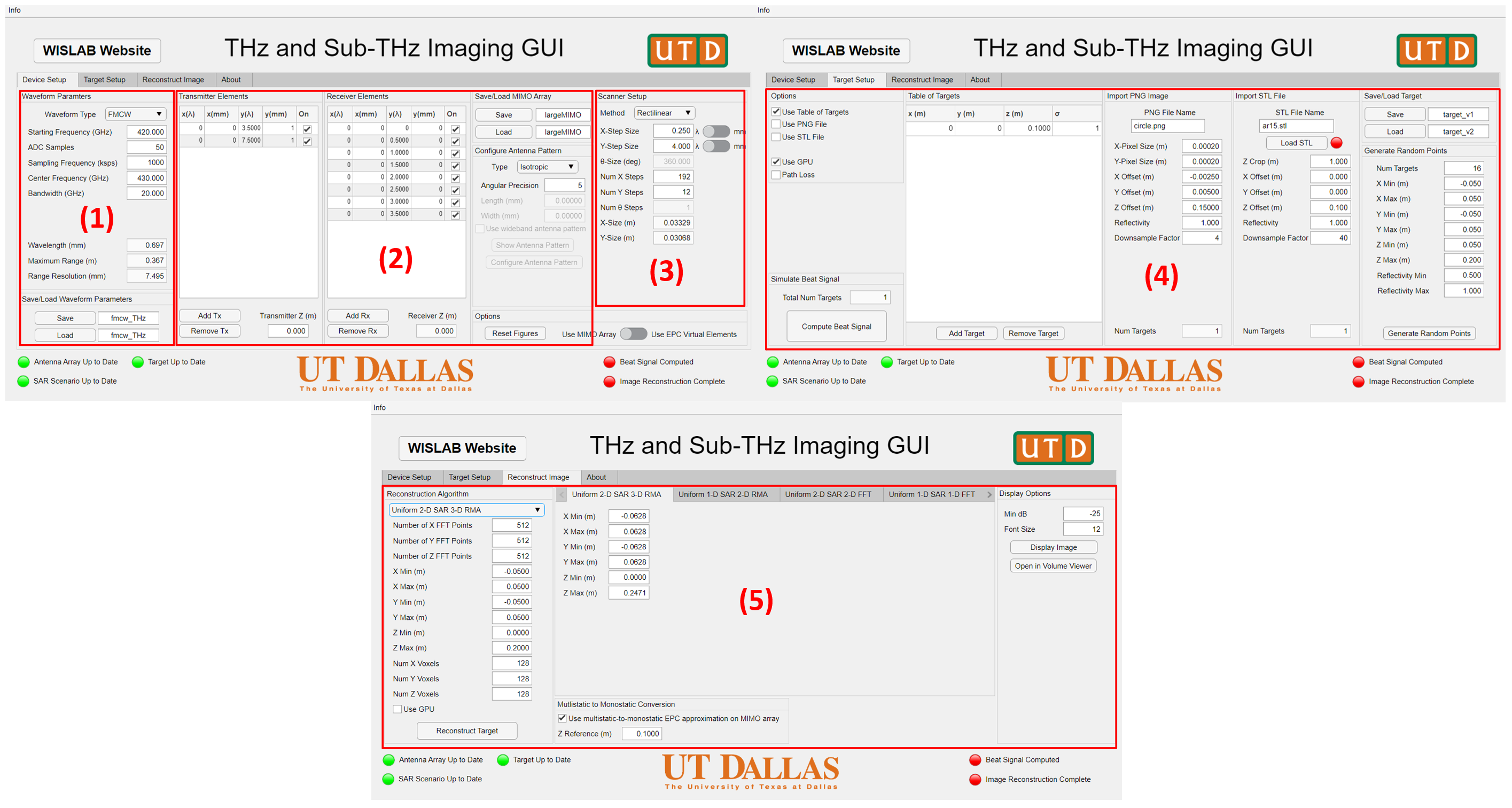} 
    \caption{Five step process in the interactive GUI.}
    \label{fig:gui_steps}
\end{figure}

The toolbox comprises five primary modules: (1) setting up the desired waveform parameters, (2) creating the MIMO antenna array, (3) configuring the scanning pattern, (4) constructing the target scene and simulating the beat signal, and (5) reconstructing and displaying the image. 
The functionality required for each step is encapsulated in five MATLAB classes for the toolbox and displayed within the three tabs of the interactive GUI, as shown in Fig. \ref{fig:gui_steps}. 
Each step is discussed in detail in the following subsections.
Further details on each step of the simulation process and the underlying methodology of the various classes can be found in the documentation of the toolbox, as discussed in Appendix \ref{app:documentation}. 

\subsection{Setting the Waveform Parameters}
\label{subsec:fmcw}
To set up the waveform parameters in the interactive GUI, navigate to the ``Device Setup'' tab (see (1) in Fig. \ref{fig:gui_steps}) and enter the desired parameters.
Although the FMCW modulation scheme is the default waveform, the user can easily employ alternatives such as OFDM or PMCW \cite{roos2019radar,zhang2015ofdm,schweizer2018stepped,sarieddeen2020next,bourdoux2016PMCW}.
As the values are changed by the user in the GUI, the dependent parameters (wavelength, maximum unambiguous range, range resolution, etc.) are automatically updated based on the user-defined fields. 
The descriptions each parameter are detailed below and in the documentation. 

For use with the toolbox API without the GUI, an object of class \texttt{THzWaveformParameters} is first initialized. 
Once the object has been constructed, the user can set the desired chirp parameters by changing the properties of the object, whose names are given below. 
It is important to note that the values given to the parameters of the \texttt{THzWaveformParameters} object must be in units of s, Hz, and m and must be scaled appropriately rather than the inputs to the interactive GUI, which are given in different units.

\begin{itemize}
    \item Starting Frequency (GHz): $f_0$/\texttt{f0} - Starting frequency of the chirp signal in GHz.
    \item ADC Samples: $N_k$/\texttt{Nk} - Number of samples by the ADC, corresponding to the number of frequency steps in the FMCW chirp.
    \item Sampling Frequency (ksps): $f_S$/\texttt{fS} - Sampling frequency of the beat signal in ksps.
    \item Center Frequency (GHz): $f_C$/\texttt{fC} - Center frequency of the device, used to calculate the wavelength $\lambda = c/f_C$, in GHz.
    \item Bandwidth (GHz): $B$/\texttt{B} - Device bandwidth in GHz.
\end{itemize}

Hence, the user has complete control over the waveform, including important parameters such as the number of samples and bandwidth, with immediate feedback to relevant parameters such as the range resolution and maximum resolvable range. 
In addition, the underlying software is thoroughly documented and straightforward to allow the user to implement custom waveforms of many varieties for simulation and image reconstruction.
For system design, having effortless access to such a tool allows for the virtual prototyping of many systems without the need for full hardware prototype construction.
Alternatively, the interactive GUI and toolbox both provide an option to save the current waveform parameters for later reuse or reference.

\subsection{Setting Up the Antenna Elements and Array}
\label{subsec:ant}
The second step is to configure an antenna array with transmitters and receivers at the desired locations and optionally use a nonideal antenna pattern. 
The interactive GUI allows the user to easily set the desired locations of the Tx and Rx elements of the array (see (2) in Fig. \ref{fig:gui_steps}). 
The MIMO array elements are assumed to be coplanar in an $x$-$y$ plane, whose $z$-coordinate can be specified in the GUI. 
The $x$ and $y$ positions of each element are set as follows: 
For example, in the $x$ direction, the value in the column labeled $x(\lambda)$ is scaled by the wavelength, as calculated above and added to the value in the column labeled $x(\text{mm})$ to compute the $x$ position of each antenna element. 
This functionality is due to the common practice of spacing the antennas by a factor of the wavelength as a result of the spatial sampling criteria \cite{yanik2019sparse}. 
Additionally, the user has complete control over the topology of the radar device and can easily prototype many different array types with various targets and image reconstruction algorithms, as discussed later. 

As discussed in Chapter \ref{ch:signal_processing}, under certain conditions, it is desirable to treat a MIMO array as a virtual monostatic array whose elements are located at the midpoints of the MIMO transceiver pairs. 
The GUI allows the user to specify whether to use the multistatic MIMO array elements, which can later be approximated as a virtual monostatic array, or to continue the simulation using a monostatic simulation algorithm with ideal SISO elements at the equivalent phase center (EPC) locations of the MIMO array. 

\begin{figure}[h]
    \centering
    \includegraphics[width=0.65\textwidth]{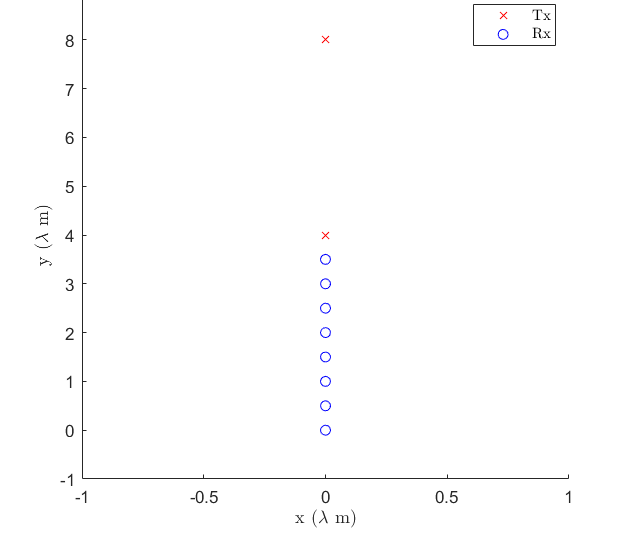} 
    \caption{Antenna array configuration of a 2 Tx, 8 Rx colinear MIMO array.}
    \label{fig:ant}
\end{figure}

As the user changes the location or status (On/Off) of the antenna elements, the interactive GUI again provides immediate feedback on the ``Graphical Window'' showing the array with either the Tx and Rx element or virtual monostatic transceiver elements (Vx), as shown in Fig. \ref{fig:ant}.

Optionally, the user can employ a nonideal antenna pattern.
By default, an ideal isotropic antenna pattern of uniform gain is employed to compute the beat signal.
However, using the interactive GUI, the user can select several built-in antenna patterns or import a custom antenna pattern from HFSS.
The two built-in options, Patch and Dipole antennas employ the MATLAB Antenna Toolbox and are limited to frequencies below 200 GHz.
Hence, the option is provided to import a custom antenna pattern simulated in HFSS for use in beat signal computation at THz frequencies.

The toolbox API employs a custom class called \texttt{THzAntennaArray} to compute the relevant antenna parameters from the user inputs. 
Once the \texttt{THzAntennaArray} object is instantiated, the user can set various parameters, discussed thoroughly in the documentation, determining whether the EPC Vx elements should be used or MIMO Tx/Rx elements (\texttt{isEPC}, a Boolean), the $z$ coordinate of the array (\texttt{z0\_m}), the antenna locations as an array in the same format as the GUI (\texttt{tableTx} and \texttt{tableRx}), and the antenna pattern properties. 
Documentation provides thorough instructions and illustrative examples of class usage, methods, and properties.
Again, both the interactive GUI and the toolbox allow the user to save the antenna array for later reuse or reference. 

\begin{figure}[h]
    \centering
    \includegraphics[width=0.55\textwidth]{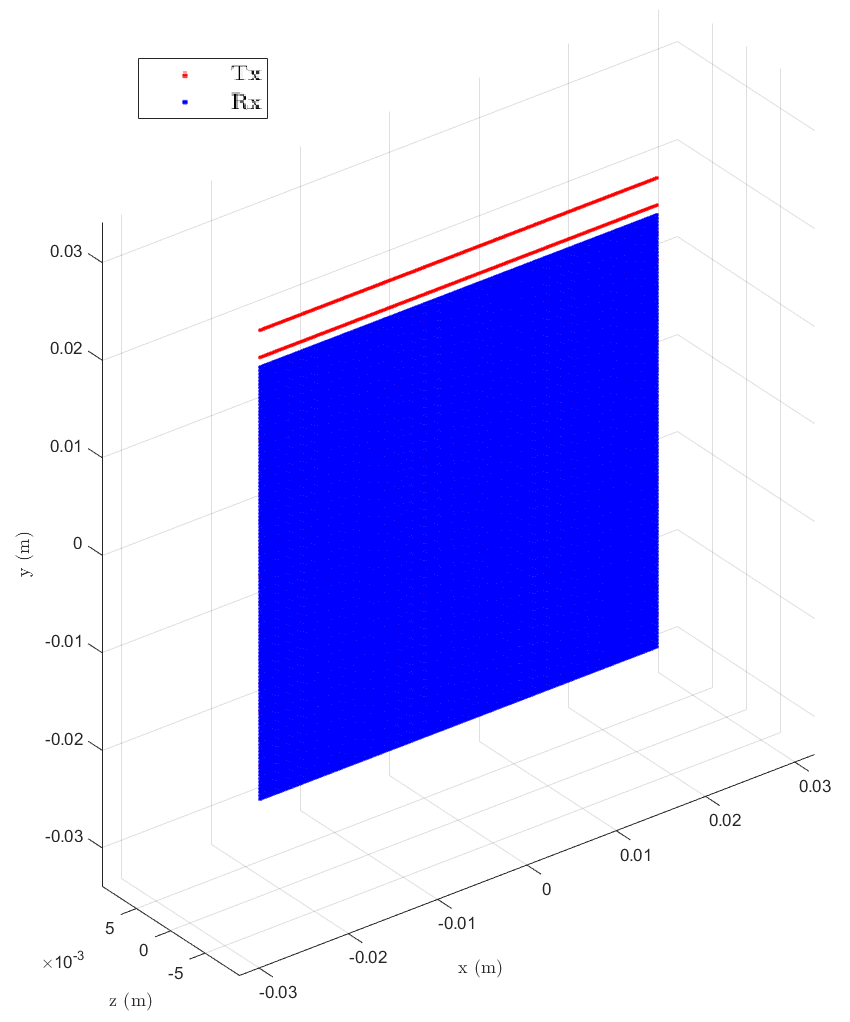} 
    \caption{Rectilinear scanning geometry with $\Delta_x = \lambda/4$, $\Delta_y = 4\lambda$, $N_x = 256$, $N_y = 16$ using the colinear MIMO array shown in Fig. \ref{fig:ant}.}
    \label{fig:scanner}
\end{figure}

\subsection{Configuring the SAR Scanning Scenario}
\label{subsec:scanner}
After the antenna array is configured, the scanning pattern can be configured using four different scanning modes: linear, rectilinear, circular, and cylindrical.
The user can select the desired scan type under the ``Method'' drop-down menub and the corresponding fields will be enabled or disabled accordingly depending on the scanning mode. 
The user has full control over the step size and the number of steps in the $x$ and $y$ directions, denoted previously as $\Delta_x$, $\Delta_y$, $N_x$, and $N_y$, respectively, in addition to the size of the angular scan $\theta_{max}$ and number of angular steps $N_\theta$. 
As the user changes the parameters in the GUI, the ``Graphical Window'' updates in real-time showing the current SAR scanning scenario, as shown in Fig. \ref{fig:scanner}.

Similarly, using the toolbox API requires initiating an object of the class \texttt{THzScanner}. 
Table \ref{tab:scanner_corr} shows the properties of the \texttt{THzScanner} class corresponding to the scanning parameters discussed previously. 
For more information, please refer to the documentation.
Once the user sets the desired SAR scanning parameters, the target scene can be constructed as follows.

\begin{table}[h]
    \centering
    \caption{SAR scanning parameters in the interactive GUI and their corresponding properties in the \texttt{THzScanner} class.}
    \begin{tabular}{c||c}
        Parameter in GUI & Property in Class \\ 
        \hline
        \hline
        Scanning Method & \texttt{method} \\
        \hline
        X-Step Size - $\Delta_x$ & \texttt{xStep\_m} \\
        \hline
        Y-Step Size - $\Delta_y$ & \texttt{yStep\_m} \\
        \hline
        $\theta$-Size (deg) - $\theta_{max}$ & \texttt{thetaMax\_deg} \\
        \hline
        Num X Steps - $N_x$ & \texttt{numX} \\
        \hline
        Num Y Steps - $N_y$ & \texttt{numY} \\
        \hline
        Num $\theta$ Steps - $N_\theta$ & \texttt{numTheta} \\
    \hline
    \hline
    \end{tabular}
    \label{tab:scanner_corr}
\end{table}

\subsection{Creating a Target Scene and Simulating the Beat Signal}
\label{subsec:target}
To create the target scene, the user must switch to the ``Target Setup'' tab (see (4) in Fig. \ref{fig:gui_steps}). 
As the user switches tabs, the interactive GUI automatically shows the corresponding figure in the ``Graphical Window.'' 
On the ``Target Setup'' tab, the user has three options for creating the target scenario: Table of Targets, Import PNG Image, Import STL File. 
The toolbox and GUI can create target scenes using any combination of these methods using the options on the left-hand side of this tab. 
Additionally, the user can choose to use a graphics processing unit (GPU) to compute the beat signal, which drastically decreases the computation time for large targets, and choose to include two-way path loss in the computation. 
As the user changes the parameters for any of the target methods, the target scene is updated and shown in real-time in the ``Graphical Window.''

\begin{figure}[h]
    \centering
    \includegraphics[width=0.95\textwidth]{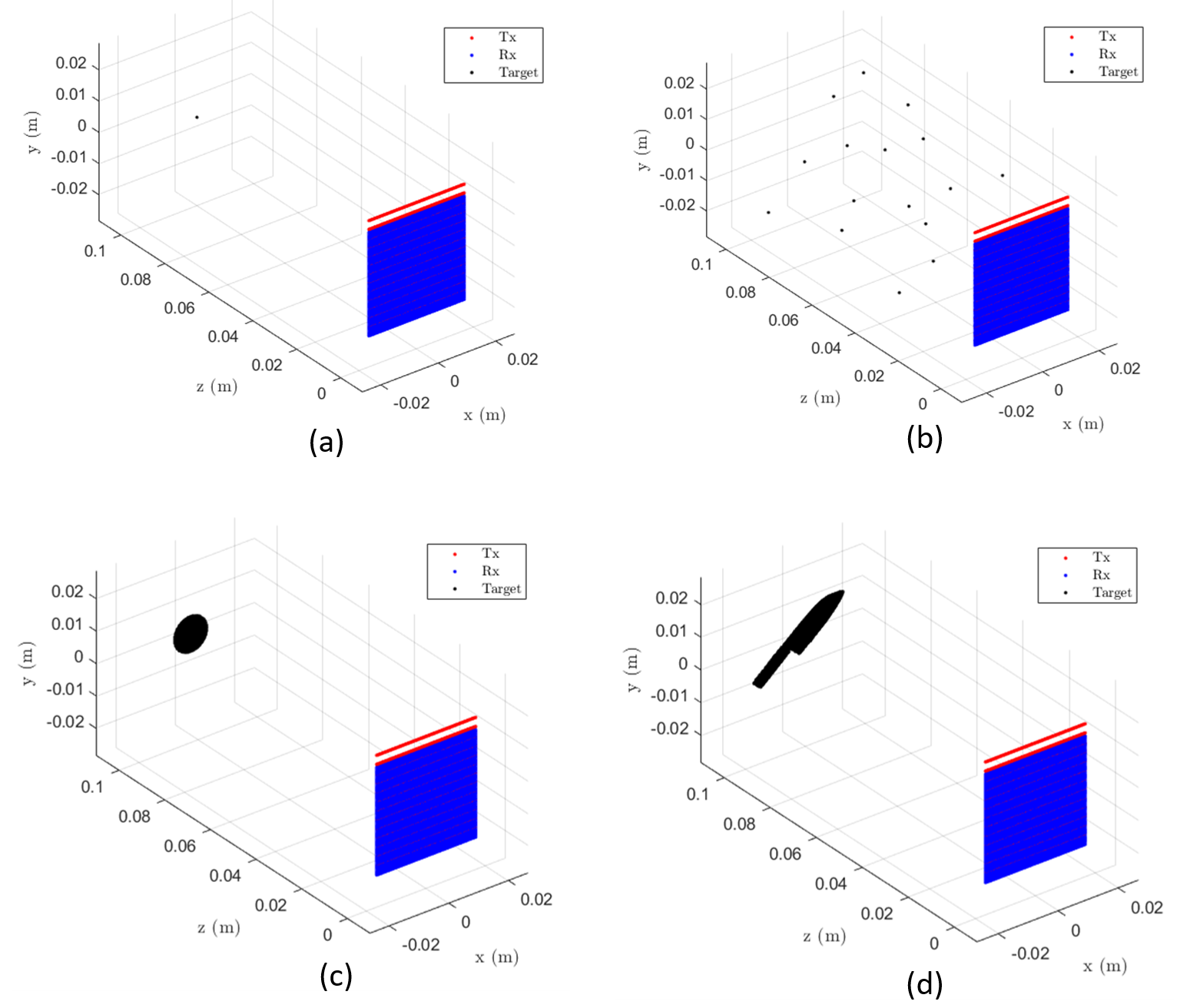} 
    \caption{Target scenes displayed by the interactive GUI with a rectilinear scanning pattern. a) Single point reflector located at $(0,0,0.1$ m$)$, 
    b) Random points generated by the GUI, c) \texttt{circle.png} with the default positioning parameters, d) \texttt{knife.stl} with the default positioning parameters.}
    \label{fig:rectilinearTargets}
\end{figure}

Using the Table of Targets method, the user can enter the \mbox{$x$-$y$-$z$} location and reflectivity of any number of desired points, which are treated as ideal point reflectors. 
Additionally, the user can generate randomly placed point targets that will be filled into the Table of Targets using the panel on the right; more information can be found in the documentation. 
Figs. \ref{fig:rectilinearTargets}a-b show two scenarios displayed by the interactive GUI using the Table of Targets method, the first with a user-entered point location and reflectivity, and the second with a randomly generated set of points given the user-specified parameters.

The user can also choose to load a PNG image, treated as a \mbox{2-D} target, in addition to or instead of the point targets entered into the Table of Targets. 
The user can import various images included in the toolbox or their own custom PNG files if desired. 
The PNG file is treated as a \mbox{2-D} $x$-$y$ image consisting of a set of point targets corresponding to the pixels of the PNG file, and the user has full control over the position and scaling of the PNG target. 
Additionally, the user can change the reflectivity of the PNG target and downsample the target to reduce the computation time of beat signal. 
For example, the included circle PNG file has been loaded with a downsampling factor of 1, meaning no downsampling, as shown in Fig. \ref{fig:rectilinearTargets}c. 

The third method for creating a target scene is to import a Standard Triangle Language (STL) file. 
An STL file is a standard file type for \mbox{3-D} computer-aided design (CAD) modeling. 
With the functionality to import \mbox{3-D} targets from STL files, our toolbox is capable of simulating highly customizable and complex target scenarios. 
The STL file is imported as a \mbox{3-D} target whose relative location, reflectivity, and downsampling factor can be specified by the user. 
Each vertex of the STL file is treated as an ideal point reflector, meaning that STL targets can comprise thousands to millions of point reflectors. 
Additionally, the user can specify the maximum value after which to crop the target along the $z$-direction. 
Given the geometry of the linear and rectilinear scans in this toolbox, cropping a dense \mbox{3-D} image after some $z$-plane is due to the rationale that the points after that $z$-plane will not be reflected back to the array as they are blocked by the portion of the target closer to the array. 
Fig. \ref{fig:rectilinearTargets}d shows the imaging scenario of a knife CAD model loaded into the toolbox as an STL file. 
The target scene can also be saved by the user in either the interactive GUI or the toolbox for reference or later reuse.

Generating a target scene using the toolbox API requires initializing a \texttt{THzTarget} object. 
The settings for each of the parameters in the GUI are laid out in detail in the documentation with examples but are omitted here due to length and repetition considerations. 

Simulation of the THz beat signal for various scanning regimes can be accomplished using a variety of means. 
To ease the computational load, we adopt a simple approach to simulate the MIMO and EPC-SISO beat signals. 
First, the target is discretized into a finite number of voxels, $N_{target}$, to be simulated independently as point reflectors using superposition. 
The locations of the MIMO array elements at each scanning iteration are computed using the \texttt{THzAntennaArray} and \texttt{THzScanner} objects, where $N_{s}$ is the number of synthetic elements in the array. 
Thus, the computation time can be reduced by precomputing the two-way distances between the antenna elements and the target voxels as an \mbox{$N_{s} \times N_{t}$} array $\mathbf{R}$, where $N_{s}$ is the number of antennas in the SAR array, and $N_{t}$ is the number of target voxels. 
Similarly, the discrete reflectivity function $p(x,y,z)$ is treated as a column vector $\mathbf{p}$ of size \mbox{$1 \times N_{t}$}. 
The antenna pattern is modeled using a complex weight matrix $\mathbf{W}$ of size \mbox{$N_s \times N_t \times N_k$} or \mbox{$N_s \times N_t$}.
The weight matrix is computed if the user employs a non-isotropic antenna; otherwise, it is ignored.
If the user employs a nonideal antenna pattern, the matrix $\mathbf{W}$ is computed by determining the angles between each antenna element and each target voxel.
Optionally, the user can simulate the antenna pattern across frequencies and $\mathbf{W}$ is computed for each frequency. 
The simulation can now be generalized for any scan type as
\begin{equation}
\label{eq:beat_sig_sim}
    s[n_{s},n_k] = \sum_{n_{t} = 1}^{N_{t}} \frac{\mathbf{p}_{1,n_{t}} \mathbf{W}_{n_{s},n_t,n_k}}{\mathbf{R}_{n_{s},n_{t}}} e^{-j (k_0 + \Delta_k n_k) \mathbf{R}_{n_{s},n_{t}}},
\end{equation}
where $n_{s}$ is the SAR antenna index, and $n_{t}$ is the target voxel index. 
Equation (\ref{eq:beat_sig_sim}) can be efficiently computed by vectorizing across $n_k$, $N_t$, or $n_{s}$. 
In most cases, $N_k << N_t,N_{s}$; thus, it is desirable to first attempt to vectorize across $n_t$, $n_{s}$ for computational efficiency, although it may be the most memory intensive.
In our investigations, the costliest operation, computing the complex exponential, can be computed quickly on a GPU because it is highly parallelizable. 

Other attempts to improve the computational complexity of this process include writing a custom implementation in a more efficient language, such as C/C++ or CUDA, or calculating the complex exponential more efficiently.
The latter method is employed in our implementation in MATLAB when no GPU is available. 
The procedure leverages the presence of any repeated values in the quantity $(k_0 + \Delta_k n_k) \mathbf{R}_{n_{s},n_{t}} (\text{mod} \ 2\pi)$ by finding unique elements and computing the complex exponential only at these values. 
However, finding the unique elements of the $\mathbf{R}$ matrix can be computationally prohibitive when $N_t$ and $N_{s}$ are large.
On the other hand, it is often more efficient than computing the entire complex exponential on the CPU for the same large target.

To perform the routine described above, the user can press the ``Compute Beat Signal'' button on the interactive GUI. 
During the simulation, the user will be provided with a dialog box showing the simulation progress and estimated time to completion. 
Once the simulation is complete, the ``Beat Signal Computed'' lamp will be green, indicating the simulation success. 
If the user changes the target scene, SAR scanning scenario, antenna array, or waveform parameters, the indicator will return to red, as the current imaging scenario has not been simulated. 
If the user is using the toolbox API, this process is equivalent to calling the \texttt{Compute()} method of the \texttt{THzTarget} object, e.g \texttt{target.Compute()}. 
Similarly, a dialog box appears, showing the progress and estimated time until completion.

Creating a target scene using the imaging and simulation toolbox is both flexible and efficient, allowing the user to import target scenes in three different manners and efficiently compute the beat signal. 
Thus, our freely-available toolbox provides easy access to many researchers wanting to learn more about near-field THz and sub-THz imaging domains, prototype a system design, develop custom image reconstruction algorithms, or programmatically generate large imaging datasets.

\subsection{High-Resolution Image Reconstruction}
\label{subsec:im}
The final step in the imaging process is to recover the image from the simulated signal. 
Numerous image reconstruction algorithms have been investigated in the literature  \cite{sheen2016three,sheen2001three,sheen2018simulation,yanik2020development,zhuge2010sparse,gao2016efficient,smith2020nearfieldisar,yanik2019sparse,gao2018_1D_MIMO,gao2018cylindricalMIMO,yanik2019cascaded,jaeschke20143d}.
Our toolbox includes 11 built-in documented image reconstruction algorithms for the four SAR scanning modes.
The provided algorithms have been made public to increase the accessibility of the implementation to the research community. 
In this subsection, we focus on using the interactive GUI and toolbox API to execute the included image reconstruction algorithms.

After navigating to the ``Reconstruct Image'' tab in the included GUI (see (5) in Fig. \ref{fig:gui_steps}) the user can enter the desired parameters in the panel to the left and select the desired image reconstruction algorithm from the drop-down menu. 
The application automatically verifies the scan type and parameters to ensure that the selected reconstruction algorithm can be applied to the simulated imaging scenario. 
The GUI may inform the user of errors in the image reconstruction parameters, allowing the user to change the parameters to recover the desired image. 
If the user has selected to use the MIMO array rather than the monostatic-equivalent SISO virtual array, there is an option to use the multistatic-to-monostatic approximation detailed in (\ref{eq:mult-to-mono}) at a specific $z$-reference plane, as discussed in Chapter \ref{ch:signal_processing}. 
Some parameters on the left are unused for some image reconstruction algorithms. 
For example, the back-projection algorithm (BPA) does not compute a fast Fourier transform (FFT) and thus does not use the first three fields containing the number of FFT points along the $x$-, $y$-, and $z$-dimensions. 
The remaining parameters refer to the reconstructed image domain, which is referred to as the imaging domain, specifically its bounds and the number of voxels along each dimension. 

For the included RMA, BPA, and FFT reconstruction algorithms, the theoretical limits on each imaging domain dimension are shown to inform the user before image reconstruction. 
Once the desired parameters are entered, the user can reconstruct the image using the desired technique. 
The image reconstruction algorithm is executed, and a dialog box appear to display the computation progress. 
As expected, any BPA reconstruction requires an exceptional amount of computation time, as discussed in \cite{soumekh1999synthetic,gao2018_1D_MIMO,gao2018cylindricalMIMO}. 
However, the efficient Fourier-based algorithms are capable of reconstructing images much more quickly. 
Once the image has been reconstructed, it is automatically displayed in the ``Graphical Window.'' 
The user also has the option of displaying the image again with a different threshold (in dB) by modifying the parameters on the right or opening the image in the MATLAB built-in Volume Viewer application. 

\begin{table}[ht]
    \centering
    \caption{Image reconstruction parameters in the interactive GUI and their corresponding properties in the \texttt{THzImageReconstruction} class.}
    \begin{tabular}{c||c}
    \hline
        Parameter in GUI & Property in Class \\ 
        \hline
        \hline
        Reconstruction Algorithm Name & \texttt{method} \\
        \hline
        Number of X FFT Points & \texttt{nFFTx} \\
        \hline
        Number of Y FFT Points & \texttt{nFFTy} \\
        \hline
        Number of Z FFT Points & \texttt{nFFTz} \\
        \hline
        X Min (m) & \texttt{xMin\_m} \\
        \hline
        X Max (m) & \texttt{xMax\_m} \\
        \hline
        Y Min (m) & \texttt{yMin\_m} \\
        \hline
        Y Max (m) & \texttt{yMax\_m} \\
        \hline
        Z Min (m) & \texttt{zMin\_m} \\
        \hline
        Z Max (m) & \texttt{zMax\_m} \\
        \hline
        Num X Voxels & \texttt{numX} \\
        \hline
        Num Y Voxels & \texttt{numY} \\
        \hline
        Num Z Voxels & \texttt{numZ} \\
        \hline
        Use GPU & \texttt{isGPU} (Boolean) \\
        \hline
        Multistatic-to-Monostatic Conversion & \texttt{isMult2Mono} (Boolean) \\
        \hline
        Z Reference (m) & \texttt{zRef\_m} \\
        \hline
        Min dB & \texttt{dBMin} \\
        \hline
        Font Size & \texttt{fontSize} \\
    \hline
    \hline
    \end{tabular}
    \label{tab:im_corr}
\end{table}

Using the toolbox API, a \texttt{THzImageReconstruction} object must be initialized. 
Table \ref{tab:im_corr} summarizes the properties of the \texttt{THzImageReconstruction} class and their corresponding parameters in the interactive GUI. 
Further discussions and examples of the \texttt{THzImageReconstruction} class are provided in the documentation. 
To compute the image reconstruction algorithm, the user must simply call the \texttt{Compute()} method of the \texttt{THzImageReconstruction} class, e.g. \texttt{im.Compute()}. 
The user can then display the image by calling the \texttt{Display()} method with the option of setting the minimum threshold in dB as the property \texttt{dBMin}.
Alternatively, the user can open the image in MATLAB Volume Viewer by calling \texttt{volumeViewer(im.imXYZ)}. 

Our open-source simulation and image reconstruction software offers a user-friendly experience for image reconstruction, providing the user with documented, efficient implementations of common image reconstruction algorithms and a seamless platform for algorithm comparison and development. 
Our toolbox streamlines the process for researchers by providing documented tools with proper interconnections for waveform parameter setup at any frequency, bandwidth, etc., creating any MIMO antenna array with an antenna gain pattern, utilizing four different scanning regimes, importing target scenes from a list of point targets, PNG image files, and high-quality CAD models, and image reconstruction. 
Thus, the entire simulation process is encapsulated into a simple tool designed for a host of applications.

\section{Overcoming Array Perturbation using Vision Transformers}
\label{sec:hffh_vit}
In this section, we develop a novel super-resolution algorithm for near-field SAR under array perturbation. 
As 5G mmWave devices become increasingly affordable and available, high-resolution SAR imaging is feasible for end-user applications and non-laboratory environments. 
Emerging applications such as freehand imaging \cite{alvarez2021towards}, unmanned aerial vehicle (UAV) imaging \cite{garcia20203DSARProcessing}, and automotive SAR \cite{iqbal2021realistic} face several unique challenges for high-resolution imaging \cite{smith2022efficient,alvarez2021freehandsystem}.
First, recovering a SAR image requires knowledge of array positions throughout the scan. 
Although recent work has introduced camera-based positioning systems capable of adequately estimating the position, recovering the algorithm efficiently is a requirement to enable edge and IoT technologies.
Efficient algorithms for near-field SAR have been developed; however, they suffer from image defocusing under position estimation error and can only produce medium-fidelity images when the array is perturbed.
In this section, we introduce a mobile-friend vision transformer (ViT) architecture to perform SAR image super-resolution and artifact mitigation under array position errors.
The proposed algorithm, Mobile-SRViT, is the first to employ a ViT approach for SAR image enhancement and is validated through simulation and empirical studies.

Traditional SAR imaging requires high-precision laboratory equipment for exact positioning of the antennas throughout the scan.
Efficient SAR imaging algorithms have been explored extensively in the literature \cite{sheen2001three,yanik2020development,smith2020nearfieldisar} leveraging the fast Fourier transform (FFT) to recover images from radar data. 
These efficient algorithms strictly require specific synthetic aperture geometries, such as planar \cite{sheen2001three,yanik2020development}, cylindrical \cite{smith2020nearfieldisar}, etc. to achieve high-resolution imaging. 
However, with the emergence of 5G and IoT technologies, near-field SAR sensing has received attention at both the system and algorithm levels \cite{alvarez2021towards,smith2022efficient}.
Although these applications operate at similar frequencies to traditional laboratory SAR \cite{yanik2019cascaded}, they suffer from two primary constraints: 1) the resulting synthetic aperture has some errors in the estimated position of each array element, and 2) because the image computation typically takes place on a low-power or mobile device, the computational load must be reduced compared to conventional imaging. 
Consequently, recovering high-fidelity images under such conditions remains an open challenge.

Previous research efforts \cite{smith2022efficient,alvarez2021freehandsystem} do not consider errors in position estimation present in practical implementations \cite{alvarez2021towards}.
In \cite{yanik2020development}, an extensive investigation into practical system design challenges was presented, and a method was developed for synchronizing SAR equipment for highly precise positioning. 
However, sophisticated scanners are infeasible for many emerging applications and positioning is degraded owing to noise in positioning or estimation. 
Such array perturbations cause defocusing and distortion of SAR images as the imaging algorithms improperly compute the matched filter weights based on noisy position estimates. 
Without knowledge of the exact positions, removing the distortion present in SAR images from the position errors remains an open challenge.
For many practical systems, these algorithms can efficiently reconstruct only medium-fidelity images. 

Recently, deep learning approaches for optical image super-resolution have been extended to the radar domain for SAR image super-resolution \cite{gao2018enhanced,dai2021imaging,jing2022enhanced,smith2021An}.
Using convolutional neural network (CNN) architectures, previous efforts have been successful in improving SAR resolution \cite{gao2018enhanced,jing2022enhanced} and removing multistatic artifacts \cite{dai2021imaging}.
However, these techniques operate on SAR images collected using traditional techniques in laboratory environments, and do not address the issues that arise because of erroneous array perturbations explored in this study.
Nevertheless, deep learning has seen tremendous success in both the optical domain, for image restoration \cite{liang2021swinir} and super-resolution \cite{lim2017enhanced}, and on conventional SAR images \cite{gao2018enhanced,dai2021imaging,jing2022enhanced}.
Hence, deep learning may be a suitable solution for near-field SAR position error artifact mitigation and super-resolution. 

Recent advances in computer vision have seen a shift from CNN-based architectures towards the attention mechanism \cite{vaswani2017attention} using Vision Transformer (ViT) techniques \cite{dosovitskiy2020image_ViT} to achieve performance gains with smaller model sizes \cite{mehta2021mobilevit,sandler2018mobilenetv2,liu2021swin}. 
In \cite{mehta2021mobilevit}, the MobileViT architecture is presented leveraging a transformer architecture for image classification. 
Later, the transformer architecture was employed for optical image super-resolution and artifact mitigation \cite{liang2021swinir}. 
Transformer techniques have appeared in recent work on radar image classification \cite{dong2021exploring} and gesture recognition \cite{zheng2021dynamic}; however, transformers have yet to be employed for SAR image super-resolution. 
In this section, we introduce a novel transformer-based architecture for SAR image super-resolution under array perturbation, called Mobile-SRViT. 
The proposed algorithm operates on images recovered by the range migration algorithm (RMA) \cite{yanik2020development} and produces high-fidelity images of intricate targets. 
We validate our mobile-friendly algorithm using simulation and empirical data from a near-field SAR scenario with positioning errors. 

\begin{figure}[ht]
    \centering
    \includegraphics[width=0.35\textwidth]{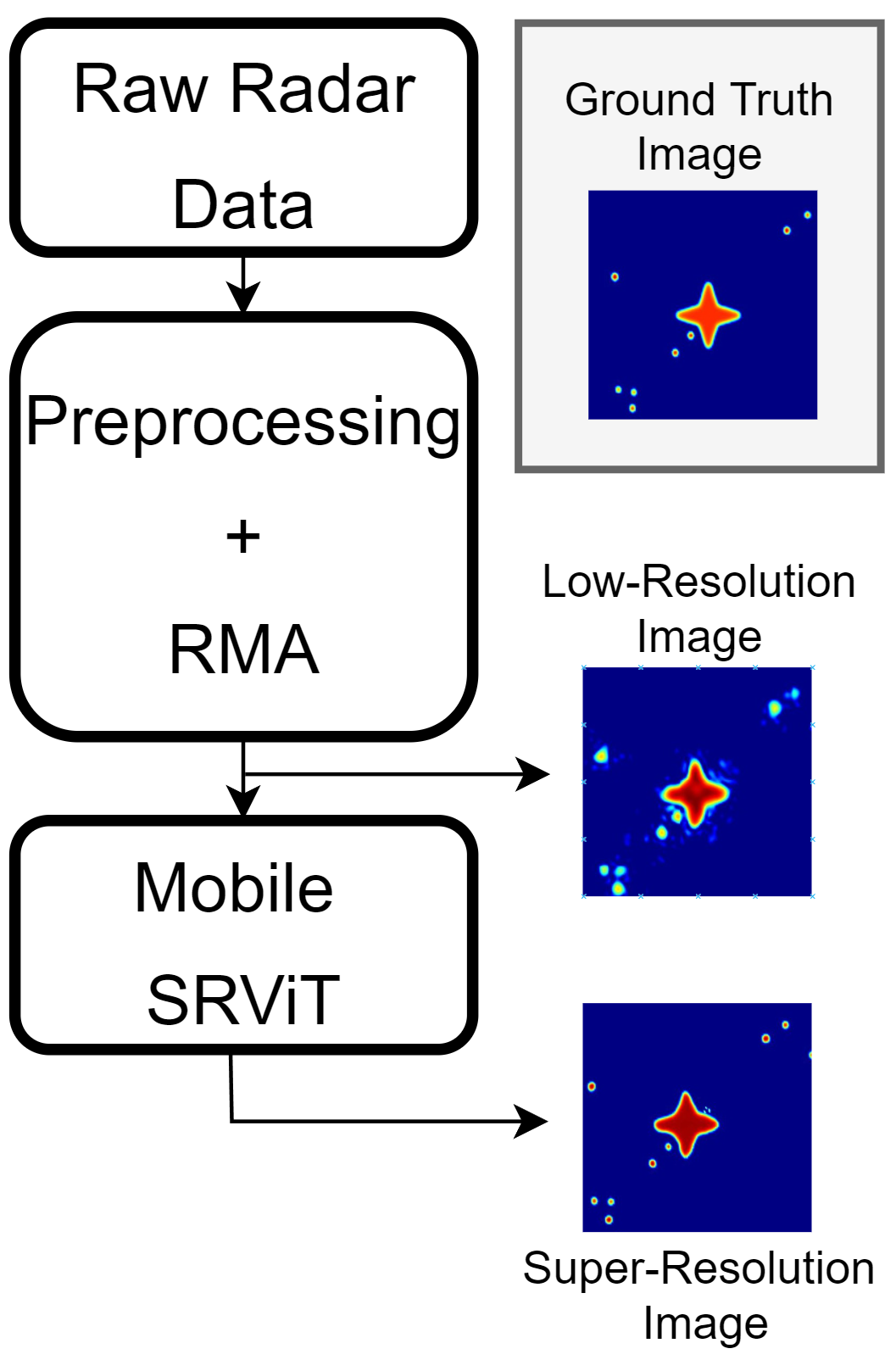}
    \caption{Operation of the Mobile-SRViT: the low resolution image produced by the RMA is restored by the Mobile-SRViT algorithm. The ground truth image is shown for reference.}
    \label{fig:hffh_vit_flow}
\end{figure} 

\subsection{ViT-Based Near-Field SAR Super-Resolution}
\label{sec:hffh_vit_methods}
In this section, we detail the proposed transformer-based approach for near-field SAR super-resolution and artifact mitigation. 
An overview of the Mobile-SRViT algorithm is shown in Fig. \ref{fig:hffh_vit_flow}.
The raw radar data are preprocessed to a planar monostatic scenario, and the RMA is applied to produce a low-resolution image with distortion, blur, and defocusing caused by imaging non-idealities.
The proposed Mobile-SRViT operates on this image to produce a super-resolution image and restore the image quality, while preserving the intricate high-frequency details of the targets.

\begin{figure}[ht]
    \centering
    \includegraphics[width=\textwidth]{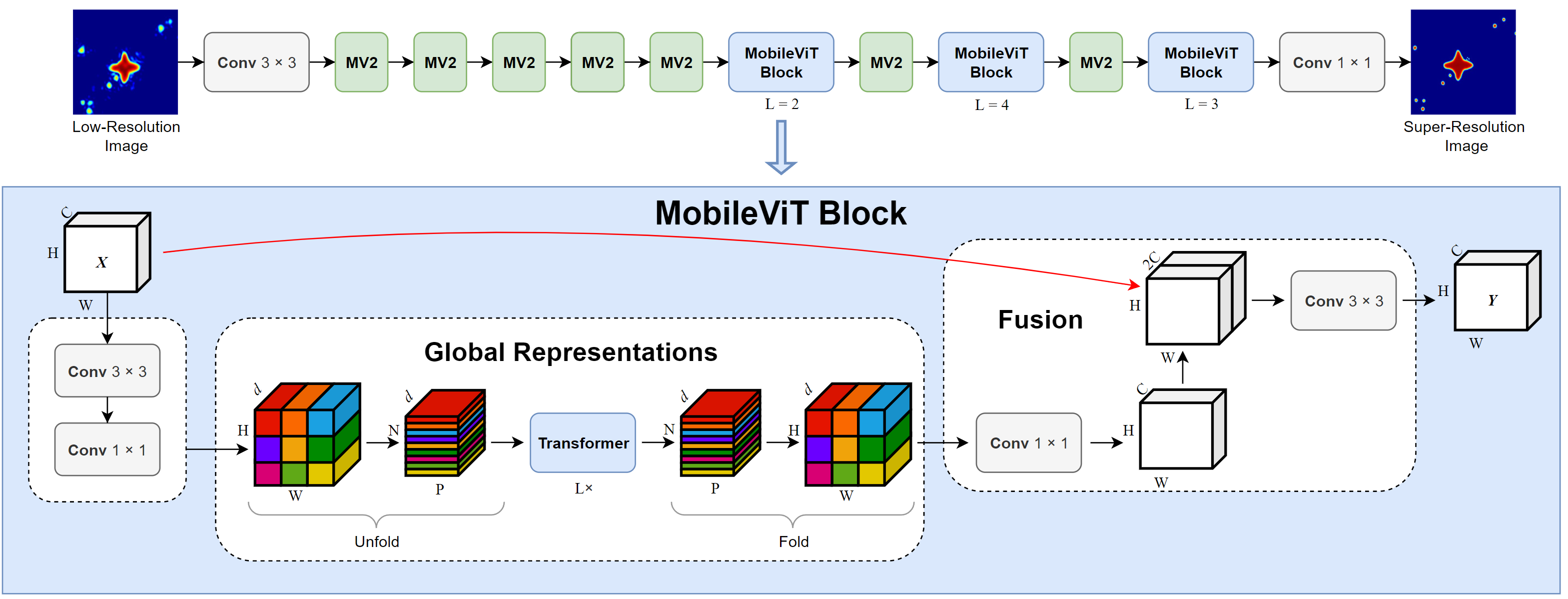}
    \caption{Mobile-SRViT architecture.}
    \label{fig:hffh_vit_net}
\end{figure}

The Mobile-SRViT architecture is based on the MobileViT network employed for image classification \cite{mehta2021mobilevit}. 
Because our network is designed for image-to-image processing, the convolution layers are modified to adhere to a fully convolutional neural network (FCNN) framework, similar to the network proposed in \cite{smith2021An}.
Fig. \ref{fig:hffh_vit_net} shows the implementation of the Mobile-SRViT algorithm, where ``MV2'' refers to the MobileNetV2 block proposed in \cite{sandler2018mobilenetv2}.
Our algorithm adopts the approach of \cite{mehta2021mobilevit} such that the image is first processed by several MobileNetV2 convolution blocks before alternating between MobileViT and MobileNetV2 operations. 
The MobileViT block is intended to model the global and local information of the input data with fewer parameters than the traditional ViT \cite{dosovitskiy2020image_ViT}. 
Given an input tensor $\mathbf{X} \in \mathbb{R}^{H \times W \times C}$, where $H$ is the height, $W$ is the width, and $C$ is the number of channels, the MobileViT block first applies a $3 \times 3$ convolution layer, followed by a $1 \times 1$ or pointwise convolution layer to produce a tensor $\mathbf{X}_L \in \mathbb{R}^{H \times W \times d}$. 
The last $1 \times 1$ convolution layer reduces the number of channels to match the input image.
To learn global representations, $\mathbf{X}_L$ is unfolded into $N$ non-overlapping patches $\mathbf{X}_U \in \mathbb{R}^{P \times N \times d}$, where $P = 4$ and $N = HW/4$.
Each of the $P$ patches is processed using a transformer architecture to encode the inter-patch relationships, yielding 
\begin{equation}
    \label{eq:transformer_mobilevit}
    \mathbf{X}_G(p) = \text{Transformer}(\mathbf{X}_U(p)), \quad p \in [1, \dots, P].
\end{equation}
Whereas most ViT implementations lose the positional location of each patch \cite{dosovitskiy2020image_ViT,liu2021swin}, the MobileViT retains the patch order and pixel order within each patch. 
Consequently, $\mathbf{X}_G \in \mathbb{R}^{P \times N \times d}$ can be directly folded to obtain $\mathbf{X}_F \in \mathbb{R}^{H \times W \times d}$. 
The resulting tensor $\mathbf{X}_F$ is projected onto a low $C$-dimensional space via a $1 \times 1$ convolution layer before being concatenated with $\mathbf{X}$, yielding $\mathbf{X}_O \in \mathbb{R}^{H \times W \times 2C}$.
Finally, using a $3 \times 3$ convolution, $\mathbf{X}_O$ is fused to form an output tensor $\mathbf{Y}$ of identical size to $\mathbf{X}$.
Interestingly, the receptive field of the MobileViT block is $H \times W$ because $\mathbf{X}_U(p)$ encodes local information from a $3 \times 3$ region via convolutions, and each pixel in $\mathbf{X}_G(p)$ encodes global information over $P$ patches \cite{mehta2021mobilevit}. 
Our implementation maintains $C = 16$ until the last MobileViT block, where $C = 32$, $d = 2C$ for each MobileViT block, and $L = \{2, 4, 3\}$. 
The images are of size $256 \times 256$, and the patch size employed is $16 \times 16$.
With this architecture, the proposed Mobile-SRViT has 69,122 parameters.
Loss is computed using the pixel-to-pixel L1 metric as
\begin{equation}
\label{eq:L1_loss}
    L_{p2p} = ||\mathbf{X}_{SR} - \mathbf{X}_{HR}||_1.
\end{equation}
Prior attempts at near-field SAR super-resolution have been purely CNN-based \cite{gao2018enhanced,dai2021imaging,jing2022enhanced,smith2021An}; however, the Mobile-SRViT detailed in this section is the first to leverage a transformer architecture for SAR imaging. 

\subsubsection{Training the Mobile-SRViT}
\label{subsubsec:hffh_vit_training}
The proposed algorithm was trained for 50 epochs using an ADAM optimizer on a single RTX3090 GPU with 24 GB of memory using 4096 samples for the training process and 1024 samples for evaluation. 
Training data were generated using the procedure detailed in \cite{gao2018enhanced,smith2022efficient,smith2021An}.
Each sample was generated with additive white Gaussian noise (AWGN) with a signal-to-noise ratio (SNR) in the range $[-10, 50]$ dB and includes AWGN positioning errors with a standard deviation of 1 mm along the $x$-, $y$-, and $z$-directions to emulate a practical scenario. 
The training process lasted approximately 6 h with an inference time of $10$ ms per sample during validation. 

\subsection{Mobile-SRViT Experimental Results}
\label{subsec:hffh_vit_results}
In this section, we conduct simulation and empirical experiments to verify the proposed algorithm.
To evaluate the performance of the SRViT algorithm, we first compare the numerical performance of the Mobile-SRViT with the BPA and RMA. 
The gold-standard BPA has no requirements for SAR array geometry and is well suited for any scanning geometry.
However, it is computationally prohibitive, particularly for mobile applications, as it computes the pixel-wise matched filter for every sampling location and frequency and is prone to distortion under array perturbations.
The RMA, on the other hand, is more efficient than the BPA by computing the matched filter using Fourier relations. 
The proposed Mobile-SRViT algorithm attempts to compensate for the distortion present in the RMA images owing to imaging non-idealities. 

\begin{table}[h]
\caption{Quantitative performance of the Mobile-SRViT compared to the BPA and RMA.} 
\centering
  \begin{tabular}{ c||c|c|c|c } 
    Metrics & Mobile-SRViT & {BPA} & {RMA} \\
    \hline \hline
    PSNR (dB) & $\mathbf{37.608}$ & $28.73$ & $27.94$\\ 
    \hline
    RMSE & $\mathbf{0.017}$ & $0.057$ & $0.061$\\
    \hline
    Time (s) & $1.113$ & $1324.8$ & $\mathbf{1.103}$\\
    \hline \hline
  \end{tabular}
\label{table:hffh_vit_performance}
\end{table}

Using a test dataset consisting of 1024 samples similar to those in the training dataset but never seen by the network, we apply the Mobile-SRViT, BPA, and RMA to the samples and measure the peak signal-to-noise ratio (PSNR), root-mean-square error (RMSE), and computation time per sample.
The results are presented in Table \ref{table:hffh_vit_performance}, where the best evaluations marked in boldface.
All experiments were conducted on a desktop PC equipped with a 12-core AMD Ryzen 9 3900X running at 4.6 GHz with 64 GB of memory.
The RMA achieves an efficient computation time of 1.103 s but falls short of the BPA in terms of RMSE and PSNR.
However, the BPA boasts the highest PSNR and lowest RMSE of the classical algorithms but requires a significantly large amount of computation time, 1324 s.
The Mobile-SRViT is superior to the other algorithms, even outperforming the BPA in PSNR and RMSE, with a total computation time of 1.113 s required to compute the RMA and pass the image through the network. 
This qualitative analysis demonstrates the superiority of the proposed method in comparison with previous techniques in terms of both computational efficiency and image quality.

\begin{figure*}[ht]
\centering
    \begin{subfigure}[b]{0.3\textwidth}
         \centering
         \includegraphics[width=\textwidth]{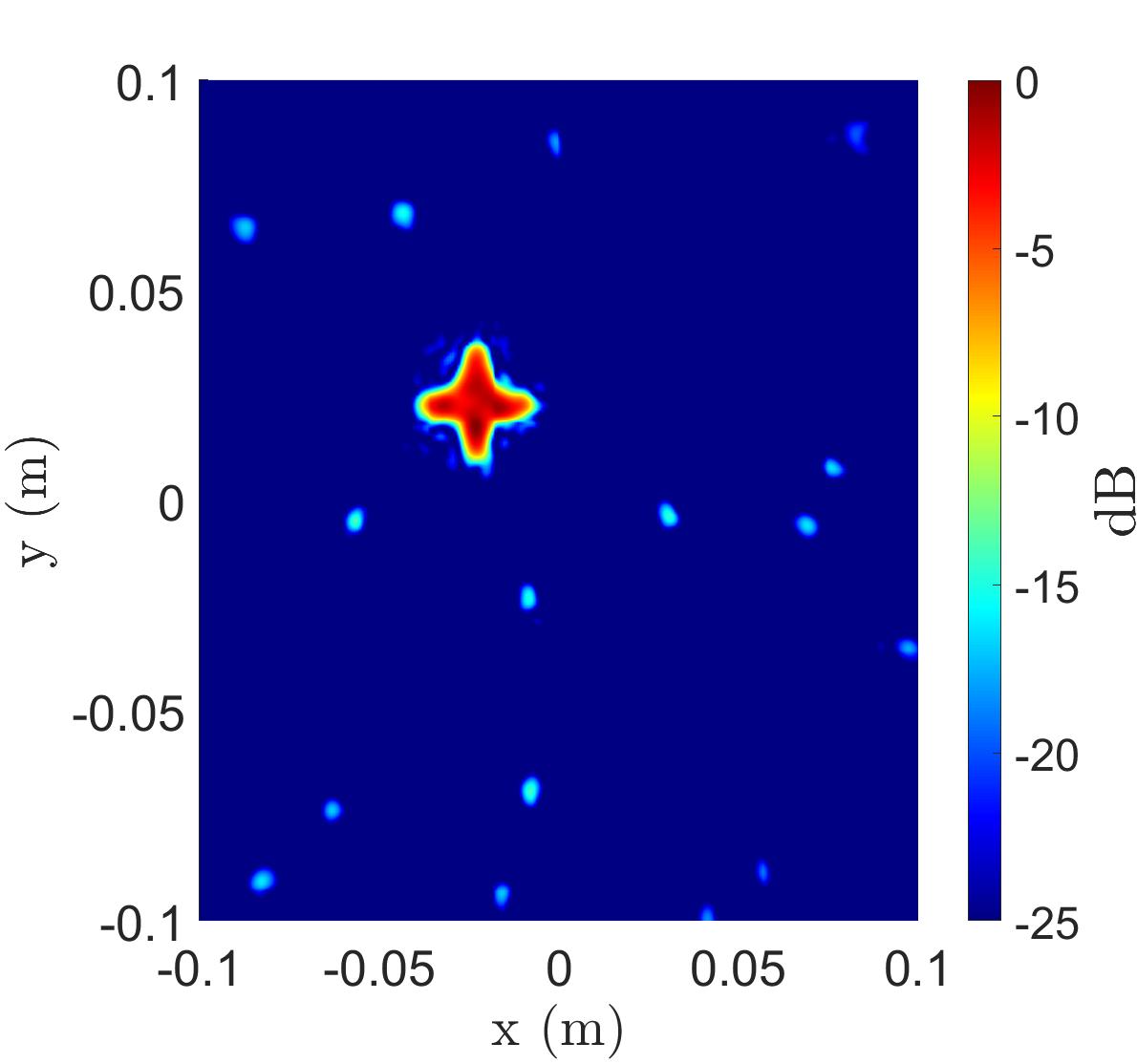} 
         \caption{}
         \label{fig:test270_lr}
    \end{subfigure}
    \begin{subfigure}[b]{0.3\textwidth}
         \centering
         \includegraphics[width=\textwidth]{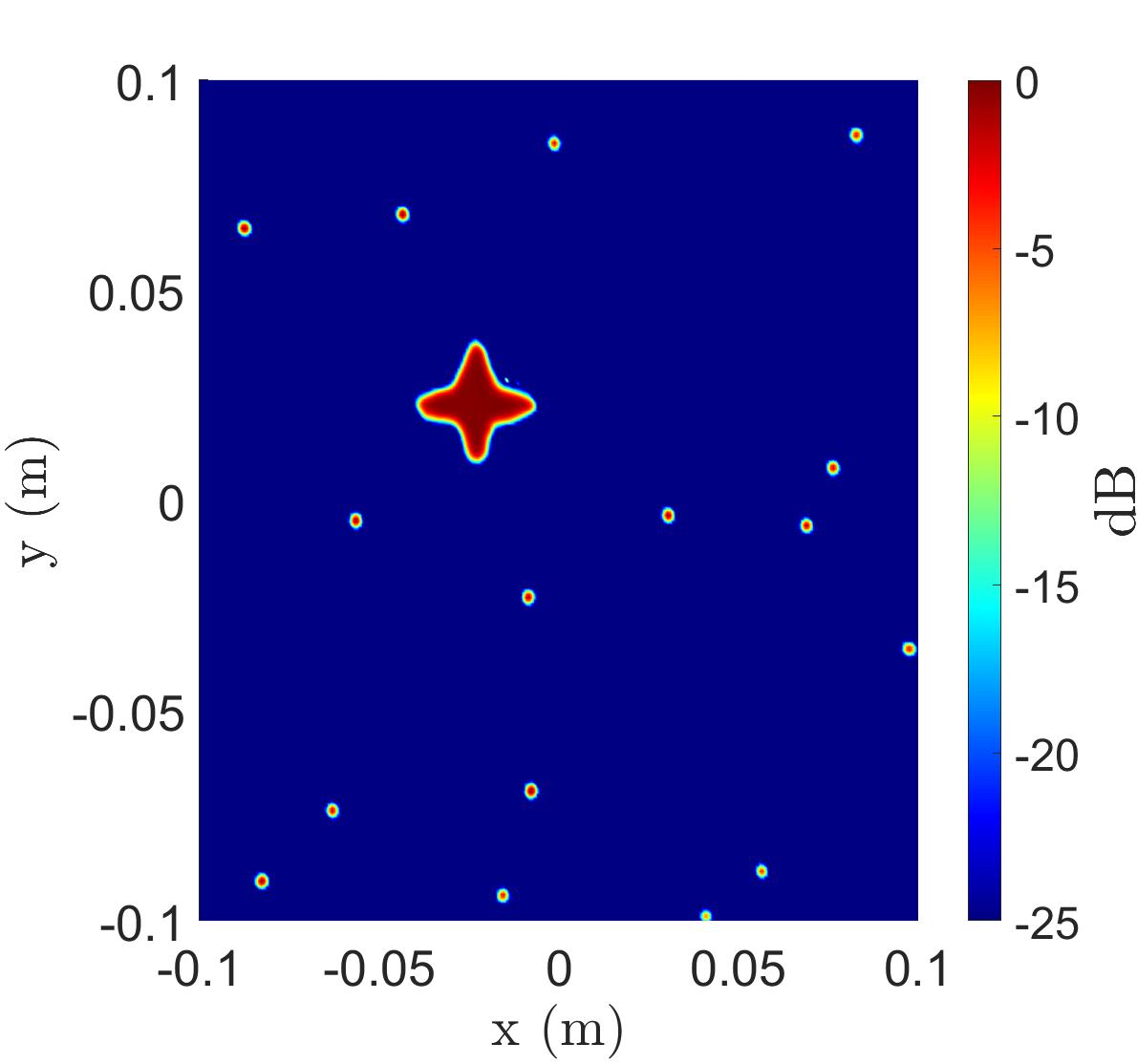} 
         \caption{}
         \label{fig:test270_sr}
    \end{subfigure}
    \begin{subfigure}[b]{0.3\textwidth}
         \centering
         \includegraphics[width=\textwidth]{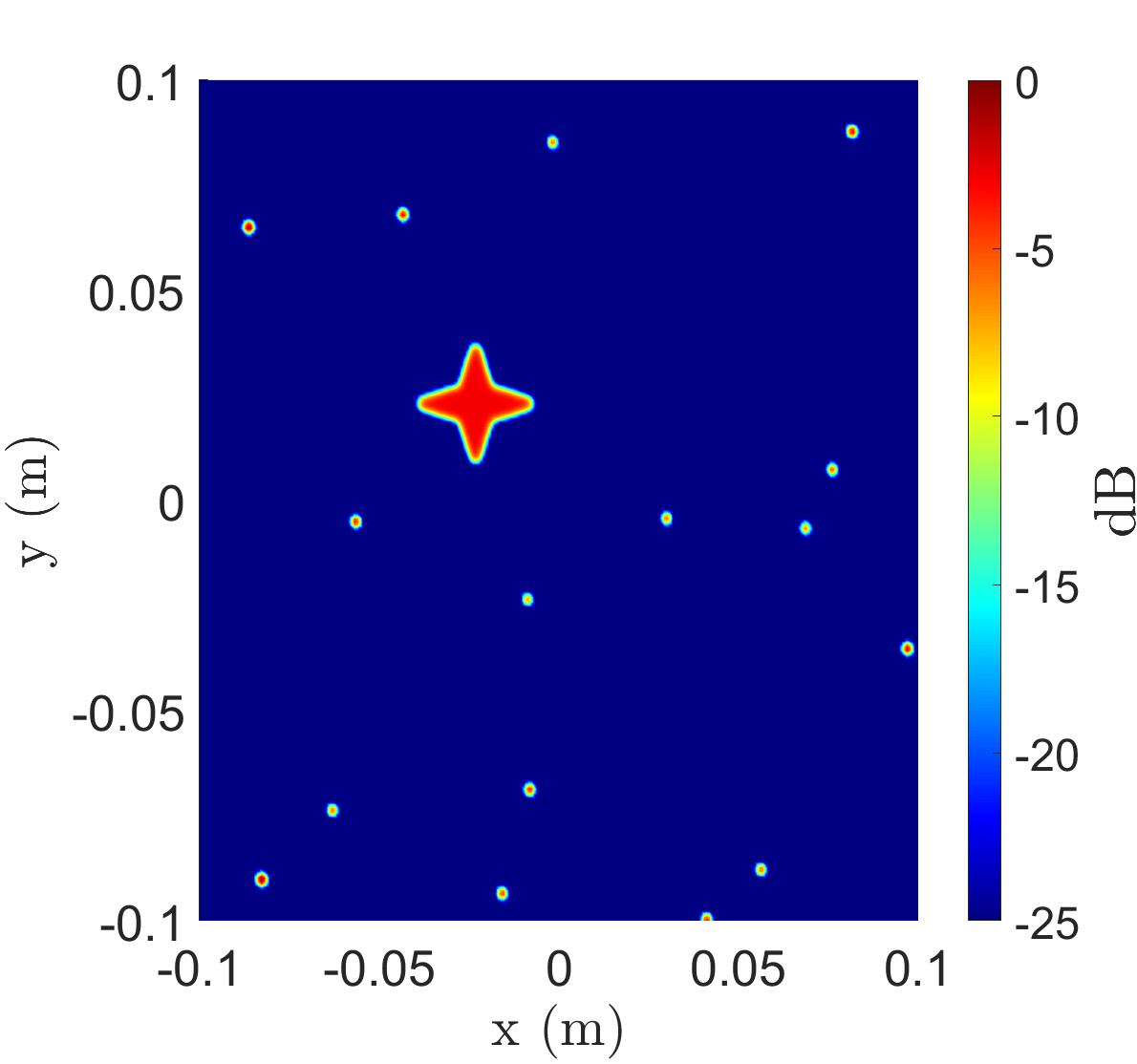} 
         \caption{}
         \label{fig:test270_hr}
    \end{subfigure}
    \vskip\baselineskip
    \begin{subfigure}[b]{0.3\textwidth}
         \centering
         \includegraphics[width=\textwidth]{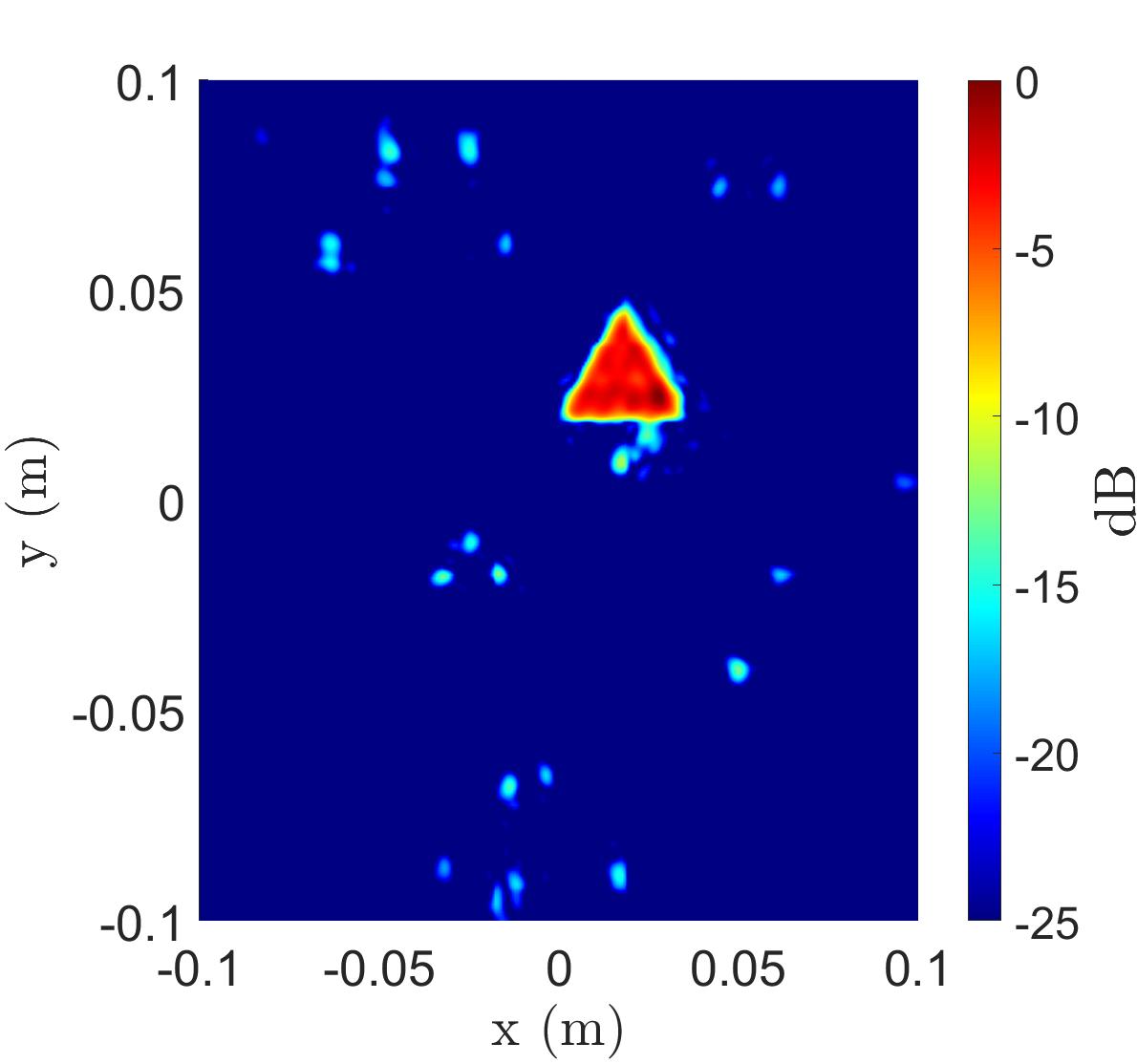} 
         \caption{}
         \label{fig:test707_lr}
    \end{subfigure}
    \begin{subfigure}[b]{0.3\textwidth}
         \centering
         \includegraphics[width=\textwidth]{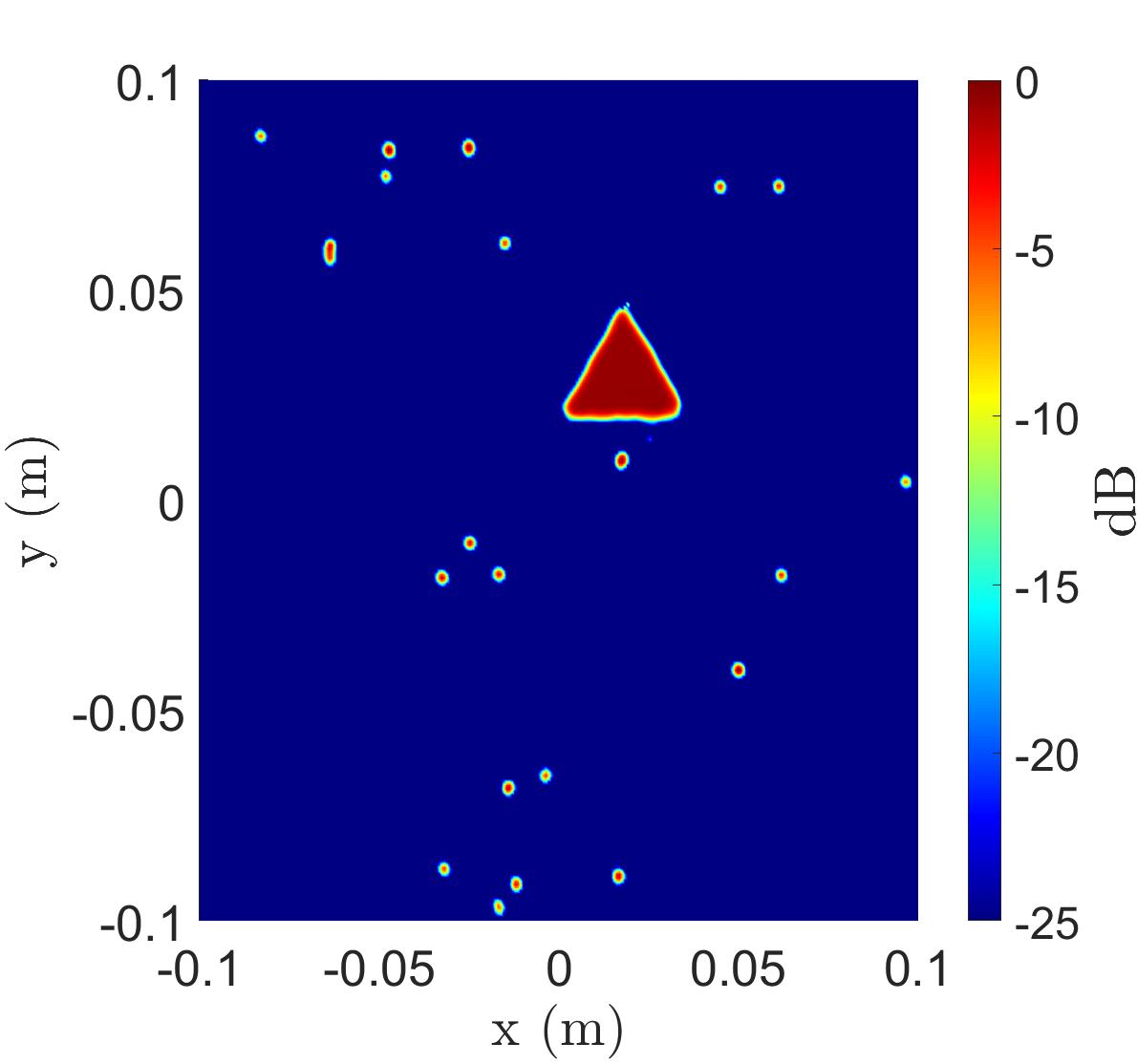} 
         \caption{}
         \label{fig:test707_sr}
    \end{subfigure}
    \begin{subfigure}[b]{0.3\textwidth}
         \centering
         \includegraphics[width=\textwidth]{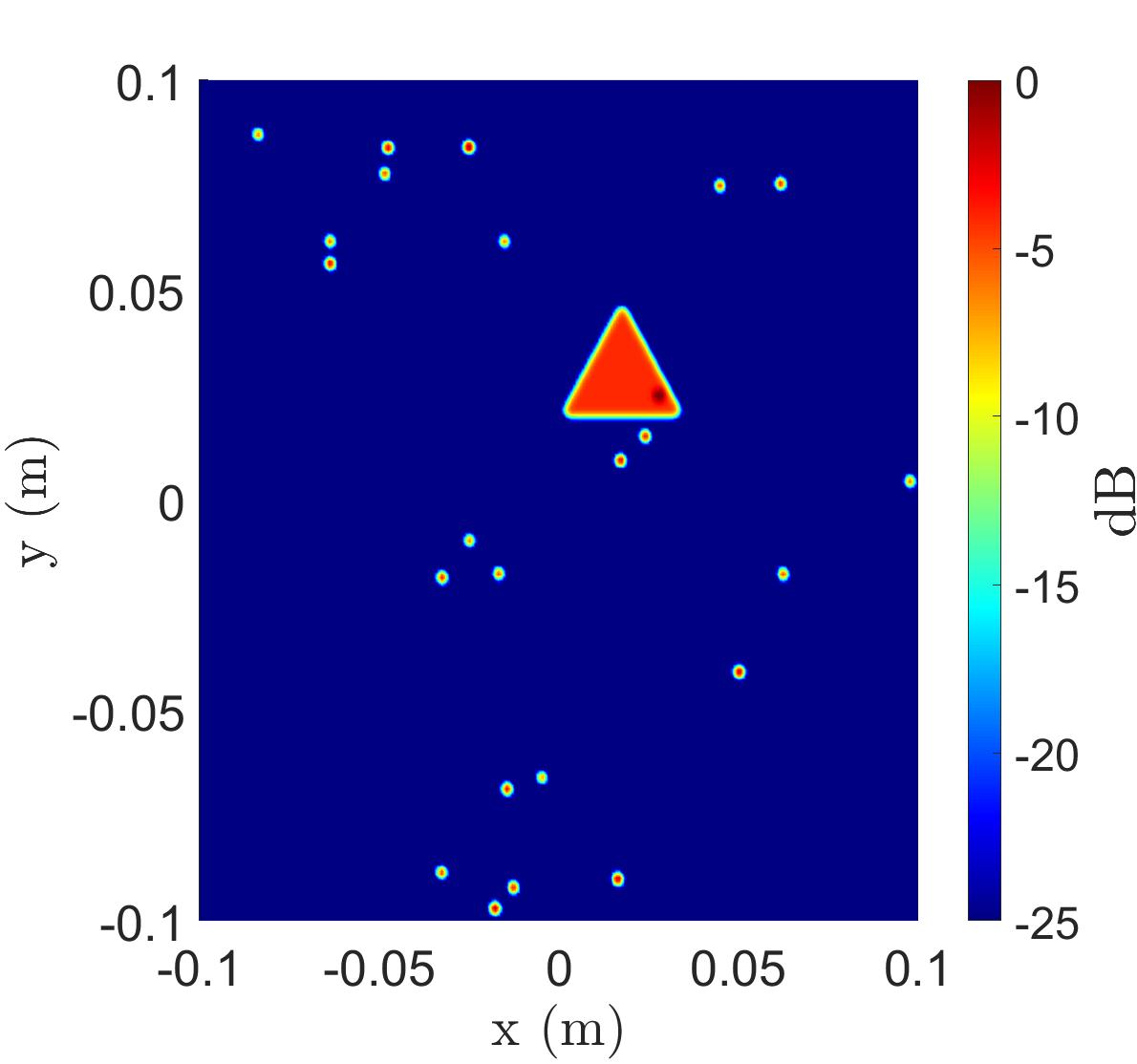} 
         \caption{}
         \label{fig:test707_hr}
    \end{subfigure}
\caption{Imaging results using the Mobile-SRViT on synthetic data. The images in the first column, (a) and (d), are produced by the RMA and input to the Mobile-SRViT. The images in the second column, (b) and (e), are the super-resolution images output from the Mobile-SRViT. The images in the third column, (c) and (f), are the ground truth images.}
\label{fig:hffh_vit_results_sim}
\end{figure*}

We further validate the performance of the proposed algorithm via visual inspection of both the simulated and empirical data.
Two samples from the testing dataset are compared in Fig. \ref{fig:hffh_vit_results_sim}.
For each sample, the proposed SAR super-resolution network recovers the solid object in addition to the point scatterers and mitigates the distortion caused by position estimation errors and system limitations.
The super-resolution images (Figs. \ref{fig:test270_sr} and \ref{fig:test707_sr}) are quite similar to the ideal images (Figs. \ref{fig:test270_hr} and \ref{fig:test707_hr}) showing an improvement over the medium-fidelity images recovered by the RMA.

\begin{figure}[ht]
\centering
    \begin{subfigure}[b]{0.45\textwidth}
         \centering
         \includegraphics[width=\textwidth]{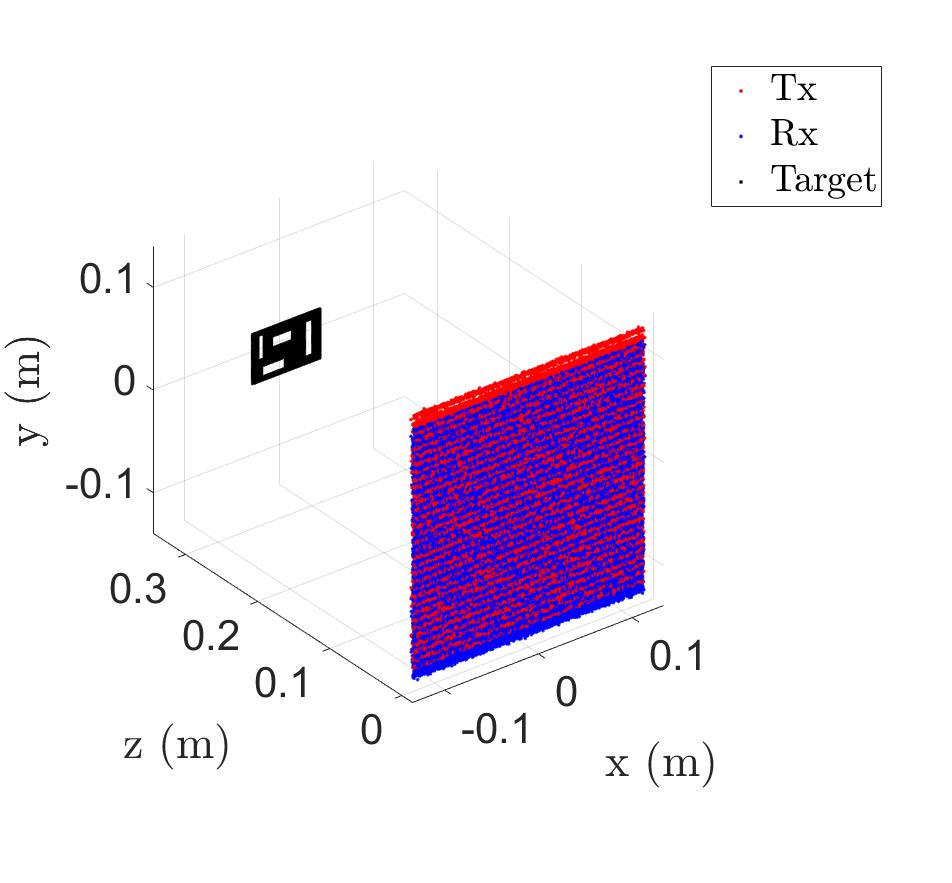}
         \caption{}
         \label{fig:perturbation_geometry_example}
    \end{subfigure}
    \vskip\baselineskip
    \begin{subfigure}[b]{0.3\textwidth}
         \centering
         \includegraphics[width=\textwidth]{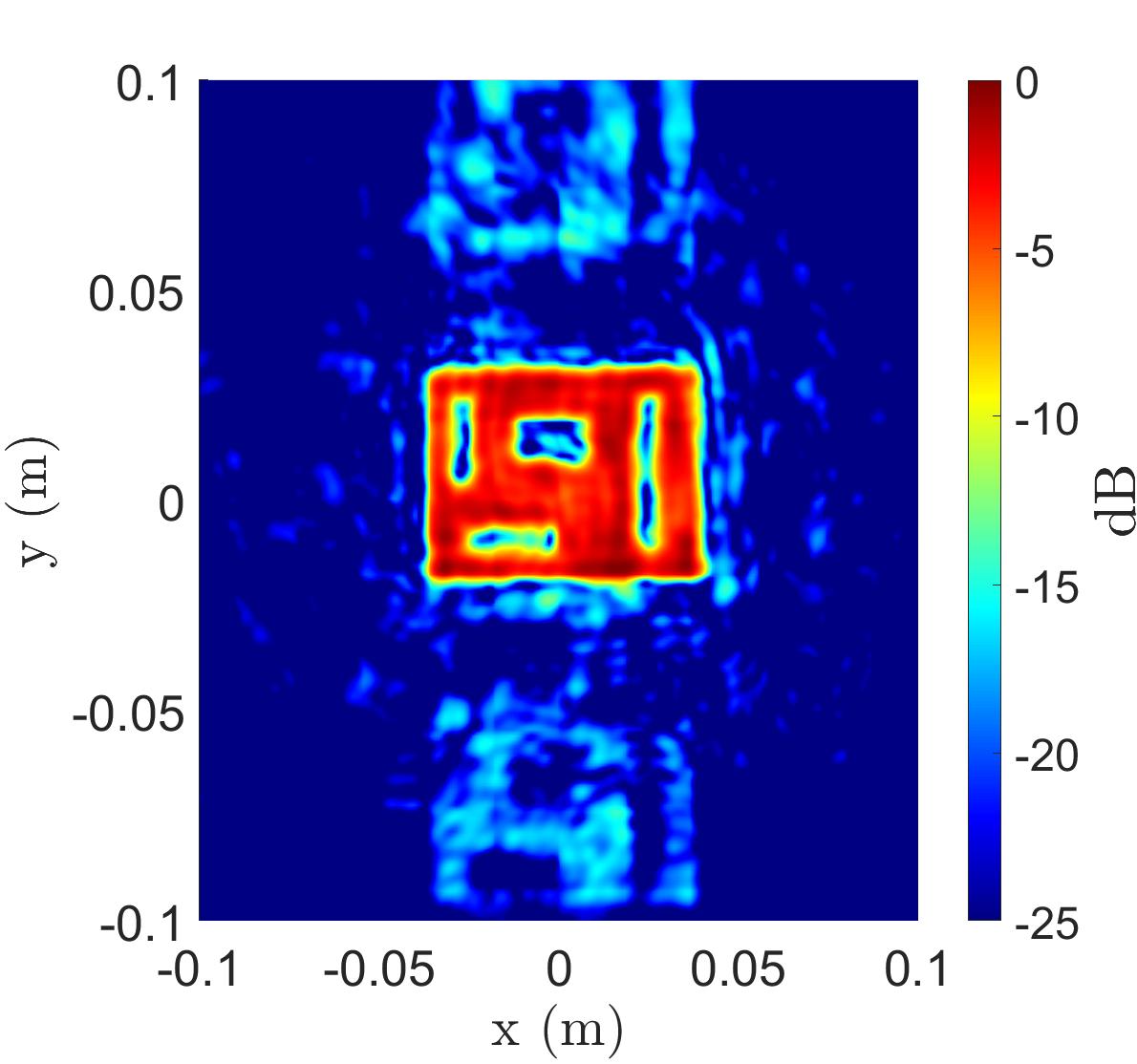}
         \caption{}
         \label{fig:exp1_lr}
    \end{subfigure}
    \begin{subfigure}[b]{0.3\textwidth}
         \centering
         \includegraphics[width=\textwidth]{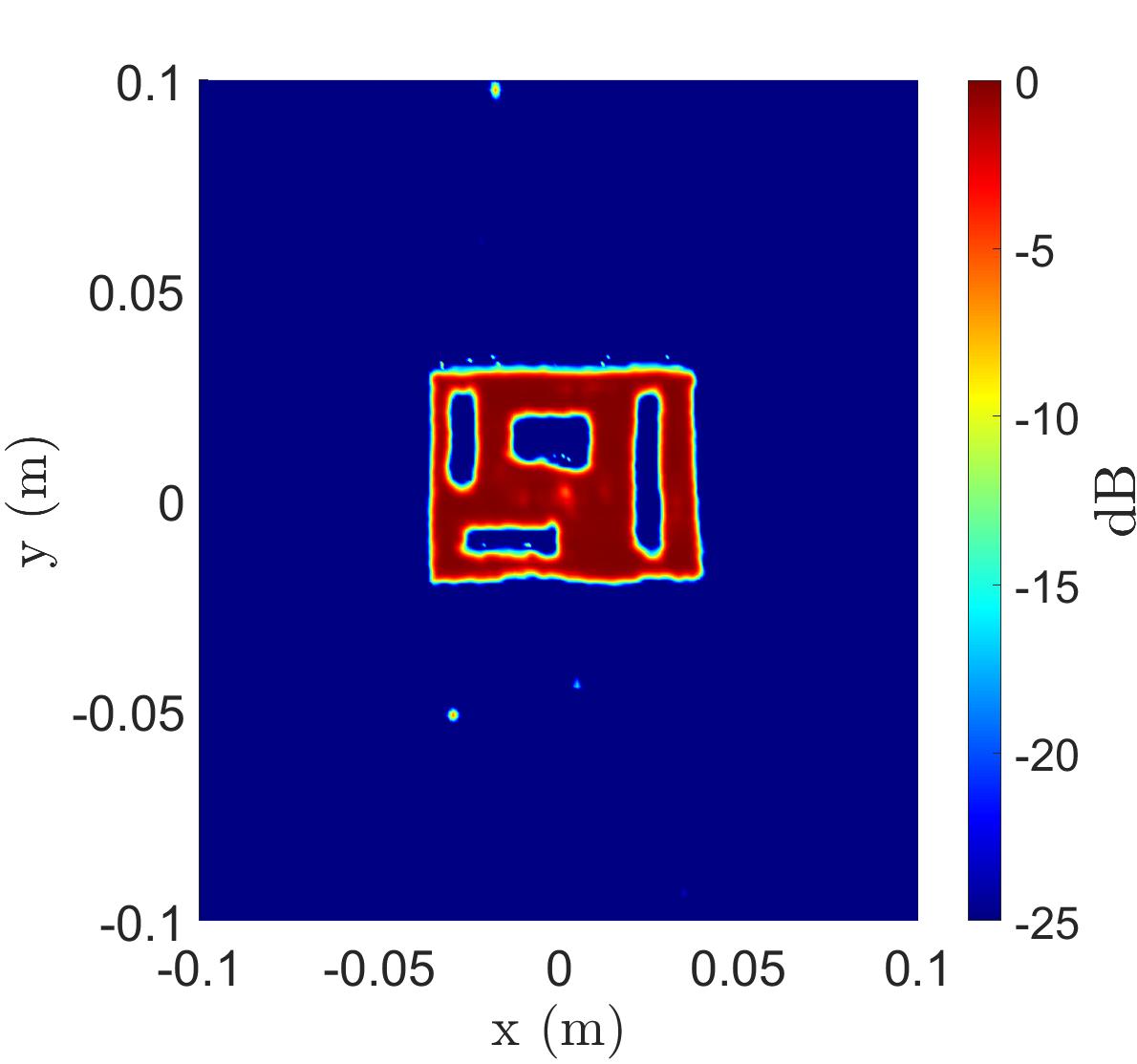}
         \caption{}
         \label{fig:exp1_sr}
    \end{subfigure}
\caption{Imaging results using the Mobile-SRViT on empirical data from a near-field SAR scenario with array perturbations: (a) SAR sampling geometry, (b) the image reconstructed by the RMA and (c) the super-resolution image produced by the Mobile-SRViT.}
\label{fig:hffh_vit_results_sim_real}
\end{figure}

To evaluate the performance of our proposed algorithm on empirical data, we first perform a SAR scan with array perturbations, as shown in Fig. \ref{fig:perturbation_geometry_example}.
After reconstructing the image with the RMA, as shown in Fig. \ref{fig:exp1_lr}, the Mobile-SRViT is applied to achieve the super-resolution image shown in Fig. \ref{fig:exp1_sr}.
The proposed algorithm not only recovers a better-resolved image but also mitigates the multistatic artifacts visible in the RMA image \cite{yanik2019sparse,smith2022efficient}. 

\section{Lightweight Deep Learning SAR Super-Resolution for Irregular Scanning Geometries}
\label{sec:hffh_cnn}
Extending the EMPM preprocessing algorithm proposed in Section \ref{sec:ffh} for efficiently recovering MIMO-SAR images from irregular scanning geometries, we propose a novel CNN-based super-resolution architecture suitable for mobile applications.
The EMPM suffers from increased sidelobes and distortion when data are collected across a large volume.
Additionally, the results in Section \ref{sec:ffh} assume that the array positions are known without error; however, in the case of position estimation errors, which are common for realistic implementations \cite{alvarez2021towards}, the image is further defocused for both the BPA and EMPM. 
In this section, we address the position estimation errors by introducing a novel data-driven super-resolution algorithm suitable for mobile near-field SAR imaging.
Whereas previous CNN-based techniques for near-field SAR super-resolution are constrained to regular geometries \cite{gao2018enhanced,wang2021tpssiNet,dai2021imaging,jing2022enhanced,cheng2020compressive,huang2019through}, we propose the first super-resolution algorithm designed for non-cooperative geometries.
Furthermore, the proposed super-resolution CNN (SRCNN) is suitable for use in computationally constrained mobile applications. 
Additionally, previous studies on SRCNNs for near field radar imaging are limited by their datasets of exclusively randomly generated point scatterers \cite{gao2018enhanced,jing2022enhanced}. 
Using the toolbox designed in the previous section, our algorithm is trained on sophisticated, realistic targets rather than simple point scatterers. 
The proposed algorithm, referred to as ``Mobile-SRGAN,'' is trained using a generative adversarial network (GAN) framework \cite{goodfellow2014generative} and utilizes an efficient depth-wise convolution-based U-Net architecture \cite{ronneberger2015unet} to obtain high fidelity SAR images with mobile-friendly computational complexity.

\begin{figure}[h]
    \centering
    \includegraphics[scale=0.6]{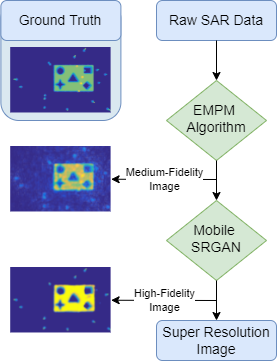}
    \caption{Flowchart of proposed design. After the raw SAR data are processed by the EMPM algorithm, the Mobile-SRGAN is applied to achieve a super-resolution image.}
    \label{fig:flowchart}
\end{figure}

The proposed algorithm is summarized in Fig. \ref{fig:flowchart}. 
The raw freehand SAR data are processed by the EMPM algorithm, yielding a medium-fidelity image that suffers from image distortion due to positional errors and the small array perturbation assumption in the EMPM.
The example image in Fig. \ref{fig:flowchart} consists of a solid object with cavities and randomly placed point targets. 
The proposed Mobile-SRGAN processes the distorted image reconstructed using the EMPM and outputs an enhanced super-resolution image.

The main contributions of this section are:

\begin{itemize}
\item[1)] Generation of near-field SAR training datasets with sophisticated targets and irregularly sampled geometries.
\item[2)] Development of an efficient SRCNN model on the complex shapes and realistic datasets.
\item[3)] Qualitative and quantitative evaluation of the SRCNN model on both real and synthetic images.
\end{itemize}

\subsection{Mobile Super-Resolution GAN}
\label{subsec:mob-srgan}

\begin{figure}[h]
    \centering
    \includegraphics[scale=0.2]{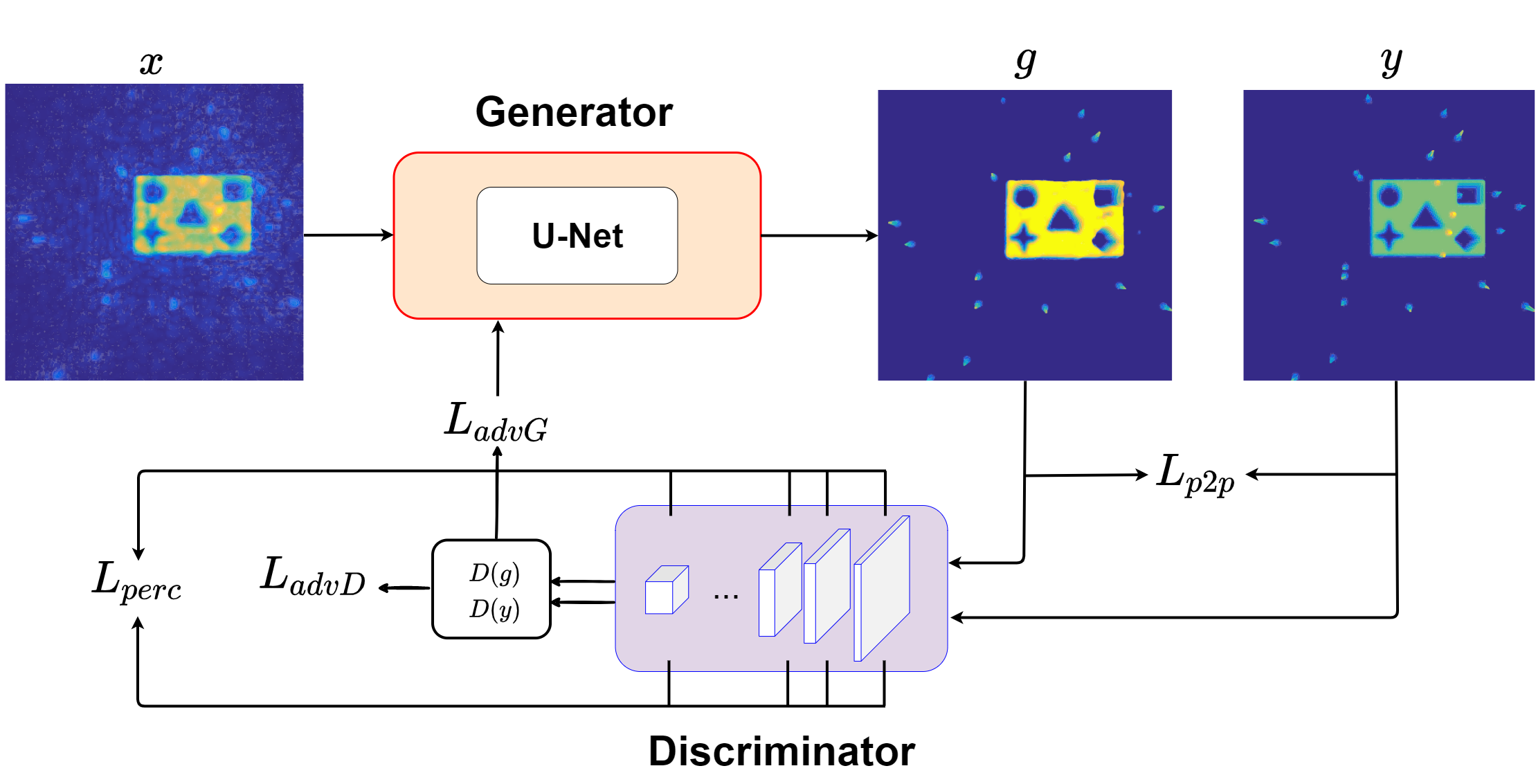}
    \caption{Mobile-SRGAN Overview}
    \label{fig:gan_overview}
\end{figure}

Using the reconstructed SAR images from the EMPM, we are tasked with mitigating the distortions in these images to generate super-resolution images that are reflective of the original target images. 
For this purpose, we propose a cGAN-based \cite{goodfellow2014generative,mirza2014conditional} Mobile-SRGAN framework that trains an encoder-decoder model to aggregate features for denoising, thereby generating a high quality SAR image. 
This framework consists of two CNNs: the Generator (G) and Discriminator (D), which work in tandem to optimize the model for achieving super-resolution. 
This particular training framework has previously been employed for medical MRI imaging \cite{Armanious2020MedGAN}, optical image super-resolution \cite{Armanious2019An_adversarial_SR} and image-to-image translation \cite{Wang2018PAN,isola2018imagetoimage}.
We adapt this architecture with depth-wise convolution operations to develop an efficient methodology that is suitable for mobile applications. 
The Mobile-SRGAN framework is illustrated in Fig. \ref{fig:gan_overview}.

The training dataset consists of synthetic (reconstructed) SAR images and their corresponding ideal images, generated using the methodology in \cite{gao2018enhanced,smith2021An}, provided to the generator-discriminator pair. 
The objective of the generator is to map the distribution of the low-resolution SAR image to the high-resolution ideal image by learning to bridge the gap between these distributions. 
This learning is guided by the discriminator, which is a classifier established to quantify the probabilistic differences between the ``clean'' image ($g$) produced by the generator and the ideal target ($y$). 
The generator attempts to produce enhanced images by transforming the reconstructed SAR image, which is then processed by the discriminator. 
This information is then fed back to the generator to guide the training towards the distribution of the ideal image. 
The network is trained in an adversarial manner, as shown in Algorithm \ref{alg:one}, until the discriminator is unable to distinguish between the ideal and clean images, implying that the generator has adequately learned to generate images akin to the high-resolution images.

\begin{algorithm}[h]
\DontPrintSemicolon
\caption{Training Process for Mobile-SRGAN}
\label{alg:one}
\KwData{$D_{train}, D_{test}$ \hfill\tcp{Train-Test Set}}
\myData{$x$, $g$ \hfill\tcp{in and out of G}}
\myData{$y$ \hfill\tcp{HR target images}}
\KwResult{Trained Model}
    \nl\For{$epoch \leftarrow 0 $ \KwTo $ MaxEpochs$}{
        \nl\For{$x \leftarrow 0 $ \KwTo $ batches $ in $ D_{train}$}{
            \nl $g \longleftarrow G(x)$\;
            \nl \tcc{Train Discriminator}
            \nl $L_{advD} \longleftarrow log(D(y))+log(1-D(g))$\;
            \nl $L_{perc} \longleftarrow \sum_{i=0}^{L_D}\lVert F_{D_i}(y) - F_{D_i}(g)\rVert_{1}$\;
            \nl backprop D       \hfill\tcp{Update D}
            \nl \tcc{Train Generator}
            \nl $L_{advG} \longleftarrow log(D(g))$\;
            \nl $L_{p2p} \longleftarrow \lambda_{p2p}\lVert y-g \rVert_{1}$\;
            \nl backprop G       \hfill\tcp{Update G}
        }
        \nl\tcc{Evaluate Generator}
        \nl PSNR$(G(D_{test}),y_{D_{test}})$\;
        \nl RMSE$(G(D_{test}),y_{D_{test}})$\;
    }
    \nl return $G$\;
\end{algorithm}
\DecMargin{0em}

The adversarial training for the generator-discriminator architecture is based on the Binary Cross-Entropy Loss ($BCE$), as described in equations \eqref{five} and \eqref{eight}.
The loss terms, $L_{adv}$, $L_{perc}$ and $L_{p2p}$, are combined to train the generator and the discriminator as illustrated in Fig. \ref{fig:gan_overview}.
The role of the discriminator is to differentiate between clean (generated) and ideal images based on their features, thereby labeling them as clean (0) or ideal (1), using the perceptual loss as calculated in \eqref{six}.
Additionally, the pixel-wise loss term penalizes the discrepancy between the exported image and the clean target \eqref{seven}.

\begin{gather}
    L_{advD} = BCE(D(y), 1) + BCE(D(g), 0), \label{five}\\
    L_{perc}= \sum_{i=0}^{L_D}\lVert F_{D_i}(y) - F_{D_i}(g)\rVert_{1}, \label{six}\\
    L_{p2p}= \lambda_{p2p}\lVert y-g \rVert_{1}, \label{seven}\\
    L_{advG} = BCE(D(g), 1), \label{eight}
\end{gather}

where $F_{D_i}$ is the intermediate activation map extracted from the $i$-th convolution layer of the discriminator, $L_D$ is the number of convolution layers in the discriminator, and $\lambda_{p2p}$ is a constant which weights the contribution of the pixel-wise L1 discrepancy. 

\begin{figure*}[h]
    \centering
    \includegraphics[scale=1, width=0.9\textwidth]{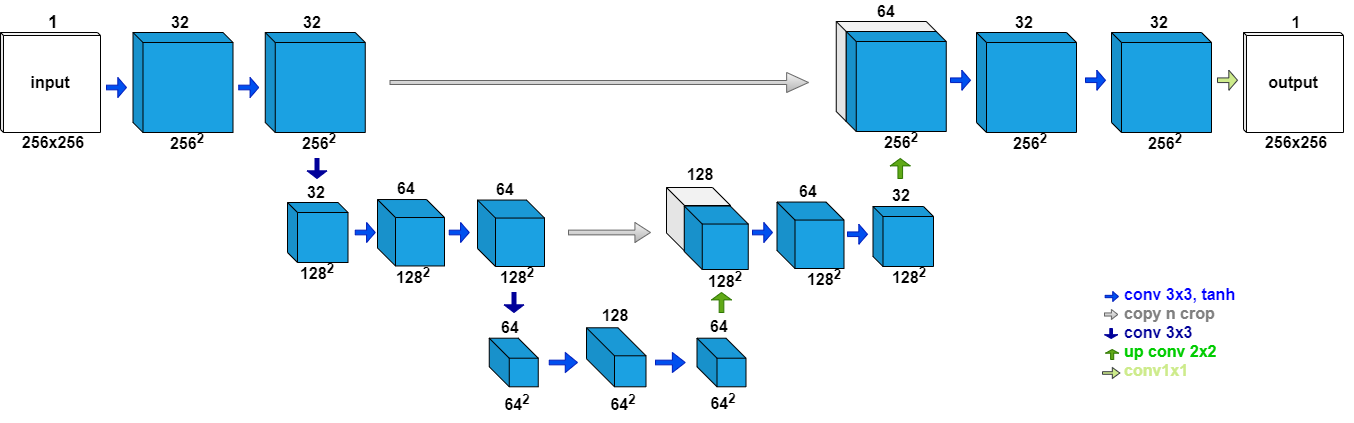}
    \caption{Mobile-SRGAN Generator Architecture} 
    \label{fig:mobu-net}
\end{figure*}

We implemented the generator using an encoder-decoder variant \cite{ronneberger2015unet}, as illustrated in Fig. \ref{fig:mobu-net}, by applying a spatial reduction step after every convolutional bottleneck on the decoder and a resolution recovery step in the encoder. 
On each bottleneck output, a corresponding residual connection forwards the required features, thereby avoiding loss of information. 
The novelty of this encoder-decoder architecture is that it factorizes the standard $3 \times 3$ convolution into a $3 \times 3$ depthwise convolution (DWC) and a $1 \times 1$ pointwise convolution (PWC) to drastically improve the efficiency \cite{guo2019depthwise}. 

DWC and PWC play different roles in generating new features: the former captures spatial correlation information, whereas the latter quantifies channel-wise correlations.
Employing DWC results in a substantial reduction in the CNN model size, with a reduction of up to 75\% by applying a DWC on the first layer consisting of 32 feature channels. 
Hence, the proposed Mobile-SRGAN architecture with 79,233 model parameters is suitable for many mobile and smartphone applications \cite{howard2017mobilenet}.
By incorporating the DWC, resulting parameterization of the spatial reduction step occurs. 
This influences performance, allowing the training process to utilize spatial reduction as desired.
The discriminator is based on the patch discriminator architecture introduced in \cite{isola2018imagetoimage}, which divides the input into $16 \times 16$ patches and classifies each patch. 
The classification score is averaged for all patches \cite{Armanious2020MedGAN}. 

The efficient mobile SAR super-resolution framework was trained for 50 epochs using the ADAM optimizer on a single TESLA P100 GPU with 16 GB of memory using 4096 samples for the training process and 1024 for the evaluation. 
The training dataset consists of SAR images containing randomly placed point scatterers, solid objects, and hollow objects, which improves over previous studies that only include images from point scatterers \cite{gao2018enhanced,jing2022enhanced,smith2021An}.
The training process was approximately 8.5 h an inference time during validation process of 14 ms per sample.

\begin{figure}[h]
  \centering
  \resizebox{\columnwidth}{!}{%
    \begin{tabular}{ccc}
    \includegraphics[width=0.33\linewidth]{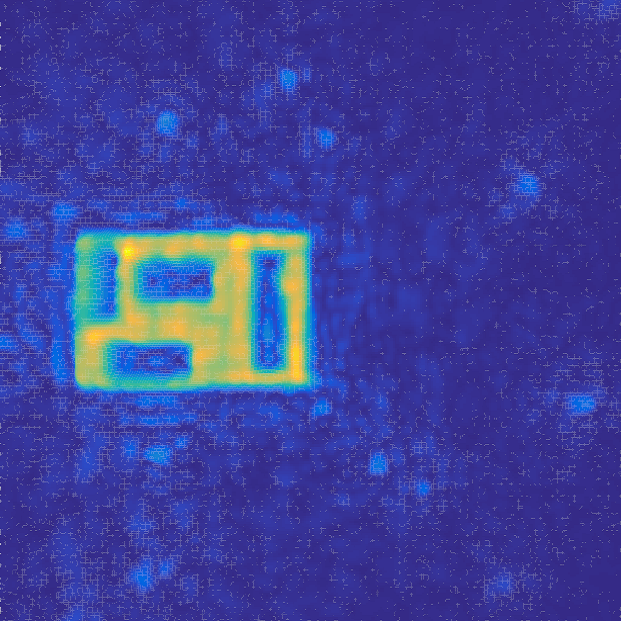}    
    &\includegraphics[width=0.33\linewidth]{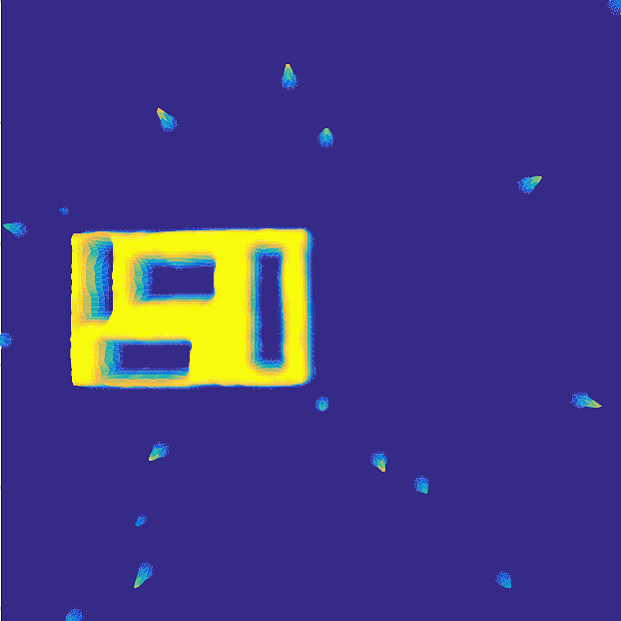}
    &\includegraphics[width=0.33\linewidth]{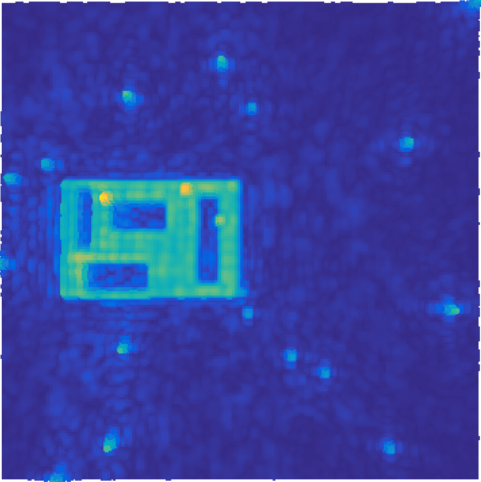}
    \\
    \scalebox{-1}[1]{\includegraphics[width=0.33\linewidth]{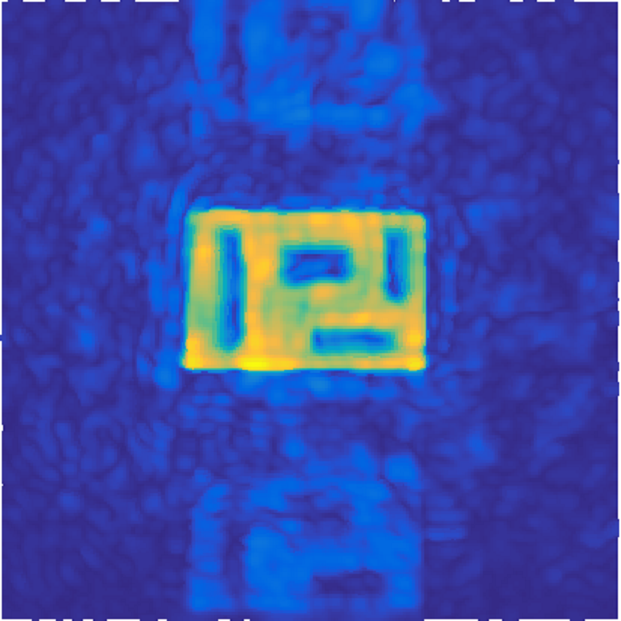}}
    &\scalebox{-1}[1]{\includegraphics[width=0.33\linewidth]{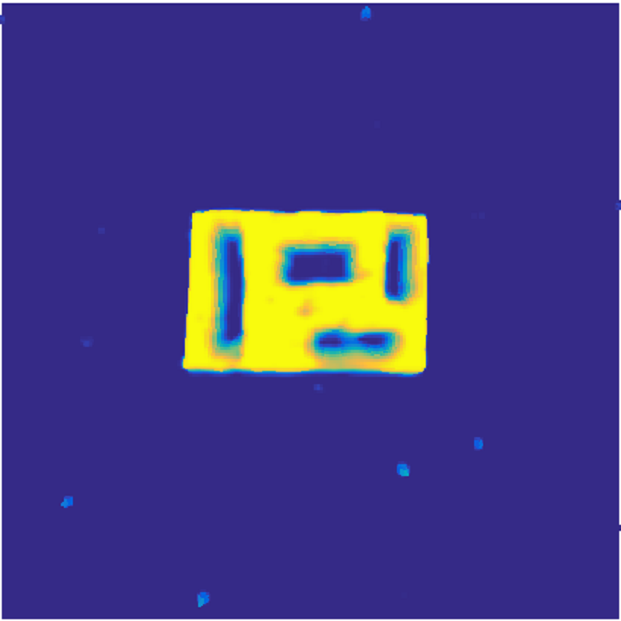}}
    &\scalebox{-1}[1]{\includegraphics[width=0.33\linewidth]{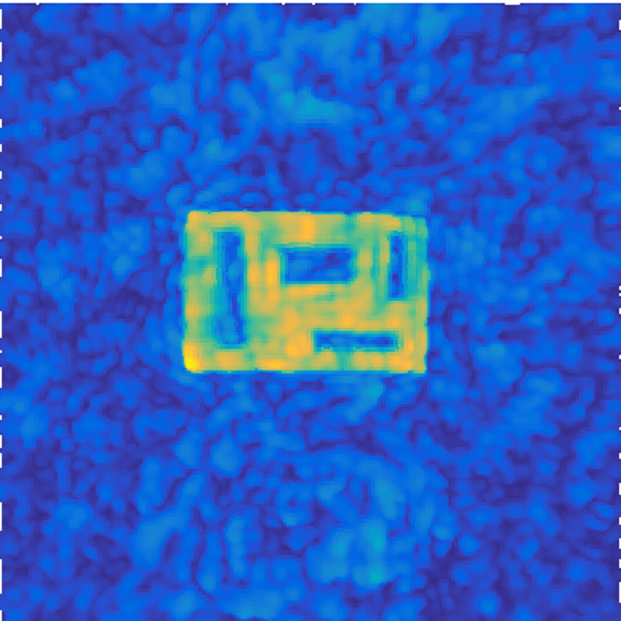}}
    \\
    (a) EMPM & (b) Mobile-SRGAN & (c) BPA
    \end{tabular}
  }
\caption{Qualitative Results: Comparison of clean images (a) EMPM, (b) Mobile-SRGAN, and (c) BPA. Results from synthetic images are shown on the first row, while results from real images are shown in the second row.}
  \label{fig:qualitative_res}
\end{figure}

\subsection{Performance Analysis of Mobile-SRGAN on Synthetic and Empirical Data}
\label{subsec:hffh_cnn_results}
We evaluate the adversarial framework with a test set consisting of 1027 ``never-seen-before'' images (1024 synthetic images and 3 real images) excluded during the model training.
The qualitative results are shown in Fig. \ref{fig:qualitative_res}. 
The network can produce rigid high-resolution images when processing new low-resolution data. 
By learning the joint distribution mapping between low- and high-resolution images, the generator can create consistent and well-structured objects by mitigating the distortion around the image while enhancing the appropriate local structure. 
The quantitative performance of the proposed framework can be measured by computing the peak signal-to-noise ratio (PSNR) and root mean square error (RMSE) between the generated images and ideal images of the synthetic testing dataset.
As shown in Table \ref{table:Table1}, the proposed method outperforms existing methods in terms of both computational efficiency and image quality. 
In particular, our algorithm achieves a higher PSNR and lower RMSE than the gold-standard BPA with only a slight increase in computation time for a single image compared to the EMPM and RMA. 


\begin{table}[h]
\centering
\caption{Quantitative performance of the Mobile-SRGAN compared to the BPA, EMPM, and RMA.} 
  \begin{tabular}{ c||c|c|c|c } 
    {Metrics} & {Mobile-SRGAN} & {BPA} & {EMPM} & {RMA} \\
    \hline \hline
    PSNR (dB) & $\mathbf{34.926}$ & $26.33$ & $20.20$ & $10.158$\\ 
    \hline
    RMSE & $\mathbf{0.019}$ & $0.044$ & $0.105$ & $0.276$\\
    \hline
    Time (s) & $1.117$ & $1324.8$ & $\mathbf{1.103}$ & $\mathbf{1.103}$\\
    \hline \hline
  \end{tabular}
\label{table:Table1}
\end{table}

The Mobile-SRGAN is the first CNN-based super-resolution algorithm for mobile freehand SAR imaging in the near-field. 
Rather than training on randomly placed point targets, we improve upon previous work by incorporating solid, intricate objects in the simulation that are more representative of real-world scenarios \cite{gao2018enhanced,jing2022enhanced,smith2021An}. 
The proposed CNN algorithm is applied to images recovered by the EMPM algorithm \cite{smith2022efficient}, yielding high-resolution low-noise SAR images and outperforming previous techniques. 
The Mobile-SRGAN is the first SAR super-resolution algorithm developed for freehand radar imaging, which is a more difficult task than traditional SAR super-resolution, and efficiently recovers high-resolution images with low computational cost, deeming it suitable for computationally constrained applications. 

In this chapter, we introduced a novel software platform for accelerating research into data-driven mmWave and THz imaging algorithms.
Several deep learning-based algorithms were developed to overcome the non-idealities present in mmWave imaging applications, such as array positioning errors and image distortion from irregular scanning geometries.
Furthermore, a focus is placed on computational efficiency, as many applications of interest are computationally limited. 
Exploring both front-end signal processing techniques and data-driven algorithms for improving mmWave imaging, we observe a lack of integration of signal processing and machine learning methods. 
The typical signal chain employs signal preprocessing prior to a data-driven algorithm, but separating these operations suffers from several limitations. 
The signal processing algorithms are unable to provide ``context'' that data-driven techniques employ to perform information extrapolation tasks such as super-resolution \cite{lim2017enhanced}. 
Additionally, most machine learning algorithms are developed and optimized for optical images and do not take into account the inherent characteristics of the FMCW signals, discussed in Chapter \ref{ch:fmcw_signal_model}. 
To overcome these limitations, we introduce a novel framework for posing solutions to radar signal processing problems that leverages the strengths of both signal processing and machine learning, called hybrid-learning algorithms. 
By developing custom data-driven algorithms to exploit the properties of the signals, we notice a significant improvement in network training convergence and final performance. 
Using signal processing techniques throughout the machine learning pipeline is shown to improve performance for several tasks including localization \cite{smith2021An} and multiband image fusion \cite{smith2023dual_radar}; however, the concept of hybrid-learning can be applied to a host of applications in signal processing and sensing \cite{xie2017aggregated_resnext,xie2021contaminated} to improve numerical and computational performance. 

\chapter{Hybrid-Learning Techniques for Contactless Musical Instrument Interface}
\label{ch:rmi}
In this chapter, we extend the efforts in Chapters \ref{ch:signal_processing} and \ref{ch:data_driven} to a more balanced approach to mmWave imaging and deep learning. 
Rather than approaching a problem from a particular perspective, we adopt a fusion of signal processing and machine learning techniques to develop end-to-end hybrid-learning algorithms, allowing the data-driven algorithm to offer insight throughout the image signal processing chain. 
Part of the following work was previously published in \cite{smith2021An}\footnote{\copyright 2021 IEEE. Reprinted, with permission, from J. W. Smith, O. Furxhi, M. Torlak, ``An {FCNN}-based super-resolution {mmWave} radar framework for contactless musical instrument interface,'' in \textit{IEEE Trans. Multimedia}, vol. 24, pp. 2315--2328, May 2021.}.

Using the interleaved hybrid-learning methodology, we design and implement a real-time system employing deep learning-based localization, classical signal processing algorithms, and a modified computer vision tracking algorithm. 
Although the proposed techniques are suitable for a host of tracking applications, this section focuses on their application as a musical interface to demonstrate the robustness of the gesture sensing pipeline and deep learning signal processing chain. 

We apply a novel fully convolutional neural network (FCNN) to preserve the geometry of the image and perform super-resolution for improved localization.
Radar signal processing using FCNNs is advantageous over other CNN techniques as it allows for data-driven ``enhancement'' rather than dimensionality reduction, as in classification.
Hence, rather than suffering from information loss, the regressive FCNN provides additional ``context'' learned during the training phase to enhance the radar data.
The enhanced data offer several advantages, such as improved SNR, clutter removal, near-field image correction, aliasing suppression, and higher-resolution peaks.
In this section, traditional radar signal processing algorithms are shown to achieve considerable performance gains when applied to enhanced data.
Our novel approach unifies FCNN-based super-resolution with near-field imaging, which requires more difficult spherical-wave compensation, on a small (8-channel) array and is shown to improve hand-tracking performance significantly.
This study is the first documented effort towards near-field radar image super-resolution using an FCNN approach for improved localization.
Incorporating our enhancement FCNN in the signal processing chain enables fine motion tracking that is unattainable by existing techniques.
Additionally, a particle filter tracking algorithm is presented to further improve tracking robustness by employing the Doppler effect. 
Compared to prior work on gesture tracking using optical solutions \cite{sun2019visual,polfreman2011multi}, our approach offers precise hand-tracking using a single mmWave sensor, offering higher depth resolution with superior privacy. 
This section proposes a novel hand-tracking method for musical interface by fusing spatiotemporal algorithms, deep learning-enhanced feature extraction, and robust position tracking algorithms. 

\section{System Model for Radar Musical Interface}
\begin{figure}[h]
	\centering
	\includegraphics[width=0.5\textwidth]{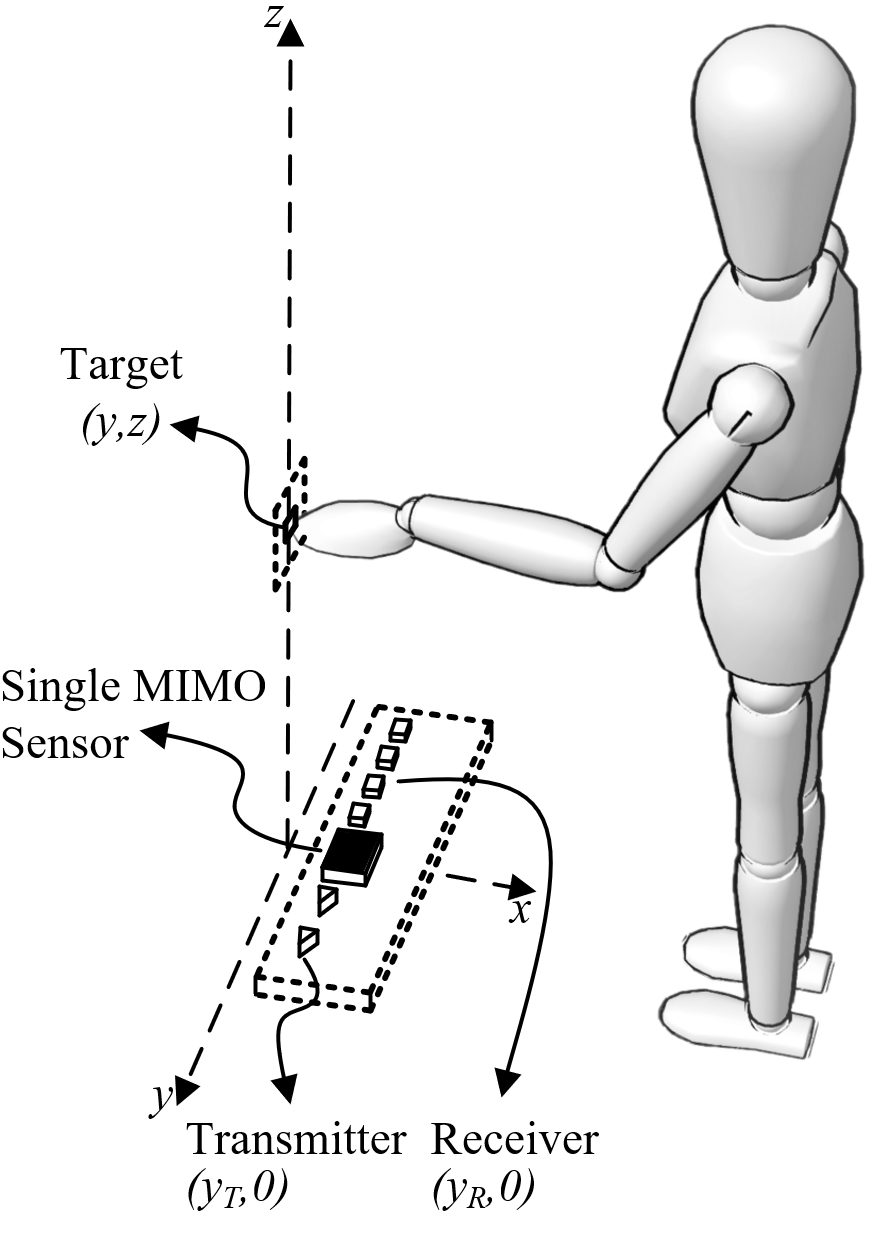}
	\caption{The imaging geometry, where the linear MIMO array faces vertically and the musician moves their hand throughout the $y$-$z$ plane.}
	\label{fig:MIMO_radar_musical_instrument_setup}
\end{figure}

In this section, we provide an overview of the system model employed by the FCNN-based musical interface and examine the spatiotemporal features of a target in motion. 
The musician's hand is modeled as a point reflector located at the point $(y,z)$, as shown in Fig. \ref{fig:MIMO_radar_musical_instrument_setup}.
To achieve high-fidelity \mbox{2-D} localization, we employ the range migration algorithm (RMA) over traditional range-angle FFT methods \cite{TI:rao2017intro, kim2020aziumth}, whose localization accuracy is known to be inferior \cite{kim2018joint}. 

Furthermore, the aforementioned Doppler principle can be leveraged to extract the velocity of the target by Fourier analysis over successive chirps, as described in Chapter \ref{ch:signal_processing}. 
The velocity is extracted from the recovered image, which is denoted by $\hat{p}(y,z)$, given the geometry in Fig. \ref{fig:MIMO_radar_musical_instrument_setup}.

\section{Classical Spatiotemporal Feature Extraction Techniques}
\label{sec:classical_gesture_tracking}
In this section, we introduce a simple approach to spatiotemporal sensing for contactless musical instrument interface.
While our system generally tracks the \mbox{2-D} position and velocity of the user's hand, we have identified three underlying features to achieve fine control of the musical interface: range, cross-range oscillation, and velocity. 
Based on the geometry shown in Fig. \ref{fig:MIMO_radar_musical_instrument_setup}, we define the range as the position of the hand along the $z$-axis, that is, the vertical displacement between the sensor and the user's hand. 
Similarly, the cross-range direction is defined as the position of the hand along the $y$-axis.
Subsequently, the cross-range oscillation is the rate at which the hand oscillates in the cross-range direction.
The velocity is given by the velocity of the hand with respect to the range $z$-axis. 
These parameters are selected such that the output musical interface is controlled primarily by the range of the musician's hand and secondarily by the cross-range oscillation and velocity.
Throughout the remainder of this section, we refer to these parameters as features extracted from the radar beat signal.

Under the simple gesture tracking regime, the \mbox{2-D} location and velocity $(\hat{y},\hat{z},\hat{v}_d)$ are extracted from the reconstructed image and buffer of recent images.
In the next section, the three parameters extracted from the raw data are treated as a vector called the noisy measurement vector $\mathbf{r}$. 
Even in the ideal case, the spatial resolution of our system along the $y$ and $z$-directions is $\delta_y = 7.5$ cm and $\delta_z = 3.75$ cm, respectively \cite{yanik2020development}.
Several other factors are not considered in the classical direct tracking method, including beam pattern, residual phase errors, and antenna coupling. 
To address these issues, we present a novel data-driven approach that employs an FCNN for super-resolution and image enhancement. 

\section{FCNN-Based Super-Resolution Feature Extraction and Particle Filter \\ Tracking Methods}
\label{sec:enhanced_gesture_tracking}
\begin{figure}[h]
	\centering
	\includegraphics[width=0.75\textwidth]{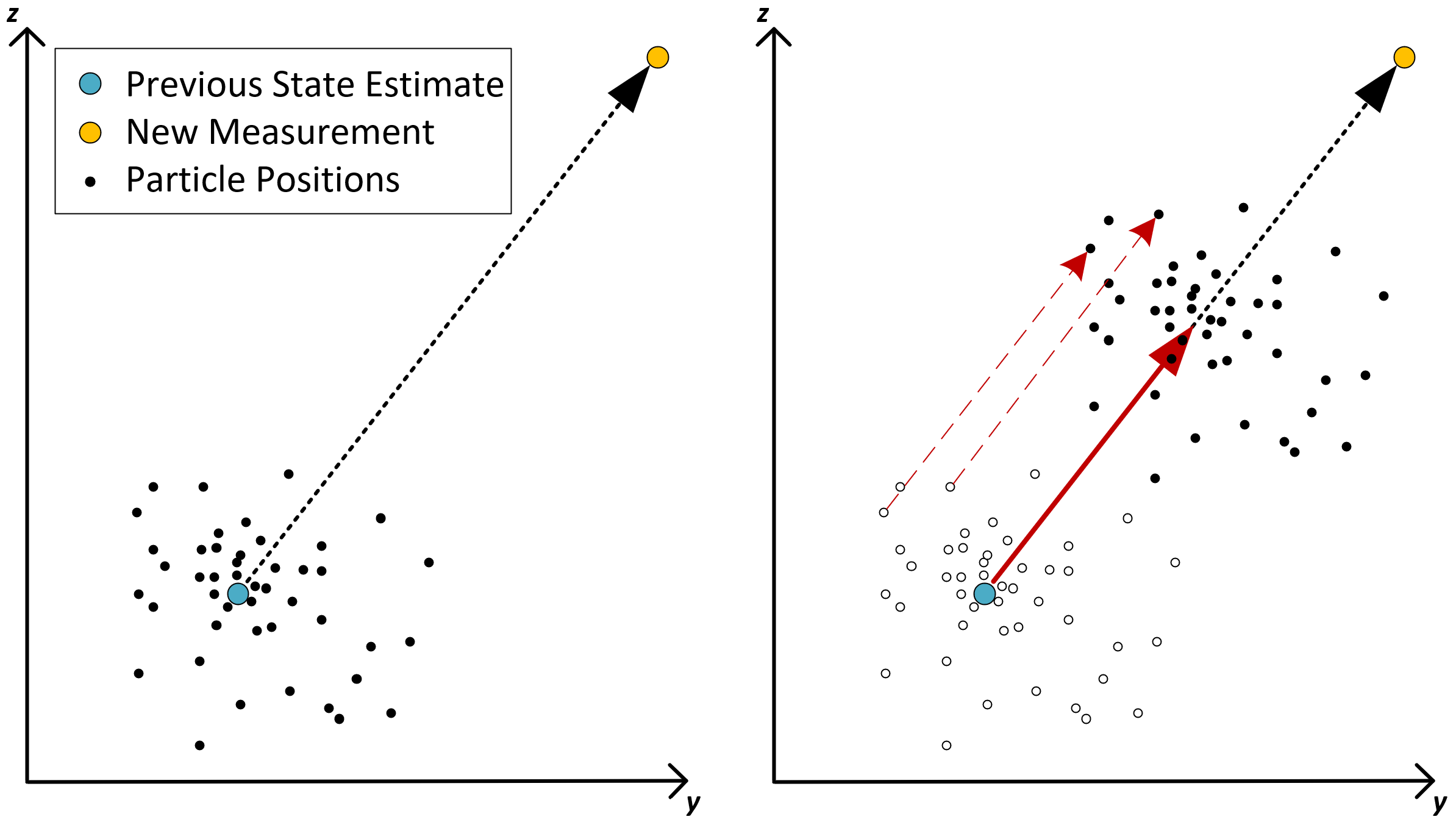}
	\caption{A visual example of the modified particle filter algorithm resampling process. The particle locations are resampled by a shift transformation towards the new measurement according to the weight vector $\mathbf{a}$, where $a_y = a_z = 0.5$.}
	\label{fig:particle_filter}
\end{figure}

In this section, we improve upon the simple tracking techniques to overcome noise and foundational non-idealities in the imaging scenario, yielding a more robust algorithm. 
First, we develop a novel algorithm based on the well-known particle filter \cite{garcia2013tracking}. 
During the particle drift step of the algorithm, rather than shifting by a constant value, our method employs a variable weight factor depending on the corroboration of the new measurement with the Doppler velocity. 
An illustration of this process is provided in Fig. \ref{fig:particle_filter}, and additional details can be found in \cite{smith2021An}.

Second, we propose a novel hybrid-learning technique to address issues such as instrumentation delay, ambient/device noise, multistatic effects, and non-spherical beam patterns. 
We present a novel FCNN-based technique for image enhancement that improves \mbox{2-D} position estimation, subsequent tracking accuracy, and Doppler spectrum SNR. 
To train the enhancement FCNN, we construct a dataset consisting of both real human hand data and synthetically generated data using the proposed MATLAB toolbox.
During training, the FCNN learns the highly nonlinear relationship between distorted, blurred RMA images and the ideal images.
Our novel training technique results in a robust and generalizable FCNN that improves the image SNR and localization by fitting to the non-ideal imaging constraints. 

Additionally, by isolating the peak corresponding to the human hand,the clutter and noise at other positions are mitigated, thereby improving the Doppler spectrum SNR and subsequent velocity estimation.
Thus, the FCNN enhances both the spatial and temporal features extracted from the radar beat signal before applying the particle filter.
Uniting the proposed particle filter and enhancement FCNN, the range, cross-range oscillation, and velocity are robustly tracked by our novel algorithms and mapped to musical interface controls.

\begin{figure}[h]
	\centering
	\includegraphics[width=0.75\textwidth]{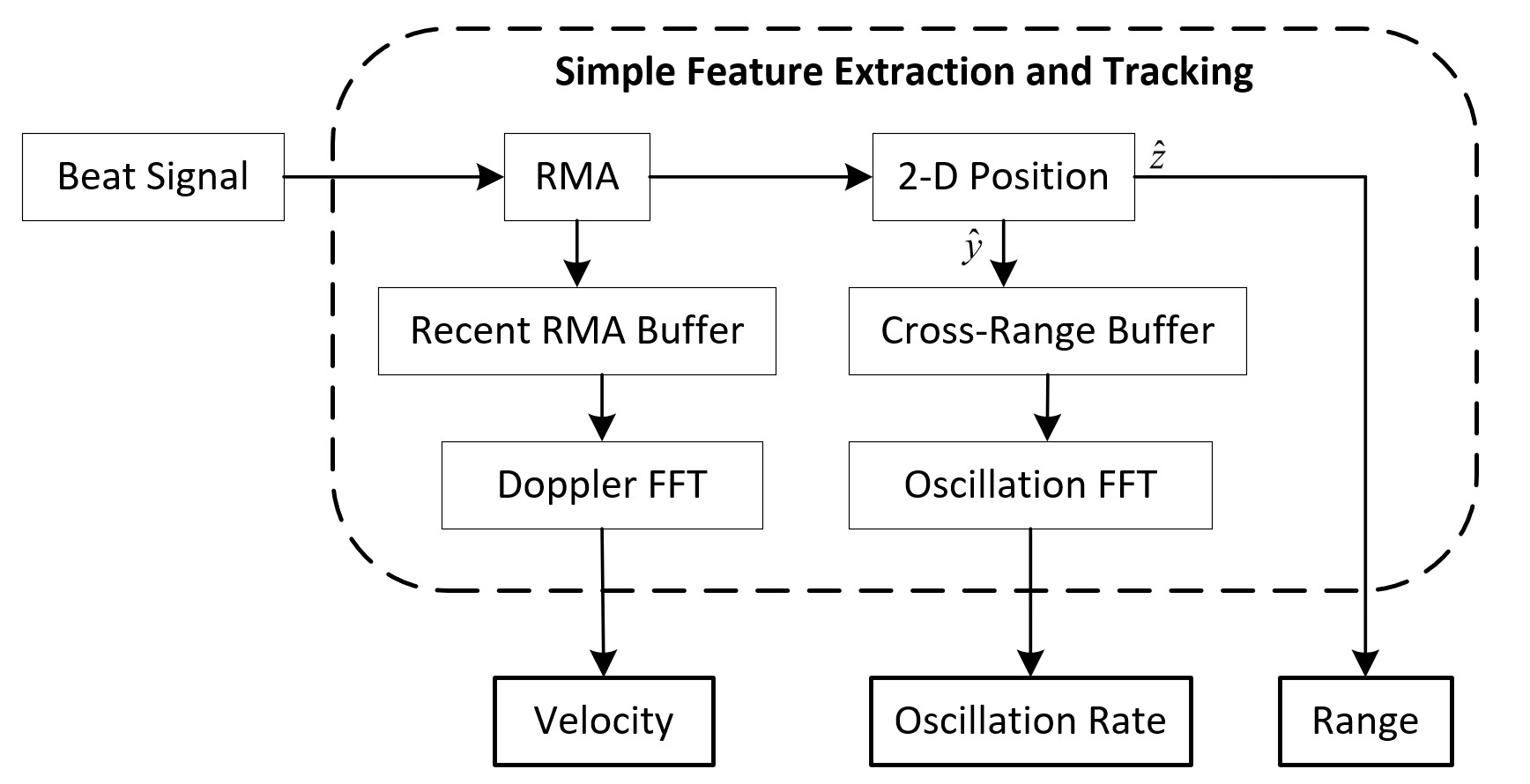}
	\caption{Simple tracking signal processing chain. After RMA is performed on the beat signal, features are extracted directly from the raw RMA image.}
	\label{fig:simple_signal_chain}
\end{figure}

\section{Simple and Enhanced Methods Compared}
\label{sec:rmi_compare}
\begin{figure}[h]
	\centering
	\includegraphics[width=0.85\textwidth]{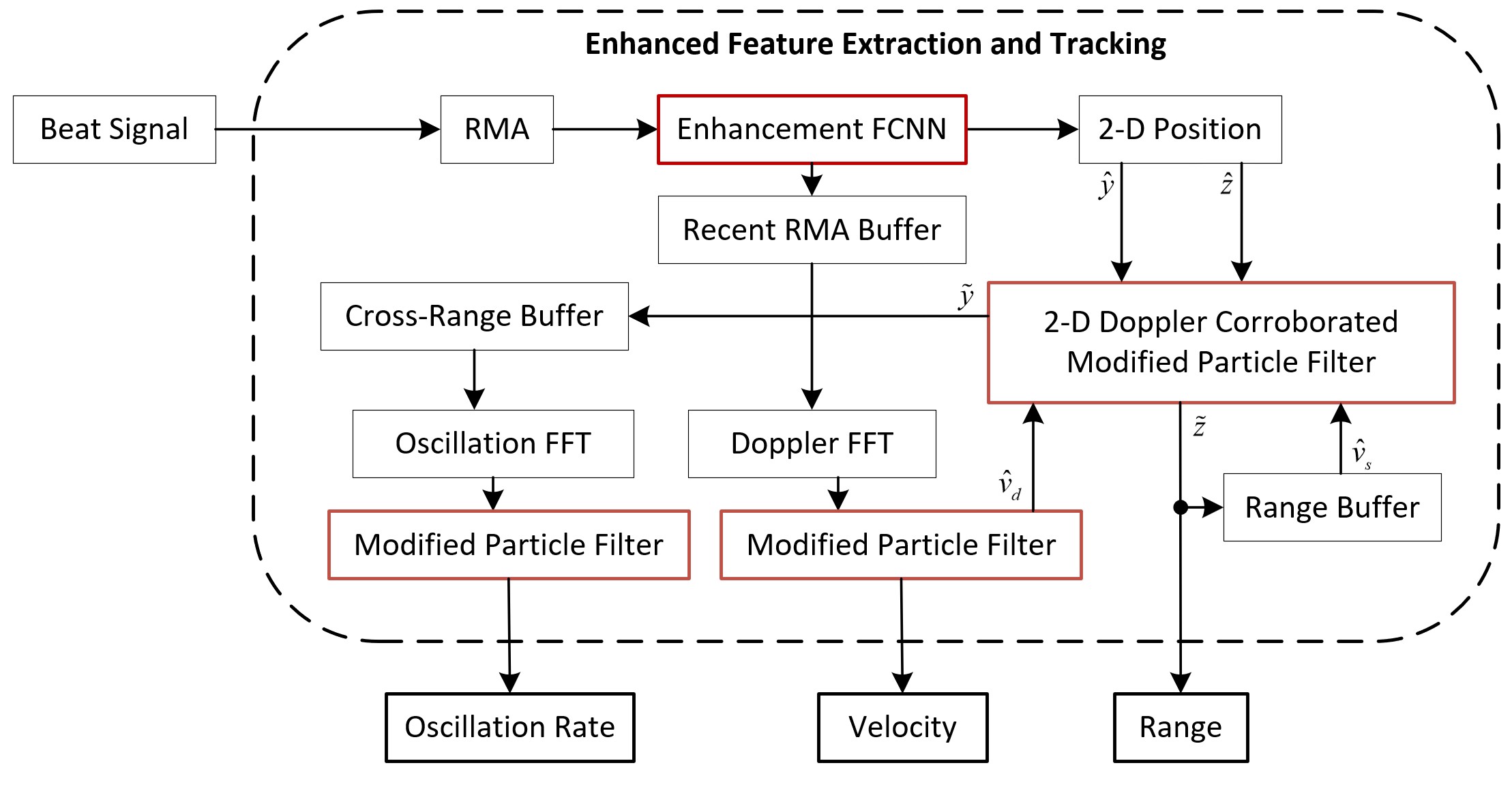}
	\caption{Enhanced tracking signal processing chain. Key elements to the enhanced methods are highlighted in red.}
	\label{fig:enhanced_signal_chain}
\end{figure}

The signal processing chain for the simple feature extraction and tracking method is illustrated in Fig. \ref{fig:simple_signal_chain}. 
The beat signal is loaded into MATLAB, where the preprocessing discussed in the previous section is performed (RMA and peak finding).
The reconstructed RMA image and raw features extracted by the classical techniques can be utilized by the particle filter algorithm and super-resolution FCNN to improve the tracking performance.
The signal processing chain for the enhanced feature extraction and tracking method is shown in Fig. \ref{fig:enhanced_signal_chain}.
Spatiotemporal features are output from the algorithm and can be used for many tracking applications.

\section{Spatiotemporal Feature Extraction and Tracking Results}
\label{sec:rmi_results}
In this section, we provide an overview of the results of our novel tracking and feature enhancement algorithms, beginning with the simple, classical techniques, and compare the performance to our proposed methods.
Our enhanced tracking regime demonstrates considerable performance improvement compared with traditional methods and allows for robust super-resolution tracking on a small radar platform unattainable by existing methods.

\subsection{Ground Truth - Ideal Motion Profile}
\label{subsec:ideal_motion_profile}

\begin{figure}[h]
	\centering
	\includegraphics[width=0.8\textwidth]{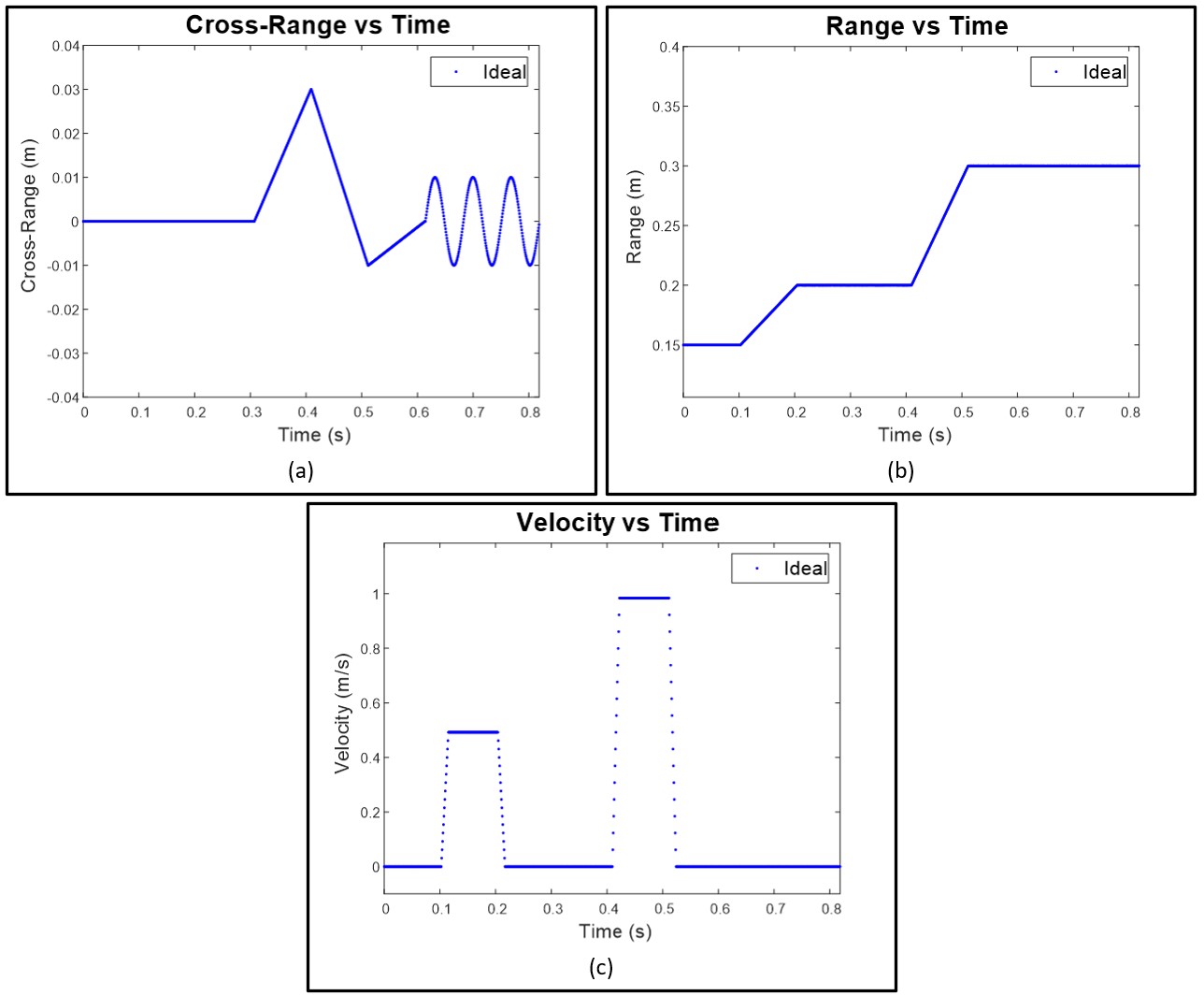}
	\caption{Ideal motion profile of the target in the (a) cross-range and (b) range directions as well as the (c) range velocity profile against time.}
	\label{fig:ideal_motion}
\end{figure}

A virtual prototyping approach is adopted to verify the feature estimation techniques. 
A point target is simulated in motion with $y$-$z$ location and velocity, as shown in Fig. \ref{fig:ideal_motion}, using
\begin{equation}
\label{eq:mimo_beat_signal_continuous}
    s(y_T,y_R,k) = \frac{p}{R_T R_R} e^{-jk(R_T + R_R)},
\end{equation}
where $R_T$ and $R_R$ are the distances from the transmitter and receiver to the point target, respectively.
This ideal motion profile is employed to compare the tracking performance of our proposed methods with that of the traditional techniques. 
Empirical noise collected from the radar with an empty scene is added to each synthetic beat signal as
\begin{equation}
\label{eq:beat_sim_with_noise}
	\tilde{s}(y_T,y_R,k) = \frac{p}{R_T R_R}e^{-jk(R_T + R_R)} + \alpha \tilde{ \omega}(y_T,y_R,k),
\end{equation}
where $\tilde{ \omega}$ is a complex-valued noise sample corrupting the amplitude and phase of the ideal simulated beat signal and $\alpha$ controls the SNR.

The motion profile in Fig. \ref{fig:ideal_motion} shows the ideal range ($z$), cross-range ($y$), and velocity ($v$) of the target. 
The motion profile includes independent and joint movements in the range and cross-range domains, in addition to sinusoidal cross-range oscillation.
For our simulations, $4096$ time samples are generated using $p \in [0.5,1]$ to simulate the variance in the hand's empirical radar cross-section (RCS), as observed empirically from prior hand data, and $\alpha \in [1,3]$ to vary the SNR among samples. 
Values for $p$ and $\alpha$ are selected randomly within the specified intervals for each time sample and provide a level of stochastic realism to the simulated data.

\subsection{Classical Spatiotemporal Imaging Results}
\label{subsec:simple_gesture_tracking_results}
First, the simple tracking methods discussed in Section \ref{sec:classical_gesture_tracking} are implemented to provide baseline performance metrics.
The signal processing chain shown in Fig. \ref{fig:simple_signal_chain} is performed to extract the spatiotemporal features. 
At each iteration, the features are extracted directly from the raw RMA images and are therefore prone to erratic behavior. 

\begin{figure}[h]
	\centering
	\includegraphics[width=0.8\textwidth]{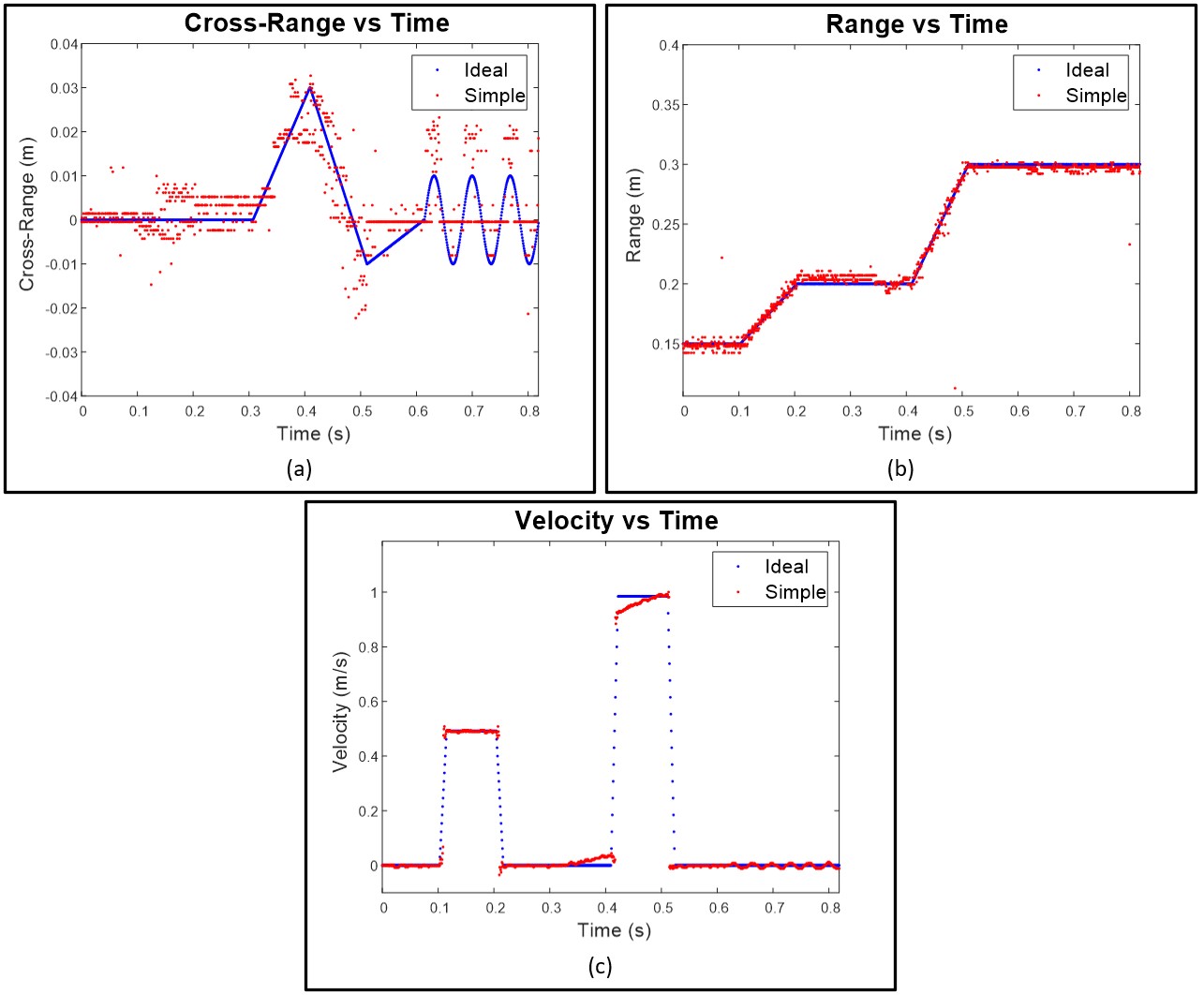}
	\caption{Motion profile using simple feature extraction techniques on each frame for every time step (red) compared with the ideal motion and velocity profiles (blue). The (a) cross-range and (b) range are measured directly from the peak of the RMA image of each frame and the (c) velocity is measured using the Doppler FFT of the raw RMA images.}
	\label{fig:simple_motion}
\end{figure}

Fig. \ref{fig:simple_motion} shows the features estimated from the data generated by (\ref{eq:beat_sim_with_noise}) using the simple methods. 
The real radar noise and varying reflectivity result in outliers and errors in the estimated location and velocity of the target, particularly in the cross-range domain.
Without more robust feature extraction and tracking techniques, the performance leaves much to be desired.
In the following sections, the performance of the simple tracking methods is quantitatively compared to the enhanced tracking methods and design considerations are discussed.

\subsection{FCNN-Based Super-Resolution Tracking Results}
\label{subsec:enhanced_gesture_tracking_results}
Assuming the motion profile in Fig. \ref{fig:ideal_motion}, our proposed particle filter algorithm is employed in an attempt to robustly track the 2-D position and Doppler velocity of the target across time, significantly improving the user's control over the interface.

\begin{figure}[h]
	\centering
	\includegraphics[width=0.8\textwidth]{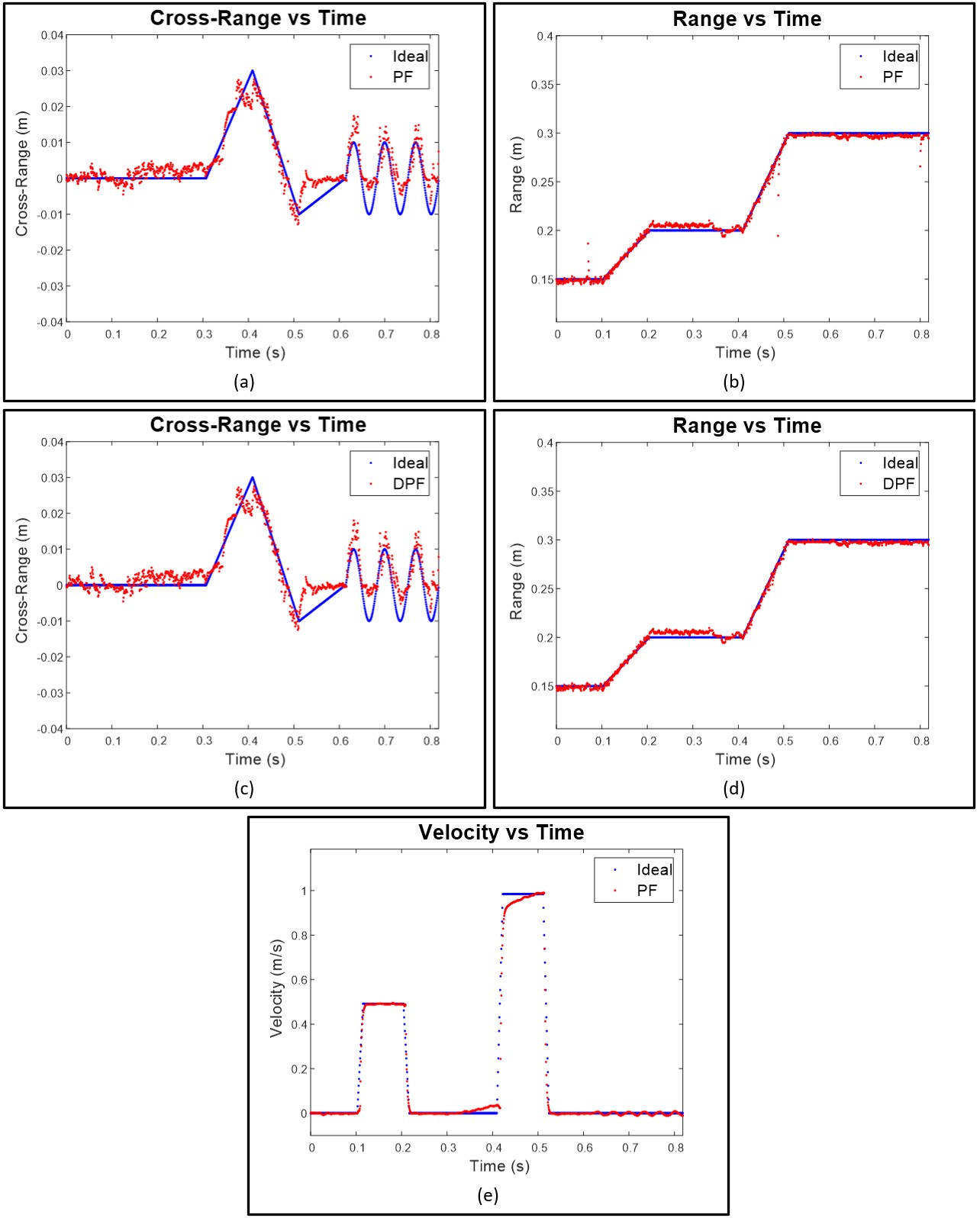}
	\caption{Particle filter (PF) and Doppler-corroborated particle filter (DPF) tracking. Improved tracking of the (a)/(c) cross-range and (b)/(d) range positions versus time using the PF/DPF with $N_z =$ 16, and (e) Doppler velocity versus time using a PF approach.}
	\label{fig:pf_motion}
\end{figure}

First, the particle filter algorithm (PF) without Doppler corroboration is implemented using the data in Fig. \ref{fig:simple_motion} as elements of the noisy measurement vector $\mathbf{r}$. 
The PF reduces the effect of noise on position estimation and improves spatiotemporal tracking performance, as shown in Fig. \ref{fig:pf_motion}.
Cross-range position tracking is the most improved compared to the traditional methods.
Next, the Doppler-corroborated particle filter (DPF) is applied to the same set of data, further improving the estimation of the range. 
The outliers in Fig. \ref{fig:pf_motion}b are mitigated by the DPF in Fig. \ref{fig:pf_motion}d because the outlying samples result in a sample velocity $\hat{v}_s$ contradicted by the Doppler velocity $\hat{v}_d$ and are weighted as unimportant in the resampling process.
The DPF algorithm improves the user experience of our interface by providing a robust and consistent tracking algorithm to smoothly estimate the 2-D position and spatiotemporal signatures of the user's hand.
However, the PF and DPF can be further improved by implementing the proposed enhancement FCNN.

\begin{figure}[h]
	\centering
	\includegraphics[width=0.85\textwidth]{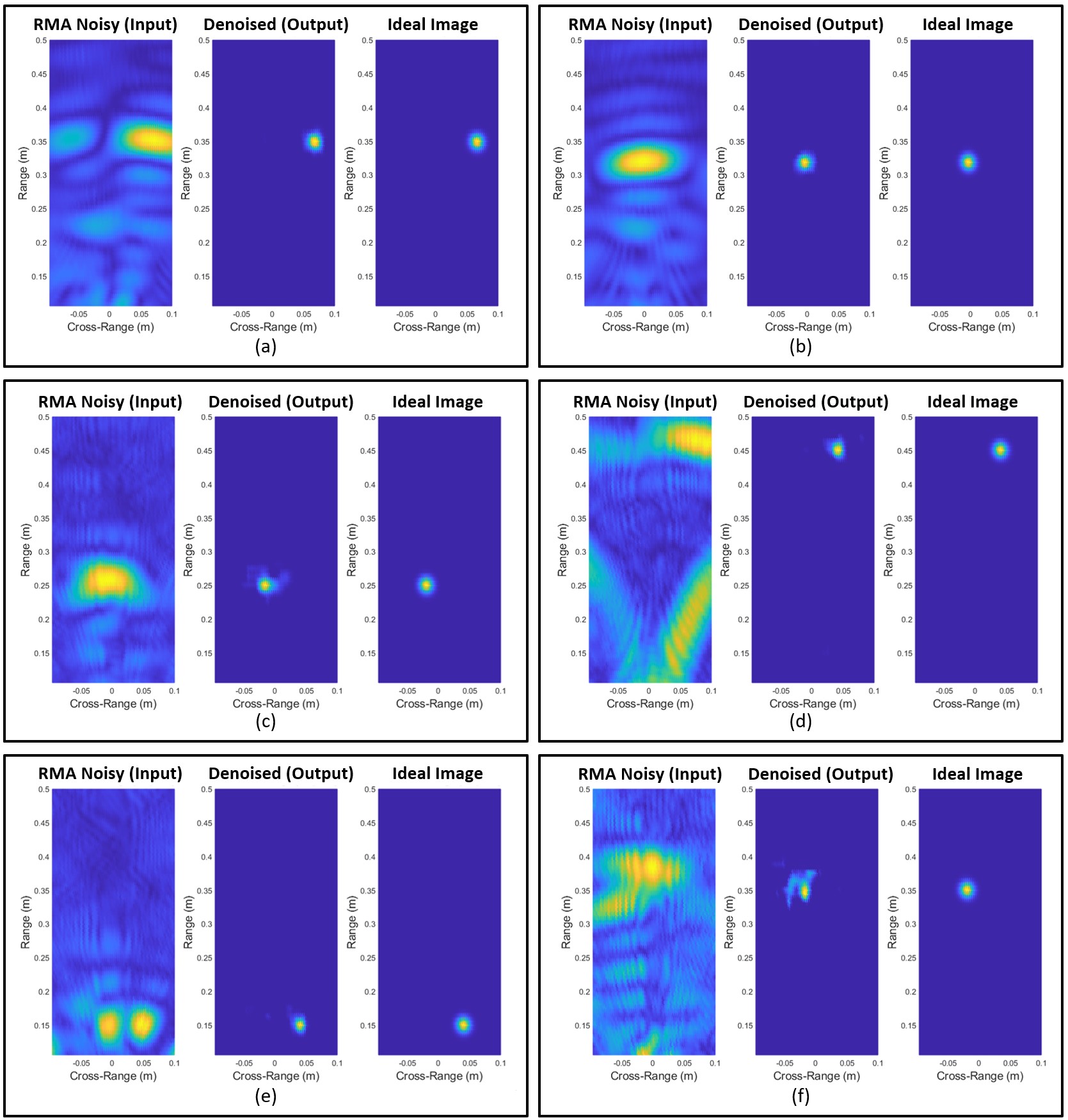}
	\caption{Enhancement FCNN applied to simulated (a,b) and real hand (c-f) RMA images for image enhancement and improved localization.}
	\label{fig:fcnn_enhancement_demo}
\end{figure}

After the super-resolution FCNN is trained, a validation dataset identical in size to the training set is collected.
Fig. \ref{fig:fcnn_enhancement_demo} shows images enhanced by the enhancement FCNN, demonstrating the robustness of the network. 
Figs. \ref{fig:fcnn_enhancement_demo}a and \ref{fig:fcnn_enhancement_demo}b show simulated point targets enhanced by the FCNN, resulting in localization super-resolution. 
Fig. \ref{fig:fcnn_enhancement_demo}c shows an RMA image reconstructed from a real hand capture close to the middle of the cross-range domain. 
The 2-D position of the hand is located more accurately, as compared with the ideal image. 
Similarly, Figs. \ref{fig:fcnn_enhancement_demo}d-\ref{fig:fcnn_enhancement_demo}f demonstrate the network's ability to enhance images degraded by small hand RCS in comparison to noise, ghosting due to non-ideal beam patterns, ambient and device noise, and other non-idealities.
The proposed enhancement FCNN simultaneously enables localization super-resolution and overcomes device and environmental issues.
Hence, the features extracted from the enhanced images are much improved compared to the raw RMA images before the FCNN and result in superior tracking performance.

\begin{table} [h]
	\caption{Simple vs Enhanced Localization RMSE}
	\centering
	\begin{tabular}{c || c |  c }
		& $y$ (m) & $z$ (m) \\
		\hline\hline
		Simple & 0.0154 & 0.023 \\ 
		\hline
		Enhanced & \textbf{0.0085} & \textbf{0.0083} \\ 
		\hline\hline
	\end{tabular}
	\label{table:fcnn_position_rmse}
\end{table}

To quantitatively compare the localization improvement of the enhancement FCNN compared to the simple method, the RMSE for the range and cross-range position are computed on the validation dataset using the two techniques and are shown in Table \ref{table:fcnn_position_rmse}, where the best evaluation is denoted in bold face. 
The enhancement FCNN improves both the resolution of the RMA images and the localization accuracy for both simulated and real data. 

\begin{figure}[h]
	\centering
	\includegraphics[width=0.8\textwidth]{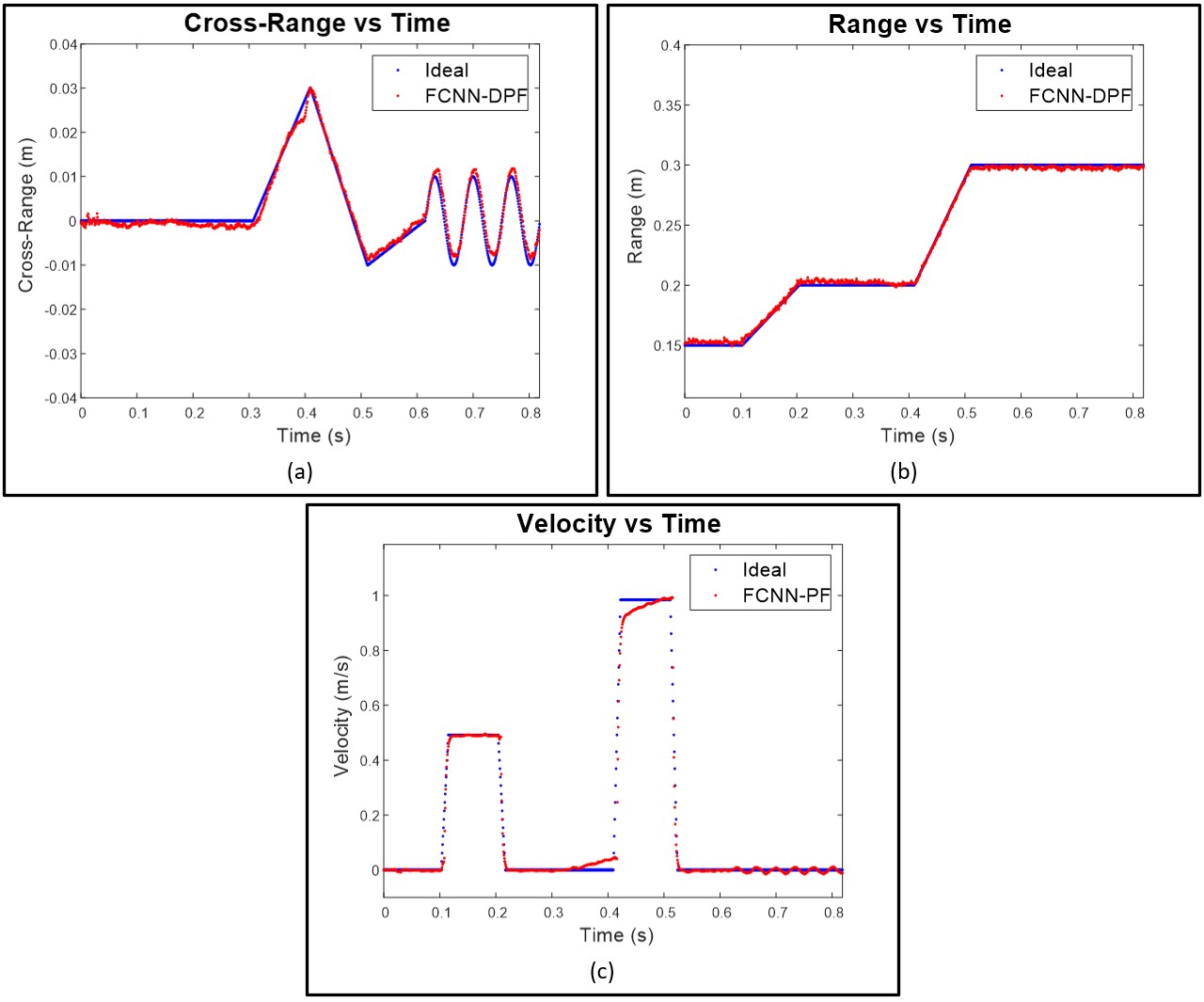}
	\caption{Spatiotemporal tracking with the FCNN-enhanced Doppler-corroborated modified particle filter algorithm.}
	\label{fig:fcnn_dpf_motion}
\end{figure}

Fig. \ref{fig:fcnn_dpf_motion} demonstrates the tracking performance of the FCNN and DPF (FCNN-DPF) on the same data as the previous tracking examples, an improvement over the DPF alone. 
By applying the FCNN-DPF, the range and cross-range tracking of the target is nearly identical to the ideal motion profile and an improvement in the velocity estimation.
Using identical sporadic data resulting in the poorly estimated cross-range positions in Fig. \ref{fig:simple_motion}a, the FCNN-DPF yields an estimation nearly identical to the ideal motion profile.
Similarly, the cross-range estimates in Fig. \ref{fig:pf_motion}a and Fig. \ref{fig:pf_motion}c are outperformed by the FCNN-DPF in Fig. \ref{fig:fcnn_dpf_motion}a.
Compared with the classical techniques and PF/DPF alone, the localization performance of the FCNN-DPF is considerably superior.

\begin{figure}[h]
	\centering
	\includegraphics[width=0.65\textwidth]{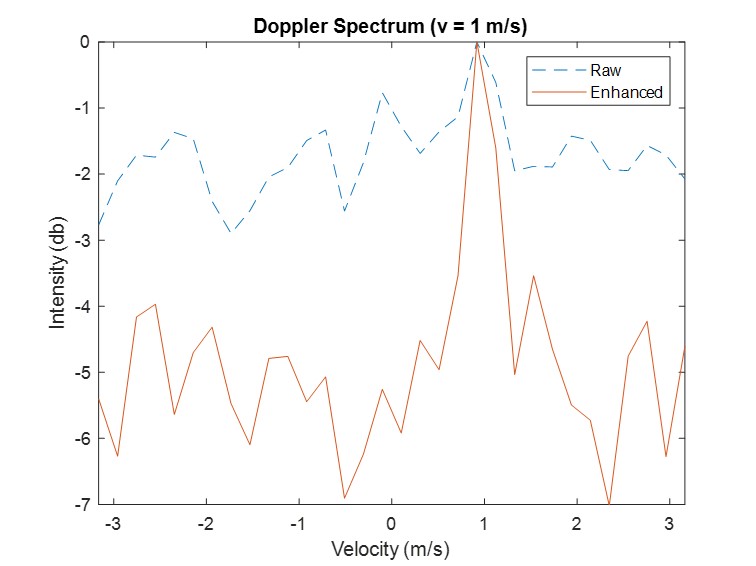}
	\caption{Comparison of the Doppler velocity spectrum when the Doppler FFT and video pulse integration steps are performed on the raw RMA images compared to the enhanced RMA images. The simulated data contains 128 frames and uses $\alpha = $ 3 for every capture to simulate a low SNR scenario.}
	\label{fig:doppler_snr}
\end{figure}

Furthermore, the FCNN is shown to improve the Doppler estimation robustness. 
As shown in Fig. \ref{fig:doppler_snr}, the Doppler spectrum SNR is improved when Doppler processing is performed on the enhanced RMA images as compared to Doppler processing on the raw RMA images.
Hence, the enhancement network improves the reliability of the Doppler velocity estimation, aiding spatiotemporal tracking.

To quantitatively compare the tracking performance of the various proposed methods, $4096$ unique motion profiles are generated, and the corresponding tracking RMSE is computed for the cross-range, range, and velocity. 
As presented in Table \ref{table:tracking_rmse2}, the RMSE for the cross-range ($y$), range ($z$), and velocity ($v$) improved with the novel algorithms proposed in this section. 

As expected, the baseline simple method yields the greatest error for the three features. 
Comparing PF and DPF, the cross-range and velocity RMSE are identical between the two techniques, but the range RMSE is improved owing to the dynamic weighting technique. 
The FCNN alone outperforms the simple method but can be improved by including the PF and DPF after image enhancement. 
Finally, the FCNN-PF and FCNN-DPF yield identical results for the cross-range and velocity RMSE, as expected, but a significant improvement is noted in the range error. 
The results in Table \ref{table:tracking_rmse2} demonstrate the considerably superior tracking performance of the enhanced tracking methods, namely the FCNN-DPF, compared with the simple tracking method.
The performance gain realized by implementing the super-resolution FCNN demonstrates the ability of the network to learn system noise and ambiguities during the training phase using both real and synthetic data. 

\begin{table} [h]
	\caption{Performance Comparison of Tracking Methods (RMSE)}
	\centering
	\begin{tabular}{ c || c | c | c | c }
		& $y$ (mm) & $z$ (mm) & $v$ (mm/s) & $\Bar{\tau}$ (ms) \\
		\hline \hline
		Simple & 7.86 & 22.0 & 72.4 & 2.29 \\
		\hline
		PF & 5.27 & 13.6 & 52.9 & 2.36 \\
		\hline
		DPF & 5.27 & 6.85 & 52.9 & 2.41 \\
		\hline
		FCNN & 7.74 & 12.3 & 58.4 & 2.67 \\
		\hline
		FCNN-PF & 3.70 & 7.44 & 44.5 & 3.92 \\
		\hline
		FCNN-DPF & 3.70 & 3.07 & 44.5 & 3.96 \\
		\hline \hline
	\end{tabular}
	\label{table:tracking_rmse2}
\end{table}

The average latency of each method, $\Bar{\tau}$, is measured as the time duration between the new sample being captured and the estimation process being completed on that sample. 
The resulting estimates are streamed across the MIDI port or sent to the built-in audio signal generation tool. 
The addition latency contributed by the subsequent synthesis engine is highly dependent on the software used and device under test; thus, it is not considered as part of the latency owing to our methods. 
A thorough analysis and comparison of the performance of the algorithms is provided in \cite{smith2021An}. 

The contactless musical instrument interface employs an interleaved hybrid-learning approach as insights from the mechanics of the problem are leveraged throughout the algorithm development for optimal localization and tracking. 
This hybrid approach is employed in the following analysis for multiband signal fusion demonstrating superior performance to previous signal processing and machine learning methods.

\chapter{Deep Learning-Based Multiband Signal Fusion for 3-D SAR Super-Resolution}
\label{ch:dual_radar}

In this chapter, we propose a novel deep learning-based solution for multiband signal fusion to achieve high-resolution synthetic aperture radar (SAR) imaging. 
This approach, called $kR$-Net, employs a hybrid, dual-domain complex-valued convolutional neural network (CV-CNN) to fuse multiband signals. 
By exploiting the relationships in both the wavenumber-domain and wavenumber spectral-domain, the framework overcomes the drawbacks of existing multiband imaging techniques for realistic scenarios and achieves high-resolution imaging of intricate targets, enabling technologies such as concealed weapon detection and occluded object classification. 
The dual-domain architecture demonstrates improved performance over a network operating exclusively in the wavenumber-domain or wavenumber spectral-domain.
Furthermore, a fully integrated multiband imaging system is developed using commercially available millimeter-wave (mmWave) radars for efficient multiband imaging. 
The proposed $kR$-Net is employed to achieve an effective bandwidth of 21 GHz from two radars each with a bandwidth of 4 GHz operating at 60 GHz and 77 GHz. 
Extensive numerical simulations and empirical experiments are conducted to demonstrate the superiority of our approach over existing methods for a diverse set of realistic 3-D SAR imaging scenarios.
Part of the following work was previously published in \cite{smith2023dual_radar}\footnote{\copyright 2023 IEEE. Reprinted, with permission, from J. W. Smith and M. Torlak, ``Deep learning-based multiband signal fusion for 3-D SAR super-resolution,'' in \textit{IEEE Trans. Aerosp. Electron. Syst.}, Apr. 2023.}.

\section{Multiband Signal Model}
\label{sec:dri_signal_model}
\begin{figure}[ht]
    \centering
    \includegraphics[width=0.65\textwidth]{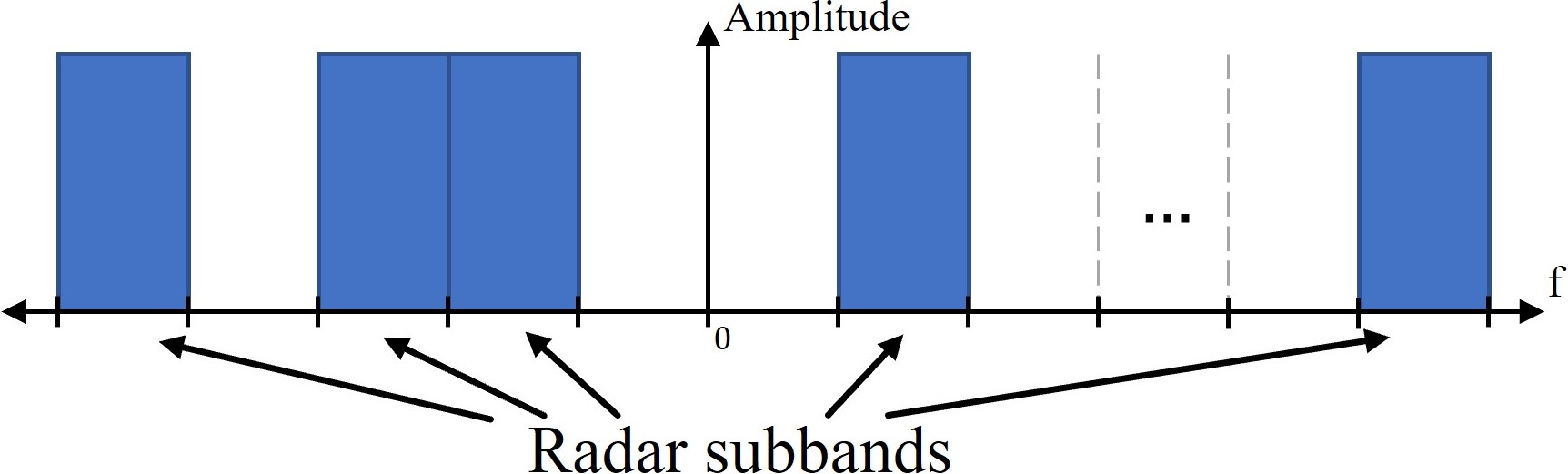}
    \caption{A baseband representation of multiband radar signaling where samples are gathered at multiple subbands separated by frequency gaps. The goal of multiband signal fusion is to recover the equivalent wideband signal spanning the entirety of the subbands.}
    \label{fig:general_multiband} 
\end{figure}

In this section, we formulate a signal model for multiband radar signaling. 
In the multiband sensing scenario, samples are taken across multiple subbands separated by frequency gaps, as shown in Fig. \ref{fig:general_multiband}, where the radar subbands represent the operating frequency ranges of the radars. 
It is important to note that the subsequent analysis and proposed algorithm assume a weak or constant relationship between the scattering properties and frequency across the entire bandwidth spanned by the subbands. 
To achieve the desired resolution, multiband signal fusion methods are applied to recover the unoccupied frequency bins and obtain the equivalent wideband signal spanning the entirety of the subbands. 

For the remainder of this study, we focus on the dual subband scenario emphasized in the existing literature \cite{cuomo1999ultrawide,tian2014sparse,zou2016matrix,wang2018wavenumber,zhang2014coherent,zhang2017multiple,tian2013multiband,li2008mft,sarkar1995mpa}.
However, the following analysis and proposed solution apply to the generalized multiband fusion problem portrayed in Fig. \ref{fig:general_multiband}, as detailed in Section \ref{subsubsec:dri_generalizability_study}. 
Additionally, although extrapolation of the signal beyond the highest and lowest subband frequencies has been proposed for previous techniques \cite{zhang2014coherent,zhang2017multiple}, we focus on developing an algorithm to reliably impute the missing signal in the frequency gap. 

In a generic near-field SAR or inverse-SAR (ISAR) scenario, two radars are mounted on a platform that scans a target scene for high-resolution imaging. 
Hence, provided proper system design, the synthetic aperture elements of both radars can overlap to produce a virtual monostatic element operating in the frequency ranges of both radars. 
Alternatively, the following fusion signal model can be achieved through other means, such as a monostatic wideband system sampled at several subbands to reduce sampling bandwidth, a MIMO dual-band system whose virtual elements of each subband overlap, or modeling the problem as fusion in the angular spatial wavenumber domain \cite{wang2018wavenumber}, provided the radars are sufficiently close to each other during scanning. 

Suppose the subbands start at frequencies $f_1$ and $f_2$, respectively, and illuminate a target in $x$-$y$-$z$ Cartesian space, where $z$ represents the downrange or range direction, and $x$-$y$ are known as the cross-range directions. 
Consider the monostatic element operating at both subbands and located at $(x',y',z')$ illuminating $N_t$ targets modeled as point scatterers, where the $i$-th target is located at $(x_i,y_i,z_i)$ with reflectivity $\alpha_i$. 
The wavenumber domain, or $k$-domain, response to a chirp signal at the first and second radars can be written as
\begin{align}
    \label{eq:radar1_response}
    s_1(n) &= \sum_{i=0}^{N_t-1} \alpha_i e^{-j2(k_1 + \Delta_k n)R_i}, \ n = 0, \dots, N_k-1, \\
    s_2(n) &= \sum_{i=0}^{N_t-1} \alpha_i e^{-j2(k_2 + \Delta_k n)R_i}, \ n = 0, \dots, N_k-1, 
    \label{eq:radar2_response}
\end{align}
where $k_1$ and $k_2$ are the wavenumbers corresponding to the starting frequencies $f_1$ and $f_2$, respectively, $k = 2\pi f/c$, $\Delta_k$ is the wavenumber sampling interval, $n$ is the time sample index, $N_k$ is the number of samples in each subband, and $R_i$ is the distance from the radar to the $i$-th scatterer, which is expressed as 
\begin{equation}
    \label{eq:R_i}
    R_i = \left[ (x' - x_i)^2 + (y' - y_i)^2 + (z' - z_i)^2 \right]^\frac{1}{2}.
\end{equation}
Although the sampling conditions, sampling rate $\Delta_k$ and number of samples $N_k$, are considered identical across subbands for simplicity, this is not a strictly necessary condition as the subbands could have different sampling conditions. 
While different values of $N_k$ raises a trivial issue, different sample rates among subbands will need to be compensated such that the spectral domains are coincident. 
In addition, the proposed signal model assumes that the scattering parameters, $\alpha_i$, are frequency-independent. 
However, in a real scenario, scattering properties of the various materials in the target scene vary across subbands, a phenomenon that may not be adequately modeled by (\ref{eq:radar1_response}) and (\ref{eq:radar2_response2}), depending on the material properties and frequency ranges of the subbands. 
Since the proposed training data scheme assumes the frequency-independence of $\alpha_i$, networks trained on these data are limited because of this assumption, as discussed in Section \ref{subsec:dri_training_details}. 

Both $s_1(\cdot)$ and $s_2(\cdot)$ are considered multisinusoidal signals because they are composed of a superposition of scaled complex exponential functions, whose frequencies are determined by the ranges $R_i$. 
Hence, the wavenumber spectral domain, known as the range domain or $R$-domain, exhibits peaks at positions corresponding to the ranges $R_i$.
Defining $\Delta_B \triangleq k_2 - k_1$, the difference between the starting wavenumbers, the signal at the second subband can be rewritten with respect to $k_1$ and different indexing as
\begin{equation}
    \label{eq:radar2_response2}
    s_2(n') = \sum_{i=0}^{N_t-1} \alpha_i e^{-j2(k_1 + \Delta_k n')R_i}, \ n' = \Tilde{N}, \dots, N,
\end{equation}
such that $\Tilde{N} \triangleq \Delta_B/\Delta_k$ is the offset between subbands 1 and 2, where $N \triangleq \Tilde{N} + N_k - 1$, $n' = n + \Tilde{N}$, and $\Tilde{N} > N_k$.
We assume that $\Tilde{N}$ is an integer based on the choices of $\Delta_k$ and $\Delta_B$, although the derivation is valid regardless.

\begin{figure}[th]
    \centering
    \includegraphics[width=0.75\textwidth]{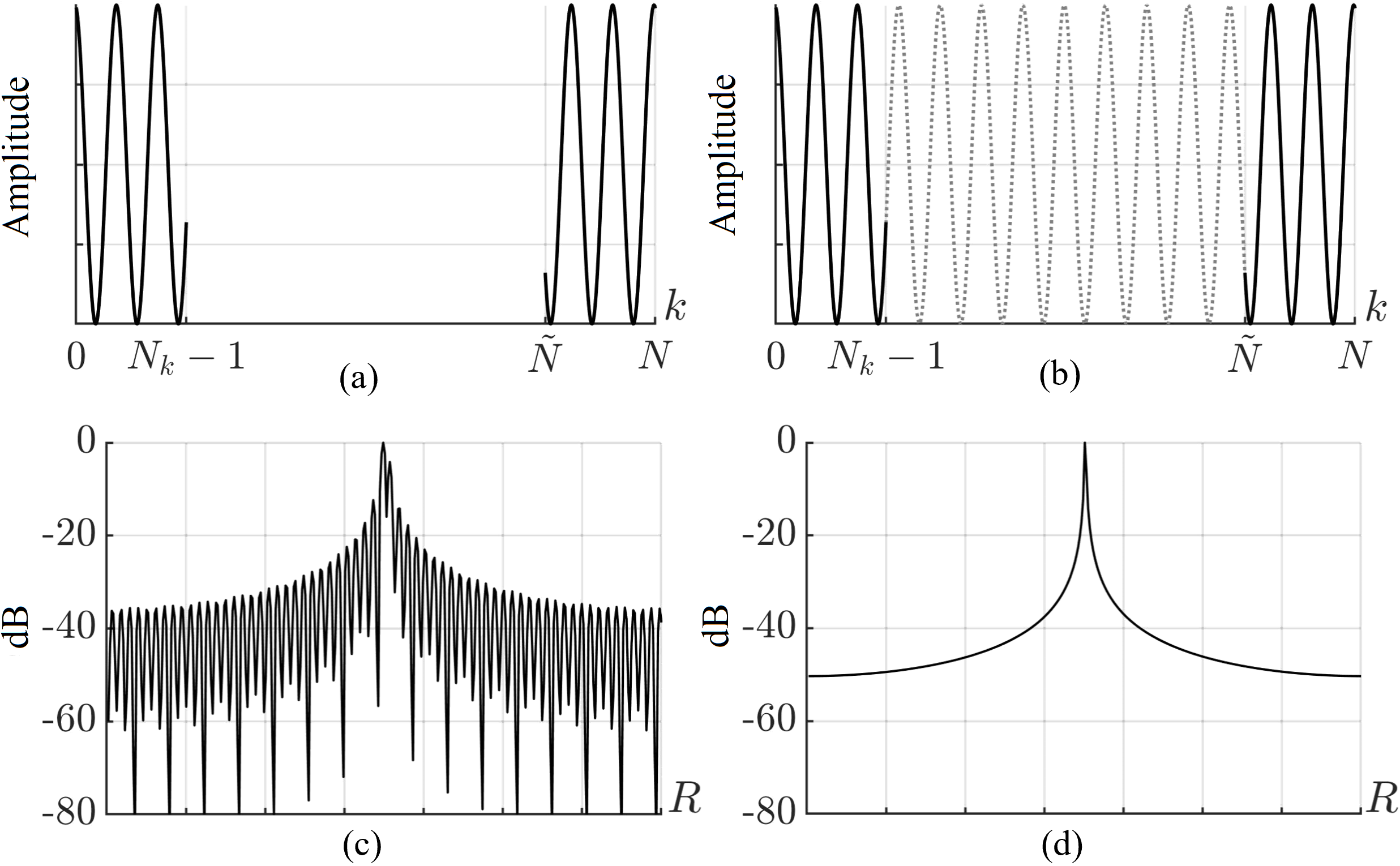}
    \caption{Multiband scenario with two subbands in the $k$-domain and $R$-domain. (a) $k$-domain non-contiguous dual-band signal, (b) $k$-domain ideal full-band signal, (c) $R$-domain spectrum of the non-contiguous dual-band signal with zero-padding, (d) $R$-domain spectrum of the ideal full-band signal.}
    \label{fig:multiband_scenario} 
\end{figure}

From (\ref{eq:radar1_response}) and (\ref{eq:radar2_response2}), the multiband scenario can be understood as a multisinusoidal signal sampled across several disjoint regions offset by $\Tilde{N}$.
We refer to the two subband scenario as \textit{non-contiguous dual-band}, as shown in Fig. \ref{fig:multiband_scenario}a.
As the two subbands are not coherent in a practical implementation, we implement the algorithm developed in \cite{wang2018wavenumber} to efficiently estimate the ICP and compensate each subband accordingly. 
Additional details on the mutual coherency among subbands can be found in \cite{tian2013multiband,tian2014sparse,zou2016matrix,wang2018wavenumber}. 
The signal in the $k$-domain can be represented in the $R$-domain, or wavenumber spectral domain, by taking the Fourier transform. 
Limiting the sampling in each subband to $N_k$ corresponds to a convolution in the $R$-domain with a discrete sinc or Dirichlet kernel of width $1/N_k$, resulting in smearing of the spectral information and causing closely spaced peaks to bend together \cite{izacard2021datadriven}.
Given the structure of the multiband signal, the $R$-domain spectrum is the sum of the spectra for each subband if the frequency gap is ignored. 
Because of the sinc-effect and phase shift in the $R$-domain corresponding to the $\Tilde{N}$ sample shift in the $k$-domain for each of the $N_t$ reflectors, the non-contiguous dual-band signal in the $R$-domain suffers from artifacts/sidelobes as the frequency gap between the subbands is neglected \cite{wang2018wavenumber}, as shown in Fig. \ref{fig:multiband_scenario}c.
This analysis is identical to the MFT \cite{li2008mft}, which results in images degraded by increased sidelobes in the range direction. 
Comparatively, the $R$-domain spectrum of the ideal \textit{full-band} signal (Fig. \ref{fig:multiband_scenario}d) does not contain spurious peaks that would distort the images recovered from the fused signal, thereby achieving an improved resolution compared with each subband and the MFT approach.

\subsection{Existing Methods for Multiband Signal Fusion}
\label{subsec:dri_existing_methods}
The objective of multiband signal fusion is to impute the bandwidth between the subbands (interpolating between the subbands or extrapolating the missing samples) in the $k$-domain to acquire the ideal full-band signal shown in Fig. \ref{fig:multiband_scenario}b, where the dotted portion represents the signal in the frequency gap. 
Methods for recovering the missing wavenumber domain data from $N_k$ to $\Tilde{N}-1$ apply MUSIC \cite{tian2014sparse} or MPA \cite{zou2016matrix,wang2018wavenumber} to estimate the signal poles in an all-pole model. 
However, these approaches assume that the estimated model order of $s_1(\cdot)$ and $s_2(\cdot)$, $\hat{N}_t \approx N_t$, is small compared to $N_k$.
From the analysis in \cite{sarkar1995mpa}, $\hat{N}_t$, the estimated number of targets in the scene, must be chosen such that $\hat{N}_t < \text{round}(N_k/3)$ for the MPA \cite{wang2018wavenumber}.
After the $\hat{N}_t$ signal poles and coefficients are computed, the missing samples can be estimated. 
However, high-resolution near-field SAR often requires imaging of intricate, continuous objects modeled by thousands or millions of point scatterers, or $N_t \gg N_k$ \cite{smith2020nearfieldisar,batra2021short}. 
As a result, traditional approaches such as the MPA assume simplistic targets, thereby neglecting high-frequency features of the target, and are unable to faithfully recover the multiband signals. 

For clarity in the remainder of this chapter, we propose new terminology to describe the spectral composition of radar target scenes based on the portion of the baseband bandwidth occupied by the reflected signal. 
A target consisting of fewer reflectors than the number of frequency samples $(N_t < N_k)$ over the specified frequency range has a \textit{low-bandwidth} relative to the bandwidth of the system. 
Low-bandwidth targets have a low model order and a low-rank sample covariance matrix, allowing conventional algorithms to adequately approximate signal poles and coefficients. 
In contrast, a target consisting of a large number of reflectors relative to the frequency sampling $(N_t \gg N_k)$ is a \textit{high-bandwidth} target, which is typical in most security and industrial applications of near-field SAR imaging. 
High-bandwidth targets contain intricate, high-frequency spatial features and have not been addressed for multiband signal fusion in previous studies. 

\begin{figure*}[t]
    \centering
    \includegraphics[width=\textwidth]{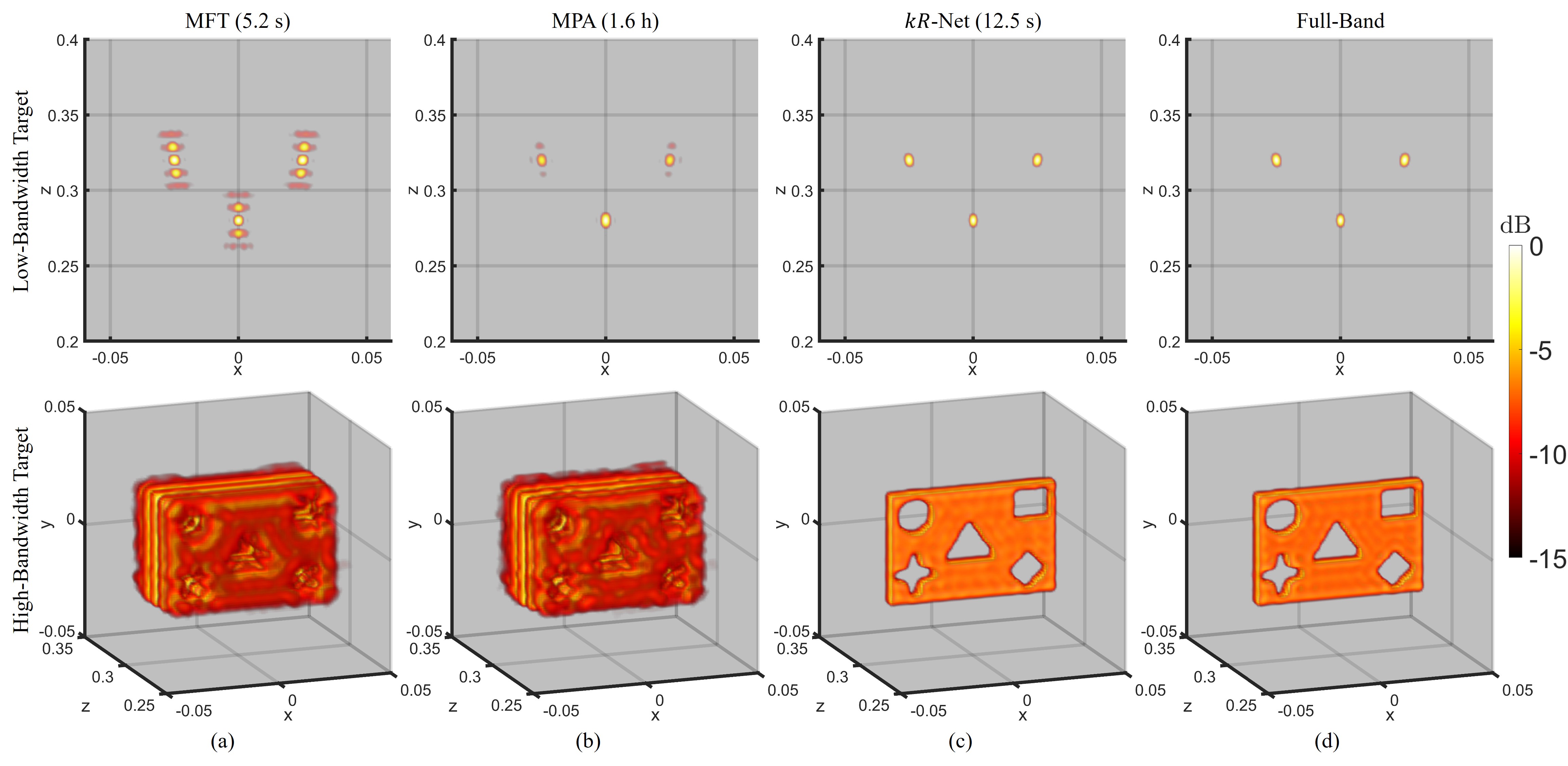}
    \caption{Demonstration of the limitations of the matrix Fourier transform (MFT) and matrix-pencil algorithm (MPA) for multiband signal fusion using two targets in simulation: (top) three point targets and (bottom) rectangle with various cutout shapes. (a) The MFT, which requires 5.2 s, suffers from significant sidelobes since it does not compensate for the missing samples in the frequency gap. (b) Although the MPA, which requires 1.6 h for a GPU implementation, achieves adequate reconstruction for the low-bandwidth target (top), its performance degrades for a high-bandwidth target (bottom). (c) The proposed $kR$-Net, which requires 12.5 s, recovers an image that closely resembles the (d) ideal reference image corresponding the full-band scenario as in Figs. \ref{fig:multiband_scenario}b and \ref{fig:multiband_scenario}d.}
    \label{fig:ex_mft_mpa_ours} 
\end{figure*}

To illustrate this phenomenon, we consider two cases: (1) a low-bandwidth target consisting of 3 point reflectors and (2) a \mbox{3-D} model of a rectangle with various cutout shapes to constitute a high-bandwidth target, as shown in the first and second rows of Fig. \ref{fig:ex_mft_mpa_ours}, respectively. 
All images are reconstructed using the RMA \cite{yanik2020development,sheen2001three,smith2022efficient,vu2022fourier} with the SAR scanning parameters detailed in Section \ref{subsubsec:dri_sim2_cutout1}. 
The MFT \cite{li2008mft} is applied to fuse the multiband data, yielding the images in the left column (Fig. \ref{fig:ex_mft_mpa_ours}a). 
As expected, since the MFT does not account for the frequency gap between the subbands, increased sidelobes are observed along the $z$-direction. 
However, the quality of images recovered using the MPA \cite{wang2018wavenumber} varies significantly. 
For the simple low-bandwidth target in the top row $(N_t = 3, N_k = 64)$, the MPA recovers each point with minimal undesirable sidelobes compared to the MFT. 
However, although its performance is better than the MFT, the MPA is plagued by considerable degradation for the high-bandwidth target, where $N_t$ is on the order of thousands and $N_t \gg N_k$, as the high-bandwidth features of the target are not adequately modeled by the MPA. 
For an equitable comparison, a parallelized GPU implementation of the MPA is employed \cite{zou2016matrix,wang2018wavenumber}. 
Hence, the computation time required for the GPU-implemented MPA is still 1.6 h for this example, deeming it unfit for many applications demanding rapid imaging, such as packing and security screening.
By comparison, the MFT and $kR$-Net boast computation times of 5.2 s and 12.5 s, respectively, enabling many common mmWave imaging solutions. 
In the right column (Fig. \ref{fig:ex_mft_mpa_ours}c), the imaging results obtained using the proposed $kR$-Net demonstrate robustness for both targets by achieving focusing performance comparable to the ideal, full-band scenario. 
Notably, the intricate features of the high-bandwidth target are retained, and the resolution in the $z$-direction is significantly improved compared with the MFT and MPA. 

Multiband signal fusion can be posed as a spectral super-resolution/restoration problem in the $R$-domain, the dual to imputation in the $k$-domain.
As shown in Figs. \ref{fig:multiband_scenario}c and \ref{fig:multiband_scenario}d, $R$-domain super-resolution of the $N_t$ peaks corresponds to imputation of the full bandwidth \cite{wang2019multi}. 
Deep learning-based solutions have proven successful in similar spectral-enhancement problems on radar images \cite{dai2021imaging,jing2022enhanced,smith2021An,vasileiou2022efficient,zhang2019target} and multisinusoidal line spectra \cite{izacard2021datadriven,pan2021complexFrequencyEstimation}, achieving resolutions exceeding the theoretical limitations. 
However, data-driven approaches have not been applied to multiband signals to achieve joint $k$-domain imputation and $R$-domain super-resolution. 
Because the multiband fusion problem has distinct features in the $k$-domain and $R$-domain, we propose a hybrid approach that operates in both domains. 


\section{Proposed Architecture of $kR$-Net for Improved 3-D Multiband SAR Imaging}
\label{sec:dri_methods}

In this section, we introduce a novel dual-domain CV-CNN architecture, referred to as $kR$-Net, to perform efficient multiband fusion for improved \mbox{3-D} near-field SAR imaging. 
The proposed framework alternates between operating in the $k$-domain and $R$-domain, allowing the network to learn the unique characteristics inherent to each domain. 
Compared with learning in only one domain, $kR$-Net demonstrates superior convergence and quantitative performance, as discussed in Section \ref{sec:dri_results}.
Additionally, the proposed algorithm is robust for low- and high-bandwidth imaging scenarios, which are common in many realistic applications.

\begin{figure*}[t]
    \centering
    \includegraphics[width=\textwidth]{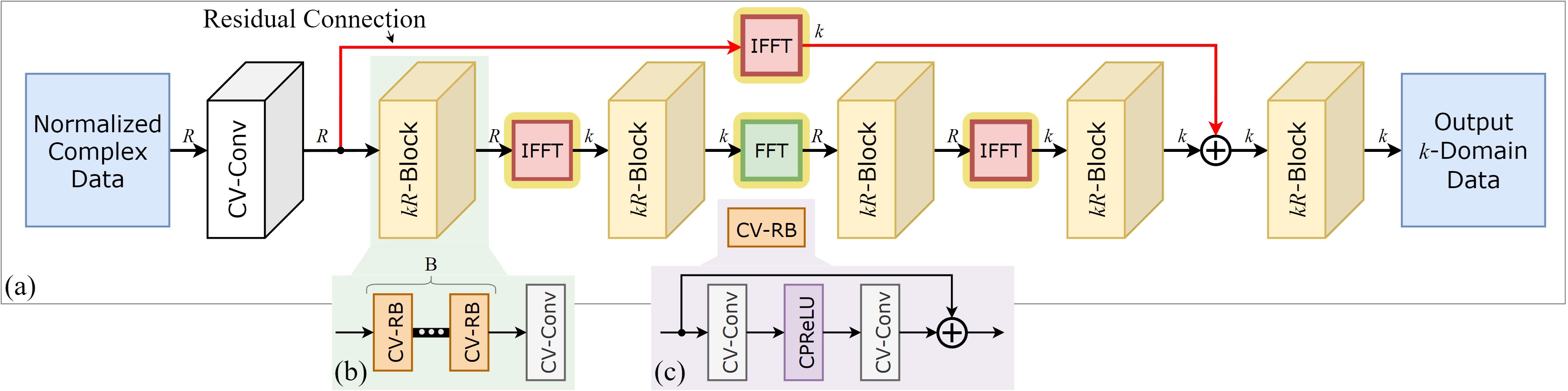}
    \caption{Architecture of the (a) hybrid, dual-domain $kR$-Net. Proposed addition of the domain transformation blocks (highlighted: FFT and IFFT) allows the network to learn important features of the signal in both the $k$- and $R$-domains improving on conventional CNN approaches. The novel, hybrid approach using FFT and IFFT blocks throughout the network, achieves superior multiband signal fusion and spectral super-resolution compared with a model without the Fourier operations. Architecture of the (b) $kR$-Block and (c) CV-RB. The $kR$-block consists of $B$ CV-RBs in cascade followed by a CV-Conv layer. Each CV-RB is a residual convolution block with a single CPReLU activation.}
    \label{fig:kRNet_overview} 
\end{figure*}

\subsection{Framework of $kR$-Net}
The architecture of $kR$-Net is shown in Fig. \ref{fig:kRNet_overview}a, where the signal domain is denoted at each connection as $k$ or $R$ for the wavenumber domain or wavenumber spectral domain, respectively, and the forward and inverse fast Fourier transform are denoted by FFT and IFFT, respectively. 
The input is given as the $R$-domain spectrum of the wavenumber domain samples and is processed in both domains by residual $kR$-blocks. 
After the residual connection, an additional $kR$-block and complex-valued convolution (CV-Conv) layer processes the signal before it is output in the $k$-domain.
The specific designs of each module are detailed as follows.

\subsubsection{Input Layer and Residual Connection}
\label{subsubsec:input_layer}
Rather than layering the real and complex parts of the signal \cite{smith2021sterile,izacard2021datadriven} or employing a two-path network \cite{wang2021tpssiNet}, the inputs to $kR$-Net are complex-valued signals of length $N$. 
The complex-valued input signals are normalized in the $R$-domain by the magnitude min-max norm before being passed to the first CV-Conv layer as
\begin{equation}
    \label{eq:min_max_norm}
    \mathbf{z} = \frac{\mathbf{x} - ||\mathbf{x}||_\text{min}}{||\mathbf{x}||_\text{max} - ||\mathbf{x}||_\text{min}},
\end{equation}
where $||\mathbf{x}||_\text{min}$ and $||\mathbf{x}||_\text{max}$ are the minimum and maximum values of the magnitude of $\mathbf{x}$. 
Hence, the phase of $\mathbf{x}$ remains unmodified, while the magnitude is scaled to be consistently between 0 and 1. 

Since the network expects an input in the $R$-domain, the FFT of the $k$-domain data is computed prior to the input to the network with zero-padding between the subbands.
Complex-valued convolution extends the convolution operation employed by CNNs to complex input data. 
To implement a CV-Conv layer, the convolution kernel matrix must be complex-valued. 
However, because the convolution between two complex-valued tensors is generally unsupported by deep learning software, we decompose the input signal $\mathbf{x} = \mathbf{x}_R + j\mathbf{x}_I$ into real and imaginary parts.
Similarly, by expressing the kernel as $\mathbf{M} = \mathbf{M}_R + j\mathbf{M}_I$, the complex-valued convolution can be written, neglecting the bias terms, as
\begin{equation}
    \label{eq:cv_conv}
    \mathbf{x} \circledast \mathbf{M} = \mathbf{x}_R \circledast \mathbf{M}_R - \mathbf{x}_I \circledast \mathbf{M}_I + j(\mathbf{x}_R \circledast \mathbf{M}_I + \mathbf{x}_I \circledast \mathbf{M}_R),
\end{equation}
where $\circledast$ denotes the real-valued convolution operation. 
By decomposing the convolution in this manner, complex-valued convolution can be computed using existing techniques operating on the real and imaginary parts of the input signal and kernel. 
The real and imaginary parts of the weight matrix $\mathbf{M}$ can be implemented as real-valued matrices according to (\ref{eq:cv_conv}), and their values are determined by complex-valued backpropagation following the convention for CV-CNNs \cite{zhang2017complex,jing2022enhanced}. 

The CV-Conv layer is a general-purpose complex-valued convolution layer defined with a kernel size $K$ and zero-padding such that the signal length of $N$ is preserved at the output, $C_\text{in}$ input channels, and $C_\text{out}$ output channels.
The first layer of $kR$-Net is a CV-Conv layer with 1 input channel and $F$ output channels, where $F$ is the number of feature channels and is constant throughout the network.
After the first CV-Conv layer, the intermediate representations are fed through an IFFT block in the residual pass-forward connection, as shown in red in Fig. \ref{fig:kRNet_overview}a. 
The residual connection preserves the information at the known subbands, and the network demonstrates superior empirical performance with the proposed configuration than without the pass-forward connection. 

\subsubsection{$kR$-Blocks and Domain Transformation Blocks}
\label{subsubsec:kR_block}
The $kR$-Block is composed of a cascade of complex-valued residual blocks (CV-RBs) followed by a single CV-Conv layer, as shown in Fig. \ref{fig:kRNet_overview}b. 
Each $kR$-Block operates on the signal in either the $k$-domain or $R$-domain, as the signal alternates between the two domains throughout the $kR$-Net. 
Furthermore, because the domain transformation blocks (FFT and IFFT) are fully differentiable, they can be treated as conventional layers in the network, and gradient backpropagation can be easily implemented \cite{jing2022enhanced}.
The Fourier operations are performed across each activation map and normalized to make the FFT and IFFT orthonormal.
Based on the convolution properties of the Fourier transform, convolution in one domain can be viewed as multiplication in the other domain. 
In this sense, applying a CV-Conv layer in the $k$-domain can be considered a fully connected layer in the $R$-domain. 
However, as illustrated later in Section \ref{subsubsec:dri_ablation_study}, our hybrid, dual-domain approach outperforms a network operating exclusively in the $k$-domain or $R$-domain in terms of convergence and numerical performance. 
Hence, compared with conventional CNN models, the addition of domain transformation blocks throughout the network is key to improving multiband fusion performance. 

The CV-RB architecture is shown in Fig. \ref{fig:kRNet_overview}c. 
Inspired by \cite{lim2017enhanced}, the residual block consists of two convolution layers separated by an activation function: the complex parametric rectified linear unit (CPReLU) \cite{jing2022enhanced}. 
Compared with the original ResNet residual block \cite{he2016deep_resnet}, the proposed residual block removes batch normalization and empirically outperforms a bottleneck residual architecture \cite{liu2022convnext} for multiband signal fusion. 

The CPReLU activation function is selected over alternatives, such as the complex ReLU (CReLU) \cite{gao2018enhanced}, which computes the sum of the ReLU operation on the real and imaginary values separately, as the CPReLU has an improved activation over the complex domain. 
Using the notation employed in (\ref{eq:cv_conv}), the CPReLU can be expressed as 
\begin{align}
    \label{eq:CPReLU}
    \begin{split}
        \text{CPReLU}(\mathbf{x}) &= \max(0, \mathbf{x}_R) + \eta_R \min(0, \mathbf{x}_R) \\ &+ j\left( \max(0, \mathbf{x}_I) + \eta_I \min(0, \mathbf{x}_I) \right) ,
    \end{split}
\end{align}
where the parameters $\eta_R$ and $\eta_I$ are learned during the training stage of the network for each CPReLU instance \cite{jing2022enhanced}. 
Because $\eta_R$ and $\eta_I$ are learned independently, different layers of the network may learn different representations of the signal in amplitude and phase, aiding network robustness. 
The CPReLU can be understood as a complex domain parametric rectified linear unit (PReLU), which extends the traditional ReLU into the negative input domain to overcome gradient saturation for negative activation values. 
In the CPReLU, the real and imaginary parts of $\mathbf{x}$ are independently processed by a PReLU and the output is complex-valued. 
In the complex domain, this corresponds to retaining information in all four quadrants, corresponding to all combinations of positive and negative real and imaginary activation values, as detailed in \cite{jing2022enhanced}. 

After the first four $kR$-Blocks, the residual connection is made in the $k$-domain followed by another $kR$-Block and CV-Conv layer before being output, as shown in Fig. \ref{fig:kRNet_overview}a.
The number of $kR$-Blocks and domain transformations was investigated empirically, and the proposed configuration yielded the optimal numerical performance. 
However, further investigation of alternate architectures and deeper neural networks is a promising route for future research. 
In the spirit of \cite{liu2022convnext}, multiple values for the convolution kernel size $K$ were investigated, and the optimal value was determined empirically to be $K = 5$.
The number of feature maps throughout the network is chosen as $F = 32$, and the number of CV-RBs for each $kR$-block is set as $B = 8$.
$kR$-Net comprises 86 CV-Conv layers and 866324 learnable parameters. 
Multiband signal fusion is performed by $kR$-Net on a signal of length $N$, yielding a fused signal in the $k$-domain of equivalent size. 
As discussed in Section \ref{subsubsec:dri_generalizability_study}, a network is trained for a specific multiband signal fusion scenario and must be retrained for application to alternate subband configurations (placement and range of subbands, sampling conditions, length $N$ of full-band signal, etc.).


\subsection{Training Details}
\label{subsec:dri_training_details}
The weights of the network are calibrated using an Adam optimizer with a learning rate of $1 \times 10^{-4}$, $\beta_1 = 0.9$, and $\beta_2 = 0.999$.
Training is performed on a single RTX3090 GPU with 24 GB of memory with a batch size of 1024 and L1 loss criterion. 
The complex-valued loss term is defined as 
\begin{equation}
    \label{eq:l1_loss_cv}
    \mathcal{L} = \sum_{\ell=0}^{N-1} \biggr[ |\hat{s}(\ell)_R - s(\ell)_R| + |\hat{s}(\ell)_I - s(\ell)_I| \biggr],
\end{equation}
where $\hat{s}(\ell)$ are the predicted signals output from the network corresponding to the full-band ground-truth vectors $s(\ell)$ and the subscripts denote the real and imaginary parts of the signals. 
The complex components are processed separately, similar to the approaches in \cite{jing2022enhanced,zhang2017complex}, by traditional L1 distance metrics, and the real-valued result is the sum of the two L1 losses from the real and imaginary parts of the predicted signal with the ground-truth signal. 

More advanced loss functions were investigated, such as the L1/L2 difference between the ground-truth signal and intermediate representations throughout the network or loss between sample covariance matrices. 
These loss functions were tested in conjunction with alternate configurations, such as varying the number of $kR$-Blocks or removing the residual connection. 
However, the architecture detailed in Fig. \ref{fig:kRNet_overview}a demonstrated superior numerical performance in both training and testing with real multiband SAR data. 
Nevertheless, future investigations into neural network design based on statistical signal processing principles will likely facilitate additional insights and promising results for signal processing problems. 
Incorporating a hybrid, data-driven signal processing approach is a promising direction for similar future efforts. 

\subsubsection{Training and Testing Datasets}
\label{subsubsec:dri_dataset}
Since there is no publicly available dataset for near-field multiband SAR imaging, we generate training and testing datasets by simulating the response to a multiband LFM radar.
The ideal noiseless full-band signals spanning both subbands and the frequency gap in the $k$-domain are generated as 
\begin{equation}
    \label{eq:fb_sim}
    s(\ell) = \sum_{i=0}^{N_t - 1} \alpha_i e^{-j 2(k_1 + \Delta_k \ell) R_i}, \ell = 0, \dots, N - 1,
\end{equation}
where $\alpha_i$ values are selected from a complex normal distribution and $R_i$ values are chosen from a uniform distribution spanning the unambiguous range of the radar. 
After computing the full-band signals $s(\cdot)$ the multiband signals $\hat{s}(\cdot)$ are obtained by nullifying the samples in the frequency gap as
\begin{equation}
    \hat{s}(\ell)=\begin{cases}
          s(\ell) \quad & \, \ell \in [0, N_k - 1] \cup [\Tilde{N}, N - 1] \\
          0 \quad & \, \ell \in [N_k, \Tilde{N} - 1] \\
     \end{cases}.
\end{equation}
The multiband signals are then corrupted with complex additive Gaussian white noise (AWGN) in each subband. 
The noisy multiband signals are used as the input to $kR$-Net after taking the FFT and employing the normalization process detailed earlier.
Each noisy multiband signal is treated as a feature vector with a corresponding label vector consisting of the ideal full-band signal in (\ref{eq:fb_sim}). 

To train the network, 1048576 samples are independently generated with $N_t$ target reflectors, where $N_t$ is randomly selected between 1 and 200. 
The SNR for each sample is selected on a continuous uniform distribution from -10 to 30 dB. 
A validation set of 2048 samples is generated using the same procedure. 
Assuming a realistic scenario with two radars with starting frequencies $f_1 = 60$ GHz and $f_2 =$ 77 GHz, where each radar has a bandwidth of $B =$ 4 GHz, we set $N_k =$ 64 and $\Delta_f =$ 62.5 MHz. 
Hence, $\Tilde{N} = 272$ and $N = 336$, and the low-rank assumption of the MPA, $\hat{N}_t < \text{round}(N_k/3)$, will often be invalid if the target is high-bandwidth and consists of many reflectors. 
Although this study employs 60 GHz and 77 GHz radars, the proposed algorithm can easily be extended to other multiband configurations. 
For practical implementation, federal communications commission (FCC) licensing limits certain combinations of subbands based on application, but the algorithm and concepts derived in this study are applicable across various subband configurations. 
By training on this dataset, the proposed algorithm learns to perform multiband fusion for high-bandwidth targets. 

It is important to note that the proposed signal model and dataset generation scheme impose limitations on the model and its generalizability. 
In a practical implementation, imperfections that are not modeled in (\ref{eq:fb_sim}), such as frequency dependence of the scattering parameters, $\alpha_i$, device non-linearity, clutter, non-Gaussian noises, and different antenna distortions among subbands, may impact the reconstruction quality. 
Hence, networks trained on these data may display different robustness depending on the imaging scenario, placing additional importance on the system design and calibration. 
Although a model trained for a specific multiband scenario will not generalize well to other subband configurations, the proposed model architecture demonstrates the ability to generalize to various multiband imaging schemes when trained with appropriate datasets, as detailed in Section \ref{subsubsec:dri_generalizability_study}. 

An alternative dataset was considered consisting of extended, solid targets, such as those shown in Figs. \ref{fig:ex_mft_mpa_ours} and \ref{fig:dri_sim6}. 
However, because there is no sufficiently diverse dataset of such multiband SAR data, the network did not generalize well across different shapes. 
Additionally, a network was first trained on a dataset of randomly placed point reflectors, as in (\ref{eq:fb_sim}), and then fine-tuned on data from solid targets, but a performance improvement was not observed. 
Further development of diverse mmWave datasets will be essential to the advancement of joint signal processing and data-driven algorithms and is a promising future direction. 

\begin{figure}[h]
    \centering
    \includegraphics[width=0.55\textwidth]{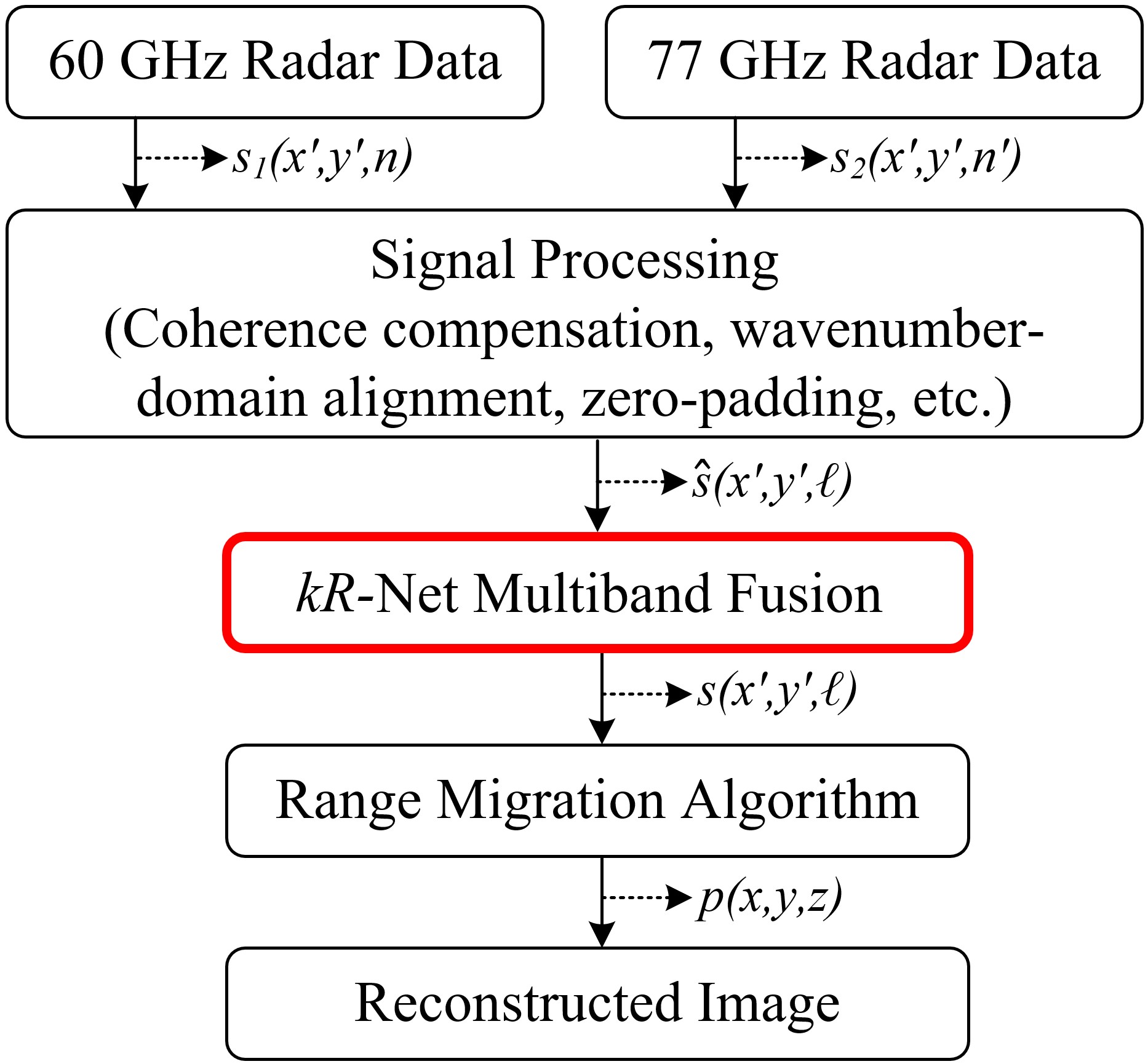}
    \caption{Multiband signal fusion pipeline for high-resolution \mbox{3-D} near-field imaging. The data from both subbands are fused using the proposed $kR$-Net producing a high-fidelity reconstruction for both low- and high-bandwidth targets.}
    \label{fig:dri_flow} 
\end{figure}

\subsection{Imaging Implementation}
\label{subsec:dri_rma_algo}
After multiband fusion is performed using $kR$-Net, the SAR image is reconstructed from the fused data. 
A summary of the imaging pipeline is presented in Fig. \ref{fig:dri_flow}. 
After the data are collected from both radars, preprocessing steps are necessary to ensure signal coherence and align the data in the $k$-domain. 
We implement the ICP compensation algorithm detailed in \cite{wang2018wavenumber} and set consistent sampling parameters across radars. 
The proposed algorithm is valid for both collocated and noncollocated antennas if the spatial wavenumber domains of all radars are sufficiently coincident, implying that both radars have similar illumination of the target. 
Applying $kR$-Net to the multiband signal is advantageous compared with classical signal processing algorithms as $kR$-Net is highly parallelizable and can efficiently perform signal fusion for many samples. 
The RMA is applied after the signal fusion step to produce a high-resolution \mbox{3-D} image \cite{yanik2020development,smith2022efficient,yanik2019cascaded}.
Compared to conventional signal processing-based algorithms for multiband fusion, the proposed $kR$-Net yields superior imaging performance and demonstrates robustness for the realistic case of high-bandwidth, intricate targets.

\section{Multiband Imaging System}
\label{sec:dri_system}
This section provides an overview of the implementation of the multiband imaging prototype using commercially available mmWave radars. 
Whereas prior research on near-field multiband radar imaging has employed sophisticated laboratory equipment \cite{wang2018wavenumber,tian2013multiband}, which is not suitable for many practical applications, we introduce a highly-integrated system that employs commercially available equipment for multiband near-field SAR. 
The proposed testbed uses two mmWave radars operating at distinct subbands and introduces a synchronization strategy to achieve efficient data collection.

\begin{figure}[h]
    \centering
    \includegraphics[width=0.75\textwidth]{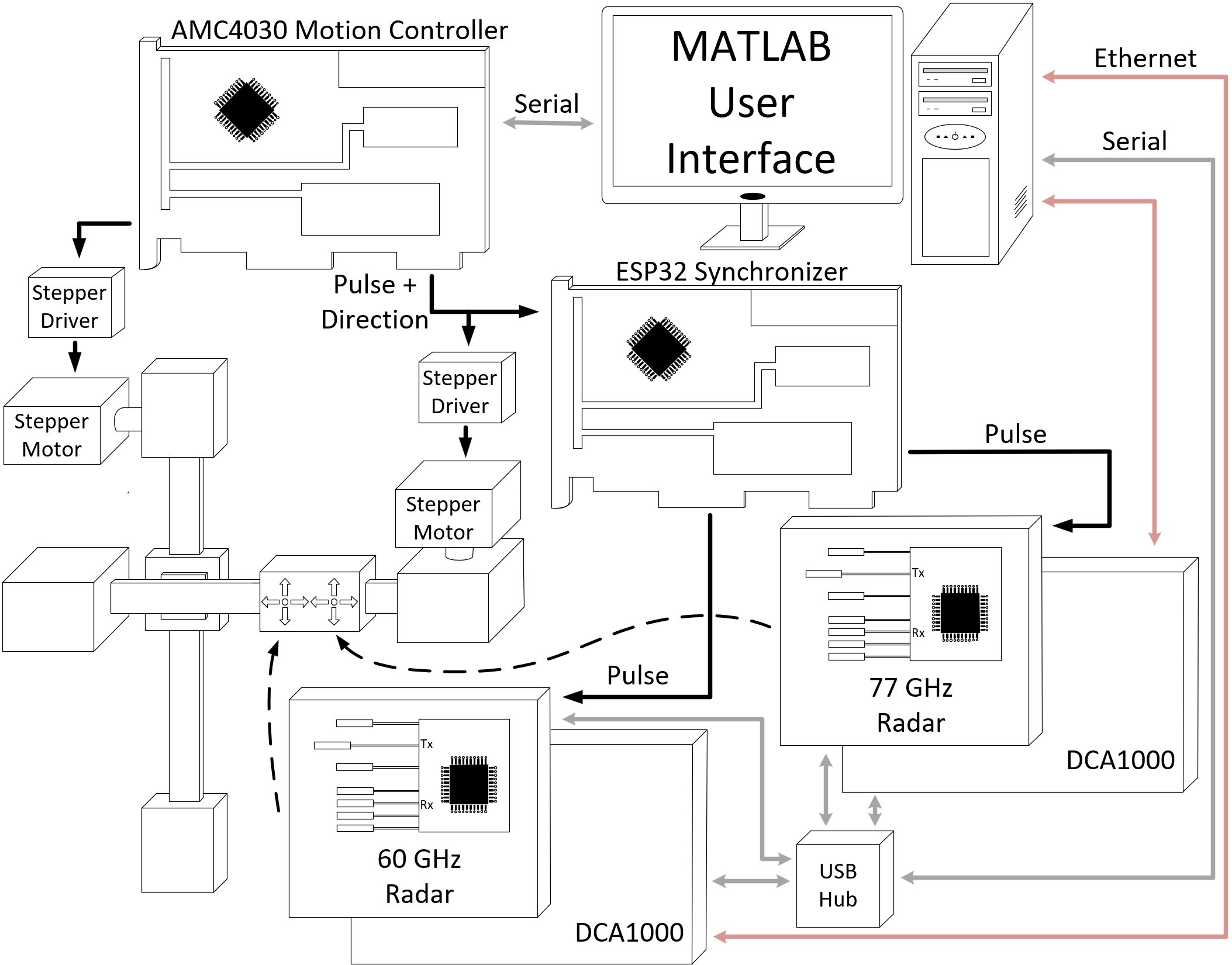}
    \caption{System architecture of the proposed multiband imaging testbed prototype.}
    \label{fig:dri_system} 
\end{figure}

An overview of the system architecture is presented in Fig. \ref{fig:dri_system}. 
The system consists of a 60 GHz radar, 77 GHz radar, two-axis mechanical scanner, motion controller, synchronization module, and a host PC.
The Texas Instruments (TI) IWR6843ISK and IWR1642BOOST are chosen as the single chip 60 GHz and 77 GHz radars, respectively.
Both radars have an operating bandwidth of $B = 4$ GHz; hence, the subbands span 60--64 GHz and 77--81 GHz. 
In addition, the LFM radars are configured using the parameters detailed in Section \ref{subsec:dri_training_details}. 
The data are captured in real-time by the TI DCA1000 evaluation module for each radar and streamed to the host PC over Ethernet. 
Both radars are mounted onto a belt-driven two-axis mechanical scanner, as shown in Fig. \ref{fig:dri_system}, such that the lowest Rx antennas on each radar are aligned, and the radars are separated horizontally by a distance of $\Delta_x$. 

The two-axis mechanical scanner is driven by stepper motors that receive pulses from a motion controller, and the entire system is controlled using a custom MATLAB user interface running on the host PC. 
The radars are scanned in the $x$- and $y$-directions at the spatial sampling Nyquist rate of $\lambda / 4$ in both the horizontal and vertical directions \cite{yanik2019sparse,sheen2001three}. 
Extending the synchronization approach in \cite{yanik2020development}, we design a novel multi-radar synchronizer for the precise positioning of both radars while operating at high scanning speeds. 
The proposed prototype achieves a speed of 500 mm/s, allowing for short scanning times required by applications such as security screening and packaging. 
However, scanning multiple physical radar modules at these speeds while achieving precise synthetic element positioning required to mitigate distortion is challenging and has not been addressed previously in the literature. 
The synchronizer monitors the stepper driver pulses that determine the position of the platform, to which the radars are mounted, as it both accelerates and decelerates. 
Additionally, the synchronizer tracks the positions of each radar and fires them independently to account for the non-uniform timing required to maintain a uniform synthetic aperture and ensure equivalent illumination of the target scene from both radars. 
Additional hardware-specific details of mmWave imaging testbeds can be found in \cite{yanik2020development}.

\section{Experimental Results}
\label{sec:dri_results}
In this section, the superiority of the proposed $kR$-Net is demonstrated using numerical simulations and empirical experiments. 
The matrix Fourier transform (MFT) algorithm \cite{li2008mft} and matrix-pencil algorithm (MPA) \cite{zou2016matrix,wang2018wavenumber} are adopted as comparison baselines for the following experiments. 
After $kR$-Net is trained using the procedure detailed in Section \ref{subsec:dri_training_details}, we conduct experiments on both synthetic and empirical multiband data to validate the performance of $kR$-Net compared with traditional signal processing approaches. 
We consider a dual-band system with the radar signaling parameters discussed previously. 

\subsection{Visual Comparison of Simulation Results}
\label{subsec:dri_qual_sim}
First, we detail various simulation results obtained using the proposed $kR$-Net algorithm for multiband signal fusion. 

\begin{figure}[h]
    \centering
    \includegraphics[width=0.7\textwidth]{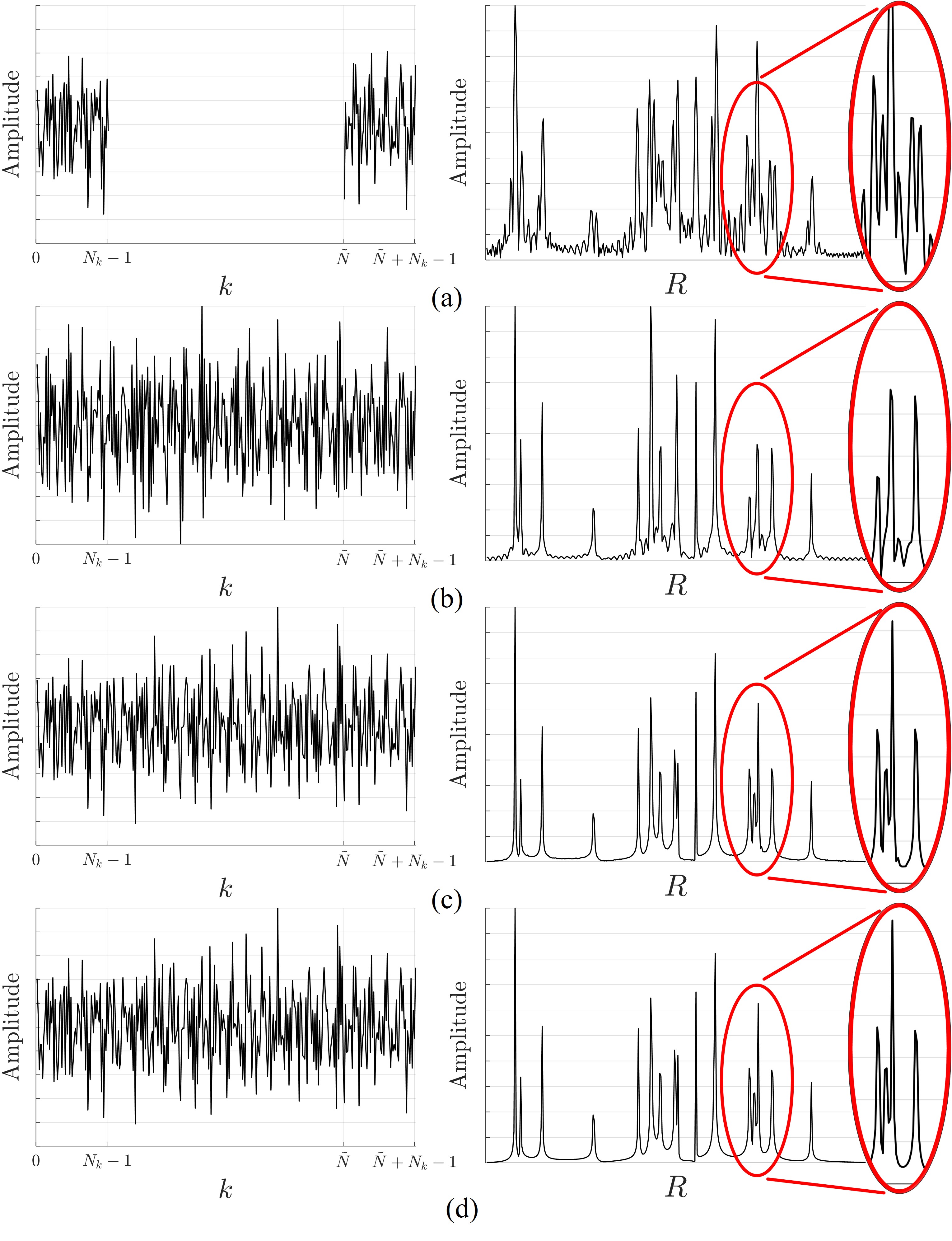}
    \caption{Closely spaced peaks resolved using $kR$-Net. Comparison of multiband signal fusion performance using (a) MFT, (b) MPA, and (c) $kR$-Net compared to (d) the ideal full-band scenario for a single simulated signal consisting of randomly placed point scatters. Left: The real part of the $k$-domain signals. Right: The magnitude of the $R$-domain spectra, demonstrating the super-resolution capability of the proposed $kR$-Net.}
    \label{fig:dri_demo3} 
\end{figure}

\subsubsection{Single Multiband Signal}
\label{subsubsec:demo3}
First, we consider the simple case of a single multiband signal captured at the two subbands used throughout the experiments with a target consisting of randomly placed points. 
The $k$-domain signals and $R$-domain spectra for the MFT, MPA, and $kR$-Net are shown in Fig. \ref{fig:dri_demo3} and compared with the corresponding ideal full-band signal. 
Each subband is corrupted with AWGN at an SNR of 20 dB. 
The MPA imputes the lost signal between the two subbands to recover a signal of length $N$; however, the resulting wideband signal deviates from the ideal signal owing to the assumptions in the MPA. 
Although it outperforms the MFT, the MPA is unable to recover every peak in the $R$-domain, and $kR$-Net yields the most accurate reconstruction of the full-band signal. 

\begin{figure}[th]
    \centering
    \includegraphics[width=0.5\textwidth]{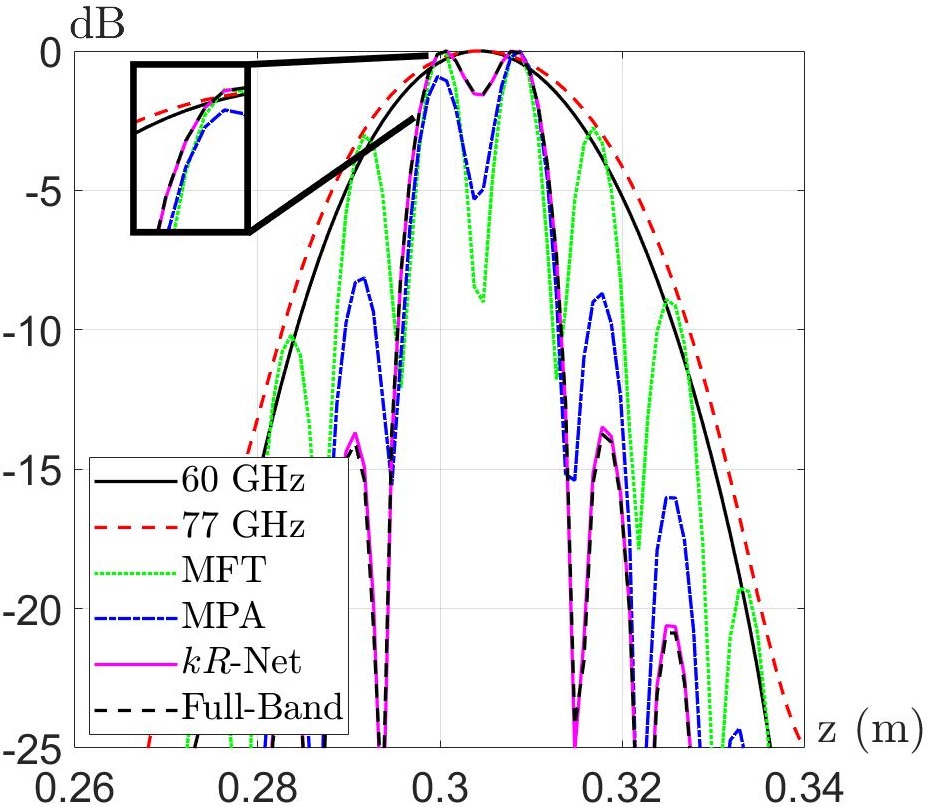}
    \caption{Comparison of imaging results from various scenarios sliced along $y = 0$ m with two simulated point scatterers spaced by $\Delta_z = 7.1$ mm, which corresponds to the minimum resolvable distance for an equivalent bandwidth of 21 GHz.}
    \label{fig:dri_res} 
\end{figure} 

\subsubsection{Effective Bandwidth Study}
\label{subsubsec:dri_res}
Given that the purpose of multiband imaging is to achieve a finer resolution by synthetically increasing the bandwidth, we consider the resolution capability of the proposed algorithm and its corresponding effective bandwidth. 
The resolution of a radar system in the downrange direction is given by $\delta_z = c/2B$, which determines the ability of the system to resolve two closely spaced reflectors. 
Hence, we compare the imaging results for various cases with two closely spaced peaks to evaluate the resolution limit of the algorithms. 
We simulate a scenario with two point scatterers located at $Z_r$ and $Z_r + \Delta_z$ from the radar boresight with AWGN at an SNR of 20 dB. 
Using $Z_r =$ 300 mm, we evaluate the performance for $\Delta_z =$ 7.1 mm, which corresponds to an effective bandwidth of 21 GHz. 
As shown in Fig. \ref{fig:dri_res}, the proposed $kR$-Net achieves a nearly identical response to the ideal full-band signal. 
Since the 60 GHz and 77 GHz radars, subbands 1 and 2, each have a bandwidth of 4 GHz, the two closely spaced reflectors are blurred into a single peak. 
The MFT resolves the two peaks but has severely increased sidelobes compared to the full-band signal.
Because the target consists of two signal components, the MPA resolves both peaks without significant distortion, but demonstrates some minor deviations from the ideal signal.
In contrast, $kR$-Net achieves a more accurate signal with lower sidelobes. 
Thus, the proposed algorithm achieves an effective bandwidth of 21 GHz because the two peaks are clearly resolved. 

\begin{figure*}[th]
    \centering
    \includegraphics[width=\textwidth]{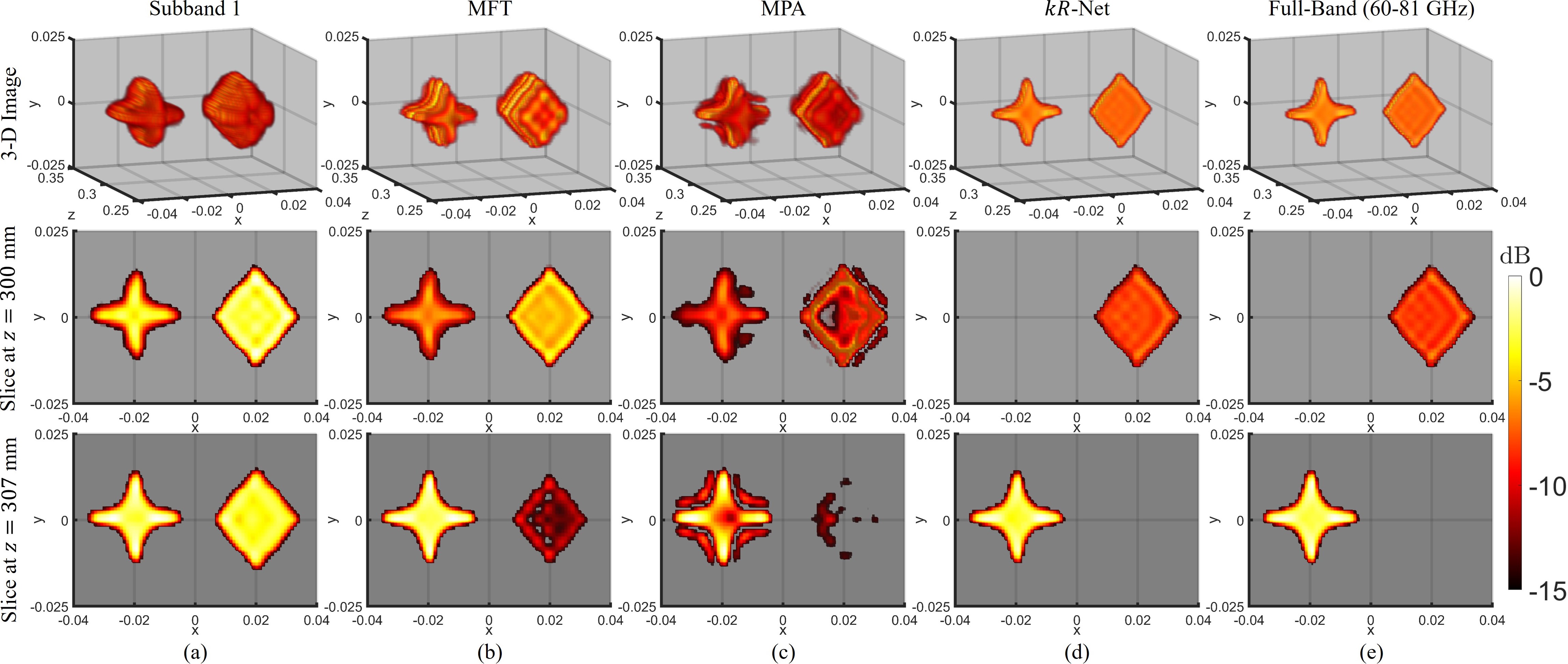}
    \caption{High-bandwidth target consisting of a diamond located at $z = 300$ mm and star located at $z = 307$ mm. The proposed $kR$-Net is able to separate the two shapes to their respective $z$-planes, exhibiting comparable imaging performance to an ideal full-band transceiver without feature loss or visible sidelobes present in the MFT and MPA images. Imaging results for simulated high-bandwidth target using (a) subband 1  (60--64 GHz), (b) MFT, (c) MPA, (d) $kR$-Net, (e) ideal full-band (60--81 GHz). First row: \mbox{3-D} image. Second row: slice at $z = 300$ mm, corresponding to the location of the diamond shape. Third row: slice at $z = 307$ mm, corresponding to the location of the star shape.}
    \label{fig:dri_sim6} 
\end{figure*}

\subsubsection{High-Bandwidth Target}
\label{subsubsec:dri_sim2_cutout1}

To visually evaluate the imaging performance of $kR$-Net and demonstrate the deficiencies of the MFT and MPA on a high-bandwidth target, two shapes are considered under the multiband scenario in the near-field with a planar array with dimensions of 0.125 m $\times$ 0.125 m, satisfying the spatial Nyquist criterion \cite{yanik2019sparse}. 
A star shape is placed on the left at the plane $z = 307$ mm and a diamond to the right at the plane $z = 300$ mm, with the synthetic array at the plane $z = 0$ m. 
For many high-resolution imaging tasks, objects must be localized, classified, or counted to identify concealed weapons, ensure correct packaging, or detect defects. 
To localize closely spaced targets, a transceiver with a bandwidth of 4 GHz may not be adequate as the low range resolution yields images stretched in the $z$-direction beyond the physical dimensions of the objects.
Fig. \ref{fig:dri_sim6} shows the \mbox{3-D} reconstructed images and slices along $z = 300$ mm and $z = 307$ mm to demonstrate the distortion along the $z$-direction that can contaminate the images and degrade system performance.  
As the objects are located in distinct $z$-planes, the second and third rows should show only the diamond and star, respectively, as shown in Fig. \ref{fig:dri_sim6}e, for the ideal full-band case.
The image recovered from subband 1 (60--64 GHz) is shown in Fig. \ref{fig:dri_sim6}a and demonstrates the limitations of low bandwidth, as both shapes are clearly visible in both slices.
Similarly, the MFT results in powerful sidelobes, as shown in Fig. \ref{fig:dri_res}, which correspond to ghost shapes along the $z$-direction for solid targets, as shown in Fig. \ref{fig:dri_sim6}b.
Hence, images recovered using the MFT are not suitable for localization or object counting tasks because they are obscured by spurious sidelobes. 
For this high-bandwidth target scenario, the image recovered using the MPA, shown in Fig. \ref{fig:dri_sim6}c, suffers from a loss of fidelity due to the simplistic multiband fusion model and the assumption of a small number of reflectors. 
As a result, the MPA images are not only contaminated with sidelobes that degrade the slices corresponding to each object but also fail to retain the high-bandwidth features of the objects. 
Comparatively, the images recovered using $kR$-Net, shown in Fig. \ref{fig:dri_sim6}d, closely resemble the ideal full-band images. 
This aligns with the conclusions drawn from the numerical experiments and demonstrates the superior performance of $kR$-Net compared to existing algorithms for realistic high-resolution imaging tasks.

\subsection{Quantitative Investigations}
\label{subsec:dri_quant}
First, we compare the imaging performance of the MFT, MPA, and $kR$-Net numerically, considering a near-field SAR scenario with a planar aperture of $200 \times 200$ synthetic elements. 
It is important to note that the MFT and MPA are classical signal processing approaches, whereas $kR$-Net employs a hybrid approach that combines data-driven techniques with signal processing algorithms. 
Fifty Monte Carlo trials are conducted for each experiment. 
We compare the various algorithms using the structural similarity index measure (SSIM), peak signal-to-noise ratio (PSNR), and normalized root mean square error (NRMSE), where SSIM is defined as
\begin{equation}
    \label{eq:ssim}
    \text{SSIM} = \frac{(2 \mu_\mathbf{x} \mu_\mathbf{y} + C_1)(2 \sigma_{\mathbf{x}\mathbf{y}} + C_2)}{(\mu_\mathbf{x}^2 + \mu_\mathbf{y}^2 + C_1)(\sigma_\mathbf{x}^2 + \sigma_\mathbf{y}^2 + C_2)},
\end{equation}
where $\mathbf{x}$ and $\mathbf{y}$ are the reconstructed image and reference image, respectively; $\mu_\mathbf{x}$, $\sigma_\mathbf{x}$ and $\mu_\mathbf{y}$, $\sigma_\mathbf{y}$ are their corresponding mean values and standard deviations, respectively; and $L$ is the dynamic range of the pixel values.
For stability $C_1 = (k_1 L)^2$ and $C_2 = (k_2 L)^2$ where $k_1 = 0.01$ and $k_2 = 0.03$ by default \cite{wang2021tpssiNet}. 
SSIM quantifies the similarity between $\mathbf{x}$ and $\mathbf{y}$, with a larger value indicating better performance and a maximum value of 1 for a perfect reconstruction. 
Similarly, higher PSNR values (dB) \cite{lim2017enhanced} and lower NRMSE values \cite{wang2021tpssiNet} indicate more accurate image reconstruction.

\begin{table}[h]
\centering
\caption{Comparison of SSIM, PSNR, and NRMSE across different number of targets ($N_t$)  using MFT, MPA, and $kR$-Net}
\label{tab:dri_sim_numerical_Nt}
\resizebox{0.75\textwidth}{!}{%
\begin{tabular}{c||ccc|ccc|ccc}
\multirow{2}{*}{$N_t$} & \multicolumn{3}{c|}{MFT} & \multicolumn{3}{c|}{MPA} & \multicolumn{3}{c}{$kR$-Net} \\ \cline{2-10} 
 
& \multicolumn{1}{c|}{SSIM} & \multicolumn{1}{c|}{PSNR} & NRMSE & \multicolumn{1}{c|}{SSIM} & \multicolumn{1}{c|}{PSNR} & NRMSE & \multicolumn{1}{c|}{SSIM} & \multicolumn{1}{c|}{PSNR} & NRMSE \\ 
\hline \hline

3 & \multicolumn{1}{c|}{0.9970} & \multicolumn{1}{c|}{48.91} & 1.145 & \multicolumn{1}{c|}{0.9997} & \multicolumn{1}{c|}{66.92} & 0.1727 & \multicolumn{1}{c|}{\textbf{0.9999}} & \multicolumn{1}{c|}{\textbf{90.90}} & \textbf{0.01779} \\ 
\hline
10 & \multicolumn{1}{c|}{0.9934} & \multicolumn{1}{c|}{45.81} & 1.142 & \multicolumn{1}{c|}{0.9959} & \multicolumn{1}{c|}{48.03} & 0.7904 & \multicolumn{1}{c|}{\textbf{0.9997}} & \multicolumn{1}{c|}{\textbf{59.36}} & \textbf{0.2439} \\ 
\hline
100 & \multicolumn{1}{c|}{0.9703} & \multicolumn{1}{c|}{39.56} & 1.108 & \multicolumn{1}{c|}{0.9684} & \multicolumn{1}{c|}{39.17} & 1.160 & \multicolumn{1}{c|}{\textbf{0.9809}} & \multicolumn{1}{c|}{\textbf{41.60}} & \textbf{0.8779} \\ 
\hline
400 & \multicolumn{1}{c|}{0.9220} & \multicolumn{1}{c|}{34.07} & 1.097 & \multicolumn{1}{c|}{0.9240} & \multicolumn{1}{c|}{33.56} & 1.164 & \multicolumn{1}{c|}{\textbf{0.9425}} & \multicolumn{1}{c|}{\textbf{35.44}} & \textbf{0.9378} \\ 
\hline
700 & \multicolumn{1}{c|}{0.8989} & \multicolumn{1}{c|}{32.28} & 1.089 & \multicolumn{1}{c|}{0.9061} & \multicolumn{1}{c|}{31.94} & 1.130 & \multicolumn{1}{c|}{\textbf{0.9250}} & \multicolumn{1}{c|}{\textbf{33.72}} & \textbf{0.9208} \\ 
\hline
1000 & \multicolumn{1}{c|}{0.8874} & \multicolumn{1}{c|}{31.39} & 1.068 & \multicolumn{1}{c|}{0.8902} & \multicolumn{1}{c|}{30.82} & 1.143 & \multicolumn{1}{c|}{\textbf{0.9094}} & \multicolumn{1}{c|}{\textbf{32.45}} & \textbf{0.9488} \\ 
\hline
1300 & \multicolumn{1}{c|}{0.8678} & \multicolumn{1}{c|}{30.03} & 1.055 & \multicolumn{1}{c|}{0.8714} & \multicolumn{1}{c|}{29.50} & 1.122 & \multicolumn{1}{c|}{\textbf{0.8922}} & \multicolumn{1}{c|}{\textbf{31.14}} & \textbf{0.9302} \\ 
\hline
Avg. & \multicolumn{1}{c|}{0.9338} & \multicolumn{1}{c|}{37.44} & 1.100 & \multicolumn{1}{c|}{0.9364} & \multicolumn{1}{c|}{40.14} & 0.9644 & \multicolumn{1}{c|}{\textbf{0.9499}} & \multicolumn{1}{c|}{\textbf{46.37}} & \textbf{0.6967} \\ 
\hline
\hline
\end{tabular}%
}
\end{table}

Images are computed using the MFT, MPA, and $kR$-Net for multiband signal fusion of the two subbands, 60--64 GHz and 77-81 GHz, and compared against the image recovered from the ideal full-band scenario spanning the entire bandwidth of 60-81 GHz as in (\ref{eq:fb_sim}). 
First, \mbox{3-D} images are generated with an SNR of 20 dB for the full-band and multiband cases, as discussed in Section \ref{subsec:dri_training_details}. 
The robustness of the various algorithms is compared over a varying number of targets $N_t$, which corresponds to the target bandwidth.
The results are presented in Table \ref{tab:dri_sim_numerical_Nt}. 
The MFT suffers from increased sidelobes because it does not account for the frequency gap. 
As expected, the MPA exhibits tremendous performance for low-bandwidth targets but is plagued by imaging degradation as the target bandwidth and $N_t$ increase. 
In contrast, the proposed $kR$-Net demonstrates robustness compared with classical algorithms, particularly when applied to high-bandwidth targets scenarios. 
This is expected as $kR$-Net is trained on a dataset containing both low- and high-bandwidth targets. 
The best evaluations are shown in boldface in Table \ref{tab:dri_sim_numerical_Nt}, indicating the superiority of $kR$-Net over the MFT and MPA for realistic \mbox{3-D} imaging scenarios. 

\begin{table}[h]
\centering
\caption{Comparison of SSIM, PSNR, and NRMSE across values of SNR using MFT, MPA, and $kR$-Net}
\label{tab:dri_sim_numerical_snr}
\resizebox{0.75\textwidth}{!}{%
\begin{tabular}{c||ccc|ccc|ccc}
\multirow{2}{*}{SNR} & \multicolumn{3}{c|}{MFT} & \multicolumn{3}{c|}{MPA} & \multicolumn{3}{c}{$kR$-Net} \\ \cline{2-10} 
 
& \multicolumn{1}{c|}{SSIM} & \multicolumn{1}{c|}{PSNR} & NRMSE & \multicolumn{1}{c|}{SSIM} & \multicolumn{1}{c|}{PSNR} & NRMSE & \multicolumn{1}{c|}{SSIM} & \multicolumn{1}{c|}{PSNR} & NRMSE \\ 
\hline \hline

20 dB & \multicolumn{1}{c|}{0.95414} & \multicolumn{1}{c|}{33.64} & 1.089 & \multicolumn{1}{c|}{0.9743} & \multicolumn{1}{c|}{39.72} & 0.5510 & \multicolumn{1}{c|}{\textbf{0.9839}} & \multicolumn{1}{c|}{\textbf{42.39}} & \textbf{0.4064} \\ 
\hline
15 dB & \multicolumn{1}{c|}{0.9487} & \multicolumn{1}{c|}{33.08} & 1.153 & \multicolumn{1}{c|}{0.9733} & \multicolumn{1}{c|}{38.72} & 0.6138 & \multicolumn{1}{c|}{\textbf{0.9801}} & \multicolumn{1}{c|}{\textbf{41.06}} & \textbf{0.4680} \\ 
\hline
10 dB & \multicolumn{1}{c|}{0.9465} & \multicolumn{1}{c|}{32.324} & 1.150 & \multicolumn{1}{c|}{0.9560} & \multicolumn{1}{c|}{37.41} & 0.6580 & \multicolumn{1}{c|}{\textbf{0.9816}} & \multicolumn{1}{c|}{\textbf{41.47}} & \textbf{0.4129} \\ 
\hline
5 dB & \multicolumn{1}{c|}{0.9485} & \multicolumn{1}{c|}{34.32} & 1.109 & \multicolumn{1}{c|}{0.9573} & \multicolumn{1}{c|}{37.25} & 0.8203 & \multicolumn{1}{c|}{\textbf{0.9800}} & \multicolumn{1}{c|}{\textbf{41.00}} & \textbf{0.5375} \\ 
\hline
0 dB & \multicolumn{1}{c|}{0.9457} & \multicolumn{1}{c|}{32.48} & 1.131 & \multicolumn{1}{c|}{0.9610} & \multicolumn{1}{c|}{37.61} & 0.6880 & \multicolumn{1}{c|}{\textbf{0.9780}} & \multicolumn{1}{c|}{\textbf{40.76}} & \textbf{0.5009} \\ 
\hline
Avg. & \multicolumn{1}{c|}{0.9487} & \multicolumn{1}{c|}{33.17} & 1.1265 & \multicolumn{1}{c|}{0.9644} & \multicolumn{1}{c|}{38.14} & 0.6662 & \multicolumn{1}{c|}{\textbf{0.9807}} & \multicolumn{1}{c|}{\textbf{41.34}} & \textbf{0.4651} \\ 
\hline
\hline
\end{tabular}%
}
\end{table}

Next, images with 200 randomly distributed point scatterers in addition to one solid object, selected from a set of 10 basic shapes including circle, square, triangle, etc., are generated for performance evaluation.
Table \ref{tab:dri_sim_numerical_snr} presents the results for the different algorithms evaluated with SNR values ranging from 0 dB to 20 dB in increments of 5 dB. 
Because $N_t > 200$ for every image, the targets are considered high-bandwidth, and the MPA is unable to adequately model the target intricacies. 
In this experiment, the MPA outperforms the MFT because of the addition of the solid target, as the MPA slightly reduces the sidelobes from the solid target, which contribute a larger amount of power than the point scatterers. 
However, $kR$-Net demonstrates robustness across low and high SNR, yielding high SSIM and PSNR in conjunction with low NRMSE. 

These analyses validate the superiority of the proposed algorithm for both low- and high-bandwidth target scenarios. 
$kR$-Net overcomes deficiencies in the MPA due to a simplistic model that deems the MPA unsuitable for many near-field imaging applications that require high-resolution imaging of intricate objects without feature loss. 
Furthermore, since $kR$-Net is highly parallelizable, its computation time per image, 12.5 s, is significantly less than that of the iterative MPA \cite{zou2016matrix,wang2018wavenumber}, which requires 1.6 h for \mbox{3-D} imaging. 
As described in Section \ref{subsec:dri_existing_methods}, the same GPU hardware is used to implement all three algorithms. 
An efficient, parallelized implementation of the MPA is employed; however, the computation time remains excessive. 
Alternatively, although the MFT requires only 5.2 s to recover an image \cite{li2008mft}, it does not account for the frequency gap between subbands and consistently demonstrates inferior performance to that of $kR$-Net in terms of image quality.

\subsubsection{Ablation Study}
\label{subsubsec:dri_ablation_study}
To demonstrate the effectiveness of the hybrid, dual-domain approach, we compare two baseline networks with $kR$-Net. 
First, we remove the FFT and IFFT blocks from the network such that it operates exclusively in the $k$-domain. 
In contrast to Fig. \ref{fig:kRNet_overview}a, the signals at the input are in the $k$-domain and remain in the $k$-domain throughout the entire network. 
For imputation problems, such as image completion or multiband signal fusion, leveraging the information positioned throughout the input signal, which in our case is the signal at each subband, to complete missing regions is challenging for a CNN. 
Each pixel of an intermediate representation, that is, the output of a convolution layer, in a CNN depends on a region of the representation in the previous convolution layer. 
Hence, a pixel in a given representation depends on a certain region of the input signal, which is known as the effective receptive field. 
However, the effective receptive field typically grows slowly due to small kernel sizes, implying that the inferred signal in the frequency gap will not be aware of the subband signals until later in the network \cite{iizuka2017globally}. 
This is an issue for multiband fusion because samples between subbands require dependence on subband signals for robust estimation. 
Hence, the performance of wavenumber domain network, called $k$-Net, degrades towards the center of the frequency gap \cite{luo2016understanding} as the information from the subband signals does not adequately impact the prediction of the frequency gap. 

\begin{figure}[h]
    \centering
    \includegraphics[width=0.65\textwidth]{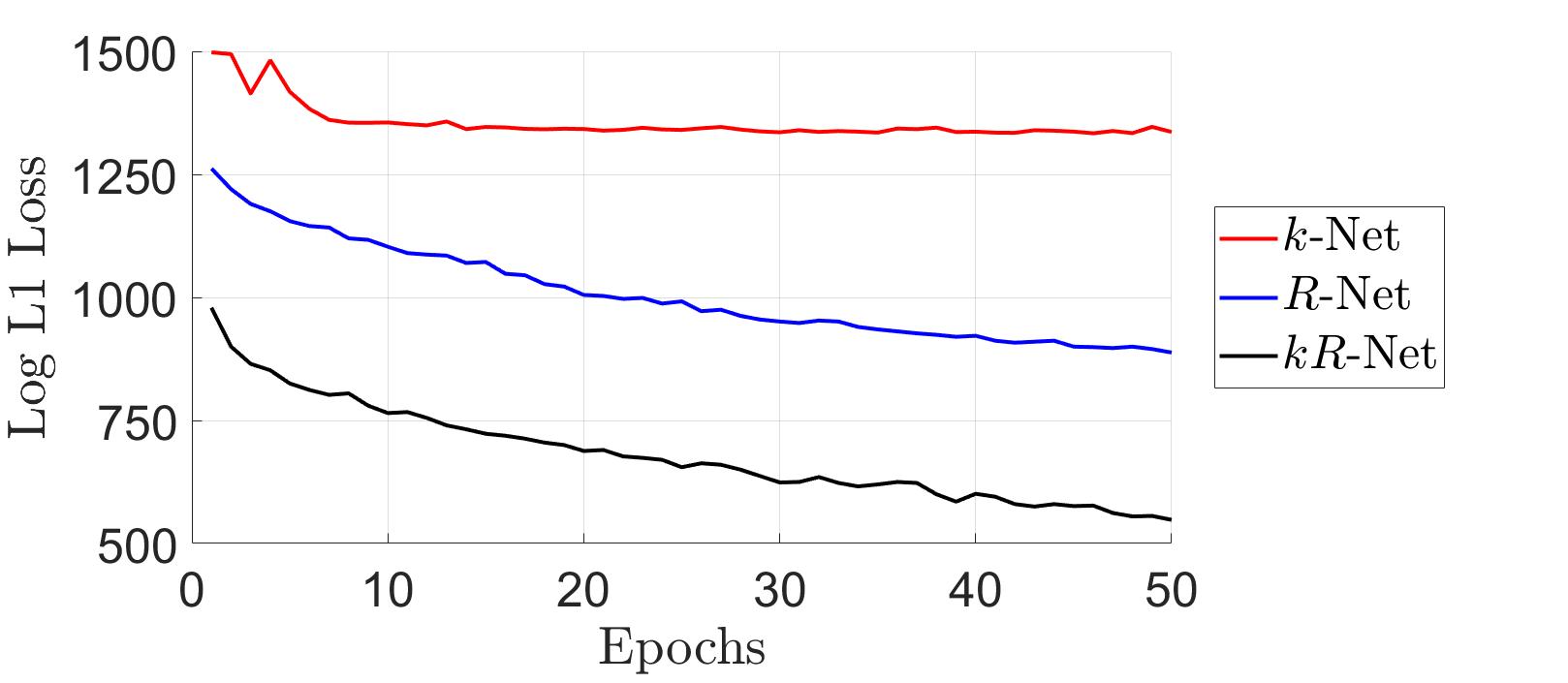}
    \caption{The proposed dual-domain $kR$-Net architecture outperforms conventional CNN models operating on $k$- or $R$-domain signals for training convergence and final performance. Comparison of log-scaled validation loss from 2048 samples during training.}
    \label{fig:dri_convergence} 
\end{figure}

Alternatively, the second baseline employs an architecture identical to that of $k$-Net but performs signal fusion in the $R$-domain. 
This technique, called $R$-Net, represents a spectral super-resolution approach to multiband signal fusion, similar to data-driven line spectra super-resolution algorithms \cite{izacard2021datadriven,pan2021complexFrequencyEstimation}.
However, despite being trained and validated on identical signals, there is a noticeable performance difference as each network operates in a different domain.

First, we compare the convergence of $kR$-Net against the baseline networks $k$-Net and $R$-Net during training, as shown in Fig. \ref{fig:dri_convergence}. 
The three algorithms are trained for 50 epochs on the same dataset, and the same dataset is used for validation.
We observe that the dual-domain architecture of $kR$-Net accelerates network training and improves the final performance. 
Comparatively, $k$-Net converges at a considerably slower rate and the signal fusion performance of $R$-Net is inferior to that of $kR$-Net. 

\begin{table}[h]
\centering
\caption{Comparison of SSIM, PSNR, and NRMSE across different number of targets ($N_t$)  using $k$-Net, $R$-Net, and $kR$-Net}
\label{tab:dri_ablation_Nt}
\resizebox{0.75\textwidth}{!}{%
\begin{tabular}{c||ccc|ccc|ccc}
\multirow{2}{*}{$N_t$} & \multicolumn{3}{c|}{$k$-Net} & \multicolumn{3}{c|}{$R$-Net} & \multicolumn{3}{c}{$kR$-Net} \\ \cline{2-10} 
 
& \multicolumn{1}{c|}{SSIM} & \multicolumn{1}{c|}{PSNR} & NRMSE & \multicolumn{1}{c|}{SSIM} & \multicolumn{1}{c|}{PSNR} & NRMSE & \multicolumn{1}{c|}{SSIM} & \multicolumn{1}{c|}{PSNR} & NRMSE \\ 
\hline \hline

3 & \multicolumn{1}{c|}{0.9972} & \multicolumn{1}{c|}{50.05} & 1.01 & \multicolumn{1}{c|}{0.9995} & \multicolumn{1}{c|}{69.98} & 0.1026 & \multicolumn{1}{c|}{\textbf{0.9999}} & \multicolumn{1}{c|}{\textbf{90.90}} & \textbf{0.01779} \\ 
\hline
10 & \multicolumn{1}{c|}{0.9946} & \multicolumn{1}{c|}{47.15} & 0.9809 & \multicolumn{1}{c|}{0.9994} & \multicolumn{1}{c|}{56.65} & 0.3308 & \multicolumn{1}{c|}{\textbf{0.9997}} & \multicolumn{1}{c|}{\textbf{59.36}} & \textbf{0.2439} \\ 
\hline
100 & \multicolumn{1}{c|}{0.9780} & \multicolumn{1}{c|}{40.92} & 0.9475 & \multicolumn{1}{c|}{0.9780} & \multicolumn{1}{c|}{41.00} & 0.9411 & \multicolumn{1}{c|}{\textbf{0.9809}} & \multicolumn{1}{c|}{\textbf{41.60}} & \textbf{0.8779} \\ 
\hline
400 & \multicolumn{1}{c|}{0.9380} & \multicolumn{1}{c|}{35.13} & 0.9715 & \multicolumn{1}{c|}{0.9351} & \multicolumn{1}{c|}{34.76} & 1.015 & \multicolumn{1}{c|}{\textbf{0.9425}} & \multicolumn{1}{c|}{\textbf{35.44}} & \textbf{0.9378} \\ 
\hline
700 & \multicolumn{1}{c|}{0.9168} & \multicolumn{1}{c|}{33.23} & 0.9746 & \multicolumn{1}{c|}{0.9167} & \multicolumn{1}{c|}{33.05} & 0.9955 & \multicolumn{1}{c|}{\textbf{0.9250}} & \multicolumn{1}{c|}{\textbf{33.72}} & \textbf{0.9208} \\ 
\hline
1000 & \multicolumn{1}{c|}{0.9035} & \multicolumn{1}{c|}{32.17} & 0.9767 & \multicolumn{1}{c|}{0.8999} & \multicolumn{1}{c|}{31.75} & 1.029 & \multicolumn{1}{c|}{\textbf{0.9094}} & \multicolumn{1}{c|}{\textbf{32.45}} & \textbf{0.9488} \\ 
\hline
1300 & \multicolumn{1}{c|}{0.8863} & \multicolumn{1}{c|}{30.96} & 0.9500 & \multicolumn{1}{c|}{0.8816} & \multicolumn{1}{c|}{30.40} & 1.012 & \multicolumn{1}{c|}{\textbf{0.8922}} & \multicolumn{1}{c|}{\textbf{31.14}} & \textbf{0.9302} \\ 
\hline
Avg. & \multicolumn{1}{c|}{0.9449} & \multicolumn{1}{c|}{38.52} & 0.9732 & \multicolumn{1}{c|}{0.9444} & \multicolumn{1}{c|}{42.51} & 0.7751 & \multicolumn{1}{c|}{\textbf{0.9499}} & \multicolumn{1}{c|}{\textbf{46.37}} & \textbf{0.6967} \\ 
\hline
\hline
\end{tabular}%
}
\end{table}

We repeat the experiment comparing performance across different numbers of targets $N_t$, for each of the three networks, and the results are shown in Table \ref{tab:dri_ablation_Nt}, where the best evaluation is marked in boldface.
Although $R$-Net outperforms the MPA in this experiment, it is unable to achieve numerical performance comparable to that of $kR$-Net. 
As $N_t$ increases, the performance of $R$-Net begins to degrade. 
This is likely due to the increased likelihood of closely spaced peaks in the $R$-domain and the resulting spectral blur. 
On the other hand, $kR$-Net demonstrates improved performance compared with both baselines and substantiates the superior ability of a hybrid approach to overcome spectral blur. 
The results demonstrate the superiority of $kR$-Net owing to its hybrid, dual-domain approach and verify that learning in both the $k$- and $R$-domains improves the quantitative performance for multiband signal fusion. 

\subsubsection{Generalizability Study}
\label{subsubsec:dri_generalizability_study}
To investigate the generalizability of the proposed technique, we consider three unique multiband scenarios with different frequency ranges, subband ranges, and number of subbands. 
Since each configuration requires a new network trained on data specific to those subbands, we simulate datasets with the same parameters as detailed in Section \ref{subsubsec:dri_dataset} for each configuration. 
Configuration $\mathcal{A}$ as two subbands with equivalent bandwidths of 2 GHz starting at 30 GHz and 38 GHz, respectively, resulting in a corresponding full-band of 30 GHz to 40 GHz. 
In Configuration $\mathcal{B}$, two subbands are used with different bandwidths, 6 GHz and 10 GHz, operating with starting frequencies of 180 GHz and 210 GHz, respectively, resulting in a full-band equivalent of 180 GHz to 220 GHz. 
Finally, a full-band range of 400 GHz to 460 GHz is achieved in Configuration $\mathcal{C}$ using three subbands with bandwidths of 8 GHz, 10 GHz, and 6 GHz and starting frequencies of 400 GHz, 428 GHz, and 454 GHz, respectively. 
\begin{table}[th]
\centering
\caption{Comparison of average PSNR and imaging time across different multiband configurations results across values of $N_t$ and SNR using MFT, MPA, and the $kR$-Net trained for the respective configuration. }
\label{tab:dri_generalizability_study}
\resizebox{0.75\textwidth}{!}{%
\begin{tabular}{c||ccc|ccc|ccc}
\multirow{2}{*}{ } & \multicolumn{3}{c|}{Configuration $\mathcal{A}$} & \multicolumn{3}{c|}{Configuration $\mathcal{B}$} & \multicolumn{3}{c}{Configuration $\mathcal{C}$} \\ \cline{2-10} 
 
& \multicolumn{1}{c|}{MFT} & \multicolumn{1}{c|}{MPA} & $kR$-Net$_\mathcal{A}$ & \multicolumn{1}{c|}{MFT} & \multicolumn{1}{c|}{MPA} & $kR$-Net$_\mathcal{B}$ & \multicolumn{1}{c|}{MFT} & \multicolumn{1}{c|}{MPA} & $kR$-Net$_\mathcal{C}$ \\ 
\hline \hline

$\bar{N_t}$ & \multicolumn{1}{c|}{31.60} & \multicolumn{1}{c|}{34.12} & \textbf{43.57} & \multicolumn{1}{c|}{31.10} & \multicolumn{1}{c|}{37.41} & \textbf{48.50} & \multicolumn{1}{c|}{32.67} & \multicolumn{1}{c|}{34.07} & \textbf{40.52} \\ 
\hline
$\bar{\text{SNR}}$ & \multicolumn{1}{c|}{33.50} & \multicolumn{1}{c|}{38.82} & \textbf{43.82} & \multicolumn{1}{c|}{33.64} & \multicolumn{1}{c|}{35.18} & \textbf{47.86} & \multicolumn{1}{c|}{33.48} & \multicolumn{1}{c|}{39.19} & \textbf{44.79} \\ 
\hline
\hline
Time (s) & \multicolumn{1}{c|}{\textbf{4.412}} & \multicolumn{1}{c|}{4362} & 9.468 & \multicolumn{1}{c|}{\textbf{4.632}} & \multicolumn{1}{c|}{5276} & 9.673 & \multicolumn{1}{c|}{\textbf{5.086}} & \multicolumn{1}{c|}{8773} & 11.39 \\ 
\hline
\hline
\hline
\end{tabular}%
}
\end{table}

Three unique networks are trained on datasets corresponding to each multiband scenario and named $kR$-Net$_\mathcal{A}$, $kR$-Net$_\mathcal{B}$, $kR$-Net$_\mathcal{C}$. 
To evaluate the numerical performance of these networks, we repeat the experiments by comparing the performance across different numbers of targets $N_t$, and SNR for each configuration. 
The average results across $N_t$ and SNR are shown in Table \ref{tab:dri_generalizability_study} along with the computation time for each algorithm, where the best evaluation is marked in boldface. 
For all three configurations, the proposed approach achieves the best numerical performance, demonstrating the ability of the proposed network to generalize to various realistic conditions. 
In addition, the MFT has the lowest computation time but the least robust image focusing, and the GPU-implemented MPA requires an excess of 1 h to impute the full-band signal with inferior reconstruction to the proposed algorithm. 
Since the MPA is only capable of multiband fusion in the case of two subbands, it must be computed twice for Configuration $\mathcal{C}$ to estimate the signal in the two frequency gaps. 
However, this increases the required computation time for the MPA to 2.4 h for each SAR image for this configuration. 
In contrast, the proposed algorithm only needs to be run once to perform a more robust reconstruction, requiring 11.4 s for Configuration $\mathcal{C}$. 
The proposed $kR$-Net demonstrates generalizability to various multisinusoidal imputation problems across a variety of frequency ranges and subband configurations. 
Provided adequate system design and hardware capable of producing multisinusoidal signals consistent with the pre-designed dataset, our technique can be trained for robust fusion across chip designs, vendors, SAR scanning patterns, etc. 
However, as described in Section \ref{sec:dri_signal_model}, the assumption of frequency-independent scattering properties may not be valid for every application and could limit the robustness of the proposed approach depending on the spectral material characteristics of the expected targets at the operating frequencies. 

\begin{figure}[h]
    \centering
    \includegraphics[width=0.65\textwidth]{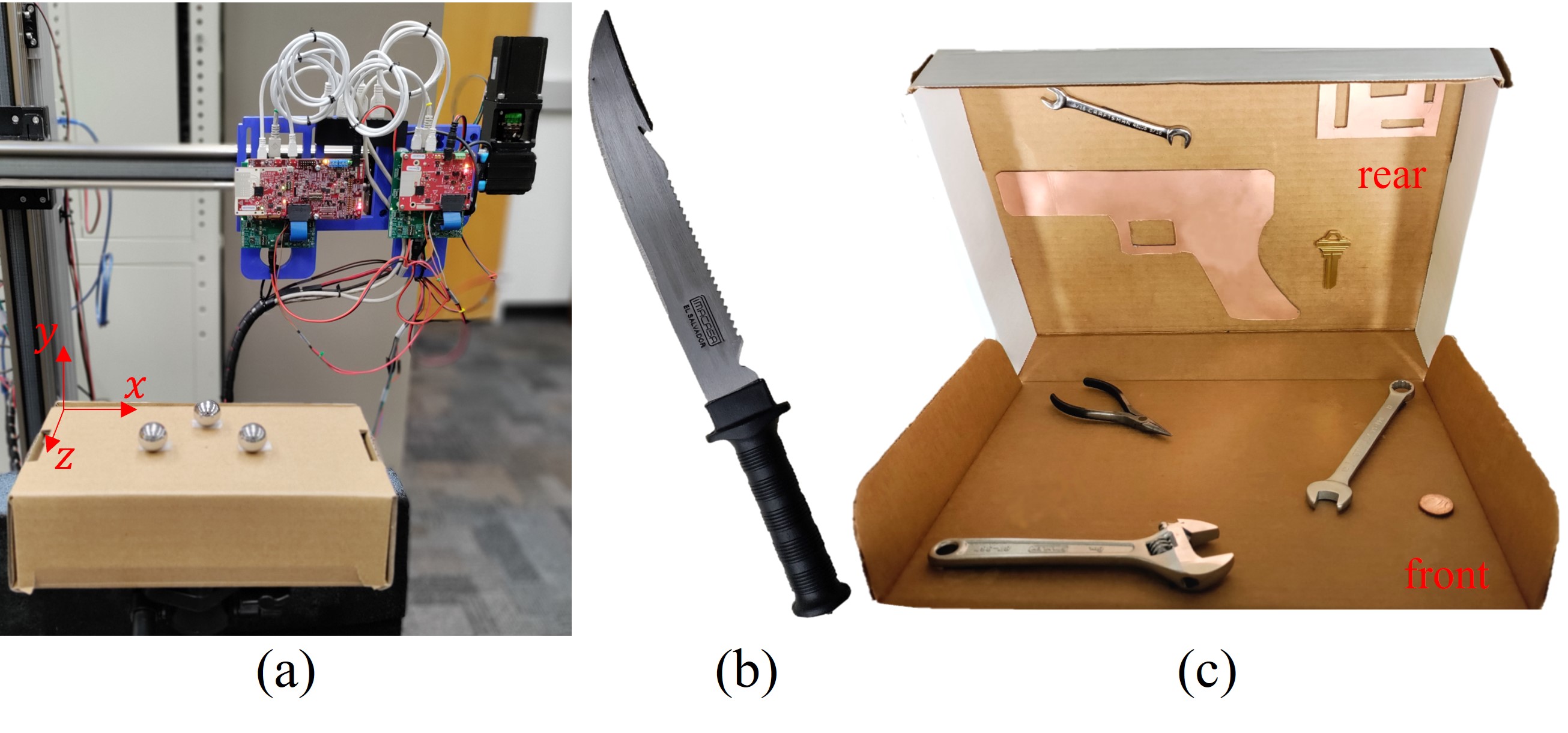}
    \caption{Various experimental targets: (a) metallic sphere targets, (b) large knife with serrated edge and notch near handle, and (c) hidden objects inside a cardboard box.}
    \label{fig:dri_targets} 
\end{figure}

\subsection{Empirical Results}
\label{subsec:dri_qual_exp}
Using the multiband imaging system detailed in Section \ref{sec:dri_system}, we acquire radar data of several objects at the two aforementioned subbands and compare the imaging results of the various multiband fusion algorithms. 

\begin{figure*}[t]
    \centering
    \includegraphics[width=\textwidth]{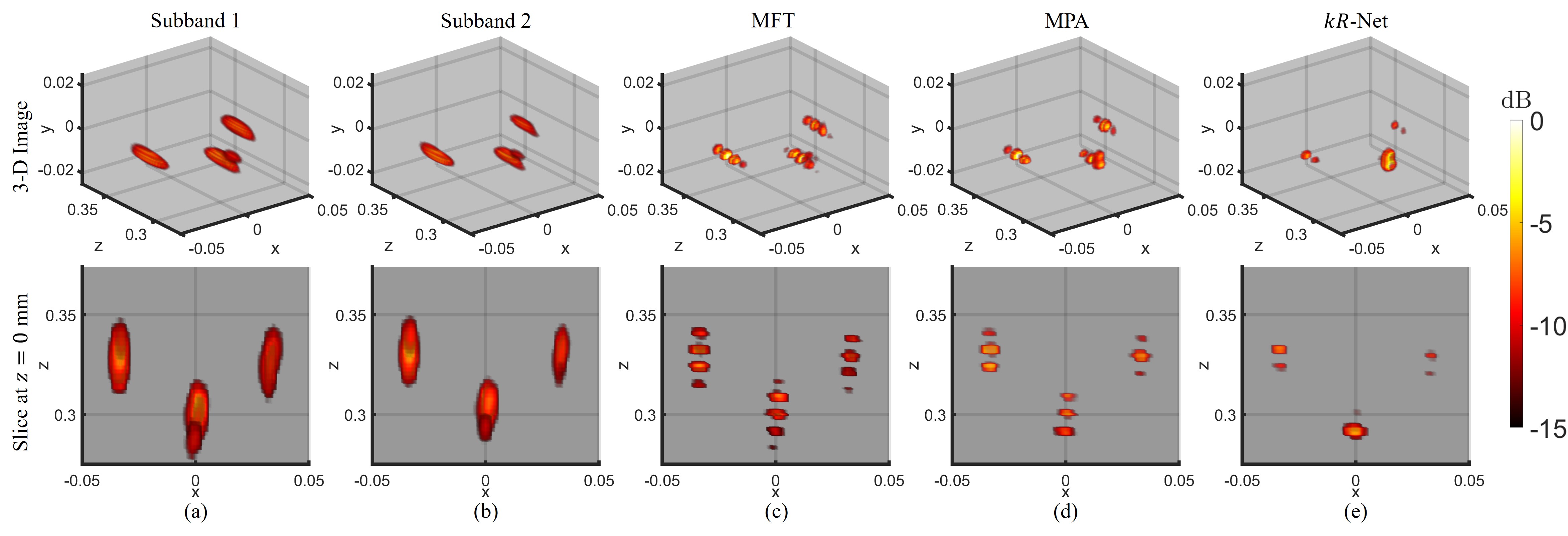}
    \caption{Whereas the MFT and MPA demonstrate ghost artifacts in the range direction, $kR$-Net resolves the three peaks with minimal sidelobes. Imaging results from metallic sphere targets, shown in Fig. \ref{fig:dri_targets}a, using (a) subband 1 (60--64 GHz), (b) subband 2 (77--81 GHz), (c) MFT, (d) MPA, (e) $kR$-Net.}
    \label{fig:dri_exp7} 
\end{figure*}

First, three metallic spheres, each with a diameter of 1.5 cm, are placed in front of the center of the array, as shown in Fig. \ref{fig:dri_targets}a. 
The sphere nearest to the planar scanner is separated by 3 cm from the other two spheres in the $z$-direction. 
The two further spheres are separated by 6 cm such that they are centered around the nearer sphere along the horizontal direction. 
Along the cross-range directions, the images are well resolved and focused; however, the range resolution varies depending on the approach. 
After sampling a planar array with dimensions of 0.125 m $\times$ 0.125 m, the recovered images are computed, as shown in Fig. \ref{fig:dri_exp7}. 
The images recovered from the first and second subbands are shown in Figs. \ref{fig:dri_exp7}a and \ref{fig:dri_exp7}b, respectively, and have low resolution in the $z$-direction because of the bandwidth of 4 GHz. 

Applying the MFT to the collected data yields the image shown in Fig. \ref{fig:dri_exp7}c. 
The MFT image is plagued by sidelobes in the $z$-direction, which obscure the location of each metallic sphere. 
This is due to the fact that the MFT does not account for the missing $k$-domain data in the signal fusion process. 
In contrast, the MPA attempts to fill the frequency gap and is relatively successful because the number of targets is small compared with the number of samples for each subband. 
By applying $kR$-Net, the high-resolution image in Fig. \ref{fig:dri_exp7}e is recovered. 
Compared with the image reconstructed using the MPA, the $kR$-Net image has decreased sidelobes for each of the three spheres. 
Even with a simple target scene, the proposed method demonstrates superior focusing performance compared with the conventional MFT and MPA approaches. 


\begin{figure*}[t]
    \centering
    \includegraphics[width=\textwidth]{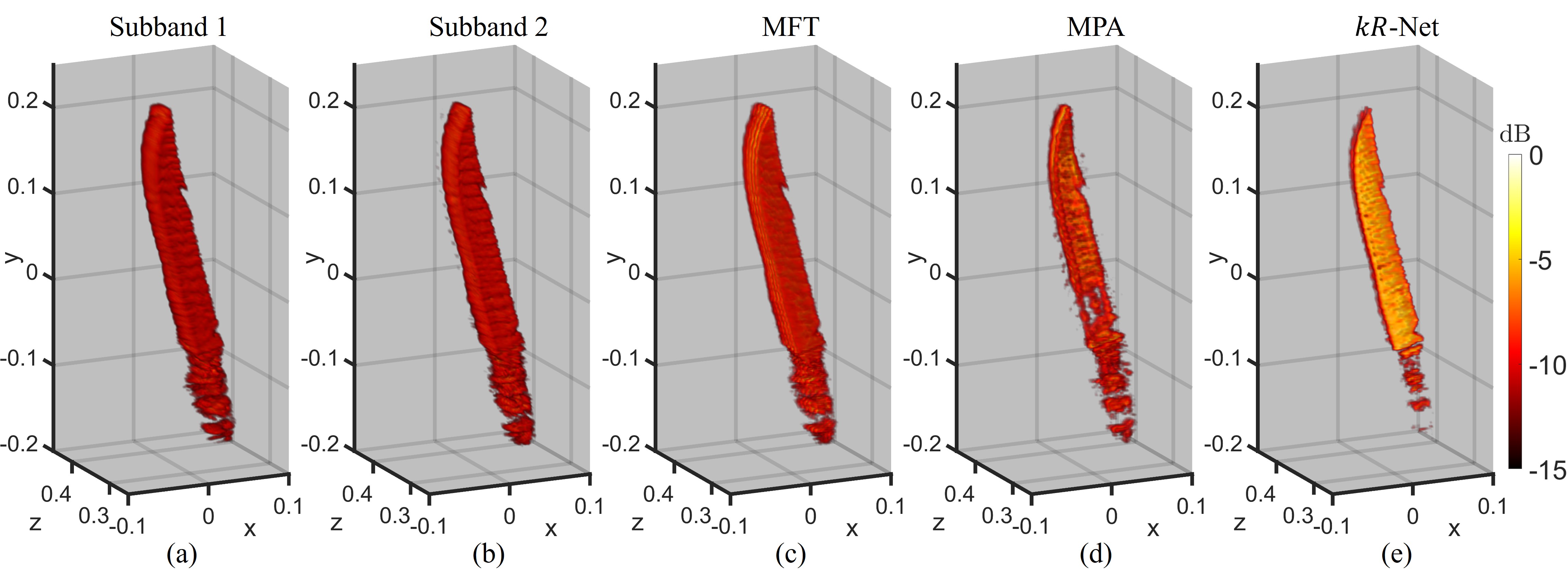}
    \caption{$kR$-Net recovers a high-fidelity image of the knife recovering the thin blade with high depth-resolution, while retaining the shape of the knife, which is lost when applying the MPA. Empirical imaging results from large knife, shown in Fig. \ref{fig:dri_targets}b, using (a) subband 1 (60--64 GHz), (b) subband 2 (77--81 GHz), (c) MFT, (d) MPA, (e) $kR$-Net.}
    \label{fig:dri_exp5} 
\end{figure*} 

Next, we consider a large knife, as shown in Fig. \ref{fig:dri_targets}b. 
A large array is synthesized with dimensions of 0.45 m $\times$ 0.8 m to scan the knife. 
The collected data are processed using the MFT, MPA, and $kR$-Net to improve the resolution, and the results are shown in Fig. \ref{fig:dri_exp5}. 
Again, the images from each subband demonstrate comparable focusing performance but are poorly resolved in the range direction because of the limited bandwidth. 
Without accounting for the missing data in the frequency gap, the MFT image has considerable sidelobes, as shown in Fig. \ref{fig:dri_exp5}c, with a similar appearance to the single-radar images in Figs. \ref{fig:dri_exp5}a and \ref{fig:dri_exp5}b. 
Although the MPA is capable of reducing the sidelobes moderately compared with the MFT, as shown in Fig. \ref{fig:dri_exp5}d, the structure of the knife is not retained because of the simplistic model employed by the MPA. 
As a result, the knife blade is distorted and ghosting is observed along the range direction. 
For concealed weapon detection or occluded item recognition, the poor reconstruction quality of the MPA for high-bandwidth targets, such as this knife, may prohibitively degrade the system performance. 
However, $kR$-Net demonstrates the best focusing performance by achieving a fine resolution in the $z$-direction while retaining the intricate features of the target. 
The serrated edge and notch on the knife are clearly visible in the recovered image shown in Fig. \ref{fig:dri_exp5}e, and the handle closely resembles the physical dimensions shown in Fig. \ref{fig:dri_targets}b. 

\begin{figure*}[t]
    \centering
    \includegraphics[width=\textwidth]{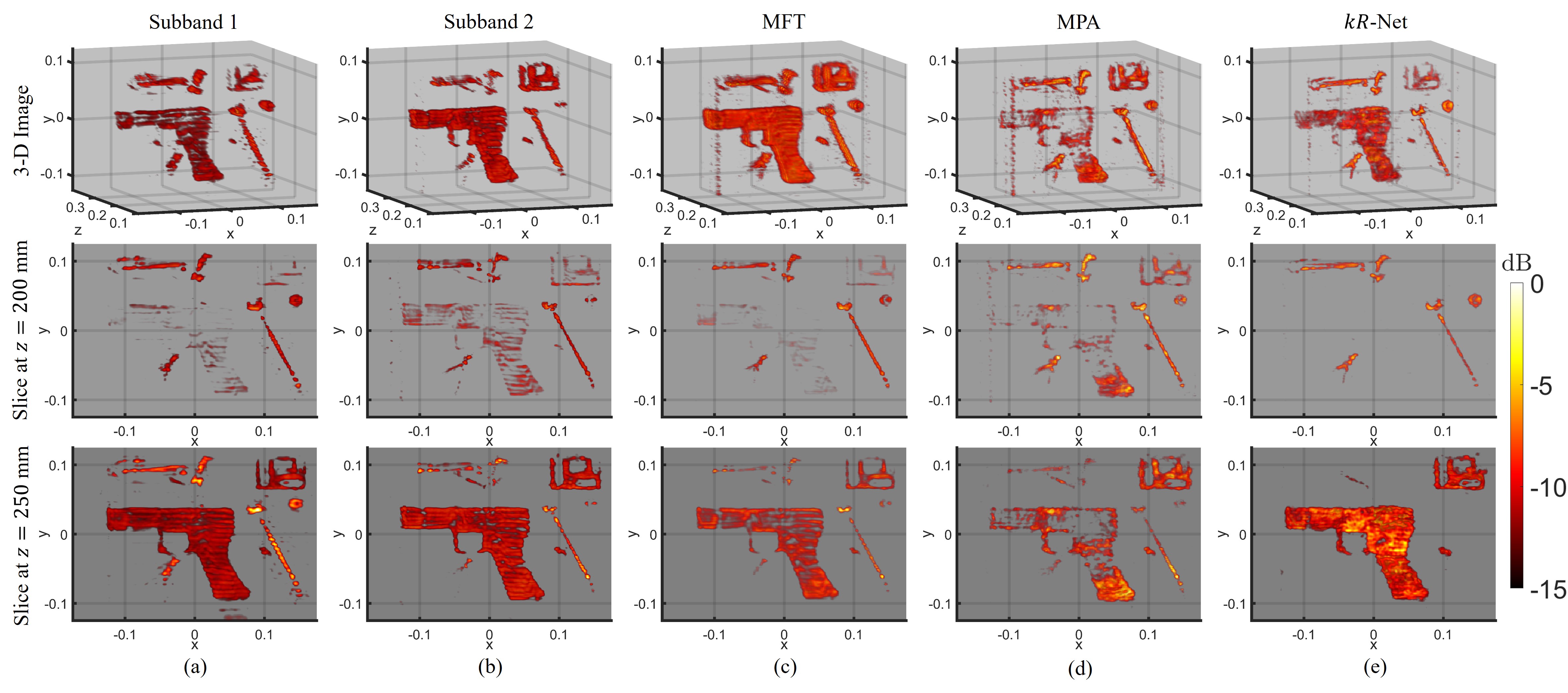}
    \caption{The proposed $kR$-Net separates the items at the front and rear of the box demonstrating high-fidelity super-resolution for a realistic hidden item scenario and superior performance compared with the MFT and MPA. 
    Empirical imaging results from hidden objects target, shown in Fig. \ref{fig:dri_targets}c, using (a) subband 1 (60--64 GHz), (b) subband 2 (77--81 GHz), (c) MFT, (d) MPA, (e) $kR$-Net. First row: \mbox{3-D} image. Second row: slice at $z = 200$ mm, corresponding to the location of the front of the box. Third row: slice at $z = 250$ mm, corresponding to the location of the rear of the box.}
    \label{fig:dri_exp11} 
\end{figure*}

Finally, we consider a hidden object scenario with several items inside a cardboard box, as shown in Fig. \ref{fig:dri_targets}c. 
The box is positioned such that the front of the box is located 200 mm from the radar boresight and parallel to the synthetic array. 
The items attached to the front of the box are separated from those in the rear by 5 cm and the box is illuminated by an array with dimensions of 0.25 m $\times$ 0.125 m. 
For more a closer spacing along the $z$-direction, the imaging results are expected to be further degraded. 
Fig. \ref{fig:dri_exp11} shows the reconstructed \mbox{3-D} images and slices at $z = 200$ mm and $z = 250$ mm, corresponding to the front and rear of the box, respectively. 
For accurate high-resolution imaging, the slice at $z = 200$ mm should contain only the objects at the front of the box and the slice at $z = 250$ mm should contain only the objects at the rear of the box. 

For the two subbands with bandwidths of 4 GHz, the recovered images are spread across the $z$-direction, and both slices shown are contaminated by objects from the front and rear, as shown in Figs. \ref{fig:dri_exp11}a and \ref{fig:dri_exp11}b. 
The small wrench and key, both located at the rear of the box, suffer from weaker reflections and occlusion and are not well resolved by the algorithms tested. 
A cylindrical SAR approach \cite{smith2020nearfieldisar,gao2016efficient} or image enhancement algorithm  \cite{dai2021imaging,jing2022enhanced,smith2021An,vasileiou2022efficient,zhang2019target,smith2022ffh_vit,yin2021study} may improve image quality in the case of occlusion. 
The image recovered using the MFT, shown in Fig. \ref{fig:dri_exp11}c, exhibits the expected behavior, as sidelobes along the range direction cause ghosting, which results in the objects being visible in both range slices. 
Similarly, the MPA reduces the sidelobes moderately compared to the MFT, as shown in Fig. \ref{fig:dri_exp11}d. 
However, as expected from prior experiments and the inherent limitations of the MPA, the sidelobes are not mitigated, and some features of the objects are lost. 
The image recovered using $kR$-Net is shown in Fig. \ref{fig:dri_exp11}e and demonstrates improved performance in two key respects. 
First, the $kR$-Net image retains the high-fidelity features of the target, which are necessary for a host of applications, including image segmentation and object classification. 
The image quality of the wrenches is particularly notable, as the jaw of each wrench is more clearly visible compared with the images recovered from the existing approaches. 
Secondly, the ghosting along the $z$-direction is significantly reduced, and the objects at the front side of the box are visible only in the $z = 200$ mm slice. 
Likewise, the objects at the rear side of the box are only visible in the $z = 250$ mm slice. 
Without the contamination observed using the MFT or MPA, the objects can be more easily localized and classified, enabling super-resolution for a host of imaging applications.  
The proposed hybrid, dual-domain algorithm yields high-resolution, high-fidelity images without feature loss and demonstrates improved performance over existing techniques in realistic scenarios. 

Through numerical simulation and empirical analysis, the proposed algorithm demonstrates superior performance to the MFT and MPA in terms of efficiency and image quality.
The proposed $kR$-Net offers improved robustness for low- and high-bandwidth target scenarios in addition to low SNR conditions. 
For practical imaging of complex, sophisticated targets, $kR$-Net achieves spatial super-resolution by improved multiband signal fusion without compromising the intricate features of the target.
Hence, $kR$-Net is better suited for high-resolution multiband imaging applications. 

\section{Conclusion}
\label{sec:dri_conclusion}
In this chapter, we introduce a novel deep learning-based algorithm for multiband signal fusion for \mbox{3-D} SAR super-resolution. 
By approaching the signal fusion problem from the wavenumber domain, we observe that imputation in the $k$-domain signal is equivalent to super-resolution in the wavenumber spectral domain. 
Hence, the proposed network employs a hybrid, dual-domain residual architecture that leverages the relationships in the $k$-domain and $R$-domain for improved performance. 
We develop a novel residual CV-CNN framework with domain transformation blocks interspersed throughout the network, resulting in superior performance compared with conventional CNN models. 
Compared with the MPA, which assumes a small number of reflectors in the scene or a low-bandwidth target, the proposed $kR$-Net is robust for imaging scenarios containing intricate targets consisting of many reflectors. 
Through simulation and empirical validation, $kR$-Net demonstrates superior imaging performance for multiband signal fusion for both low- and high-bandwidth targets. 
The hybrid architecture outperforms the equivalent single-domain networks operating in either the $k$-domain or $R$-domain. 
Extensive numerical investigations validate the superiority of $kR$-Net compared with the conventional MFT and MPA for signal fusion, in addition to single-domain CNN models. 
Using a custom multi-radar mechanical prototype built from commercially available mmWave radars, we conduct imaging experiments on various targets and observe significantly improved performance of $kR$-Net in terms of image focusing and efficiency. 

The hybrid-learning approach demonstrates superior numerical and computational performance compared to previous signal processing and machine learning-based solutions for multiband signal fusion. 
However, the hybrid concept presented in the chapter can be applied to a host of problems in sensing and signal processing. 
Specifically, we demonstrated the improvement by using a dual-domain CNN architecture, implying that the learned features in both domains contribute to improve performance. 
This conclusion alone can have a large impact on a host of signal processing problems. 
More generally, by exploiting the properties of the signal in conjunction with a data-driven approach, hybrid-learning algorithms can outperform traditional signal processing and machine learning methods. 
\chapter{Summary and Future Directions}
\label{ch:conclusion}

\section{Summary}
\label{sec:summary}

In this dissertation, we presented a novel approach to near-field mmWave imaging problems by leveraging machine learning and signal processing techniques in a fusion methodology, allowing for the exploitation of the advantages of both conventional and data-driven algorithms. 
We examined the impact of front-end signal processing techniques on deep learning perception. 
We investigated optimal signal processing methodologies for static and dynamic gesture recognition and demonstrated a novel sterile training technique to improve hand gesture classification rates. 
We developed an efficient algorithm for irregular SAR scanning geometries in the near-field, enabling technologies such as smartphone SAR imaging, near-field automotive SAR, and UAV imaging.
We presented an efficient high-fidelity dataset generation platform and discussed several near-field SAR super-resolution efforts. 
Using the proposed framework and knowledge from previous algorithmic and system-level investigations, we developed two novel deep learning-based algorithms to overcome non-idealities in common and emerging near-field SAR imaging modalities. 
We designed and implemented a real-time deep learning-based super-resolution contactless interface for musical instrumentation. 
We presented a novel hybrid, dual-domain architecture for multiband signal fusion. 
The proposed framework not only outperforms existing traditional techniques, but also demonstrates superiority over single-domain networks for the same tasks, indicating the advantages of hybrid-learning algorithms for RF signal processing problems compared to exclusively signal processing or deep learning-based solutions. 
In simulations and experiments, we demonstrated the efficacy of hybrid-learning algorithms for improved mmWave perception, sensing, tracking, and imaging. 

\section{Future Directions}
\label{sec:proposed_work}

In this dissertation, we presented hybrid-learning algorithms for improved mmWave imaging; however, there are many future directions for hybrid-learning algorithms applied to imaging and sensing problems across the frequency spectrum. 
Localization and classification problems using RF sensors were explored in this dissertation by applying CNN architectures directly to radar data. 
Extending these analyses to include a hybrid approach is a promising direction for further improving computational efficiency and system performance. 

In this dissertation, we explored mobile handheld SAR imaging systems applying a conventional technique for efficient imaging and proposed a deep learning algorithm to operate on the resulting images. 
Uniting these approaches would require designing a data-driven technique for image super-resolution at the data-level, rather than the image level.
Such an end-to-end network could benefit from an interleaved hybrid-learning approach by applying neural processing at various stages of the image recovery process. 

As demonstrated in this dissertation, an interleaved learning approach can drastically improve the multiband signal fusion performance. 
Similar problems in imaging and wireless communications may witness success with hybrid algorithms, including bandwidth extrapolation and line-spectra super-resolution. 

Finally, as transformer architectures, an alternative to CNNs, have drastically influenced the computer vision community, they are also of interest for hybrid-algorithms in RF signal processing. 
The transformer can employ multiple-scale architecture, which can be understood from a signal processing perspective as a varying window size, similar to wavelet processing. 
This technique may be used to learn different aspects of the frequency content of signals to improve sensing, imaging, and communication performance. 

The field of deep learning for RF signal processing is ripe with potential and should be explored using a hybrid approach to optimally leverage signal processing theory and data-driven methodologies. 


\appendix 

\chapter{Multivariate Taylor Series Expansion}
\label{app:taylor_series_ffh}
Consider an infinitely differentiable real-valued function and an open neighborhood around $(u,v,w) = (u_0,v_0,w_0)$. 
Let $\mathbf{x} = [u \ v \ w]^T$, and $\mathbf{x}_0 = [u_0 \ v_0 \ w_0]^T$. Hence, the multivariate Taylor series expansion of $f(\mathbf{x})$ in the neighborhood of $\mathbf{x}_0$ can be written as
\begin{equation}
\label{eq:taylor_series}
    f(\mathbf{x}) = f(\mathbf{x}_0) + (\mathbf{x}-\mathbf{x}_0)^T \nabla \mathbf{f}(\mathbf{x}_0) 
    + \frac{1}{2!}(\mathbf{x}-\mathbf{x}_0)^T \mathbf{H}(\mathbf{x}_0) (\mathbf{x}-\mathbf{x}_0) + \dots ,
\end{equation}
where $\nabla \mathbf{f}$ is the vector of first derivatives
\begin{equation}
    \nabla \mathbf{f}(\mathbf{x}) = 
    \begin{bmatrix}
    f_u(\mathbf{x}) \\
    f_v(\mathbf{x}) \\
    f_w(\mathbf{x})
    \end{bmatrix},
\end{equation}
and $\mathbf{H(x)}$ is the Hessian matrix of the second derivatives as
\begin{equation}
    \mathbf{H(x)} = 
    \begin{bmatrix}
    f_{uu}(\mathbf{x}) & f_{uv}(\mathbf{x}) & f_{uw}(\mathbf{x}) \\
    f_{vu}(\mathbf{x}) & f_{vv}(\mathbf{x}) & f_{vw}(\mathbf{x}) \\
    f_{wu}(\mathbf{x}) & f_{wv}(\mathbf{x}) & f_{vv}(\mathbf{x}) 
    \end{bmatrix}.
\end{equation}

\section{Taylor Series Expansion of Round-Trip Distance for Irregular Scanning Geometries}
\label{app:proof_ffh}
The round-trip distance between the $\ell$-th Tx/Rx pair, whose transmitter and receiver elements are located at $(x_T,y_T,z_\ell)$ and $(x_R,y_R,z_\ell)$, respectively, and the scatterer located at $(x,y,z)$ is expressed in (\ref{eq:Rl_of_xT_xR_yT_yR}). 
Substituting (\ref{eq:xT_xR_yT_yR_to_virtual}) and (\ref{eq:zl_to_virtual}) into (\ref{eq:Rl_of_xT_xR_yT_yR}), $R_\ell^{RT}$ can be expressed as a function of the distances between the Tx and Rx elements along the $x$- and $y$-directions, $d_\ell^x$ and $d_\ell^y$, respectively, and displacement along the $z$-direction, $d_\ell^z$:
\begin{align}
\begin{split}
    \label{eq:Rl}
    R_\ell^{RT}(d_\ell^x,d_\ell^y,d_\ell^z) & \\
    &= \left[(x' - \frac{d_\ell^x}{2} - x)^2 + (y' - \frac{d_\ell^y}{2} - y)^2 + (Z_0 + d_\ell^z - z)^2 \right]^{\frac{1}{2}} \\
    &+ \left[(x' + \frac{d_\ell^x}{2} - x)^2 + (y' + \frac{d_\ell^y}{2} - y)^2 + (Z_0 + d_\ell^z - z)^2 \right]^{\frac{1}{2}}.
\end{split}
\end{align}
The first derivatives of (\ref{eq:Rl}), evaluated at $d_\ell^x = d_\ell^y = d_\ell^z = 0$, are
\begin{align}
\begin{split}
\label{eq:first_derivatives}
    \frac{\partial R_\ell^{RT}}{\partial d_\ell^x} \biggr\rvert_{(d_\ell^x = d_\ell^y = d_\ell^z = 0)} &= \frac{\partial R_\ell^{RT}}{\partial d_\ell^y} \biggr\rvert_{(d_\ell^x = d_\ell^y = d_\ell^z = 0)} = 0, \\
    \frac{\partial R_\ell^{RT}}{\partial d_\ell^y} \biggr\rvert_{(d_\ell^x = d_\ell^y = d_\ell^z = 0)} &= \frac{2(Z_0 - z)}{R_0},
\end{split}
\end{align}
where $R_0$ is the distance between the virtual monostatic element located at $(x',y',Z_0)$ and the point scatterer at $(x,y,z)$, as expressed in (\ref{eq:R0}).

The second derivatives of (\ref{eq:Rl}), evaluated at the point of interest, can be derived as
\begin{align}
\begin{split}
    \label{eq:second_derivates}
    \frac{\partial^2 R_\ell^{RT}}{\partial (d_\ell^x)^2} \biggr\rvert_{(d_\ell^x = d_\ell^y = d_\ell^z = 0)} &= \frac{1}{2R_0} \left[ 1 - \frac{(x'-x)^2}{R_0^2} \right], \\
    \frac{\partial^2 R_\ell^{RT}}{\partial (d_\ell^y)^2} \biggr\rvert_{(d_\ell^x = d_\ell^y = d_\ell^z = 0)} &= \frac{1}{2R_0} \left[ 1 - \frac{(y'-y)^2}{R_0^2} \right], \\
    \frac{\partial^2 R_\ell^{RT}}{\partial (d_\ell^z)^2} \biggr\rvert_{(d_\ell^x = d_\ell^y = d_\ell^z = 0)} &= \frac{2}{R_0} \left[ 1 - \frac{(Z_0-z)^2}{R_0^2} \right], \\
    \frac{\partial^2 R_\ell^{RT}}{\partial d_\ell^x d_\ell^y} \biggr\rvert_{(d_\ell^x = d_\ell^y = d_\ell^z = 0)} &= -\frac{(x'-x)(y'-y)}{2R_0^3}, \\
    \frac{\partial^2 R_\ell^{RT}}{\partial d_\ell^x d_\ell^z} \biggr\rvert_{(d_\ell^x = d_\ell^y = d_\ell^z = 0)} &= \frac{\partial^2 R_\ell^{RT}}{\partial d_\ell^y d_\ell^z} \biggr\rvert_{(d_\ell^x = d_\ell^y = d_\ell^z = 0)} = 0.
\end{split}
\end{align}

Substituting (\ref{eq:first_derivatives}) and (\ref{eq:second_derivates}) into (\ref{eq:taylor_series}), the quadratic approximation of $R_\ell$ can be expressed as
\begin{align}
\begin{split}
\label{Rl_approximation}
    R_\ell^{RT} &\approx 2R_0 + \frac{2(Z_0-z)d_\ell^z}{R_0} + \frac{(d_\ell^x)^2 + (d_\ell^y)^2 + 4(d_\ell^z)^2}{4R_0} \\
    &- \frac{\left[(x'-x)d_\ell^x + (y'-y)d_\ell^y\right]^2 + 4(Z_0-z)^2(d_\ell^z)^2}{4R_0^3}.
\end{split}
\end{align}
\chapter{Downloading the Toolbox and \\ Accessing the Documentation}
\label{app:documentation}
The proposed toolbox and interactive GUI can be downloaded via GitHub or the authors' website. 
Using the MathWorks file exchange platform, our toolbox can be downloaded and installed to your MATLAB installation by opening the \texttt{THz and Sub-THz Imaging Toolbox.mltbx} MATLAB toolbox installer. 
To use the functionality of the toolbox from the repository available on our website, simply add the the main folder to the MATLAB search path.

\begin{itemize}
    \item \href{https://github.com/josiahwsmith10/THz-and-Sub-THz-Imaging-Toolbox}{https://github.com/josiahwsmith10/THz-and-Sub-THz-Imaging-Toolbox}
    \item \href{https://labs.utdallas.edu/wislab/}{labs.utdallas.edu/wislab/}
\end{itemize}

To set up and open the directory of the toolbox, call the provided script \texttt{THzSimulator()}. 
(Note: the main folder of the toolbox must be on the MATLAB path and is included when installed via the MathWorks website.) 
To open the interactive GUI, either call \texttt{THzSimulator()} and open the application in MATLAB Application Designer or call the \texttt{THzSimulatorGUI()} script to open the stand-alone GUI. 

To access the documentation created for this toolbox, after installing the packaged MATLAB toolbox or downloading the repository and adding the main folder to the MATLAB path, navigate to the Home tab in the main MATLAB window and select the Help button indicated by a question mark icon. 
The Documentation Home window will open and the THz Imaging Toolbox will appear under the Supplemental Software section\footnote{Refer to \url{www.mathworks.com/help/matlab/matlab_prog/display-custom-documentation.html} for more detailed instructions.}. 
Additionally, the ``Info" menu and ``About" tab of the interactive GUI contain quick links to the documentation.
Inside our documentation, you will find the Getting Started Guide, User Guide, Reference, and Examples sections detailing the usage and functionality of the various tools.


\begin{thesisbib}  
\setlength{\bibsep}{12pt plus 1pt minus 1pt}
\bibliography{library.bib}
\end{thesisbib}  

\begin{biosketch}
Josiah W. Smith received the BSEE degree (\textit{summa cum laude}) in electrical engineering from The University of Texas at Dallas in 2019, where he is currently pursuing a PhD degree in electrical engineering specializing in communications engineering. 

In 2019, he was an undergraduate research intern at the Texas Analog Center of Excellence (TxACE) working on radar reception for hand gesture recognition and high resolution near-field MIMO synthetic aperture radar imaging algorithms. During the summer of 2020, he developed real-time human-computer interaction algorithms for mmWave radar with imec-USA. In the summer of 2021, he developed advanced deep learning and data-driven algorithms for user experience enhancement at Apple in the Display Technologies group.

Mr. Smith was awarded the Texas Instruments Analog Excellence Graduate Fellowship in August 2019 and the Louis Beecherl, Jr. Graduate Research Fellowship in August 2021. In 2020, he was awarded first alternate for the student paper competition in the \textit{IEEE Radar Conf. 2020}, ranking 6/143 and best poster award at the \textit{2020 TxACE Symposium}. He is a student IEEE member. His current research interests include new regime radar imaging algorithm development, ultrawideband radar imaging algorithms, terahertz radar, radar perception, computer vision, machine learning, millimeter-wave sensing, and phased array signal processing.

\end{biosketch}

\begin{vita}  


  \begin{center}
    {\LARGE\bfseries Josiah W. Smith} \\[5pt]
    March 31, 2022
  \end{center}

  \bigskip

  {\large\bfseries Contact Information:\par}
  \medskip
  \noindent\vtop{\hsize=.49\hsize
    Department of Electrical and Computer Eng.\par
    The University of Texas at Dallas\par
    800 W.~Campbell Rd.\par
    Richardson, TX 75080-3021, U.S.A.\par}
  \hfil\vtop{\hsize=.49\hsize
    \par}\par

  \bigskip

  {\large\bfseries Educational History:\par}
  \medskip
  BSEE, Electrical Engineering,
    The University of Texas at Dallas, 2019\par

  \bigskip

  {\large\bfseries Employment History:\par}
  \medskip
  Research Assistant, The University of Texas at Dallas, August 2019 -- May 2022\par
  Machine Learning Intern, Apple - EE Display Team, March 2021 -- August 2021\par
  Computational Imaging Intern, imec-USA, March 2020 -- August 2020\par
  Undergraduate Researcher, Texas Analog Center of Excellence (TxACE), January 2019 -- August 2019\par

  \bigskip

  {\large\bfseries Publications	:\par}
  \medskip
  [J1] \textbf{J. W. Smith}, S. Thiagarajan, R. Willis, Y. Makris and M. Torlak, ``Improved Static Hand Gesture Classification on Deep Convolutional Neural Networks Using Novel Sterile Training Technique,'' \textit{IEEE Access}, vol. 9, pp. 10893--10902, Jan. 2021.\par 
  [J2] \textbf{J. W. Smith}, O. Furxhi, M. Torlak, ``An {FCNN}-Based Super-Resolution {mmWave} Radar Framework for Contactless Musical Instrument Interface,'' \textit{IEEE Trans. on Multimedia}, May 2021.\par 
  [J3] \textbf{J. W. Smith} and M. Torlak, ``Efficient \mbox{3-D} Near-Field MIMO-SAR Imaging for Irregular Scanning Geometries,'' \textit{IEEE Access}, vol. 10, pp. 10283--10294. Jan. 2022.\par
  [J4] \textbf{J. W. Smith} and M. Torlak, ``Deep learning-based multiband signal fusion for 3-D SAR super-resolution,'' in \textit{IEEE Trans. Aerosp. Electron. Syst.}, Apr. 2023.\par
  [J5] \textbf{J. W. Smith} and M. Torlak, ``Survey of emerging systems and algorithms for near-field THz SAR imaging,'' \textit{Proc. IEEE}, submitted for publication.\par
  [C1] \textbf{J. W. Smith}, M. E. Yanik and M. Torlak, ``Near-Field MIMO-ISAR Millimeter-Wave Imaging,'' in \textit{Proc. IEEE Radar Conf. (RadarConf)}, Florence, Italy, Sep. 2020, pp. 1--6.\par 
  [C2] \textbf{J. W. Smith}, Y. Alimam, G. Vedula, and M. Torlak, ``A vision transformer approach for efficient near-field SAR super-resolution under array perturbation,'' in \textit{Proc. IEEE Tex. Symp. Wirel. Microw. Circuits Syst. (WMCS)}, Waco, TX, USA, Apr. 2022, pp. 1--6.\par
  [C3] C. Vasieleiou, \textbf{J. W. Smith}, S. Thiagarajan, M. Nigh, Y. Makris, and M. Torlak, ``Efficient CNN-based super resolution algorithms for mmWave mobile radar imaging,'' in \textit{Proc. IEEE Int. Conf. Image Process. (ICIP)}, Bourdeaux, France, Oct. 2022, pp. 3803--3807.\par
  \bigskip
  
  {\large\bfseries Professional Recognitions and Honors:\par}
  \medskip
  Louis Beecherl, Jr. Graduate Research Fellowship, Aug. 2021\par
  Best Poster Award, TxACE Symposium, Oct. 2020\par 
  First Alternate Student Paper (Ranked 6/143), IEEE Radar Conference 2020, Sep. 2020\par
  Texas Instruments Analog Excellence Graduate Fellowship, Aug. 2019\par 
  First Place Senior Capstone Project, UTDesign II Expo, May 2019\par 
  First Place Senior Capstone Project, UTDesign I Expo, Dec. 2018\par 
  Greater Texas Foundation Removing Educational Barriers Endowed Scholarship, Aug. 2017\par 
  Academic Excellence Full-Ride Scholarship, The University of Texas at Dallas, Aug. 2016\par
  
  \bigskip

  {\large\bfseries Professional Memberships:\par}
  \medskip
  Institute of Electrical and Electronics Engineers (IEEE), 2016 -- present\par

\end{vita}  

\end{document}